\documentclass[11pt,a4paper,oneside]{scrartcl}

\usepackage[T1]{fontenc}
\usepackage[utf8]{inputenc}
\usepackage{libertinus}
\usepackage{amsmath,amsthm}
\usepackage[libertinus,slantedGreek,varbb]{newtxmath}
\usepackage[scr=rsfs,frak=euler]{mathalpha}
\DeclareOldFontCommand{\rm}{\normalfont\rmfamily}{\mathrm}
\usepackage{microtype}
\usepackage[a4paper,margin=2.65cm]{geometry}
\usepackage{enumitem}
\usepackage{xcolor}
\usepackage{tikz}
\usepackage{placeins}
\usetikzlibrary{arrows.meta,calc}
\definecolor{figblue}{HTML}{0072B2}
\definecolor{figred}{HTML}{D55E00}
\colorlet{notecol}{black!62}
\tikzset{
 cell diagram/.style={x=1cm,y=1cm,font=\small,
   >={Latex[length=4pt,width=3pt]},line width=.5pt,
   line cap=round,line join=round},
 fig note/.style={font=\footnotesize,text=notecol},
 fig title/.style={font=\small},
 cell grid/.style={black!22,line width=.4pt},
 dim line/.style={{Latex[length=3pt,width=2.4pt]}-{Latex[length=3pt,width=2.4pt]},
   black!70,line width=.45pt},
 ext line/.style={black!40,line width=.35pt,densely dotted},
}
\usepackage{hyperref}

\hypersetup{
 hidelinks,
 bookmarksnumbered=true,
 pdftitle={Lossless propagation bounds for long-range fermionic interactions},
 pdfauthor={Stefan Teufel}
}

\setkomafont{author}{\sffamily}
\setkomafont{date}{\sffamily}
\setkomafont{sectionentry}{\normalfont\normalsize\sffamily}
\DeclareTOCStyleEntry[beforeskip=0pt,linefill=\TOCLineLeaderFill]{tocline}{section}
\DeclareTOCStyleEntry[entryformat=\normalfont\sffamily,pagenumberformat=\normalfont\sffamily]{tocline}{subsection}
\setlist[itemize]{topsep=4pt,itemsep=2pt,parsep=1pt}
\setlist[enumerate]{topsep=4pt,itemsep=2pt,parsep=1pt}
\allowdisplaybreaks

\newtheorem{theorem}{Theorem}[section]
\newtheorem{proposition}[theorem]{Proposition}
\newtheorem{lemma}[theorem]{Lemma}
\newtheorem{corollary}[theorem]{Corollary}
\theoremstyle{definition}
\newtheorem{hypothesis}[theorem]{Hypothesis}
\newtheorem{definition}[theorem]{Definition}
\theoremstyle{remark}
\newtheorem{remark}[theorem]{Remark}

\let\proofqedsymbol\qedsymbol
\newcommand{\pushStatementQED}{\renewcommand{\qedsymbol}{\ensuremath{\diamond}}\pushQED{\qed}}
\AddToHook{env/hypothesis/begin}{\pushStatementQED}
\AddToHook{env/hypothesis/end}{\popQED}
\AddToHook{env/definition/begin}{\pushStatementQED}
\AddToHook{env/definition/end}{\popQED}
\AddToHook{env/remark/begin}{\pushStatementQED}
\AddToHook{env/remark/end}{\popQED}
\AddToHook{env/proof/begin}{\let\qedsymbol\proofqedsymbol}

\makeatletter
\newcommand{\keepwithproofpointer}{\@endparpenalty=10000\relax}
\makeatother
\newcommand{\proofpointer}[3][Proof]{\par\nopagebreak[4]\noindent\hyperref[#3]{\emph{#1:}} #2, p.~\pageref{#3}.\par}

\newcommand{\A}{\mathcal A}
\newcommand{\C}{\mathbb C}
\newcommand{\R}{\mathbb R}
\newcommand{\1}{\mathbf 1}
\newcommand{\dd}{\mathrm d}
\newcommand{\I}{\mathrm{i}}
\newcommand{\E}{\mathrm{e}}
\newcommand{\dist}{\operatorname{dist}}
\newcommand{\spec}{\operatorname{spec}}
\newcommand{\supp}{\operatorname{supp}}
\newcommand{\diam}{\operatorname{diam}}
\newcommand{\ball}{\mathrm B}
\newcommand{\loc}{\mathrm{loc}}
\newcommand{\pol}{\mathrm{pol}}
\newcommand{\Z}{\mathbb Z}
\newcommand{\N}{\mathbb N}
\newcommand{\F}{\mathbb F}
\newcommand{\EE}{\mathbb E}
\newcommand{\cB}{\mathcal B}
\newcommand{\cC}{\mathcal C}
\newcommand{\cD}{\mathcal D}
\newcommand{\cG}{\mathcal G}
\newcommand{\cH}{\mathcal H}
\newcommand{\cI}{\mathcal I}
\newcommand{\cK}{\mathcal K}
\newcommand{\cM}{\mathcal M}
\newcommand{\cN}{\mathcal N}
\newcommand{\cP}{\mathcal P}
\newcommand{\cR}{\mathcal R}
\newcommand{\cS}{\mathcal S}
\newcommand{\cT}{\mathcal T}
\newcommand{\cV}{\mathcal V}
\newcommand{\cW}{\mathcal W}
\newcommand{\scrB}{\mathscr B}
\newcommand{\scrC}{\mathscr C}
\newcommand{\scrE}{\mathscr E}
\newcommand{\scrF}{\mathscr F}
\newcommand{\scrL}{\mathscr L}
\newcommand{\scrV}{\mathscr V}
\newcommand{\frakP}{\mathfrak P}
\newcommand{\frakb}{\mathfrak b}
\newcommand{\frakm}{\mathfrak m}
\newcommand{\frakt}{\mathfrak t}
\newcommand{\id}{\mathrm{id}}
\newcommand{\CAR}{\operatorname{CAR}}
\newcommand{\Br}{\operatorname{Br}}
\newcommand{\ad}{\operatorname{ad}}
\newcommand{\Ran}{\operatorname{Ran}}
\newcommand{\dvol}{d_{\rm vol}}
\newcommand{\Cvol}{C_{\rm vol}}
\newcommand{\Ccell}{C_{\rm cell}}
\newcommand{\Dcell}{D_{\rm cell}}
\newcommand{\mucell}{\mu_{\rm cell}}
\newcommand{\Kcell}{K_{\rm cell}}

\title{Lossless propagation bounds\\ for long-range fermionic interactions}
\author{Stefan Teufel\thanks{Fachbereich Mathematik, Universit\"at T\"ubingen, Auf der Morgenstelle 10, 72076 T\"ubingen, Germany.}
}
\date{October 31, 2026}
\begin{document}
\maketitle

\begin{abstract}
Propagation estimates underpin much of the rigorous analysis of extended quantum many-body systems, providing a systematic way to exploit the spatial locality of Hamiltonian interactions.
We prove propagation estimates for extensive fermionic many-body interactions with polynomial time growth, no loss of spatial decay, and an optimal threshold for the polynomial spatial exponent.

More precisely, we construct local representatives of evolved fermionic many-body interactions on metric spaces with polynomial volume growth of order \(\dvol\), under an additional assumption on cell partitions.
The Hamiltonian interaction \(\Phi_u\) and the interaction \(\Psi\) being evolved are even.
We assume bounds on the sums of term norms over all \(X\) containing a given site, uniform in the site and, for \(\Phi_u\), in time, with weight \(\E^{\kappa|X|}(1+\diam X)^a\) for \(\Phi_u(X)\) and \(\E^{\kappa|X|}(1+\diam X)^p\) for \(\Psi(X)\), where \(a>\dvol\), \(0\le p\le a\), and \(\kappa\) is sufficiently large.
For suitable constants \(c,b>0\), the interaction norms of the representatives, with weight \(\E^{c(1+|t-s|)^{-b}|X|}(1+\diam X)^p\), grow at most polynomially in the elapsed time \(|t-s|\).
The key feature is that \(p=a\) is allowed: the Hamiltonian need not have stronger spatial decay than the interaction being evolved.
The decay threshold \(a>\dvol\) is optimal in general: uniform polynomial time bounds at \(p=a\) fail for \(0<a\le\dvol\), even for smooth time-dependent two-site interactions.

Polynomial time growth permits integration against filters with sufficiently many finite absolute moments, including those used in quasi-adiabatic constructions and quantum belief propagation.
We study consequences of our lossless propagation bounds for filtered interactions and the dynamics they generate.
These results give the dynamical estimates needed in perturbation theory with extensive perturbations, including gap stability, the principle that local perturbations perturb locally (LPPL), adiabatic theory, and response theory.

All results also hold for quantum spin systems.
\end{abstract}

\clearpage
\tableofcontents
\clearpage

\section{Introduction}

The spatial locality of Hamiltonian interactions is a central ingredient in the rigorous analysis of extended quantum many-body systems.
Propagation estimates make this structure usable: they quantify how the evolution of an observable is affected by distant degrees of freedom and by changes to the Hamiltonian outside a surrounding region.
These estimates enter the construction of infinite-volume dynamics, spectral flow, and perturbative expansions, as well as the analysis of stability and response.

Lieb--Robinson bounds \cite{LR72} express this control through estimates on commutators, or graded commutators, of evolved local observables with distant test observables.
For even observables, such bounds, taken uniformly over norm-one test observables outside a region, control how well the evolved observable is approximated by its tracial conditional expectation onto that region \cite{NSY18}.
For applications to extensive perturbations, estimates on individual evolved observables must be supplemented by control of the interaction obtained by evolving every local term and decomposing each result into local observables.
For polynomially decaying interactions, summing these local observables in a weighted interaction norm can reduce the spatial decay exponent, even when each evolved term is well localized.
We construct a decomposition that preserves this exponent.

We work with the CAR algebra \(\A:=\CAR(\bigoplus_{x\in\Gamma}\C^{r_x})\) over a countable metric space \((\Gamma,d)\), with a finite number \(r_x\ge1\) of fermionic modes at each site \(x\).
We assume that \(\Gamma\) has polynomial volume growth of order \(\dvol\) and admits partitions into finite cells at every scale, with cell diameters bounded by a fixed multiple of the scale and uniformly bounded adjacency degree.
These geometric assumptions hold, for example, for every uniformly discrete set in \(\R^n\) with the induced Euclidean metric and \(\dvol=n\); see Remark~\ref{rem:euclidean-cells}.
Treating each cell as a single site provides the controlled spatial coarse graining required in our proofs.
For finite \(X\subseteq\Gamma\), let \(\A_X\subseteq\A\) be the algebra generated by the fermionic modes in \(X\) and \(\A_X^+\) its even part.
An interaction is a map \(X\mapsto\Xi(X)\in\A_X^+\) on the nonempty finite subsets of \(\Gamma\), and for \(p,k\ge0\) and \(\kappa>0\) we use the interaction norms
\begin{align}
 \|\Xi\|^{\pol}_{p;k}
 &:=\sup_{x\in\Gamma}\sum_{X\ni x}
       |X|^k(1+\diam X)^p\|\Xi(X)\|,
       \label{eq:polynomial-norm}\\
 \|\Xi\|^{\#}_{p;\kappa}
 &:=\sup_{x\in\Gamma}\sum_{X\ni x}
       \E^{\kappa|X|}(1+\diam X)^p\|\Xi(X)\|.
       \label{eq:card-norm}
\end{align}
The superscript \(\#\) refers to the exponential weight in the number \(|X|\) of sites.
The exponent \(p\) measures spatial decay, while \(k\) and \(\kappa\) control the size of the supports.
Let \(\Phi=(\Phi_u)_{u\in\R}\) be a path of Hamiltonian interactions, with \(\Phi_u(X)^*=\Phi_u(X)\) and \(u\mapsto\Phi_u(X)\) norm-continuous for every \(X\).
Let \(\kappa_0>0\) be the constant of Theorem~\ref{thm:card-propagation}, which depends only on the cell adjacency bound, and assume, for some \(a>\dvol\) and \(0\le L<\infty\),
\[
 \sup_{u\in\R}\|\Phi_u\|^{\#}_{a;\kappa_0}\le L.
\]
By Proposition~\ref{prop:polynomial-dynamics}, \(\Phi\) generates a cocycle of automorphisms \(\tau^\Phi_{s,t}:\A\to\A\), \(s,t\in\R\).

Our main result, Theorem~\ref{thm:card-propagation}, constructs complex-linear maps \(\cT^\Phi_{s,t}\) for evolving other interactions \(\Psi\), whose terms need not be self-adjoint.
The construction first decomposes each evolved term \(\tau^\Phi_{s,t}(\Psi(X))\) into local observables using linear maps chosen independently of \(\Psi\).
Summing the resulting contributions assigned to each output support \(Y\) then defines \((\cT^\Phi_{s,t}\Psi)(Y)\); see Definition~\ref{def:interaction-representatives}.
The induced interaction maps satisfy, for every \(s,t\in\R\), \(0\le p\le a\), and \(k\ge0\),
\begin{equation}
 \|\cT^\Phi_{s,t}\Psi\|^{\pol}_{p;k}
 \le C_{p,k}(1+|t-s|)^{m_{p,k}}
                 \|\Psi\|^{\#}_{p;\kappa_0},
 \label{eq:introduction-propagation}
\end{equation}
whenever the right-hand side is finite.
Here \(C_{p,k}\) depends only on \(a,L,p,k\) and the geometric constants, and \(m_{p,k}\) is given in \eqref{eq:time-exponent}.
We call this estimate \emph{lossless}: the evolved interaction is bounded with the same spatial exponent \(p\) as the input interaction.
The bounds on \(\|\Phi_u\|^{\#}_{a;\kappa_0}\) and \(\|\Psi\|^{\#}_{p;\kappa_0}\), with their exponential support-size weights, are the crucial assumptions that allow \(p=a\), i.e.\ that the Hamiltonian need not decay faster than the interaction being evolved.
For suitable constants \(c,b>0\), the exponential support-size norm \(\|\cT^\Phi_{s,t}\Psi\|^{\#}_{p;c(1+|t-s|)^{-b}}\) also grows at most polynomially in \(|t-s|\).
The results and proofs extend to quantum spin systems with finite-dimensional local algebras and to fermionic systems with additional finite-dimensional matrix algebras at the sites, with constants independent of the local dimensions (Remark~\ref{rem:spin-systems}).

The decay threshold \(a>\dvol\) is optimal in general for uniform lossless propagation bounds with polynomial time growth.
Using the state-transfer constructions of Tran et al.\ \cite{TranStateTransfer21}, we exclude an extension of these bounds at \(p=a\) to any \(0<a\le\dvol\), already for smooth time-dependent two-site interactions on \(\Z^n\) with \(\dvol=n\); see Proposition~\ref{prop:decay-optimality} in Appendix~\ref{app:decay-optimality}.

For the passage from observables to interactions, Nachtergaele--Sims--Young \cite{NSY19} give a construction for general quasi-local maps in which each transformed term is decomposed into differences of conditional expectations on growing neighborhoods of its original support.
To estimate the resulting interaction, one sums the errors in approximating each transformed term within these neighborhoods, with polynomial volume factors \cite[Section~5.4]{NSY19}.
When these errors decay polynomially with the neighborhood radius, the volume factors reduce the spatial decay exponent in the interaction estimate.

\begin{figure}[htbp]
\centering
\begin{tikzpicture}[cell diagram,x=.36cm,y=.36cm]
 \begin{scope}
  \node[fig title] at (5.5,9.6) {(a) A two-site commutator};
  \foreach \i in {0,...,11} \foreach \j in {0,...,7}
   \fill[black!28] (\i,\j) circle[radius=.85pt];
  \draw[figred,line width=1.1pt] (2,3) -- (9,4);
  \filldraw[fill=figblue!70!black,draw=white] (2,3) circle[radius=2.3pt];
  \filldraw[fill=figred,draw=white] (9,4) circle[radius=2.3pt];
  \node[font=\footnotesize,anchor=north,fill=white,inner sep=1pt] at (2,2.55) {$x$};
  \node[font=\footnotesize,anchor=north,fill=white,inner sep=1pt] at (9,3.55) {$y$};
  \node[font=\footnotesize,text=figred!85!black,anchor=south,fill=white,inner sep=1pt]
    at (5.5,3.75) {$\I[\Phi_u(\{x,y\}),A]$};
  \node[fig note,align=center,text width=4.6cm,anchor=north] at (5.5,-.95)
   {supported on $\{x,y\}$,\\ for every distance $d(x,y)$};
 \end{scope}
 \begin{scope}[shift={(14.5,0)}]
  \node[fig title] at (5.5,9.6) {(b) Assignment to a ball};
  \begin{scope}
   \clip (-.6,-.6) rectangle (11.6,7.6);
   \fill[black!8] (2,3) circle[radius=7.071];
   \draw[black!50,densely dashed] (2,3) circle[radius=7.071];
  \end{scope}
  \foreach \i in {0,...,11} \foreach \j in {0,...,7}{
   \pgfmathtruncatemacro\inball{((\i-2)*(\i-2)+(\j-3)*(\j-3)<=50.01)}
   \ifnum\inball=1 \fill[black!62] (\i,\j) circle[radius=.85pt];
   \else \fill[black!22] (\i,\j) circle[radius=.85pt];\fi}
  \draw[->,black!75,line width=.6pt,shorten >=2.5pt] (2,3) -- (9,4)
    node[midway,above=1pt,font=\footnotesize,fill=black!8,inner sep=1pt] {$d(x,y)$};
  \filldraw[fill=figblue!70!black,draw=white] (2,3) circle[radius=2.3pt];
  \filldraw[fill=figred,draw=white] (9,4) circle[radius=2.3pt];
  \node[font=\footnotesize,anchor=north,fill=black!8,inner sep=1pt] at (2,2.55) {$x$};
  \node[font=\footnotesize,anchor=north,fill=black!8,inner sep=1pt] at (9,3.55) {$y$};
  \node[fig note,align=center,text width=4.6cm,anchor=north] at (5.5,-.95)
   {a ball of radius $d(x,y)$:\\ at most $C(1+d(x,y))^{\dvol}$ sites};
 \end{scope}
 \begin{scope}[shift={(29,0)}]
  \node[fig title] at (5.5,9.6) {(c) Assignment to cells};
  \fill[figblue!42] (1.5,1.5) rectangle (3.5,3.5);
  \fill[figblue!16] (7.5,3.5) rectangle (9.5,5.5);
  \foreach \u in {-.5,1.5,...,11.5} \draw[cell grid] (\u,-.5) -- (\u,7.5);
  \foreach \v in {-.5,1.5,...,7.5} \draw[cell grid] (-.5,\v) -- (11.5,\v);
  \foreach \i in {0,...,11} \foreach \j in {0,...,7}
   \fill[black!28] (\i,\j) circle[radius=.85pt];
  \filldraw[fill=figblue!70!black,draw=white] (2,3) circle[radius=2.3pt];
  \filldraw[fill=figred,draw=white] (9,4) circle[radius=2.3pt];
  \node[font=\footnotesize,anchor=east,inner sep=1pt] at (1.65,3) {$x$};
  \node[font=\footnotesize,anchor=north,inner sep=1pt] at (9,3.55) {$y$};
  \draw[dim line] (2,2) -- (9,5)
    node[midway,sloped,below=1pt,fill=white,inner sep=1pt,font=\footnotesize] {$\diam Q(S)$};
  \draw[ext line] (7.5,5.5) -- (7.5,8.0);
  \draw[ext line] (9.5,5.5) -- (9.5,8.0);
  \draw[dim line] (7.5,7.85) -- (9.5,7.85) node[midway,above=0pt,font=\footnotesize] {$\ell$};
  \node[fig note,align=center,text width=4.6cm,anchor=north] at (5.5,-.95)
   {two cells: at most $C\ell^{\dvol}$ sites,\\ independent of $d(x,y)$};
 \end{scope}
\end{tikzpicture}
\caption{Assigning supports, illustrated on a square lattice.
(a)~For \(X=\{x\}\) and \(A\in\A_X^+\), the commutator \(\I[\Phi_u(\{x,y\}),A]\) is supported on \(\{x,y\}\).
(b)~The neighborhood decomposition of \cite{NSY19} assigns this contribution to a neighborhood of \(x\) containing \(y\).
When contributions from all input terms are summed through a fixed site, these enlarged supports introduce volume factors.
(c)~Each cell consists of the lattice sites inside its square.
The cell decomposition assigns this commutator to the union \(Q(S)\) of the two cells containing \(x,y\) (dark: the initial cell; light: the added cell).
This union contains at most \(C\ell^{\dvol}\) sites and has diameter between \(d(x,y)\) and \(d(x,y)+2\Ccell\ell\).
Here \(\Ccell\ell\) bounds the diameter of each cell, as in Hypothesis~\ref{hyp:cell-partitions}.
Its cardinality bound is independent of \(d(x,y)\); the chosen cell size grows polynomially with time (Figure~\ref{fig:scales}).}
\label{fig:cell-assignment}
\end{figure}
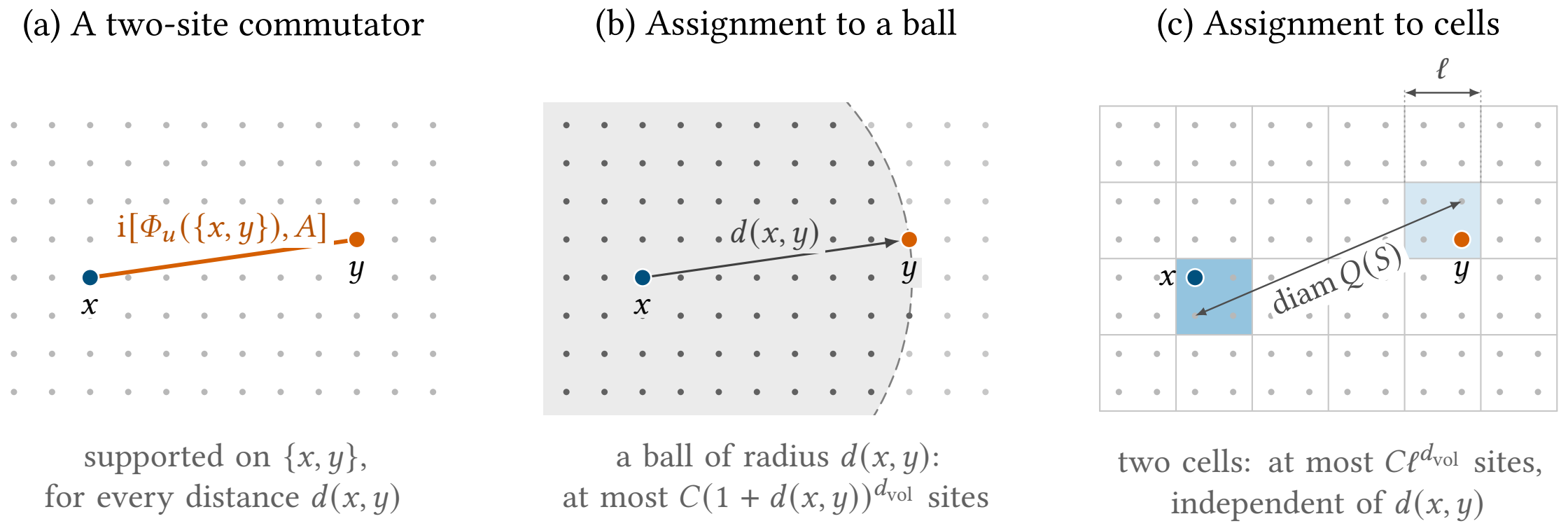

We instead assign the local terms representing each evolved observable to unions of cells.
McDonough--Yin--Lucas--Zhang \cite{McDonoughEtAl25} decompose evolved observables into operators supported on specified unions of boxes.
We adapt this construction to the cell partitions of Hypothesis~\ref{hyp:cell-partitions} at a scale~\(\ell\).
We estimate separately the number of cells in each support and the diameter of their union in the original metric.
The difference from a neighborhood decomposition is already visible for a single two-site term.
For distinct sites \(x,y\) and \(A\in\A_{\{x\}}^+\), the commutator \(\I[\Phi_u(\{x,y\}),A]\) is supported on \(\{x,y\}\), however large \(d(x,y)\) is.
The neighborhood decomposition of Nachtergaele--Sims--Young \cite{NSY19} assigns contributions reaching \(y\) to a neighborhood of \(x\) large enough to contain \(y\), and the volume of this neighborhood then enters the interaction estimate.
The cell construction instead assigns each contribution to a specified union of cells, which may contain distant cells without the region between them.
Figure~\ref{fig:cell-assignment} illustrates the difference.

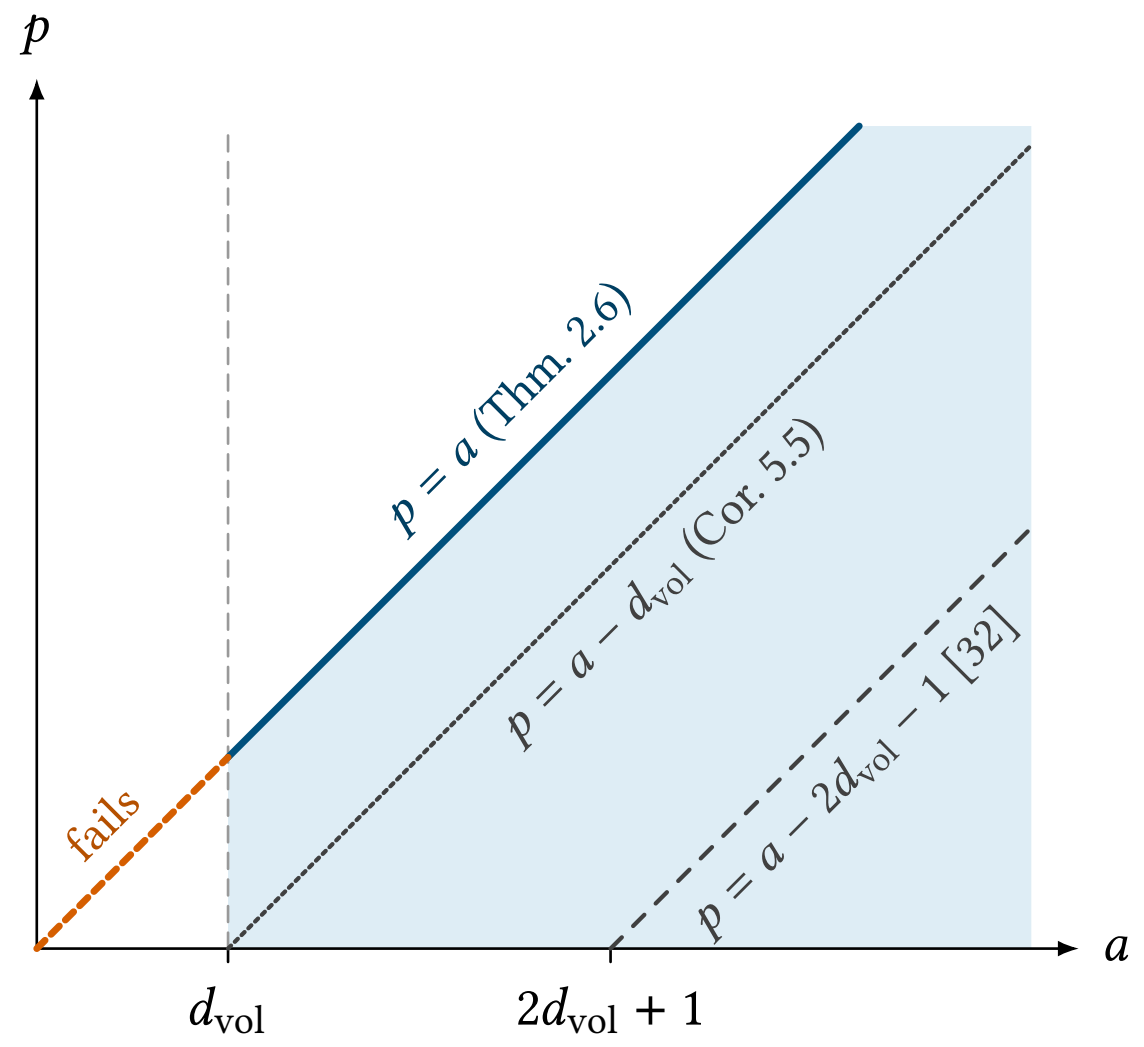
\begin{figure}[htbp]
\centering
\begin{minipage}[c]{0.485\linewidth}
\centering
\begin{tikzpicture}[cell diagram,x=1.1cm,y=1.1cm,scale=1.15,transform shape]
 \fill[figblue!13] (1,0) -- (5.2,0) -- (5.2,4.3) -- (4.3,4.3) -- (1,1) -- cycle;
 \draw[->] (0,0) -- (5.45,0);
 \draw[->] (0,0) -- (0,4.55);
 \node[font=\footnotesize,anchor=west] at (5.45,0) {$a$};
 \node[font=\footnotesize,anchor=south] at (0,4.55) {$p$};
 \foreach \t/\lab in {1/{\dvol},3/{2\dvol+1}}{
  \draw (\t,0) -- (\t,-.07);
  \node[font=\footnotesize,anchor=north] at (\t,-.07) {$\lab$};}
 \draw[black!40,densely dashed] (1,0) -- (1,4.3);
 \draw[black!75,line width=.8pt,dashed] (3,0) -- (5.2,2.2)
   node[midway,sloped,font=\scriptsize,anchor=north]
   {$p=a-2\dvol-1$ \cite{TeufelWessel25}};
 \draw[black!75,line width=.8pt,densely dotted] (1,0) -- (5.2,4.2)
   node[midway,sloped,font=\scriptsize,anchor=north]
   {$p=a-\dvol$ (Cor.~\ref{cor:polynomial-interaction-shells})};
 \draw[figblue!70!black,line width=1.3pt] (1,1) -- (4.3,4.3)
   node[midway,sloped,font=\scriptsize,text=figblue!55!black,anchor=south]
   {$p=a$ (Thm.~\ref{thm:card-propagation})};
 \draw[figred,line width=1.3pt,dash pattern=on 2.2pt off 1.6pt] (0,0) -- (1,1)
   node[midway,sloped,font=\scriptsize,text=figred!85!black,anchor=south] {fails};
\end{tikzpicture}
\end{minipage}\hfill\begin{minipage}[c]{0.51\linewidth}
\setlength{\abovecaptionskip}{0pt}
\caption{Spatial exponent \(a\) of the Hamiltonian versus spatial exponent \(p\) of the evolved interaction.
Shaded: Theorem~\ref{thm:card-propagation} preserves \(p\) for \(a>\dvol\) and \(0\le p\le a\), under exponential support-size weights and Hypothesis~\ref{hyp:cell-partitions}.
Under polynomial support-size weights and without cells, Corollary~\ref{cor:polynomial-interaction-shells} applies strictly below the dotted line, for \(0\le p<a-\dvol\).
The quasi-adiabatic construction of \cite{TeufelWessel25} on \(\Z^n\), \(\dvol=n\), applies strictly below the dashed line, for \(0<p<a-2n-1\).
Red: an extension with uniform polynomial time growth is excluded in general for \(p=a\) and \(0<a\le\dvol\) (Proposition~\ref{prop:decay-optimality}).}
\label{fig:exponent-map}
\end{minipage}
\end{figure}

To compare the neighborhood decomposition directly with our cell decomposition, we combine the Lieb--Robinson bound of Teufel--Wessel with the construction in their proof of Proposition~8 \cite{TeufelWessel25}.
For example, on finite boxes in \(\Z^n\), with \(\dvol=n\), a uniform bound on \(\|\Phi_u\|^{\pol}_{a;1}\) gives an interaction \(\Psi_{s,t}\) representing the evolution of \(\Psi\), with
\[
 \|\Psi_{s,t}\|^{\pol}_{p;1}
 \le C(1+|t-s|)^M\|\Psi\|^{\pol}_{p;2},
 \qquad p>0,\quad a>p+2\dvol+1,
\]
for suitable constants \(C,M\), uniformly in the box.
Like \eqref{eq:introduction-propagation}, this estimate preserves the spatial exponent \(p\) of the interaction being evolved and grows polynomially in time.
However, it requires the Hamiltonian to decay faster by more than \(2\dvol+1\).
Our exponential support-size assumption replaces this separation of spatial exponents and allows \(p=a>\dvol\).
Figure~\ref{fig:exponent-map} compares the admissible exponents.

\FloatBarrier
To prove the propagation bounds, we split the Hamiltonian interaction at a range cutoff.
Adapting the estimates of McDonough--Yin--Lucas--Zhang \cite{McDonoughEtAl25} to the fermionic algebra, we decompose the evolution of each \(\Psi(X)\) under the finite-range part into operators supported on unions of cells.
Lemma~\ref{lem:finite-volume-matrix-commutator} expresses the fermionic Lieb--Robinson estimate of Nachtergaele--Sims--Young \cite{NSY18} through a time-ordered series of matrices indexed by the sites.
 Rubiliani--Wesle--Wessel \cite{RubilianiWesleWessel26} obtain finer commutator bounds by summing over self-avoiding sequences of supports.
Lemma~\ref{lem:card-separated} gives a time-dependent fermionic version of the separated-cell estimates of McDonough--Yin--Lucas--Zhang \cite{McDonoughEtAl25}.
We prove this estimate by combining Lemma~\ref{lem:finite-volume-matrix-commutator} with a differential equation for nested commutators.
The matrix bound of Lemma~\ref{lem:finite-volume-matrix-commutator} also gives the observable-locality estimates without a cell assumption in Section~\ref{sec:filters}.

To recover the full dynamics, we include the Hamiltonian terms beyond the range cutoff through the interaction picture, following \cite{FossFeigEtAl15,MKN17,EMNY20}.
These long-range terms are first evolved with the finite-range dynamics; the resulting interaction generates the second evolution needed to recover the full dynamics.
We control the growth of supports in the commutator expansion of this additional evolution by decreasing the exponential support-size weight, following Abanin--De~Roeck--Ho--Huveneers \cite{ADHH17} and Yin--Lucas \cite{YinLucas25}.
For \(a>\dvol\), the range cutoff and cell size can be chosen to grow polynomially with the time horizon, yielding the propagation bound \eqref{eq:introduction-propagation}.
Section~\ref{sec:propagation} explains the construction and the choice of scales.

The polynomial time bound \eqref{eq:introduction-propagation} permits integration of the evolved interaction against filters with sufficiently many finite absolute moments, without decreasing the spatial exponent \(p\).
In Theorem~\ref{thm:card-limits}, we prove continuity and local convergence of the propagated and filtered interactions.
Under an additional stretched-exponential moment condition on the filter, the filtered interactions also admit a stretched-exponential support-size weight.

In Theorem~\ref{thm:smooth-hamiltonian-dependence}, we study dependence on the Hamiltonian in interaction norms.
For fixed \(s,t\) and an interaction \(\Psi\) with \(\|\Psi\|^\#_{p;\kappa_0}<\infty\), the map \(\Phi\mapsto\cT^\Phi_{s,t}\Psi\) is real analytic with respect to the uniform \(\|\cdot\|^{\#}_{a;\kappa_0}\)-norm on Hamiltonian paths, with values in the interaction space with norm \(\|\cdot\|^{\pol}_{p;k}\).
Its derivatives satisfy polynomial time bounds with the same spatial exponent \(p\).

Bru--de Siqueira Pedra \cite{BruPedra17} obtain norm expansions for the change in fermionic dynamics under perturbations supported in a finite region, with thermodynamic limits for both autonomous and time-dependent dynamics.
For polynomial spatial decay, their estimates require a larger decay exponent for higher expansion orders.
Under our fixed exponential support-size assumption on the Hamiltonian, perturbation directions, and input interaction, we instead control derivatives of every order for extensive perturbations in interaction norms while keeping the polynomial spatial exponents \(a\) and \(p\) fixed.

For an autonomous Hamiltonian, we also construct interactions representing repeated time filtering and commutators under the same spatial-decay assumption as Theorem~\ref{thm:card-propagation}; see Proposition~\ref{prop:iterated-filters}.
Control of these iterated filters and commutators is needed for higher-order perturbation theory and adiabatic perturbation theory, as discussed below.

The interaction estimate also yields local-approximation bounds for individual observables in the quasi-local algebra \(\A\).
For \(A\in\A_X\), Corollary~\ref{cor:observable-LR} approximates \(\tau^\Phi_{s,t}(A)\) by an observable supported in the \(R\)-neighborhood of \(X\), with an error that decays as \((1+R)^{-a}\), grows at most polynomially in \(|t-s|\), and has a prefactor proportional to \(|X|\|A\|\).
It also bounds the difference between the full and the finite-volume dynamics, with the same spatial decay and one additional time power.
The local-approximation estimates of \cite{TeufelWessel25} use spatial exponents strictly below \(a\); we obtain exponent \(a\) under our exponential support-size assumption and cell partitions.

In Theorem~\ref{thm:polynomial-local-topology}, we weaken the exponential support-size assumption to polynomial weights.
This is useful for dynamics generated by filtered interactions, since our filtering estimates do not in general give an exponential support-size bound.
We assume only an integrable \(\|\cdot\|^{\pol}_{p;1}\)-norm and a uniform \(\|\cdot\|^{\pol}_{p;0}\)-bound on the generating interaction.
The local-approximation error still decays as \((1+R)^{-p}\), but the bound grows exponentially in time when the \(\|\cdot\|^{\pol}_{p;1}\)-norm is uniformly bounded.
On fixed finite time intervals, this exponential factor remains bounded and does not obstruct the locality estimates needed for spectral flow, where the coupling parameter plays the role of time; see Corollary~\ref{cor:filtered-flow}.

Without cell partitions or exponential support-size weights, a uniform \(\|\Xi_u\|^{\pol}_{a;0}\)-bound with \(a>\dvol\) also gives polynomial time growth for individual observables, at every spatial exponent \(0<q<a\) (Proposition~\ref{prop:polynomial-time-locality}).
Under the same Hamiltonian assumption, Corollary~\ref{cor:polynomial-interaction-shells} gives polynomial time bounds for evolved interactions in \(\|\cdot\|^{\pol}_{p;0}\), assuming a finite \(\|\cdot\|^{\pol}_{p;1}\)-norm for the input interaction and \(0\le p<a-\dvol\).

All three observable estimates include odd observables.
Subsection~\ref{subsec:observable-extensions} explains the auxiliary-fermion reduction used in Corollary~\ref{cor:observable-LR} and Proposition~\ref{prop:polynomial-time-locality}, following the locality arguments of \cite{BourneOgata21,WilmingWerner22}.

Lieb--Robinson bounds with polynomial spatial tails for long-range spin and fermion systems go back to Hastings--Koma \cite{HastingsKoma06}, and the \(F\)-function approach of Nachtergaele--Ogata--Sims \cite{NOS06} covers polynomially decaying spin interactions at the price of an exponential time factor.
Polynomial light cones were subsequently proved by Foss-Feig--Gong--Clark--Gorshkov \cite{FossFeigEtAl15} and improved by Tran et al.\ \cite{TranEtAl19}.
Matsuta--Koma--Nakamura \cite{MKN17} and Else--Machado--Nayak--Yao \cite{EMNY20} developed bounds for general many-body interactions, the latter with spatial decay arbitrarily close to the exponent controlling the interaction tail.
For pair interactions decaying as \(r^{-\alpha}\) in spatial dimension \(\dvol\), Kuwahara--Saito \cite{KuwaharaSaito20} and Tran et al.\ \cite{TranHierarchy20} proved strictly linear light cones when \(\alpha>2\dvol+1\), following the one-dimensional result of Chen--Lucas \cite{ChenLucas19}.
Tran et al.\ \cite{TranEtAl21} subsequently established the sharp light-cone exponents.
For mutually commuting spin interactions, Lemm--Wessel \cite{LemmWessel25} establish enhanced long-range commutator and local-approximation bounds, with applications to clustering and to the principle that local perturbations perturb locally (LPPL).

On the fermionic side, Nachtergaele--Sims--Young \cite{NSY18} establish Lieb--Robinson bounds and conditional-expectation estimates for local approximation, Bru--de Siqueira Pedra \cite{BruPedra17} develop multi-commutator estimates, and Teufel--Wessel \cite{TeufelWessel25}, building on \cite{MKN17,EMNY20}, obtain fermionic Lieb--Robinson bounds with polynomial light cones and a prefactor linear in the smaller support size.
For two-site interactions, Hernández-Santana--Gogolin--Cirac--Acín \cite[Lemma~2]{HernandezSantanaEtAl17} prove anticommutator bounds for odd observables with polynomial light cones.

\subsection*{Applications and further consequences}

Many constructions in perturbation theory involve extensive perturbations, whose finite-volume norms need not remain bounded as the volume grows.
They therefore require estimates on the interaction terms themselves.
Under the assumptions stated above, we control time filtering, parameter derivatives, and finite iterations of filtering and commutators without decreasing the polynomial spatial exponent.
The approximation results allow us to pass from finite perturbation cutoffs to extensive interactions and from finite volumes to infinite systems.

Quasi-adiabatic continuation \cite{Hastings04,HastingsWen05} and spectral flow \cite{BMNS12,NSY19} use a filtered derivative of a differentiable Hamiltonian path \(\lambda\mapsto\Phi_\lambda\).
For each fixed \(\lambda\), this construction requires evolving \(\partial_\lambda\Phi_\lambda\) with the physical dynamics generated by \(\Phi_\lambda\) and integrating against a quasi-adiabatic filter.
The resulting interaction serves as the generator of a flow in \(\lambda\).
Our propagation and filtering estimates justify these steps when \(\Phi_\lambda\) and \(\partial_\lambda\Phi_\lambda\) satisfy the interaction-norm bounds with the same spatial exponent~\(a\).
In Corollary~\ref{cor:filtered-flow}, we prove locality of the flow generated by the resulting interaction and its convergence under approximation of the input interactions.
For smooth input paths, we also prove smooth dependence of the flow on \(\lambda\).

For ground-state transport in infinite volume, Moon--Ogata \cite{MoonOgata20} construct spectral flow under a bulk-gap assumption for finite-range spin systems with a unique gapped ground state.
Becker--Teufel--Wesle \cite{BTW26} obtain such a flow for fermionic systems with super-polynomially decaying interactions.
Our propagation and filtering estimates address the locality bounds needed to extend these bulk-gap constructions to polynomially decaying interactions.

Two companion papers apply these estimates to gapped free fermion systems, including pairing terms, perturbed by weak many-body interactions, where both the free coefficients and the interaction may decay only polynomially in space.
In \cite{Stability}, we show that the bulk gap persists and that a spectral flow, whose generator keeps the spatial decay exponent of the interaction, maps the free ground state to a ground state of the interacting bulk system.
In \cite{LPPLBdG}, we show that low-energy states are determined locally: if an interaction agrees with the bulk interaction inside a region \(U\), then every state with excitation energy below three quarters of the free gap is locally close to the bulk ground state deep inside \(U\), whatever the interaction does outside \(U\).
This yields uniqueness of the bulk ground state and a strong LPPL principle for defects of arbitrary strength.
Section~\ref{sec:applications} states the flow and convergence results used in these applications.

Higher-order perturbation theory and adiabatic perturbation theory require parameter derivatives, commutators of previously constructed terms, and repeated quasi-adiabatic filtering; see Becker--Teufel--Wesle \cite{BTWAd25}, Teufel--Wessel \cite{TeufelWessel25}, and references therein.
The locality of these successive corrections is needed both to approximate the dynamics and to justify response formulas for extensive perturbations.
In Theorem~\ref{thm:smooth-hamiltonian-dependence} and Proposition~\ref{prop:iterated-filters}, we control the derivatives and finite expressions of this form at a fixed spatial exponent.
Increasing the derivative order or the number of operations in such an expression requires higher filter moments, but no stronger spatial decay of the original interactions.

At positive temperature, time-filtered perturbations also enter quantum belief propagation \cite{Hastings07} and its applications to locality and stability of Gibbs states \cite{CMTW25}.
For infinite fermionic lattice systems, Becker \cite{BeckerQBP26} establishes quantum belief propagation directly on the underlying \(C^*\)-algebra and applies it to locality properties of KMS states.
Our estimates control the interactions representing these filtered perturbations without decreasing their spatial exponent, including for perturbations distributed throughout the system.
Together with assumptions on decay of correlations, these bounds can be used to study the local effect of changing the Hamiltonian.

The extensions in Proposition~\ref{prop:onsite-terms} and Remark~\ref{rem:spin-systems} make the propagation and filtering estimates available also with spatially unbounded onsite terms and for quantum spin systems.

The paper is organized as follows.
Section~\ref{sec:results} gives the setting, defines the interaction representatives, and states the main propagation and filtering results.
Section~\ref{sec:finite-range} proves the commutator estimates and uses them to construct the finite-range decomposition.
Section~\ref{sec:propagation} proves bounds on fixed cell partitions and combines them to obtain lossless propagation.
Section~\ref{sec:filters} develops parameter dependence, filtering, and iterated commutators.
It also proves the commutator and observable-locality estimates, including the bounds under polynomial support-size assumptions.
Section~\ref{sec:applications} treats flows generated by filtered interactions, spectral transport, and the extensions to unbounded onsite terms and spin systems.
Appendix~\ref{app:dynamics} proves the criterion for local interaction convergence and derives the basic dynamics construction from the independent integrated-norm argument in Section~\ref{sec:filters}.
Appendix~\ref{app:decay-optimality} proves the optimality of the decay assumption.

\section{Setting and main results}
\label{sec:results}

\subsection{Setting and assumptions}

Let \((\Gamma,d)\) be a countable metric space and write \(\ball_R(x):=\{y\in\Gamma\,|\,d(x,y)\le R\}\).
Except where weaker metric assumptions are stated explicitly, we assume the volume-growth bound
\begin{equation}
 |\ball_R(x)|\le C_\Gamma(1+R)^{\dvol},
 \qquad x\in\Gamma,\quad R\ge0,\qquad \dvol>0.
 \label{eq:volume-growth}
\end{equation}
Write \(X\Subset\Gamma\) for a finite subset and \(X^{(R)}:=\{y\in\Gamma\,|\,\dist(y,X)\le R\}\).
Here \(\dist(y,X):=\inf_{x\in X}d(y,x)\), \(\dist(X,Y):=\inf_{x\in X,y\in Y}d(x,y)\), and \(\diam X:=\sup_{x,y\in X}d(x,y)\) for nonempty sets.
Limits \(\Lambda\nearrow\Gamma\) are taken over finite subsets of \(\Gamma\), directed by inclusion.
The notation \(\Lambda_n\nearrow\Gamma\) denotes an increasing sequence of finite subsets with union \(\Gamma\).
For finite positive integers \(r_x\), let \(\A=\CAR(\bigoplus_{x\in\Gamma}\C^{r_x})\), with local algebras \(\A_X\).
The function \(x\mapsto r_x\) need not be bounded.
For an arbitrary region \(Y\subseteq\Gamma\), define
\[
 \A_Y:=\overline{\bigcup_{X\Subset Y}\A_X},
\]
where the closure is taken in norm.
The parity automorphism \(\Theta\) sends each creation and annihilation operator to its negative; \(\A_X^+\) denotes its fixed-point algebra.
For a parity-homogeneous observable \(A\), write \(|A|\in\{0,1\}\) for its parity.
Even observables commute with all observables on disjoint supports.
We use the CAR conventions of \cite{BR97} and the normalized tracial conditional expectations \(\F_X^{\rm tr}:\A\to\A_X\) of Araki--Moriya \cite[Theorem~4.7 and Corollary~4.8]{ArakiMoriya03}.
For averaging formulas and local-approximation estimates, we use Nachtergaele--Sims--Young \cite[Section~4]{NSY18}.

The construction of interaction representatives in Theorem~\ref{thm:card-propagation} uses one additional geometric assumption: partitions of \(\Gamma\) into cells at every scale.
It is not needed for the existence and convergence of dynamics in Proposition~\ref{prop:polynomial-dynamics} or for the observable-locality bounds in Theorem~\ref{thm:polynomial-local-topology} and Proposition~\ref{prop:polynomial-time-locality}.
Theorem~\ref{thm:polynomial-local-topology} and Remark~\ref{rem:general-spatial-weights} require only that metric balls be finite.

\begin{hypothesis}[Partitions into cells]
\label{hyp:cell-partitions}
There are constants \(\Ccell\ge1\) and \(\Dcell\in\N_0\) such that, for every \(\ell\ge1\), there exist an index set \(\cB_\ell\) and pairwise disjoint, nonempty finite cells \(Q_b^{(\ell)}\subseteq\Gamma\), \(b\in\cB_\ell\), satisfying
\[
 \Gamma=\bigcup_{b\in\cB_\ell}Q_b^{(\ell)}
 \quad\text{with}\quad
 \diam Q_b^{(\ell)}\le \Ccell\ell
 \quad(b\in\cB_\ell).
\]
On the vertex set \(\cB_\ell\), define a graph by joining distinct \(b,c\in\cB_\ell\) precisely when \(\dist(Q_b^{(\ell)},Q_c^{(\ell)})\le\ell\).
We require this graph to have maximum degree at most \(\Dcell\).
\end{hypothesis}

Hypothesis~\ref{hyp:cell-partitions} isolates the geometric properties of the box partitions used by McDonough--Yin--Lucas--Zhang \cite[Section~VI\,A]{McDonoughEtAl25}.
On doubling metric spaces with finite balls, partitions with these properties also follow from the metric-cube construction of Hyt\"onen--Kairema \cite[Theorem~2.2]{HytonenKairema12}.

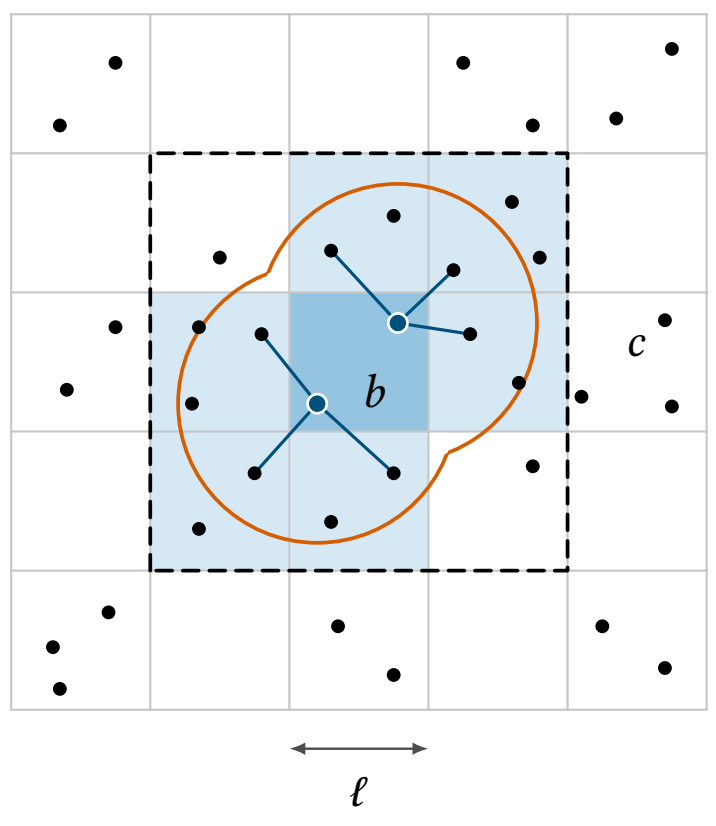
\begin{figure}[htbp]
\centering
\begin{minipage}[c]{0.36\linewidth}
\centering
\begin{tikzpicture}[cell diagram,x=.92cm,y=.92cm]
 \foreach \x/\y in {-1/-1,0/-1,-1/0,1/0,0/1,1/1}
  \fill[figblue!16] (\x,\y) rectangle (\x+1,\y+1);
 \fill[figblue!42] (0,0) rectangle (1,1);
 \draw[cell grid,step=1] (-2,-2) grid (3,3);
 \draw[densely dashed,line width=.7pt] (-1,-1) rectangle (2,2);
 \coordinate (p) at (.2,.2); \coordinate (q) at (.78,.78);
 \begin{scope}
  \clip (-2,-2) -- (-2,3) -- (3,3) -- (3,-2) -- cycle (q) circle[radius=1];
  \draw[figred,line width=.7pt] (p) circle[radius=1];
 \end{scope}
 \begin{scope}
  \clip (-2,-2) -- (-2,3) -- (3,3) -- (3,-2) -- cycle (p) circle[radius=1];
  \draw[figred,line width=.7pt] (q) circle[radius=1];
 \end{scope}
 \foreach \x/\y in {-.25/-.3,.75/-.3,-.2/.7}
  \draw[figblue!70!black,line width=.6pt] (p) -- (\x,\y);
 \foreach \x/\y in {1.3/.7,.3/1.3,1.18/1.16}
  \draw[figblue!70!black,line width=.6pt] (q) -- (\x,\y);
 \foreach \x/\y in {-1.7/-1.55,-1.3/-1.3,-1.65/-1.85,
   .35/-1.4,.75/-1.75,2.25/-1.4,2.7/-1.7,
   -.25/-.3,-.65/-.7,.3/-.65,.75/-.3,1.75/-.25,
   -1.6/.3,-1.25/.75,-.2/.7,-.7/.2,-.65/.75,
   1.65/.35,1.3/.7,2.1/.25,2.7/.8,2.75/.18,
   -.5/1.25,.75/1.55,.3/1.3,1.18/1.16,1.6/1.65,1.8/1.25,
   -1.65/2.2,-1.25/2.65,1.25/2.65,1.75/2.2,2.35/2.25,2.75/2.75}
  \fill (\x,\y) circle[radius=1.3pt];
 \foreach \c in {p,q}
  \filldraw[fill=figblue!70!black,draw=white,line width=.5pt] (\c) circle[radius=1.8pt];
 \node[font=\footnotesize] at (.62,.3) {$b$};
 \node[font=\footnotesize] at (2.5,.62) {$c$};
 \draw[dim line] (0,-2.28) -- (1,-2.28);
 \node[font=\footnotesize,anchor=north] at (.5,-2.33) {$\ell$};
\end{tikzpicture}
\end{minipage}\hfill
\begin{minipage}[c]{0.61\linewidth}
\setlength{\abovecaptionskip}{0pt}
\caption{Cubic cells and the degree bound in Remark~\ref{rem:euclidean-cells}.
Squares depict ambient cubes of side length \(\ell\); the dots inside each cube form its cell.
The dark cube contains the two sites of \(Q_b\).
The red curve encloses the \(\ell\)-neighborhood of \(Q_b\) in \(\R^2\), whose intersection with \(\Gamma\) is \(\{x\in\Gamma\,|\,\dist(x,Q_b)\le\ell\}\).
Exactly six other cells, shaded light blue, have sites in this neighborhood; the blue segments show distances at most \(\ell\).
The neighborhood lies inside the dashed block of indices \(b+\{-1,0,1\}^2\).
The two unshaded cells in the block have distance greater than \(\ell\) from \(Q_b\); \(c=b+(2,0)\) lies outside the block.
In ambient dimension \(n\), all neighbor indices lie in \(b+\{-1,0,1\}^{n}\), so the degree is at most \(3^n-1\).}
\label{fig:cell-degree-bound}
\end{minipage}
\end{figure}

\begin{remark}[Euclidean point sets]
\label{rem:euclidean-cells}
Suppose \(\Gamma\subseteq\R^n\) has the induced Euclidean distance and satisfies~\eqref{eq:volume-growth}.
For each \(\ell\ge1\), the nonempty sets
\[
 Q_b:=\Gamma\cap\ell\bigl(b+[0,1)^n\bigr),\qquad b\in\Z^n,
\]
form a partition into finite cells of diameter at most \(\sqrt n\,\ell\).
If \(|b_j-c_j|\ge2\) for some coordinate, half-openness gives \(\|x-y\|_2>\ell\) for every \(x\in Q_b\), \(y\in Q_c\).
Finiteness of the cells implies \(\dist(Q_b,Q_c)>\ell\).
Thus Hypothesis~\ref{hyp:cell-partitions} holds with \(\Ccell=\sqrt n\) and \(\Dcell=3^n-1\).
Figure~\ref{fig:cell-degree-bound} illustrates the count for \(n=2\).
Uniformly discrete Euclidean point sets, including Delone sets, satisfy \eqref{eq:volume-growth} with \(\dvol=n\), by packing disjoint balls around their sites.
\end{remark}

\begin{remark}[Doubling metric spaces]
\label{rem:doubling-cells}
The following covering condition is sufficient for Hypothesis~\ref{hyp:cell-partitions} and does not require a Euclidean embedding.
Suppose all metric balls are finite and there is \(N\in\N\) such that every ball of radius \(r>0\) can be covered by at most \(N\) balls of radius \(r/2\).
At scale \(\ell\ge1\), choose a maximal set of centers whose pairwise distances are greater than \(\ell\).
Assign each site to a nearest center, resolving ties by a fixed ordering of the centers.
Maximality puts each cell in the ball of radius \(\ell\) about its center, so the cells are finite and have diameter at most \(2\ell\).
Centers of adjacent cells have distance at most \(3\ell\).
Iterating the covering assumption three times covers each ball of radius \(3\ell\) by at most \(N^3\) balls of radius \(3\ell/8\), each containing at most one center.
Thus Hypothesis~\ref{hyp:cell-partitions} holds with \(\Ccell=2\) and \(\Dcell=N^3-1\).
\end{remark}

We use the interaction convention from the introduction and the norms defined in \eqref{eq:polynomial-norm}--\eqref{eq:card-norm}.
Hamiltonian interactions are in addition self-adjoint term by term, whereas the input interactions to be evolved need not be.
An interaction \(\Xi\) has range at most \(R\) if \(\Xi(Z)=0\) whenever \(\diam Z>R\); an interaction path has range at most \(R\) if this holds at every time.
A family of interactions \((\Psi_\xi)_{\xi\in\frakP}\), indexed by a topological parameter space \(\frakP\), is componentwise norm-continuous if \(\xi\mapsto\Psi_\xi(X)\) is continuous in norm for every fixed nonempty \(X\Subset\Gamma\).
The exponential support-size weight dominates every polynomial one: for \(p,k\ge0\) and \(\kappa>0\),
\[
 \|\Psi\|^{\pol}_{p;k}
 \le\Bigl(\sup_{n\ge1}n^k\E^{-\kappa n}\Bigr)
                         \|\Psi\|^{\#}_{p;\kappa}.
\]
Conversely, for interactions with at most \(q\) sites in each support the two kinds of norms are equivalent: if \(\Psi(X)=0\) for \(|X|>q\), then
\[
 \|\Psi\|^{\#}_{p;\kappa}
 \le \E^{\kappa q}\|\Psi\|^{\pol}_{p;0}.
\]
To measure local convergence of interactions, we use, for \(p,k\ge0\) and a finite nonempty \(K\), the local seminorm
\begin{equation}
 \|\Psi\|^{\pol,\loc}_{p;k,K}
 :=\sup_{x\in K}\sum_{X\ni x}|X|^k(1+\diam X)^p\|\Psi(X)\|.
 \label{eq:polynomial-local-seminorm}
\end{equation}
For a finite region \(\Lambda\Subset\Gamma\), the restricted interaction \(\Psi^\Lambda\) is defined on \(\Gamma\) by \(\Psi^\Lambda(X):=\1_{\{X\subseteq\Lambda\}}\Psi(X)\).
This deletes terms while keeping the underlying space \(\Gamma\) and its cell partitions fixed.
For a nonempty \(X\Subset\Gamma\) and \(A\in\A_X^+\), define the single-support interaction by
\begin{equation}
 (\iota_X A)(Z):=\1_{\{Z=X\}}A.
 \label{eq:single-support-interaction}
\end{equation}

Let \((\Phi_u)_{u\in J}\) be a componentwise norm-continuous Hamiltonian path on an interval \(J\).
Its finite-volume Hamiltonian is \(H_\Lambda(u):=\sum_{X\subseteq\Lambda}\Phi_u(X)\).
We define the propagator \(U_\Lambda(t,s)\) and the finite-volume dynamics \(\tau^{\Phi,\Lambda}_{s,t}\), for \(s,t\in J\) and \(A\in\A\), by
\[
\begin{gathered}
 \partial_t U_\Lambda(t,s)=-\I H_\Lambda(t)U_\Lambda(t,s),
 \qquad U_\Lambda(s,s)=\1,\\
 \tau_{s,t}^{\Phi,\Lambda}(A):=U_\Lambda(t,s)^*AU_\Lambda(t,s).
\end{gathered}
\]
These finite-volume dynamics obey the cocycle law \(\tau^{\Phi,\Lambda}_{s,t}\circ\tau^{\Phi,\Lambda}_{t,v} =\tau^{\Phi,\Lambda}_{s,v}\) for \(s,t,v\in J\).
Under the hypotheses of Proposition~\ref{prop:polynomial-dynamics} below, they converge in norm on every observable to the infinite-volume dynamics \(\tau^\Phi\).
The limit obeys the same cocycle law and, for every local observable \(A\), satisfies the differential equation
\begin{equation}
 \partial_t\tau_{s,t}^{\Phi}(A)
 =\I\,\tau_{s,t}^{\Phi}(\delta_{\Phi_t}(A)),
 \qquad
 \tau_{s,s}^{\Phi}(A)=A,
 \qquad
 \delta_\Phi(A):=\sum_X[\Phi(X),A],
 \label{eq:dynamics}
\end{equation}
with the derivative taken in norm.
For a constant path we write \(\tau_t^\Phi:=\tau_{0,t}^\Phi\).

\begin{proposition}[Dynamics and local limits]
\label{prop:polynomial-dynamics}
Assume \eqref{eq:volume-growth}, and let \(p>\dvol\).
A componentwise norm-continuous path of self-adjoint interactions \((\Phi_u)_{u\in J}\) on a compact interval \(J\), with \(\sup_{u\in J}\|\Phi_u\|^{\pol}_{p;0}<\infty\), generates a unique strongly continuous parity-preserving automorphism cocycle satisfying \eqref{eq:dynamics}.
As \(\Lambda\nearrow\Gamma\), the finite-volume dynamics \(\tau^{\Phi,\Lambda}\) converge in norm on every observable, uniformly in \(s,t\in J\).
If componentwise norm-continuous paths of self-adjoint interactions \(\Phi_u^{(n)}\), \(u\in J\), have a common \(\|\cdot\|^{\pol}_{p;0}\)-bound and
\[
 \int_J\|\Phi_u^{(n)}(X)-\Phi_u(X)\|\,\dd u\xrightarrow{n\to\infty}0
\]
for every nonempty finite \(X\), their cocycles converge to \(\tau^\Phi\) in norm on every observable, uniformly in \(s,t\in J\).
\keepwithproofpointer
\end{proposition}
\proofpointer{Appendix~\ref{subsec:dynamics-proof}}{proof:polynomial-dynamics}

For a componentwise norm-continuous self-adjoint interaction path \(\Phi\) on an arbitrary interval \(I\), suppose that, for some \(p>\dvol\), \(\sup_{u\in J}\|\Phi_u\|^{\pol}_{p;0}<\infty\) for every compact subinterval \(J\subseteq I\).
Applying Proposition~\ref{prop:polynomial-dynamics} on these subintervals gives cocycles that agree on overlaps by uniqueness and hence define \(\tau^\Phi_{s,t}\) for all \(s,t\in I\).

\subsection{Lossless quasi-local evolution of long-range interactions}

We begin by explicitly defining the decompositions of evolved local observables into local terms.

\begin{definition}[Local representatives of evolved interactions]
\label{def:interaction-representatives}
Let \((\Phi_u)_{u\in I}\) be a componentwise norm-continuous path of self-adjoint interactions on an interval \(I\).
For a fixed time horizon \(T\ge1\), choose a scale \(\ell=\ell(T)\ge1\) and a partition \((Q_b)_{b\in\cB_\ell}\) from Hypothesis~\ref{hyp:cell-partitions}.
For finite \(S\subseteq\cB_\ell\) and nonempty \(X\Subset\Gamma\), put
\[
 Q(S):=\bigcup_{b\in S}Q_b,
 \qquad
 S_0(X):=\{b\in\cB_\ell\,|\,Q_b\cap X\ne\varnothing\}.
\]
For \(s,t\in I\) and \(A\in\A_X\), define the fixed-scale coefficients by
\begin{equation}
 (\cT^{\Phi,T}_{s,t;X}A)(Q(S))
 :=\sum_{S_0(X)\subseteq U\subseteq S}
       (-1)^{|S|-|U|}\tau^{\Phi,Q(U)}_{s,t}(A),
 \qquad S_0(X)\subseteq S\Subset\cB_\ell,
 \label{eq:full-dynamics-mobius}
\end{equation}
and set \((\cT^{\Phi,T}_{s,t;X}A)(Y):=0\) on every other support \(Y\).
The cells are disjoint and nonempty, so \(Q(S)\) determines \(S\).

To combine the fixed-scale coefficients, set \(T_n:=2^n\), \(n\in\N_0\), and fix a scale \(\ell(T_n)\) and a cell partition for each \(n\).
Choose a smooth even function \(\chi:\R\to[0,1]\), nonincreasing on \([0,\infty)\), with \(\chi(v)=1\) for \(|v|\le1/2\) and \(\chi(v)=0\) for \(|v|\ge1\), and put
\begin{equation}
 \chi_0(v):=\chi(v),\qquad
 \chi_n(v):=\chi(v/2^n)-\chi(v/2^{n-1}),\quad n\ge1.
 \label{eq:time-cutoffs}
\end{equation}
These functions are nonnegative and satisfy \(\sum_{n=0}^N\chi_n(v)=\chi(v/2^N)\to1\).
For \(n\ge1\), \(\chi_n(v)=0\) unless \(2^{n-2}<|v|<T_n\), so the sum is locally finite.
See Figure~\ref{fig:scales}(b).
Define the local coefficients of the representative by
\begin{equation}
 (\cT^\Phi_{s,t;X}A)(Y)
 :=\sum_{n\ge0}\chi_n(t-s)
       (\cT^{\Phi,T_n}_{s,t;X}A)(Y).
 \label{eq:source-map-construction}
\end{equation}
For an interaction \(\Psi\), define the induced map on interactions by
\begin{equation}
 (\cT^\Phi_{s,t}\Psi)(Y)
 :=\sum_{\varnothing\ne X\subseteq Y}
       (\cT^\Phi_{s,t;X}\Psi(X))(Y).
 \label{eq:interaction-propagation-map}
\end{equation}
Each coefficient map is complex-linear, commutes with adjoints and parity, and takes values in \(\A_Y\).
For even \(A\), the resulting family of coefficients is an interaction.
\end{definition}

\begin{figure}[htbp]
\centering
\begin{tikzpicture}[cell diagram]
 \def\s{.62}
 \def\mobrow#1#2#3{%
  \begin{scope}[shift={(0,#1)}]
   \fill[figblue!42] (0,0) rectangle (\s,\s);
   \ifnum#2=1 \fill[figblue!16] (\s,0) rectangle (2*\s,\s);\fi
   \ifnum#3=1 \fill[figblue!16] (2*\s,0) rectangle (3*\s,\s);\fi
   \draw[black!45] (0,0) rectangle (3*\s,\s);
   \draw[black!45] (\s,0) -- (\s,\s);
   \draw[black!45] (2*\s,0) -- (2*\s,\s);
   \fill (.27*\s,.62*\s) circle[radius=1.2pt];
   \fill (.66*\s,.3*\s) circle[radius=1.2pt];
   \foreach \x/\y in {.75/.77,1.25/.30,1.67/.64,2.27/.67,2.74/.33,2.60/.82}
    \fill[black!45] (\x*\s,\y*\s) circle[radius=1.2pt];
  \end{scope}}
 \foreach \lab/\x in {b/0,c/1,d/2}
  \node[font=\footnotesize,anchor=south] at ({(\x+.5)*\s},\s+.03) {$\lab$};
 \node[fig note,anchor=south west] at (2.4,\s+.03) {coefficient at \(Q(S)\)};
 \mobrow{0}{0}{0}
 \mobrow{-.95}{1}{0}
 \mobrow{-1.9}{0}{1}
 \mobrow{-2.85}{1}{1}
 \foreach \y/\lab/\form/\note in {
   0/{\{b\}}/{\tau_{b}}/{restricted to the cell of \(X\)},
   -.95/{\{b,c\}}/{\tau_{bc}-\tau_{b}}/{correction from the adjacent cell \(Q_c\)},
   -1.9/{\{b,d\}}/{\tau_{bd}-\tau_{b}}/{zero if \(Q_b,Q_d\) are uncoupled},
   -2.85/{\{b,c,d\}}/{\tau_{bcd}-\tau_{bc}-\tau_{bd}+\tau_{b}}/{second difference in \(c\) and \(d\)}}{
  \node[font=\small,anchor=east] at (-.2,\y+.5*\s) {$S=\lab$};
  \node[font=\small,anchor=west] at (2.4,\y+.5*\s) {$\form$};
  \node[fig note,anchor=west,text width=5.6cm] at (7.0,\y+.5*\s) {\note};}
 \draw[black!60] (2.35,-3.2) -- (6.6,-3.2);
 \node[font=\small,anchor=east] at (-.2,-3.62) {sum};
 \node[font=\small,anchor=west] at (2.4,-3.62) {$\tau_{bcd}$};
 \node[fig note,anchor=west,text width=5.6cm] at (7.0,-3.62)
   {restricted to \(Q_b\cup Q_c\cup Q_d\)};
\end{tikzpicture}
\caption{The fixed-scale coefficients \eqref{eq:full-dynamics-mobius} for \(A\in\A_X^+\) with \(S_0(X)=\{b\}\).
Black dots mark \(X\); grey dots mark sites in \(\Gamma\setminus X\).
The dark cell contains \(X\), and light cells belong to \(S\setminus S_0(X)\).
Write \(\tau_U:=\tau^{\Phi,Q(U)}_{s,t}(A)\), with cell sets abbreviated by their labels, for example \(\tau_{bc}:=\tau^{\Phi,Q_b\cup Q_c}_{s,t}(A)\).
The four coefficients add up to \(\tau_{bcd}\).
The third coefficient vanishes if no Hamiltonian term meets both \(Q_b\) and \(Q_d\) at any time between \(s\) and \(t\).
Under the hypotheses and scale choices of Proposition~\ref{prop:fixed-scale}, the coefficients decay exponentially in \(|S|\), their sum is absolutely convergent, and it equals the infinite-volume evolution.}
\label{fig:mobius-coefficients}
\end{figure}
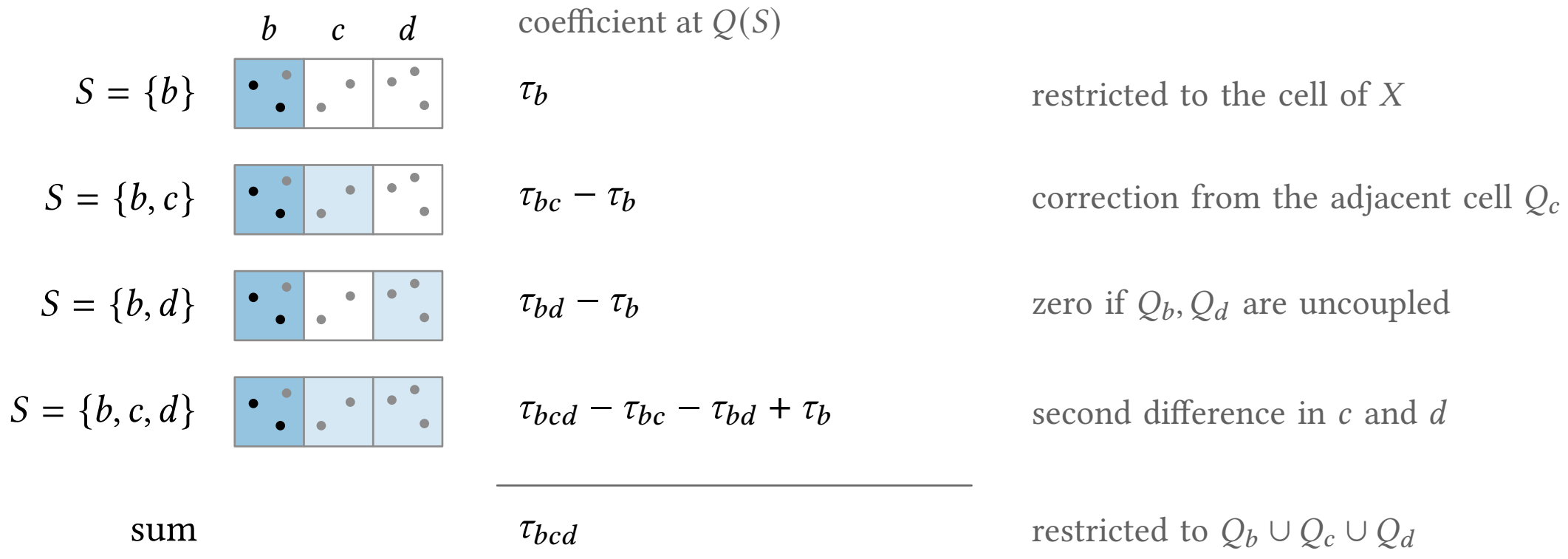

Formula~\eqref{eq:full-dynamics-mobius} extends the inclusion--exclusion construction of McDonough--Yin--Lucas--Zhang \cite[Section~VI\,A, (119)--(120)]{McDonoughEtAl25} to arbitrary initial cell sets \(S_0(X)\).
For every finite cell set \(V\supseteq S_0(X)\), summing the coefficients in \eqref{eq:full-dynamics-mobius} over \(S_0(X)\subseteq S\subseteq V\) telescopes to the evolution on \(Q(V)\):
\begin{equation}
 \sum_{S_0(X)\subseteq S\subseteq V}
       (\cT^{\Phi,T}_{s,t;X}A)(Q(S))
 =\tau^{\Phi,Q(V)}_{s,t}(A),
 \qquad S_0(X)\subseteq V\Subset\cB_\ell.
 \label{eq:card-telescope}
\end{equation}
Figure~\ref{fig:mobius-coefficients} illustrates this cancellation for an input support inside a single cell.

The interaction maps \(\cT^\Phi_{s,t}\) need not form a cocycle, since the cell decompositions used in their construction depend on the elapsed time \(|t-s|\).

We call the terms \(\Psi(X)\) of the input interaction \(\Psi\) the \emph{input terms} and their supports \(X\) the \emph{input supports}.
The sets \(Y\) assigned to the terms \((\cT^\Phi_{s,t}\Psi)(Y)\) are the \emph{output supports}.

In the following theorem, we choose the cell scales \(\ell(T_n)\) so that the local coefficients sum absolutely to the evolved observables and the induced interaction maps satisfy polynomial time bounds.

\begin{theorem}[Lossless propagation of interactions]
\label{thm:card-propagation}
Assume \eqref{eq:volume-growth} and Hypothesis~\ref{hyp:cell-partitions}.
There exists \(\kappa_0=\kappa_0(\Dcell)>0\) such that, for every \(a>\dvol\) and \(0\le L<\infty\), one can choose the cell scales \(\ell(T_n)\), the corresponding partitions, and the time cutoffs in Definition~\ref{def:interaction-representatives}, together with constants
\[
 c_*>0,\qquad C_p,C_{p,k}>0
 \quad(0\le p\le a,\ k\ge0),
\]
with the following properties.
The constants depend only on \(a,L,\dvol,C_\Gamma,\Ccell,\Dcell\) and their indicated indices.
Set
\begin{equation}
 m_{p,k}:=(1+\gamma)((k+1)\dvol+p),
 \qquad \gamma:=\frac{\dvol+1}{a-\dvol}.
 \label{eq:time-exponent}
\end{equation}
For every choice of local multiplicities \((r_x)_{x\in\Gamma}\), every interval \(I\subseteq\R\), and every componentwise norm-continuous path \((\Phi_u)_{u\in I}\) of self-adjoint interactions satisfying
\[
 \sup_{u\in I}\|\Phi_u\|^{\#}_{a;\kappa_0}\le L,
\]
the maps \eqref{eq:source-map-construction}, restricted to even inputs,
\[
 \cT^\Phi_{s,t;X}:\A_X^+
 \rightarrow\{\text{interactions on }\Gamma\},
\]
are complex-linear and have the following properties for \(s,t\in I\) and nonempty \(X\Subset\Gamma\).
\begin{enumerate}[label=(\roman*),leftmargin=*]
\item \emph{Decomposition of evolved local observables.}
For every \(A\in\A_X^+\), one has \((\cT^\Phi_{s,t;X}A)(Y)=0\) unless \(X\subseteq Y\), and
\begin{equation}
 \tau^\Phi_{s,t}(A)
 =\sum_{Y\supseteq X}(\cT^\Phi_{s,t;X}A)(Y),
 \label{eq:source-representation}
\end{equation}
where the sum converges absolutely in norm.
The maps preserve adjoints.
For fixed \(X,Y,A\), the function
\[
 (s,t)\mapsto(\cT^\Phi_{s,t;X}A)(Y)
\]
is norm-continuous.

\item \emph{The induced map on interactions.}
The induced map \eqref{eq:interaction-propagation-map} is complex-linear and preserves adjoints.
Let \(\Psi\) be an interaction, and set
\begin{equation}
 \rho(s,t):=c_*(1+|t-s|)^{-(1+\gamma)\dvol}.
 \label{eq:support-rate}
\end{equation}
For every \(0\le p\le a\) with \(\|\Psi\|^{\#}_{p;\kappa_0}<\infty\), the same family of maps satisfies
\begin{align}
 \sup_{x\in\Gamma}\sum_{Y\ni x}\E^{\rho(s,t)|Y|}(1+\diam Y)^p
       \sum_{\varnothing\ne X\subseteq Y} &
       \bigl\|(\cT^\Phi_{s,t;X}\Psi(X))(Y)\bigr\|
 \nonumber\\*
 & \le C_p(1+|t-s|)^{m_{p,0}}
                 \|\Psi\|^{\#}_{p;\kappa_0}\,.
 \label{eq:exponential-support-propagation}
\end{align}
By the triangle inequality, the right-hand side also bounds \(\|\cT^\Phi_{s,t}\Psi\|^{\#}_{p;\rho(s,t)}\).
Consequently, for every \(0\le p\le a\) and \(k\ge0\),
\begin{align}
 \sup_{x\in\Gamma}\sum_{Y\ni x}|Y|^k(1+\diam Y)^p
       \sum_{\varnothing\ne X\subseteq Y}&
       \bigl\|(\cT^\Phi_{s,t;X}\Psi(X))(Y)\bigr\|
 \nonumber\\*
 & \le C_{p,k}(1+|t-s|)^{m_{p,k}}
                 \|\Psi\|^{\#}_{p;\kappa_0},
 \label{eq:card-propagation}
\end{align}
whenever the norm on the right is finite.
The same right-hand side bounds \(\|\cT^\Phi_{s,t}\Psi\|^{\pol}_{p;k}\).
The constants can be chosen with \(C_{p,0}=C_p\) and
\[
 C_{p,k}\le C_p\max\!\left\{1,
                \left(\frac{k}{\E c_*}\right)^k\right\},
 \qquad k>0.
\]

\item \emph{Representation of the evolved interaction.}
For every interaction \(\Psi\) with \(\|\Psi\|^{\#}_{0;\kappa_0}<\infty\) and every local observable \(B\),
\begin{equation}
 \sum_Y[(\cT^\Phi_{s,t}\Psi)(Y),B]
 =\sum_X[\tau^\Phi_{s,t}(\Psi(X)),B],
 \label{eq:card-propagation-derivation}
\end{equation}
with both series converging absolutely in norm.

If additionally \(\sum_X\E^{\kappa_0|X|}\|\Psi(X)\|<\infty\), then
\begin{equation}
 \sum_Y(\cT^\Phi_{s,t}\Psi)(Y)
 =\sum_X\tau^\Phi_{s,t}(\Psi(X)),
 \label{eq:card-propagation-finite}
\end{equation}
with both series converging absolutely in norm.
\end{enumerate}
The bounds hold with the same constants for the restricted interactions \(\Phi^\Lambda,\Psi^\Lambda\) on \(\Gamma\), and after deleting terms of \(\Phi\) or \(\Psi\).
\keepwithproofpointer
\end{theorem}
\proofpointer{\S\ref{subsec:propagation-proof}}{proof:card-propagation}

For a single input term \(A\in\A_X^+\), the weighted sum over all output supports \(Y\supseteq X\) admits a smaller time exponent than \(m_{p,k}\) in \eqref{eq:card-propagation}.
Since the input support \(X\) is fixed, the proof need not count its possible positions relative to a prescribed output cell.
This saves \((1+\gamma)\dvol\) in the time exponent and reduces the filter moments required in Theorem~\ref{thm:card-limits}\ref{item:card-limits-estimates}.
Set
\begin{equation}
 \widetilde m_{p,k}:=(1+\gamma)(k\dvol+p),
 \qquad 0\le p\le a,\quad k\ge0.
 \label{eq:source-time-exponent}
\end{equation}
Under the hypotheses of Theorem~\ref{thm:card-propagation}, using its chosen partitions, cutoffs, and constants, the following estimates hold for every nonempty finite \(X\), \(A\in\A_X^+\), \(s,t\in I\), \(0\le p\le a\), and \(k\ge0\):
\begin{align}
 \sum_{Y\supseteq X}|Y|^k(1+\diam Y)^p
       \bigl\|(\cT^\Phi_{s,t;X}A)(Y)\bigr\|
 \le C_{p,k}\E^{\kappa_0|X|}(1+\diam X)^p
        (1+|t-s|)^{\widetilde m_{p,k}}\|A\|.
 \label{eq:card-propagation-total}
\end{align}
With an exponential support-size weight, we have
\begin{align}
 \sum_{Y\supseteq X}\E^{\rho(s,t)|Y|}(&1+\diam Y)^p
                         \|(\cT^\Phi_{s,t;X}A)(Y)\|
 \le C_p\E^{\kappa_0|X|}(1+\diam X)^p
                (1+|t-s|)^{\widetilde m_{p,0}}\|A\|.
 \label{eq:exponential-source-total}
\end{align}
Both estimates are proved together with Theorem~\ref{thm:card-propagation} in Section~\ref{sec:propagation}.
At \(p=k=0\), \eqref{eq:card-propagation-total} is uniform in both times; this gives the total-operator identities for \(L^1\) filters in Theorem~\ref{thm:card-limits}\ref{item:card-limits-identities} and Proposition~\ref{prop:iterated-filters} under their total-summability assumptions.
The representation \eqref{eq:source-representation} and the bounds \eqref{eq:card-propagation-total}--\eqref{eq:exponential-source-total} on the weighted sums of term norms over all output supports \(Y\supseteq X\) also hold for arbitrary \(A\in\A_X\), with the same constants; see Corollary~\ref{rem:odd-representatives}.

To compare the weighted interaction exponent \(a\) with the more familiar pointwise decay exponent, suppose that the Hamiltonian interaction \(\Phi_u\) has only two-site terms and that \(\|\Phi_u(\{x,y\})\|\le C(1+d(x,y))^{-\varpi}\) uniformly in \(u\in I\) and distinct \(x,y\).
Volume growth then gives \(\sup_{u\in I}\|\Phi_u\|^{\#}_{a;\kappa}<\infty\) for \(a<\varpi-\dvol\).
Hence \(\varpi>2\dvol\) is sufficient to choose \(a>\dvol\).

On \(\Z^n\), with \(\dvol=n\), the condition \(\varpi>2\dvol\) is precisely the threshold above which uniform norm Lieb--Robinson bounds with algebraic light cones hold for time-dependent two-site interactions \cite{FossFeigEtAl15,TranEtAl19}.
For \(\dvol<\varpi\le2\dvol\), Tran et al.\ \cite[Eq.~(3) and Section~IV]{TranStateTransfer21} construct time-dependent two-site spin Hamiltonians with transfer times growing more slowly than every positive power of the distance.
This excludes bounds with polynomial time growth and positive polynomial spatial decay in that range.
In Proposition~\ref{prop:decay-optimality}, we prove that Theorem~\ref{thm:card-propagation} cannot be extended to \(0<a\le\dvol\) with uniform polynomial time bounds at \(p=a\), even for smooth paths of fermionic two-site interactions.

Spatially unbounded onsite Hamiltonian terms, for which the full interaction need not satisfy the hypotheses of Proposition~\ref{prop:polynomial-dynamics}, are treated separately in Proposition~\ref{prop:onsite-terms}.

\subsection{Continuity, local limits, and filtering}

In applications, the Hamiltonian interaction and the interaction being evolved often depend on parameters or arise as limits of approximations. 
The evolved interaction is also integrated against time filters.
The following theorem handles all three situations.

For a Hamiltonian path \(\Phi\) generating a strongly continuous automorphism cocycle on \(\R\), a complex filter \(W\in L^1(\R)\), and \(s\in\R\), define the bounded linear map
\[
 \cI^\Phi_{W,s}(A):=\int_\R W(v)\tau^\Phi_{s,s+v}(A)\,\dd v,
 \qquad A\in\A.
\]
If \(\Phi\) satisfies the hypotheses of Theorem~\ref{thm:card-propagation}, fix the partitions and cutoffs provided there and, for an interaction \(\Psi\) with \(\|\Psi\|^{\#}_{0;\kappa_0}<\infty\), define
\begin{equation}
 \cK^\Phi_{W,s}[\Psi](Y):=\int_\R W(v)
       (\cT_{s,s+v}^\Phi\Psi)(Y)\,\dd v.
 \label{eq:card-filter-interaction}
\end{equation}
The uniform single-input bound \eqref{eq:card-propagation-total} at \(p=k=0\), summed over the finitely many input supports \(X\subseteq Y\), gives absolute convergence in norm of each coefficient integral.

\begin{theorem}[Continuity, local limits, and filtering]
\label{thm:card-limits}
Assume \eqref{eq:volume-growth} and Hypothesis~\ref{hyp:cell-partitions}.
Fix \(a>\dvol\) and \(0\le L<\infty\).
Use the constants \(\kappa_0,c_*,C_p,C_{p,k}\) and exponents \(\gamma,m_{p,k}\) from Theorem~\ref{thm:card-propagation} for the Hamiltonian bound \(L\), and \(\widetilde m_{p,k}\) from \eqref{eq:source-time-exponent}.
For every \(0\le p\le a\), \(b>0\), and \(0<\alpha\le1\), there are constants \(C_{p,b,\alpha}>0\) and \(C_{p,b,\alpha,k}>0\), \(k\ge0\), for the estimates in part~\ref{item:card-limits-stretched}, depending only on the indicated indices and the propagation data above, such that the following holds.
Let \(I\subseteq\R\) be an interval and \(\frakP\) a compact Hausdorff parameter space.
Consider componentwise norm-continuous families of Hamiltonian interactions
\[
 (\Phi_{\xi,u})_{(\xi,u)\in\frakP\times I},
 \qquad
 (\Phi^{(n)}_{\xi,u})_{(\xi,u)\in\frakP\times I},
 \quad n\in\N,
\]
whose \(\|\cdot\|^{\#}_{a;\kappa_0}\)-norms are bounded by \(L\), uniformly in all parameters and \(n\).
Suppose that, for every nonempty \(Z\Subset\Gamma\) and every compact subinterval \(J\subseteq I\),
\begin{equation}
 \lim_{n\to\infty} \sup_{\xi\in\frakP}\int_J
       \|\Phi^{(n)}_{\xi,u}(Z)-\Phi_{\xi,u}(Z)\|\,\dd u
 \;=\;0.
 \label{eq:hamiltonian-component-convergence}
\end{equation}

Choose the same cell partitions and time cutoffs for all \(\xi\in\frakP\) and \(n\in\N\), as in Theorem~\ref{thm:card-propagation} for \(a,L\).
With these choices, define the maps \(\cT^{\Phi_\xi}_{s,t;X}\) and \(\cT^{\Phi^{(n)}_\xi}_{s,t;X}\), and the induced interaction maps \(\cT^{\Phi_\xi}_{s,t}\) and \(\cT^{\Phi^{(n)}_\xi}_{s,t}\), by Definition~\ref{def:interaction-representatives}.
Let \(0\le M_\Psi<\infty\), and consider componentwise norm-continuous families of interactions
\[
 (\Psi_\xi)_{\xi\in\frakP},
 \qquad (\Psi_\xi^{(n)})_{\xi\in\frakP},
 \quad n\in\N,
\]
whose \(\|\cdot\|^{\#}_{a;\kappa_0}\)-norms are bounded by \(M_\Psi\), uniformly in \(\xi\) and \(n\).
Suppose that \(\Psi_\xi^{(n)}\to\Psi_\xi\) componentwise, uniformly in \(\xi\).
In parts \ref{item:card-limits-coefficients}--\ref{item:card-limits-stretched}, assume additionally that \(I=\R\) and \(W\in L^1(\R)\).
\begin{enumerate}[label=(\roman*),ref=(\roman*),leftmargin=*]
\item \emph{Parameter continuity and local convergence.}
\label{item:card-limits-continuity}
The interaction \(\cT_{s,t}^{\Phi_\xi}\Psi_\xi\) is componentwise norm-continuous on \(\frakP\times I^2\) and, for every \(0\le q<a\), \(k\ge0\), nonempty \(K\Subset\Gamma\), and compact subinterval \(J\subseteq I\),
\begin{equation}
  \lim_{n\to\infty} \sup_{\xi\in\frakP}\sup_{s,t\in J}
 \bigl\|\cT_{s,t}^{\Phi_\xi^{(n)}}\Psi_\xi^{(n)}
       -\cT_{s,t}^{\Phi_\xi}\Psi_\xi\bigr\|^{\pol,\loc}_{q;k,K}
 \;=\;0.
 \label{eq:card-local-topology}
\end{equation}

\item \emph{Filtered coefficients.}
\label{item:card-limits-coefficients}
Abbreviate the filtered interactions as
\[
 \cK_{\xi,s}:=\cK^{\Phi_\xi}_{W,s}[\Psi_\xi],
 \qquad
 \cK^{(n)}_{\xi,s}:=\cK^{\Phi_\xi^{(n)}}_{W,s}[\Psi_\xi^{(n)}],
\]
and for each nonempty \(X\Subset\Gamma\), define
\[
 \cK_{\xi,s;X}(Y):=\int_\R W(v)
       (\cT^{\Phi_\xi}_{s,s+v;X}\Psi_\xi(X))(Y)\,\dd v.
\]
The integrals in \eqref{eq:card-filter-interaction} defining \(\cK_{\xi,s}(Y)\) and \(\cK^{(n)}_{\xi,s}(Y)\), as well as the integral defining \(\cK_{\xi,s;X}(Y)\), exist in norm.
The interaction \(\cK_{\xi,s}\) is componentwise norm-continuous in \((\xi,s)\).
It is self-adjoint when \(\Psi_\xi\) is self-adjoint and \(W\) is real.
For every nonempty finite \(Y\) and compact interval \(J\subseteq\R\),
\[
 \lim_{n\to\infty} \sup_{\xi\in\frakP}\sup_{s\in J}
       \|\cK^{(n)}_{\xi,s}(Y)-\cK_{\xi,s}(Y)\|
\;=\;0.
\]
\item \emph{Weighted estimates.}
\label{item:card-limits-estimates}
For \(0\le p\le a\) and \(k\ge0\), set
\[
 \begin{aligned}
 \mu_{p,k}(W):=\int_\R|W(v)|(1+|v|)^{m_{p,k}}\,\dd v,\qquad
 \widetilde\mu_{p,k}(W):=\int_\R|W(v)|(1+|v|)^{\widetilde m_{p,k}}\,\dd v.
 \end{aligned}
\]
For every \(0\le p\le a\) and \(k\ge0\) with \(\mu_{p,k}(W)<\infty\), one has
\begin{equation}
 \|\cK_{\xi,s}\|^{\pol}_{p;k}
 \le C_{p,k}\|\Psi_\xi\|^{\#}_{p;\kappa_0}
       \mu_{p,k}(W).
 \label{eq:filtered-norm}
\end{equation}
For every \(0\le p\le a\) and \(k\ge0\) with \(\widetilde\mu_{p,k}(W)<\infty\), and every nonempty \(X\Subset\Gamma\), one has
\begin{align}
 \sum_{Y\supseteq X}|Y|^k(1+\diam Y)^p
                         \|\cK_{\xi,s;X}(Y)\|
 \le C_{p,k}\E^{\kappa_0|X|}(1+\diam X)^p\|\Psi_\xi(X)\|
       \widetilde\mu_{p,k}(W).
 \label{eq:filtered-source-total}
\end{align}
For each fixed \(X\), the estimate with \(p=k=0\) requires only \(W\in L^1(\R)\).

\item \emph{Derivation and total-operator identities.}
\label{item:card-limits-identities}
Write \(\cI_{\xi,s}:=\cI^{\Phi_\xi}_{W,s}\) for the filtering map on \(\A\).
If \(\mu_{0,0}(W)<\infty\), then for every local observable \(B\),
\begin{equation}
 \sum_Y[\cK_{\xi,s}(Y),B]
 =\sum_X[\cI_{\xi,s}(\Psi_\xi(X)),B] =\int_\R W(v)\sum_X[\tau^{\Phi_\xi}_{s,s+v}(\Psi_\xi(X)),B]\,\dd v,
 \label{eq:filtered-derivation}
\end{equation}
with absolute convergence of the sums and the integral of the sum of commutator norms.
If \(\sum_X\E^{\kappa_0|X|}\|\Psi_\xi(X)\|<\infty\), then already for \(W\in L^1(\R)\),
\begin{equation}
 \sum_Y\cK_{\xi,s}(Y)
 =\cI_{\xi,s}\left(\sum_X\Psi_\xi(X)\right),
 \label{eq:filtered-total-operator}
\end{equation}
with absolute norm convergence of the operator sums.

\item \emph{Local convergence of filtered interactions.}
\label{item:card-limits-filtered-convergence}
For every \(0\le q<a\) and \(k\ge0\) with \(\mu_{q,k}(W)<\infty\), every nonempty \(K\Subset\Gamma\), and every compact interval \(J\subseteq\R\),
\begin{equation}
 \lim_{n\to\infty}\sup_{\xi\in\frakP}\sup_{s\in J}
 \bigl\|\cK^{(n)}_{\xi,s}-\cK_{\xi,s}\bigr\|^{\pol,\loc}_{q;k,K}
 \;=\;0.
 \label{eq:filtered-local-topology}
\end{equation}

\item \emph{Stretched-exponential support weights.}
\label{item:card-limits-stretched}
For \(b>0\) and \(0<\alpha\le1\), set
\[
 \theta:=\frac{\alpha}{\alpha+(1+\gamma)\dvol},\qquad
 b':=c_*^\theta(b/2)^{1-\theta},\qquad
 F_{b,\alpha}(W):=\int_\R |W(v)|
                       \E^{b(1+|v|)^\alpha}\,\dd v.
\]
If \(F_{b,\alpha}(W)<\infty\), then, for every \(0\le p\le a\), \(\xi\in\frakP\), and \(s\in\R\),
\begin{align}
 \sup_{x\in\Gamma}\sum_{Y\ni x}\E^{b'|Y|^\theta}(1+\diam Y)^p
       \sum_{\varnothing\ne X\subseteq Y}\|\cK_{\xi,s;X}(Y)\|
\le C_{p,b,\alpha}F_{b,\alpha}(W)
                  \|\Psi_\xi\|^{\#}_{p;\kappa_0}.
 \label{eq:filtered-stretched-support}
\end{align}
For every \(k\ge0\), one also has
\begin{align}
 \sup_{x\in\Gamma}\sum_{Y\ni x}|Y|^k
       \E^{b'|Y|^\theta/2}(1+\diam Y)^p
       \sum_{\varnothing\ne X\subseteq Y}\|\cK_{\xi,s;X}(Y)\|
 \le C_{p,b,\alpha,k}F_{b,\alpha}(W)
                  \|\Psi_\xi\|^{\#}_{p;\kappa_0}.
 \label{eq:filtered-stretched-moments}
\end{align}
\end{enumerate}
\keepwithproofpointer
\end{theorem}
\proofpointer{\S\ref{subsec:continuity-filtering}}{proof:card-limits}

The strict inequality \(q<a\) in parts~\ref{item:card-limits-continuity} and~\ref{item:card-limits-filtered-convergence} is necessary in general; see the example following Lemma~\ref{lem:componentwise-local-continuity} in Appendix~\ref{app:dynamics}.

Theorem~\ref{thm:smooth-hamiltonian-dependence} in Subsection~\ref{subsec:differentiability} extends these results to parameter derivatives of propagated and filtered interactions.
Proposition~\ref{prop:iterated-filters} in Subsection~\ref{subsec:iterated-filtering} treats repeated filtering and commutators.

\subsection{Propagation bounds for local observables}

Theorem~\ref{thm:card-propagation} also yields propagation bounds for individual local observables, including odd ones, as stated in the following corollary.

\begin{corollary}[Local approximation with polynomial time growth]
\label{cor:observable-LR}
Assume \eqref{eq:volume-growth} and Hypothesis~\ref{hyp:cell-partitions}, and choose \(\kappa_0\) as in Theorem~\ref{thm:card-propagation}.
Fix \(a>\dvol\) and \(0\le L<\infty\).
There exists
\[
 C=C(a,L,\dvol,C_\Gamma,\Ccell,\Dcell)>0
\]
such that the following holds for every choice of local multiplicities \((r_x)_{x\in\Gamma}\) and every componentwise norm-continuous path \((\Phi_u)_{u\in\R}\) of self-adjoint interactions satisfying
\[
 \sup_{u\in\R}\|\Phi_u\|^{\#}_{a;\kappa_0}\le L.
\]
For every nonempty \(X\Subset\Gamma\), \(A\in\A_X\), \(s,t\in\R\), and \(R\ge0\),
\begin{equation}
 \|\tau_{s,t}^\Phi(A)-\F_{X^{(R)}}^{\rm tr}
                       (\tau_{s,t}^\Phi(A))\|
 \le C|X|\|A\|
        (1+|t-s|)^{m_{a,0}}(1+R)^{-a}.
 \label{eq:observable-LR}
\end{equation}
Here \(m_{a,0}\) is given in \eqref{eq:time-exponent}.
For every finite \(\Lambda\) with \(X\subseteq\Lambda\subsetneq\Gamma\),
\begin{equation}
 \|\tau_{s,t}^\Phi(A)-\tau_{s,t}^{\Phi,\Lambda}(A)\|
 \le C|X|\|A\|(1+\dist(X,\Lambda^c))^{-a}
       \frac{(1+|t-s|)^{m_{a,0}+1}-1}{m_{a,0}+1}.
 \label{eq:polynomial-time-finite-approximation}
\end{equation}
Both approximation bounds hold uniformly under finite restrictions and deletion of Hamiltonian terms.
If \(A\) is parity-homogeneous, \(Y\subseteq\Gamma\setminus X^{(R)}\), and \(B\in\A_Y\) is parity-homogeneous, then
\[
 \bigl\|\tau_{s,t}^\Phi(A)B-(-1)^{|A||B|}B\tau_{s,t}^\Phi(A)\bigr\|
 \le 2C\min\{|X|,|Y|\}\|A\|\|B\|
        (1+|t-s|)^{m_{a,0}}(1+R)^{-a}.
\]
For infinite \(Y\), the minimum is understood as \(|X|\).
\keepwithproofpointer
\end{corollary}
\proofpointer{\S\ref{subsec:observable-locality}}{proof:observable-LR}

If one is willing to give up some spatial decay in Corollary~\ref{cor:observable-LR}, the time exponent and the dependence on the support size improve: for every \(0<p\le a\),
\begin{align}
 \|\tau_{s,t}^\Phi(A)-\F_{X^{(R)}}^{\rm tr}
                       (\tau_{s,t}^\Phi(A))\|
 \le 2^{1-p/a}C^{p/a}|X|^{p/a}\|A\|
       (1+|t-s|)^{(p/a)m_{a,0}}(1+R)^{-p}.
 \label{eq:observable-interpolation}
\end{align}
This follows from \eqref{eq:observable-LR}, the trivial bound \(2\|A\|\), and \(\min\{u,v\}\le u^{1-p/a}v^{p/a}\).
The same interpolation applies to the finite-volume approximation and graded commutator bounds, using \(2\|A\|\|B\|\) as the trivial bound for the latter.
For \(0<p<a\), the time exponent \((p/a)m_{a,0}\) is smaller than \(m_{p,0}\).

Filtered interactions are themselves used as generators of dynamics, for instance in spectral flow.
The filtering estimates of Theorem~\ref{thm:card-limits}, however, do not in general provide the exponential support-size bound required by Theorem~\ref{thm:card-propagation}, so the latter cannot be applied to these dynamics directly.
The following theorem covers these dynamics under polynomial support-size assumptions and preserves the spatial decay exponent.
Its bound depends exponentially on the integrated interaction norm and hence grows exponentially in time when the interaction norm is uniformly bounded.
Its proof is independent of Theorem~\ref{thm:card-propagation}.

\begin{theorem}[Locality from integrated interaction norms]
\label{thm:polynomial-local-topology}
Let \(p>0\), let \((\Gamma,d)\) be a countable metric space with finite balls, and choose positive integer local multiplicities \((r_x)_{x\in\Gamma}\).
Let \(J\subseteq\R\) be a compact interval and \((\Xi_u)_{u\in J}\) a componentwise norm-continuous path of self-adjoint interactions satisfying
\[
 \sup_{u\in J}\|\Xi_u\|^{\pol}_{p;0}<\infty,
 \qquad
 \int_J\|\Xi_u\|^{\pol}_{p;1}\,\dd u<\infty.
\]
These interactions generate a unique strongly continuous parity-preserving automorphism cocycle \(\tau^\Xi\) satisfying \eqref{eq:dynamics}.
As \(\Lambda\nearrow\Gamma\), their finite-volume dynamics converge in norm on every observable, uniformly in \(s,t\in J\).
To quantify this convergence, set, for \(s,t\in J\),
\[
 M(s,t):=\int_{\min(s,t)}^{\max(s,t)}
                     \|\Xi_u\|^{\pol}_{p;1}\,\dd u.
\]
For every nonempty \(X\Subset\Gamma\), \(A\in\A_X\), \(s,t\in J\), \(R\ge0\), and \(\Lambda\Subset\Gamma\) with \(X^{(R)}\subseteq\Lambda\), one has
\begin{equation}
 \|\tau_{s,t}^\Xi(A)-\tau_{s,t}^{\Xi,\Lambda}(A)\|
 \le |X|\|A\|\bigl(\E^{2M(s,t)}-1\bigr)(1+R)^{-p},
 \label{eq:polynomial-finite-approximation}
\end{equation}
and
\begin{equation}
 \|\tau_{s,t}^\Xi(A)-\F_{X^{(R)}}^{\rm tr}
                         (\tau_{s,t}^\Xi(A))\|
 \le 2|X|\|A\|\bigl(\E^{2M(s,t)}-1\bigr)(1+R)^{-p}.
 \label{eq:polynomial-cocycle-locality}
\end{equation}
If \(A\) is parity-homogeneous, \(Y\subseteq\Gamma\setminus X^{(R)}\), and \(B\in\A_Y\) is parity-homogeneous, then
\begin{align}
 \bigl\|\tau_{s,t}^\Xi(A)B-(-1)^{|A||B|}B\tau_{s,t}^\Xi(A)\bigr\|
 \le 2\min\{|X|,|Y|\}\|A\|\|B\|
          \bigl(\E^{2M(s,t)}-1\bigr)(1+R)^{-p}.
 \label{eq:integrated-graded-commutator}
\end{align}
For infinite \(Y\), the minimum is understood as \(|X|\).
\keepwithproofpointer
\end{theorem}
\proofpointer{\S\ref{subsec:observable-locality}}{proof:polynomial-local-topology}

The interpolation \eqref{eq:observable-interpolation} applies also here and gives the bounds of Theorem~\ref{thm:polynomial-local-topology} at every spatial exponent \(0<q\le p\), with the support-size factors and \(\E^{2M(s,t)}-1\) raised to the power \(q/p\) and a numerical prefactor at most \(2\).
It applies as well to the locality bounds in Corollary~\ref{cor:filtered-flow}, with \(p=a\).

\begin{remark}[Other spatial weights]
\label{rem:general-spatial-weights}
Let \(w:[0,\infty)\to[1,\infty)\) be nondecreasing, with
\[
 w(0)=1,\qquad w(r+s)\le w(r)w(s)\quad(r,s\ge0),
 \qquad \lim_{R\to\infty}w(R)=\infty.
\]
For \(k\ge0\), define
\[
 \|\Xi\|_{w;k}:=\sup_{x\in\Gamma}\sum_{Z\ni x}
             |Z|^k w(\diam Z)\|\Xi(Z)\|.
\]
Under the metric and regularity hypotheses of Theorem~\ref{thm:polynomial-local-topology}, assume
\[
 \sup_{u\in J}\|\Xi_u\|_{w;0}<\infty,\qquad
 \int_J\|\Xi_u\|_{w;1}\,\dd u<\infty.
\]
All conclusions of that theorem then hold with
\[
 M_w(s,t):=\int_{\min(s,t)}^{\max(s,t)}\|\Xi_u\|_{w;1}\,\dd u
\]
in place of \(M(s,t)\), and \(w(R)^{-1}\) in place of \((1+R)^{-p}\) in the approximation and graded commutator bounds.
For example, \(w(r)=\E^{cr^\vartheta}\) satisfies the weight assumptions for \(c>0\) and \(0<\vartheta\le1\).
\keepwithproofpointer
\end{remark}
\proofpointer{\S\ref{subsec:observable-locality}}{proof:general-spatial-weights}

A uniform \(\|\Xi_u\|^{\pol}_{a;0}\)-bound with \(a>\dvol\) also gives polynomial time growth for observable locality if the spatial exponent is decreased to \(0<q<a\).
Proposition~\ref{prop:polynomial-time-locality} in Subsection~\ref{subsec:polynomial-time-locality} states and proves this bound without a cell assumption.

\section{Interaction representatives for finite-range dynamics}
\label{sec:finite-range}

We begin the proof of Theorem~\ref{thm:card-propagation} with a finite-volume commutator estimate that requires no cell partition.
We then use this estimate to control nested commutators with test observables in separated cells.
These bounds allow us to decompose evolved interaction terms into operators supported on unions of cells.
To organize these operators and estimate their weighted sums, we introduce interactions on the cell partition.
In Section~\ref{sec:propagation}, we use this decomposition to treat Hamiltonians with polynomial spatial decay.

\begin{samepage}
\subsection{A finite-volume commutator estimate}
\label{subsec:finite-volume-commutators}

We first estimate the commutator of an evolved observable with one even test observable.
We will use this estimate to control propagation from a cell to the boundary of its neighborhood and later to prove the observable bounds without a cell assumption.
The matrix estimate below follows from the fermionic Lieb--Robinson iteration of Nachtergaele--Sims--Young \cite[Lemma~3.3 and Eqs.~(3.29)--(3.31)]{NSY18}.

\par
\end{samepage}

\begin{lemma}
\label{lem:finite-volume-matrix-commutator}
Let \((\Gamma,d)\) be a countable metric space, let \(\Omega\Subset\Gamma\), and choose a finite-dimensional Hilbert space \(\mathcal H_y^{\rm aux}\) at each \(y\in\Omega\).
Put
\[
 \widetilde\A_Y:=\A_Y\otimes
       \bigotimes_{y\in Y}\cB(\cH_y^{\rm aux}),
 \qquad Y\subseteq\Omega,
\]
with parity acting on the CAR factor.
Let \((\Xi_u)_{u\in I}\) be a componentwise norm-continuous path on an interval \(I\), with even self-adjoint terms \(\Xi_u(Z)\in\widetilde\A_Z^+\), \(\varnothing\ne Z\subseteq\Omega\).
Denote by \(\tau^{\Xi,\Omega}\) the dynamics generated by \(\sum_{Z\subseteq\Omega}\Xi_u(Z)\).
Define the matrices
\[
 K_u^\Omega(x,y):=
 2\sum_{\substack{Z\subseteq\Omega\\x,y\in Z}}\|\Xi_u(Z)\|,
 \qquad x,y\in\Omega.
\]
For \(s,t\in I\), put \(T:=|t-s|\), choose \(\sigma\in\{-1,1\}\) with \(t=s+\sigma T\), and set
\[
 \Delta_n(T):=\{(u_1,\ldots,u_n)\,|\,0\le u_1\le\cdots\le u_n\le T\},
 \qquad \dd\mathbf u:=\dd u_1\cdots\dd u_n,
\]
\[
 \mathsf U_{s,t}^\Omega:=\1_\Omega+
 \sum_{n=1}^\infty\int_{\Delta_n(T)}
 K_{s+\sigma u_n}^\Omega\cdots K_{s+\sigma u_1}^\Omega\,\dd\mathbf u,
 \qquad \1_\Omega(x,y):=\delta_{xy}.
\]
The products are ordinary matrix products, and \(\mathsf U^\Omega_{s,s}=\1_\Omega\).

Then the following estimates hold for every \(s,t\in I\).
\begin{enumerate}[label=(\roman*),leftmargin=*]
\item \emph{Commutator estimate.}
For all nonempty \(X,Y\subseteq\Omega\), \(A\in\widetilde\A_X\), and \(B\in\widetilde\A_Y^+\),
\begin{equation}
 \|[\tau_{s,t}^{\Xi,\Omega}(A),B]\|
 \le2\|A\|\|B\|
       \sum_{x\in X}\sum_{y\in Y}\mathsf U_{s,t}^\Omega(x,y).
 \label{eq:polynomial-matrix-commutator}
\end{equation}
\item \emph{Weighted matrix estimate.}
If \(w:[0,\infty)\to[1,\infty)\) is finite, nondecreasing, and satisfies \(w(0)=1\) and \(w(r+s)\le w(r)w(s)\), then
\begin{equation}
 \sup_{x\in\Omega}\sum_{y\in\Omega}
 w(d(x,y))\mathsf U_{s,t}^\Omega(x,y)
 \le\exp\!\left(
 \int_{\min(s,t)}^{\max(s,t)}
 \sup_{x\in\Omega}\sum_{y\in\Omega}
 w(d(x,y))K_u^\Omega(x,y)\,\dd u\right).
 \label{eq:finite-volume-weighted-matrix-bound}
\end{equation}
\end{enumerate}
\end{lemma}

\begin{proof}
Let \(s<t\) and \(X\cap Y=\varnothing\).
We use the iterated commutator bound of Nachtergaele--Sims--Young \cite[Eqs.~(3.29)--(3.31)]{NSY18}, before estimating the support sums by a spatial decay function.
Their argument applies to the algebras \(\widetilde\A_Z\): an even operator commutes with every operator supported on a disjoint set, and auxiliary factors at distinct sites commute.
After enlarging the boundary sums to all overlapping supports and numbering the supports in increasing time order, this bound reads
\[
 \|[\tau_{s,t}^{\Xi,\Omega}(A),B]\|
  \le2\|A\|\|B\|\sum_{n=1}^{\infty}2^n
       \int_{s\le u_1\le\cdots\le u_n\le t}
 \sum_{\substack{\varnothing\ne Z_1,\ldots,Z_n\subseteq\Omega\\
                  Z_1\cap Y\ne\varnothing,\ Z_n\cap X\ne\varnothing\\
                  Z_j\cap Z_{j+1}\ne\varnothing\ (1\le j<n)}}
       \prod_{j=1}^n\|\Xi_{u_j}(Z_j)\|\,\dd\mathbf u.
\]
The contribution of order \(n\) is at most
\[
 2\|A\|\|B\|\,
 \frac{\left(2(t-s)\sup_{u\in[s,t]}
                 \sum_{Z\subseteq\Omega}\|\Xi_u(Z)\|\right)^n}{n!},
\]
so the series converges absolutely.

For each support sequence, choose
\[
 x_0\in Y\cap Z_1,\qquad
 x_j\in Z_j\cap Z_{j+1}\ (1\le j<n),\qquad
 x_n\in Z_n\cap X.
\]
Multiplying by \(2^n\) and enlarging the sum to all choices of sites and supports gives
\[
2^n
 \sum_{\substack{\varnothing\ne Z_1,\ldots,Z_n\subseteq\Omega\\
                  Z_1\cap Y\ne\varnothing,\ Z_n\cap X\ne\varnothing\\
                  Z_j\cap Z_{j+1}\ne\varnothing\ (1\le j<n)}}
       \prod_{j=1}^n\|\Xi_{u_j}(Z_j)\| \le
 \sum_{y\in Y}\sum_{x\in X}
       (K_{u_1}^\Omega\cdots K_{u_n}^\Omega)(y,x) =\sum_{x\in X}\sum_{y\in Y}
       (K_{u_n}^\Omega\cdots K_{u_1}^\Omega)(x,y).
\]
The equality transposes the entire product and uses the symmetry of each \(K_u^\Omega\).
Integration over the ordered times and summation over \(n\) prove \eqref{eq:polynomial-matrix-commutator} for disjoint supports.
For intersecting supports, the bound \(2\|A\|\|B\|\) is covered by the identity contribution in \(\mathsf U_{s,t}^\Omega\).
For \(t<s\), apply the argument to \(\check\Xi_v:=-\Xi_{-v}\), whose dynamics satisfy \(\tau_{-s,-t}^{\check\Xi,\Omega}=\tau_{s,t}^{\Xi,\Omega}\).
The resulting matrix series has \(\sigma=-1\), as in the statement.
At \(s=t\), only the identity contribution remains.

For a matrix \(F\) on \(\Omega\), define
\[
 \|F\|_{w,\mathrm{row}}:=
 \sup_{x\in\Omega}\sum_{y\in\Omega}w(d(x,y))|F(x,y)|.
\]
The triangle inequality and the assumptions on \(w\) give \(w(d(x,y))\le w(d(x,z))w(d(z,y))\).
Applying this inequality in the matrix product yields \(\|FG\|_{w,\mathrm{row}}\le\|F\|_{w,\mathrm{row}}\|G\|_{w,\mathrm{row}}\).
Also \(\|\1_\Omega\|_{w,\mathrm{row}}=1\).
Consequently, the norm of the \(n\)-th integral in the matrix series is at most
\[
 \frac1{n!}\left(
 \int_{\min(s,t)}^{\max(s,t)}\|K_u^\Omega\|_{w,\mathrm{row}}\,\dd u
 \right)^n.
\]
Summation proves convergence of the series and \eqref{eq:finite-volume-weighted-matrix-bound}.
\end{proof}

\subsection{Commutators in separated cells}
\label{subsec:separated-cell-commutators}

For the remainder of this section, assume Hypothesis~\ref{hyp:cell-partitions}.
Whenever the scale \(\ell\ge1\) is fixed, we write \(Q_b\) for \(Q_b^{(\ell)}\).
Constants may depend on \(\dvol,C_\Gamma,\Ccell,\Dcell\), in addition to the dependencies stated below.

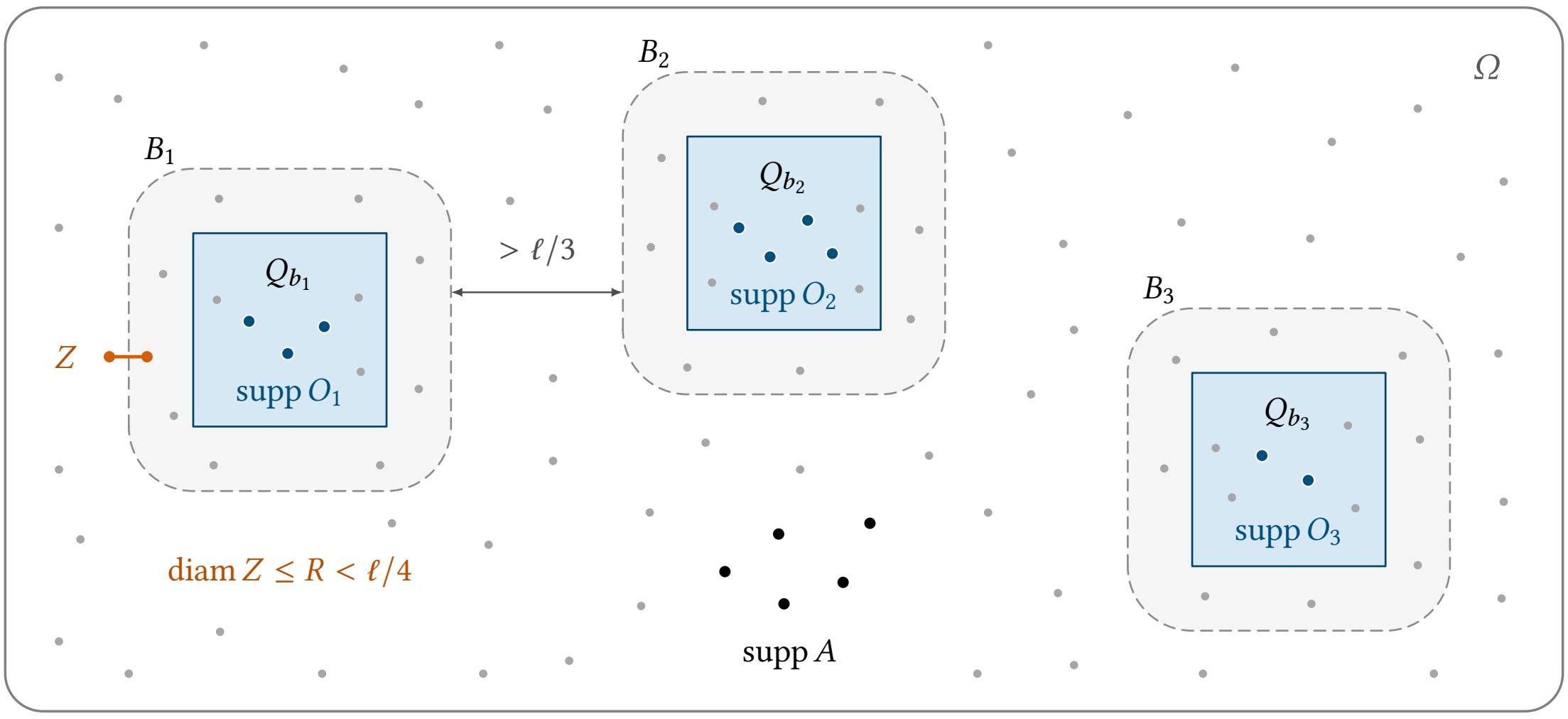
\begin{figure}[!htbp]
\centering
\begin{tikzpicture}[cell diagram]
 \draw[black!50,line width=.65pt,rounded corners=10pt]
  (.15,.25) rectangle (14.65,6.8);
 \node[font=\small,text=black!65] at (13.95,6.25) {$\Omega$};
 \foreach \x/\y/\i in {2.8/3.8/1,7.4/4.7/2,12.1/2.5/3}{
  \filldraw[fill=black!4,draw=black!45,densely dashed,rounded corners=.6cm]
   (\x-1.5,\y-1.5) rectangle (\x+1.5,\y+1.5);
  \filldraw[fill=figblue!16,draw=figblue!60!black]
   (\x-.9,\y-.9) rectangle (\x+.9,\y+.9);
  \node[font=\footnotesize] at (\x,\y+.52) {$Q_{b_\i}$};
  \node[font=\footnotesize,anchor=south west,inner sep=1pt]
   at (\x-1.4,\y+1.5) {$B_\i$};
  \node[font=\footnotesize,text=figblue!70!black] at (\x,\y-.59)
   {$\supp O_\i$};
 }
 \foreach \x/\y in {2.42/3.88,2.78/3.58,3.12/3.83,
   6.98/4.75,7.27/4.48,7.62/4.82,7.85/4.51,
   11.85/2.63,12.28/2.4}
  \filldraw[fill=figblue!70!black,draw=white,line width=.4pt]
   (\x,\y) circle[radius=1.65pt];
 \draw[dim line] (4.3,4.15) -- (5.9,4.15)
  node[midway,above=3pt,font=\footnotesize] {$>\ell/3$};
 \foreach \x/\y in {6.85/1.55,7.35/1.9,7.95/1.45,8.2/2,7.4/1.25}
  \fill (\x,\y) circle[radius=1.5pt];
 \node[font=\footnotesize] at (7.45,.78) {$\supp A$};
 \draw[figred,line width=.9pt] (1.12,3.55) -- (1.47,3.55);
 \foreach \x in {1.12,1.47}
  \fill[figred] (\x,3.55) circle[radius=1.5pt];
 \node[font=\footnotesize,text=figred!85!black,anchor=east]
  at (.96,3.55) {$Z$};
 \node[font=\footnotesize,text=figred!85!black] at (2.8,1.55)
  {$\diam Z\le R<\ell/4$};
 \foreach \x/\y in {2.12/4.08,3.44/4.1,3.46/3.41,
   6.73/4.24,8.11/4.93,6.75/4.95,8.1/4.18,
   11.42/2.7,12.65/2.91,12.72/2.14,11.57/2.24,
   1.62/4.32,2.14/5.02,3.44/5.02,4.01/4.45,
   4/3.25,3.64/2.54,2.09/2.54,1.72/3,
   6.16/4.57,6.26/5.4,7.2/5.93,8.29/5.92,
   8.66/4.73,8.58/3.9,7.55/3.42,6.5/3.45,
   10.94/2.51,11.05/3.52,11.96/3.78,13.16/3.56,
   13.32/2.78,13.3/1.61,12.31/1.25,11.32/1.32,
   .65/.9,.65/2.5,.65/4.75,.65/6.15,
   1.3/.6,3.1/.6,4.6/.6,5.4/.72,9.2/.6,10.1/.68,11.3/.6,13.45/.6,
   .85/1.85,2.15/.99,3.75/2,4.65/1.8,5.25/2.58,5.25/3.35,
   4.85/5,5.2/5.85,1.2/5.95,3.3/6.23,4/5.9,2/6.45,4.75/6.45,
   9.3/6.45,10.6/5.8,11.6/6.24,12.5/5.55,13.3/5.86,14.1/5.18,
   10/4.6,11.1/4.8,12.3/4.65,13.7/4.48,9.52/5.45,
   9.3/2.1,9.95/1.35,9.7/3.2,10.1/3.8,
   6.07/1.23,6.15/2.1,6.67/2.75,7.55/2.5,8.75/2.63,
   14.08/1.3,14.1/2.2,14.05/3.58}
  \fill[black!35] (\x,\y) circle[radius=1.1pt];
\end{tikzpicture}
\caption{Support geometry in Lemma~\ref{lem:card-separated} for \(N_{\rm sep}=3\).
The test observables \(O_i\) are supported inside the cells \(Q_{b_i}\), surrounded by the neighborhoods \(B_i=\{x\in\Gamma\,|\,\dist(x,Q_{b_i})\le\ell/3\}\).
All three neighborhoods lie in \(\Omega\); the initial observable \(A\) is supported in their complement.
Pairwise cell distances greater than \(\ell\) imply \(\dist(B_i,B_j)>\ell/3>R\) for \(i\ne j\).
Thus each Hamiltonian support of diameter at most \(R\) meets at most one neighborhood, as illustrated by \(Z\).
Gray dots indicate other sites of \(\Gamma\); the outlines are schematic.}
\label{fig:separated-cell-commutators}
\end{figure}

\begin{samepage}
The next lemma estimates nested commutators of an evolved observable with one test observable in each of several separated cells.
We prove a time-dependent fermionic version of the separated-cell estimates of McDonough--Yin--Lucas--Zhang \cite[Theorem~7, Corollaries~2 and~4, and Section~IV\,D]{McDonoughEtAl25}, using Lemma~\ref{lem:finite-volume-matrix-commutator} and a differential equation for the nested commutators.
For initial time zero, write \(\tau_t^\Xi=\tau_{0,t}^\Xi\) also for a time-dependent generator.
Figure~\ref{fig:separated-cell-commutators} illustrates the support assumptions of the commutator estimate for three test observables.
We allow additional finite-dimensional factors at each site because, in the proof of Lemma~\ref{lem:card-short}, auxiliary qubits encode restricted dynamics and express their alternating sums as nested commutators.
\par
\end{samepage}

\begin{samepage}
\begin{lemma}
\label{lem:card-separated}
There is a constant \(C>0\), depending only on the geometric constants, with the following property.
Let \(R\ge1\), \(J_{\rm fr}\ge0\), and let \(\Omega\Subset\Gamma\).
Choose the auxiliary site factors and the algebras \(\widetilde\A_Y\) as in Lemma~\ref{lem:finite-volume-matrix-commutator}.
Let \((\Xi_u)_{u\in\R}\) be a componentwise norm-continuous path with even self-adjoint terms \(\Xi_u(Z)\in\widetilde\A_Z^+\), \(\varnothing\ne Z\subseteq\Omega\), extended by zero to other supports.
Assume that this path has range at most \(R\) and satisfies \(\sup_{u\in\R}\|\Xi_u\|^{\pol}_{0;1}\le J_{\rm fr}\).
Take cells from Hypothesis~\ref{hyp:cell-partitions} with \(\ell>4R\), and choose distinct, pairwise nonadjacent cells \(Q_{b_1},\ldots,Q_{b_{N_{\rm sep}}}\).
Set \(B_i:=\{x\in\Gamma\,|\,\dist(x,Q_{b_i})\le\ell/3\}\).
Assume \(\bigcup_iB_i\subseteq\Omega\), let \(A\in\widetilde\A_{\Omega\setminus\bigcup_iB_i}^+\), and let \(O_i\in\widetilde\A_{Q_{b_i}}^+\) have norm at most one.
The dynamics generated by \(\sum_{Z\subseteq\Omega}\Xi_u(Z)\) satisfy, for every \(t\in\R\),
\begin{equation}
 2^{-N_{\rm sep}}
 \|[O_{N_{\rm sep}},[\cdots,[O_1,\tau_t^{\Xi,\Omega}(A)]\cdots]]\|
 \le \|A\|
 \left(C\ell^{\dvol}
 \E^{2\E J_{\rm fr}|t|-\ell/(4R)}\right)^{N_{\rm sep}}.
 \label{eq:card-separated}
\end{equation}
\end{lemma}
\end{samepage}

\pagebreak[0]
\begin{samepage}
To see why \eqref{eq:card-separated} is nontrivial, note that repeated use of \(\|[O_i,B]\|\le2\|B\|\), together with the isometry of the dynamics, bounds the left-hand side of \eqref{eq:card-separated} by \(\|A\|\).
For \(N_{\rm sep}\ge1\), applying the single-cell case of \eqref{eq:card-separated} to the innermost commutator and estimating the remaining commutators in this way gives the factor \(C\ell^{\dvol}\E^{2\E J_{\rm fr}|t|-\ell/(4R)}\) only once.
The gain in \eqref{eq:card-separated} is one such factor for every selected cell.
\par
\end{samepage}

\begin{proof}
We compare the full evolution with the separate evolutions inside the neighborhoods \(B_i\).
The resulting differential equation relates each nested commutator to commutators involving one fewer test observable.
Only Hamiltonian terms crossing the boundary of a neighborhood occur in the additional terms, and we estimate their contributions with Lemma~\ref{lem:finite-volume-matrix-commutator}.
We give the proof for \(t>0\); negative times follow by time reflection, and the case \(t=0\) is immediate.

\smallskip
\noindent \emph{Boundary contributions.}
For \(D\subseteq\Omega\), let \(\tau^{\Xi,D}\) denote the finite-volume dynamics generated by \(H_D(v):=\sum_{Z\subseteq D}\Xi_v(Z)\).
For \(0\le u\le t\), put
\[
 P_i(u):=\tau_{t-u,0}^{\Xi,B_i}(O_i),
 \qquad
 E_i(u):=\I[H_\Omega(t-u)-H_{B_i}(t-u),P_i(u)].
\]
The operator \(P_i(u)\) is even, belongs to \(\widetilde\A_{B_i}\), has norm at most one, and satisfies \(P_i(t)=O_i\).
Terms supported outside \(B_i\) commute with \(P_i(u)\), so
\[
 E_i(u)=\I\sum_{\substack{Z\subseteq\Omega\,|\,Z\cap B_i\ne\varnothing\\
                         Z\not\subseteq B_i}}
                 [\Xi_{t-u}(Z),P_i(u)].
\]
Define
\[
 b_i(u):=\sum_{\substack{Z\subseteq\Omega\,|\,Z\cap B_i\ne\varnothing\\
                         Z\not\subseteq B_i}}
                 \|[\Xi_{t-u}(Z),P_i(u)]\|,
 \quad \mbox{then}\quad \|E_i(u)\|\le b_i(u).
\]
For \(i\ne j\), the cell separation gives \(\dist(B_i,B_j)>\ell/3>R\).
No nonzero Hamiltonian term therefore meets two neighborhoods.
Every summand defining \(E_i(u)\) is even and supported in \(Z\cup B_i\), which is disjoint from \(B_j\) for \(j\ne i\).
Consequently, \([E_i(u),P_j(u)]=0\) for \(i\ne j\).

\smallskip
\noindent \emph{Induction on the number of test observables.}
Set
\[
 F_\varnothing(u):=\tau_{t-u,t}^{\Xi,\Omega}(A).
\]
For \(\mathcal J=\{i_1<\cdots<i_m\}\subseteq\{1,\ldots,N_{\rm sep}\}\), define
\[
 F_{\mathcal J}(u):=
 [P_{i_m}(u),[\cdots,[P_{i_1}(u),F_\varnothing(u)]\cdots]].
\]
Differentiation with respect to the initial time gives
\[
 \begin{aligned}
 F_\varnothing'(u)&=\I[H_\Omega(t-u),F_\varnothing(u)],\\
 P_i'(u)&=\I[H_{B_i}(t-u),P_i(u)]
          =\I[H_\Omega(t-u),P_i(u)]-E_i(u).
 \end{aligned}
\]
Differentiating the nested commutator \(F_{\mathcal J}(u)\) by the product rule gives one contribution from each \(P_i(u)\), \(i\in\mathcal J\), and one from \(F_\varnothing(u)\).
For operators \(H,B,C\in\widetilde\A_\Omega\), the Jacobi identity gives
\[
 [H,[B,C]]=[[H,B],C]+[B,[H,C]].
\]
Applying it successively with \(H=H_\Omega(t-u)\) shows that all contributions involving \(\I[H_\Omega(t-u),\cdot]\) sum to \(\I[H_\Omega(t-u),F_{\mathcal J}(u)]\).
For \(i=i_r\in\mathcal J\), the remaining contribution replaces \(P_i(u)\) by \(-E_i(u)\) in its position in the nested commutator.
For every \(j\ne i\) and \(B\in\widetilde\A_\Omega\), the relation \([P_j(u),E_i(u)]=0\) and the Jacobi identity imply
\[
 [P_j(u),[E_i(u),B]]
 =[[P_j(u),E_i(u)],B]+[E_i(u),[P_j(u),B]]
 =[E_i(u),[P_j(u),B]].
\]
Applying this identity in turn to \(P_{i_{r+1}}(u),\ldots,P_{i_m}(u)\) puts \(E_i(u)\) in the outermost commutator and leaves the other test observables in their original order.
The contribution from \(-E_i(u)\) therefore equals \(-[E_i(u),F_{\mathcal J\setminus\{i\}}(u)]\).
Combining the Hamiltonian and boundary contributions gives
\[
 F_{\mathcal J}'(u)
 =\I[H_\Omega(t-u),F_{\mathcal J}(u)]
   -\sum_{i\in\mathcal J}
          [E_i(u),F_{\mathcal J\setminus\{i\}}(u)].
\]
Since \(F_\varnothing(0)=A\) is supported outside every \(B_i\), we have \(F_{\mathcal J}(0)=0\) for nonempty \(\mathcal J\).
The homogeneous equation is implemented by unitaries and hence isometric.
Variation of constants therefore gives
\[
 2^{-|\mathcal J|}\|F_{\mathcal J}(u)\|
 \le\sum_{i\in\mathcal J}\int_0^u
       b_i(v)\,2^{-(|\mathcal J|-1)}
       \|F_{\mathcal J\setminus\{i\}}(v)\|\,\dd v.
\]
Starting with \(\|F_\varnothing(u)\|=\|A\|\), induction yields
\begin{equation}
 2^{-|\mathcal J|}\|F_{\mathcal J}(u)\|
 \le\|A\|\prod_{i\in\mathcal J}\int_0^u b_i(v)\,\dd v.
 \label{eq:separated-boundary-product}
\end{equation}
Indeed, substituting the bounds for \(\mathcal J\setminus\{i\}\) into the preceding integral gives \(\|A\|\) times
\[
 \sum_{i\in\mathcal J}\int_0^u
 b_i(v)\prod_{j\in\mathcal J\setminus\{i\}}
              \left(\int_0^v b_j(w)\,\dd w\right)\dd v
 =\prod_{i\in\mathcal J}\int_0^u b_i(v)\,\dd v,
\]
by differentiating the product on the right and using its value at zero.

\smallskip
\noindent \emph{Propagation from a cell to its boundary.}
Let \(K_v^{B_i}\) and \(\mathsf U_{v,w}^{B_i}\) be the matrices of Lemma~\ref{lem:finite-volume-matrix-commutator} for the interaction restricted to \(B_i\).
The range assumption implies
\[
 \sup_{x\in B_i}\sum_{y\in B_i}\E^{d(x,y)/R}K_v^{B_i}(x,y)
 \le2\E\sup_{x\in B_i}\sum_{\substack{Z\subseteq B_i\\Z\ni x}}
                         |Z|\|\Xi_v(Z)\| \le2\E J_{\rm fr}.
\]
Consequently, \eqref{eq:finite-volume-weighted-matrix-bound} gives
\[
 \sup_{x\in B_i}\sum_{y\in B_i}\E^{d(x,y)/R}
        \mathsf U_{t-u,0}^{B_i}(x,y)
 \le\E^{2\E J_{\rm fr}(t-u)}.
\]
Apply Lemma~\ref{lem:finite-volume-matrix-commutator} to \(P_i(u)\) and each Hamiltonian term \(\Xi_{t-u}(Z)\) whose support \(Z\) meets both \(B_i\) and its complement.
Here we regard the restricted interaction as an interaction on \(\Omega\), extended by zero; its matrix series has no entries connecting \(B_i\) to \(\Omega\setminus B_i\).
We obtain
\[
 b_i(u)\le2\sum_{x\in Q_{b_i}}\sum_{y\in B_i}
      \mathsf U_{t-u,0}^{B_i}(x,y)
      \sum_{\substack{Z\subseteq\Omega\,|\,y\in Z\\Z\not\subseteq B_i}}
                         \|\Xi_{t-u}(Z)\|.
\]
The inner sum is at most \(J_{\rm fr}\).
If a nonzero term in this sum contains \(y\), choose \(z\in Z\setminus B_i\).
Then \(\dist(z,Q_{b_i})>\ell/3\) and \(d(y,z)\le R\), so \(\dist(y,Q_{b_i})>\ell/3-R\).
The weighted matrix bound therefore yields
\[
 b_i(u)\le2J_{\rm fr}|Q_{b_i}|
       \E^{2\E J_{\rm fr}(t-u)-\ell/(3R)+1}.
\]
For \(J_{\rm fr}>0\), integration gives
\[
 \int_0^t b_i(u)\,\dd u
 \le |Q_{b_i}|\E^{-\ell/(3R)}
                 \bigl(\E^{2\E J_{\rm fr}t}-1\bigr).
\]
If \(J_{\rm fr}=0\), every Hamiltonian term vanishes, and \(b_i=0\).
Finally, volume growth and \(\diam Q_{b_i}\le\Ccell\ell\) give
\[
 |Q_{b_i}|\le C_\Gamma(1+\Ccell\ell)^{\dvol}
             \le C_\Gamma(1+\Ccell)^{\dvol}\ell^{\dvol}.
\]
At \(u=t\), the left-hand side of \eqref{eq:separated-boundary-product} for \(\mathcal J=\{1,\ldots,N_{\rm sep}\}\) is the left-hand side of \eqref{eq:card-separated}.
Substituting the preceding bounds proves \eqref{eq:card-separated} with \(C=\E C_\Gamma(1+\Ccell)^{\dvol}\), using \(-\ell/(3R)\le-\ell/(4R)\).
\end{proof}

\subsection{Cell interactions and local decompositions}
\label{subsec:card-finite-range}

We now use Lemma~\ref{lem:card-separated} to decompose evolved interaction terms into operators supported on unions of cells.
We first introduce cell interactions to organize these operators and compare their weighted norms with the norms of interactions on sites.
For finite \(S\subseteq\cB_\ell\), use the union \(Q(S)\subseteq\Gamma\) from Definition~\ref{def:interaction-representatives}.
Disjointness of the cells and the volume bound give
\[
 |Q(S)|=\sum_{b\in S}|Q_b|
 \le C_\Gamma(1+\Ccell\ell)^{\dvol}|S|.
\]

\begin{definition}[Interactions on a cell partition]
\label{def:cell-interactions}
Fix a cell partition \((Q_b)_{b\in\cB_\ell}\).
A cell interaction is a map
\[
 \Psi:S\mapsto\Psi(S)\in\A_{Q(S)}^+,
 \qquad \varnothing\ne S\Subset\cB_\ell\,.
\]
It is self-adjoint if \(\Psi(S)^*=\Psi(S)\) for every \(S\).
A path \((\Psi_u)_{u\in I}\) of cell interactions on an interval \(I\) is componentwise norm-continuous if \(u\mapsto\Psi_u(S)\) is norm-continuous for every \(S\).
We identify \(\Psi\) with the interaction on physical sites that assigns \(\Psi(S)\) to \(Q(S)\) and vanishes on site sets that are not unions of cells.
For \(\rho,p\ge0\), define the corresponding cell interaction norms
\begin{equation}
 \|\Psi\|^{\#,\ell}_{p;\rho}:=
 \sup_{b\in\cB_\ell}
 \sum_{S\ni b}\E^{\rho|S|}
 (1+\diam Q(S))^p\|\Psi(S)\|.
 \label{eq:card-cell-norm}
\end{equation}
The superscript \(\ell\) refers to the fixed cell partition \((Q_b)_{b\in\cB_\ell}\).
\end{definition}

To express a site interaction \(\Psi\) on the same partition, group its terms according to the cells they meet:
\begin{equation}
 \Psi^{(\ell)}(S):=
 \sum_{\substack{\varnothing\ne X\subseteq Q(S)\\S_0(X)=S}}
       \Psi(X),
 \qquad \varnothing\ne S\Subset\cB_\ell,
 \label{eq:site-to-cell-grouping}
\end{equation}
where \(S_0(X)\) is the set of cells meeting \(X\), as in Definition~\ref{def:interaction-representatives}.
For every finite \(U\Subset\cB_\ell\), grouping the site supports according to \(S_0(X)\) gives
\begin{equation}
 \sum_{\varnothing\ne S\subseteq U}\Psi^{(\ell)}(S)
 =\sum_{\varnothing\ne X\subseteq Q(U)}\Psi(X).
 \label{eq:site-to-cell-hamiltonian}
\end{equation}
The site interaction associated with a cell interaction \(\Psi\) satisfies \(\|\Psi\|^{\pol}_{p;0}\le \|\Psi\|^{\#,\ell}_{p;\rho}\).
For every fixed \(k\ge0\), volume growth also gives
\begin{equation}
 |Q(S)|^k\le C_k\ell^{k\dvol}|S|^k,\qquad
 \|\Psi\|^{\pol}_{p;k}\le C_{\rho,k}\ell^{k\dvol}
                  \|\Psi\|^{\#,\ell}_{p;\rho}\quad(\rho>0).
 \label{eq:cell-to-site-moments}
\end{equation}
The second inequality follows from \(\sup_{m\ge1}m^k\E^{-\rho m}<\infty\).

\begin{figure}[!hbp]
\centering
\begin{tikzpicture}[cell diagram]
 \begin{scope}[x=.5cm,y=.5cm]
  \node[fig title] at (6.5,7.1) {Admissible cell sets};
  \foreach \x/\y in {1/1,2/1,1/2,2/3,3/3,4/3, 9/3,10/4,11/4,10/2,11/2}
   \fill[figblue!16] (\x,\y) rectangle (\x+1,\y+1);
  \foreach \x/\y in {2/2,3/2,10/3}
   \fill[figblue!42] (\x,\y) rectangle (\x+1,\y+1);
  \fill[black!5] (6,5) rectangle (8,6);
  \draw[cell grid,step=1] (0,0) grid (13,6);
  \draw[black!55,densely dashed,line width=.6pt] (6,5) rectangle (8,6);
  \node[fig note,anchor=west,fill=white,inner sep=1.5pt] at (8.15,5.5)
    {misses $S_0$};
  \foreach \x/\y in {2.35/2.6,2.72/2.22,3.4/2.68,10.3/3.62,10.66/3.3}
   \fill (\x,\y) circle[radius=1.3pt];
  \coordinate (Xleft) at (2.35,2.6);
  \coordinate (Xright) at (10.66,3.3);
  \coordinate (Dleft) at ($(Xleft)!1.85cm!-90:(Xright)$);
  \coordinate (Dright) at ($(Xright)+(Dleft)-(Xleft)$);
  \draw[ext line] ($(Xleft)!.08cm!(Dleft)$) -- ($(Dleft)!-.08cm!(Xleft)$);
  \draw[ext line] ($(Xright)!.08cm!(Dright)$) -- ($(Dright)!-.08cm!(Xright)$);
  \draw[dim line] (Dleft) -- (Dright)
    node[midway,fill=white,inner sep=1.5pt,font=\footnotesize] {$\diam X$};
  \begin{scope}[shift={(.6,-2.1)}]
   \fill[figblue!42] (0,-.35) rectangle (.7,.35);
   \node[font=\footnotesize,anchor=west] at (.85,0) {$S_0$};
   \fill[figblue!16] (2.6,-.35) rectangle (3.3,.35);
   \node[font=\footnotesize,anchor=west] at (3.45,0) {$S\setminus S_0$};
   \fill (6.55,0) circle[radius=1.3pt];
   \node[font=\footnotesize,anchor=west] at (6.8,0) {sites of $X$};
  \end{scope}
 \end{scope}
\end{tikzpicture}
\caption{Cell sets in the local decomposition of Lemma~\ref{lem:card-short}.
Cells are obtained by intersecting Euclidean squares with \(\Gamma\).
The dots form the initial support \(X\) of the interaction term being evolved, and the dark squares are the cells \(S_0=\{b\,|\,Q_b\cap X\ne\varnothing\}\) meeting \(X\).
All blue-shaded squares together specify a cell set \(S\supseteq S_0\); its union \(Q(S)\) is the support assigned to one of the local terms in the decomposition of the evolved interaction term.
This local term vanishes if \(S\) has a connected component in the cell graph that does not meet \(S_0\), as would happen after adding the dashed component.
The set \(S\) need not be connected, so \(Q(S)\) need not contain the region between distant parts of \(X\).}
\label{fig:cell-reconstruction}
\end{figure}
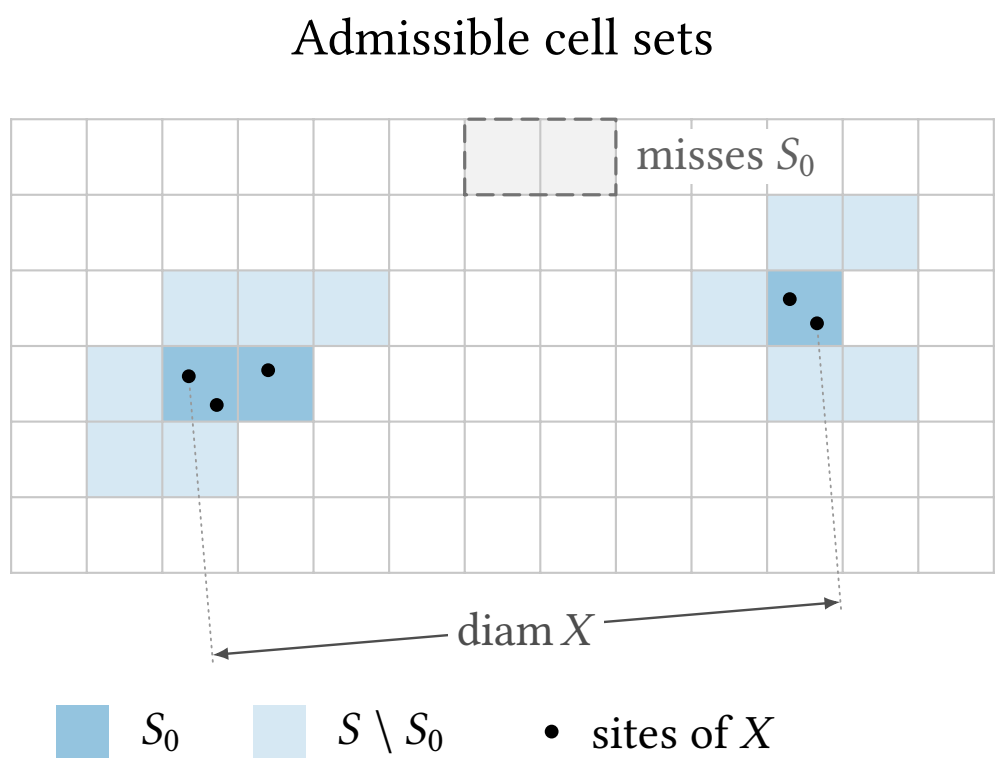

The following lemma gives the required decomposition, with the cell scale chosen according to the interaction range and the time horizon.
It adapts the box expansion of McDonough--Yin--Lucas--Zhang \cite[Theorem~9 and Lemma~8, Sections~VI\,A and~VI\,C]{McDonoughEtAl25} from a single initial box to arbitrary initial supports on the cell graphs of Hypothesis~\ref{hyp:cell-partitions}.
The estimates include the dependence on the initial support and the polynomial diameter weight.
Any prescribed exponential weight in the number of output cells can be obtained by choosing a sufficiently strong input support-size weight and a sufficiently large cell scale.
In Theorem~\ref{thm:card-propagation}, we use the following lemma with \(\rho=4\), so the resulting \(\kappa_0\) depends only on \(\Dcell\).
Figure~\ref{fig:cell-reconstruction} illustrates the cell sets carrying these local terms.

\begin{lemma}
\label{lem:card-short}
Fix \(\rho>0\).
There exists \(\kappa_0>0\), depending only on \(\rho,\Dcell\), such that, for every \(0\le J_{\rm fr}<\infty\), there are \(c>4\) and constants \(C_p,C'_p>0\), \(p\ge0\), with the following property.
The constants \(c,C_p,C'_p\) depend only on \(\rho,J_{\rm fr},\dvol,C_\Gamma,\Ccell,\Dcell\) and their indicated indices.
For every \(T\ge0\), \(R\ge1\), and every componentwise norm-continuous path \((\Xi_u)_{|u|\le T}\) of self-adjoint interactions of range at most \(R\) satisfying \(\sup_{|u|\le T}\|\Xi_u\|^{\pol}_{0;1}\le J_{\rm fr}\), set
\begin{equation}
 \ell:=\left\lceil cR\bigl(T+\log(2+R)\bigr)\right\rceil.
 \label{eq:card-cell-scale}
\end{equation}
Choose a cell partition \((Q_b)_{b\in\cB_\ell}\) from Hypothesis~\ref{hyp:cell-partitions}.
For every interaction \(\Psi\) and nonempty \(X\Subset\Gamma\), let \(S_0(X)\) be the set of cells meeting \(X\), as in Definition~\ref{def:interaction-representatives}.
Define the cell interaction \(A_X(\cdot;t)\) by the coefficient formula \eqref{eq:full-dynamics-mobius}, using \(\Phi=\Xi\), \(s=0\), \(A=\Psi(X)\), and the chosen cell partition, and set \(A_X(S;t)=0\) unless \(S\supseteq S_0(X)\).
The maps \(\Psi(X)\mapsto A_X(S;t)\) are complex-linear, preserve adjoints, and use only Hamiltonian terms \(\Xi_u(Z)\) with \(Z\subseteq Q(S)\).
For \(|t|\le T\),
\[
 \tau_t^\Xi(\Psi(X))
 =\sum_{S\supseteq S_0(X)}A_X(S;t)
\]
with absolute convergence in norm.
Moreover, \(A_X(S;t)=0\) whenever a connected component of \(S\) in the cell graph is disjoint from \(S_0(X)\).
For every \(p\ge0\), these terms satisfy
\begin{equation}
 \sup_{b\in\cB_\ell}
 \sum_{S\ni b}\E^{\rho|S|}(1+\diam Q(S))^p
 \sum_{\varnothing\ne X\subseteq Q(S)}\|A_X(S;t)\|  \le C_p\ell^{\dvol}
 \left(\|\Psi\|^{\#}_{p;\kappa_0}
       +\ell^p\|\Psi\|^{\#}_{0;\kappa_0}\right).
 \label{eq:card-short}
\end{equation}
For each fixed \(X\), the sum over all output cell sets satisfies
\begin{equation}
 \sum_{S\supseteq S_0(X)}\E^{\rho|S|}(1+\diam Q(S))^p\|A_X(S;t)\|
 \le C'_p\E^{\kappa_0|X|}\|\Psi(X)\|
       \bigl((1+\diam X)^p+\ell^p\bigr).
 \label{eq:finite-range-total-bound}
\end{equation}
\end{lemma}

\begin{proof}
Put
\[
 c_0:=\Dcell+1,\quad A_D:=4c_0,\quad
 \mucell:=\rho+3+\log A_D,\quad
 \Kcell:=2c_0(\mucell+\log2),\quad \kappa_0:=\Kcell+1.
\]
In the cell graph, \(c_0\) bounds the number of cells adjacent or equal to a given cell.
Below, \(A_D^m\) bounds the number of contributing cell sets with \(m\) cells, \(\mucell\) is the decay rate per cell, and \(\Kcell\) accounts for the cells meeting the initial support.
Fix \(J_{\rm fr}\ge0\), and let \(C_{\rm sep}\ge1\) be a constant for which Lemma~\ref{lem:card-separated} holds.
Choose \(c>4\) so that
\[
 \frac c4-\dvol\log(c+1)
 \ge 2\E J_{\rm fr}+2\dvol+\log C_{\rm sep}+\Kcell.
\]
For \(T\ge0\) and \(R\ge1\), put \(h:=T+\log(2+R)\) and \(\ell=\lceil cRh\rceil\), and define
\[
 \varepsilon:=C_{\rm sep}\ell^{\dvol}
                 \E^{2\E J_{\rm fr}T-\ell/(4R)}.
\]
Since \(h\ge1\), \(\ell\le(c+1)Rh\), \(\log R\le h\), and \(\log h\le h\), the choice of \(c\) gives
\[
 \log\varepsilon
 \le \log C_{\rm sep}+\dvol\log(c+1)
       +\left(2\E J_{\rm fr}+2\dvol-\frac c4\right)h
 \le-\Kcell.
\]
Extend \(\Xi\) to \(\R\) by \(\Xi_u:=\Xi_{-T}\) for \(u<-T\) and \(\Xi_u:=\Xi_T\) for \(u>T\).

\smallskip
\noindent \emph{Finite-volume formula and vanishing terms.}
Let \(G_\ell\) be the cell graph of Hypothesis~\ref{hyp:cell-partitions}.
Fix an interaction \(\Psi\) and a nonempty finite \(X\subseteq\Gamma\), and put
\[
 S_0:=S_0(X),
 \qquad A:=\Psi(X).
\]
For a finite cell set \(U\), define
\[
 H_U(u):=\sum_{\varnothing\ne Z\subseteq Q(U)}\Xi_u(Z),
 \qquad \tau_t^{H_U}:=\tau_{0,t}^{\Xi,Q(U)}.
\]
For finite \(S\supseteq S_0\), the coefficients in the statement are
\begin{equation}
 A_X(S;t)=
 \sum_{S_0\subseteq U\subseteq S}
 (-1)^{|S|-|U|}\tau_t^{H_U}(A)
 \in\A_{Q(S)}^+.
 \label{eq:card-mobius}
\end{equation}
This finite sum proves complex-linearity, preservation of adjoints, and dependence only on Hamiltonian terms contained in \(Q(S)\).
Equation~\eqref{eq:card-telescope} gives the sum of these coefficients in a finite volume.
Define
\[
 \scrC(S_0):=
 \{S\Subset\cB_\ell\,|\,S\supseteq S_0,
   \ C\cap S_0\ne\varnothing
   \text{ for every component }C\text{ of }G_\ell[S]\}.
\]
We now prove that \(A_X(S;t)=0\) whenever \(S\supseteq S_0\) and \(S\notin\scrC(S_0)\).
Let \(C\) be a component of \(G_\ell[S]\) disjoint from \(S_0\).
Every nonzero term \(\Xi_u(Z)\) meets only pairwise adjacent cells, since \(\diam Z\le R<\ell\).
For \(S_0\subseteq V\subseteq S\setminus C\) and \(W\subseteq C\), it follows that
\[
 H_{V\cup W}(u)=H_V(u)+H_W(u),\qquad
 [H_V(u),H_W(v)]=0,\qquad [H_W(u),A]=0
 \quad(u,v\in\R).
\]
The propagator for \(H_{V\cup W}\) therefore factors into the propagators for \(H_V\) and \(H_W\), and \(\tau_t^{H_{V\cup W}}(A)=\tau_t^{H_V}(A)\).
Substitution in \eqref{eq:card-mobius} gives
\[
 A_X(S;t)=
 \sum_{S_0\subseteq V\subseteq S\setminus C}
 (-1)^{|S\setminus C|-|V|}\tau_t^{H_V}(A)
 \sum_{W\subseteq C}(-1)^{|C|-|W|}=0.
\]
\smallskip
\noindent \emph{Exponential decay in the number of cells.}
We first select many pairwise nonadjacent cells of \(S\) far from \(X\).
We then group the terms in \eqref{eq:card-mobius} into alternating sums over the selected cells and express each of these sums as a matrix element of a nested commutator in an enlarged algebra.
Each successive commutator is taken with an operator that flips the auxiliary qubits in one selected cell.
For \(|t|\le T\), \eqref{eq:card-separated} bounds the nested commutator by a product with one factor \(2\varepsilon\) for each selected cell, where \(\varepsilon\le\E^{-\Kcell}\) by the choice of \(\ell\).
The resulting decay compensates for the number of output supports and yields \eqref{eq:card-short}, with an exponential weight in the number of cells.
Write \(m:=|S|\) and \(n_0:=|S_0|\).
For a cell set \(V\), let
\[
 \cN(V):=V\cup
 \{b\in\cB_\ell\,|\,b\text{ is adjacent to some }a\in V\}.
\]
Fix an ordering of the finite set \(S\).
Starting with \(E_0:=S\setminus\cN(S_0)\), define recursively
\[
 b_i:=\min E_{i-1},\qquad
 E_i:=E_{i-1}\setminus\cN(\{b_i\})
 \quad\text{while }E_{i-1}\ne\varnothing.
\]
Let \(N_{\rm sep}\) be the first index with \(E_{N_{\rm sep}}=\varnothing\), and put \(\cM:=\{b_1,\ldots,b_{N_{\rm sep}}\}\).
The selected cells are pairwise nonadjacent, and
\[
 c_0N_{\rm sep}\ge |E_0|\ge m-c_0n_0,\qquad
 \dist(Q_{b_i},X)>\ell\quad(1\le i\le N_{\rm sep}).
\]

To construct this enlarged algebra, we encode the choice of \(U\subseteq S\) in auxiliary qubits, one for each site of \(Q(S)\).
The qubit at \(x\) is switched on precisely when \(x\in Q(U)\), and each Hamiltonian term acts only if all qubits on its support are switched on.
For this fixed \(S\), set \(\cH_S^{\rm aux}:=\bigotimes_{x\in Q(S)}\C^2\).
Let \(q_x:=|1\rangle\langle1|\) act on the factor at \(x\), and define an interaction on \(\A\otimes\cB(\cH_S^{\rm aux})\) by
\[
 \widehat\Xi^S_u(Z):=
 \begin{cases}
 \Xi_u(Z)\otimes\displaystyle\prod_{x\in Z}q_x,
       &\varnothing\ne Z\subseteq Q(S),\\
 0,    &Z\not\subseteq Q(S),
 \end{cases}
 \qquad
 \widehat H_S(u):=\sum_{Z\subseteq Q(S)}\widehat\Xi^S_u(Z).
\]
The terms \(\widehat\Xi^S_u(Z)\) are even and self-adjoint, are supported in \(Z\), and satisfy
\[
 \|\widehat\Xi^S_u(Z)\|
 =\1_{\{Z\subseteq Q(S)\}}\|\Xi_u(Z)\|,\qquad
 \sup_{u\in\R}\|\widehat\Xi^S_u\|^{\pol}_{0;1}\le J_{\rm fr}.
\]
Put \(Y_t:=\tau_t^{\widehat H_S}(A\otimes\1)\).
For \(U\subseteq S\), define the unit vector
\[
 |e_U\rangle
 :=\bigotimes_{x\in Q(S)}
       |\1_{Q(U)}(x)\rangle\in\cH_S^{\rm aux}.
\]
We identify \(\A_{Q(S)}\otimes\cB(\cH_S^{\rm aux})\) with matrices over \(\A_{Q(S)}\) in the standard tensor-product basis of \(\cH_S^{\rm aux}\).
Then \(\1\otimes|e_U\rangle\) is a column over \(\A_{Q(S)}\), and \((\1\otimes\langle e_V|)Y(\1\otimes|e_U\rangle)\) is a matrix entry whose norm is at most \(\|Y\|\).
By the definition of \(\widehat\Xi^S\),
\[
 \widehat H_S(u)(\1\otimes|e_U\rangle)
 =(\1\otimes|e_U\rangle)H_U(u).
\]
The propagators of \(\widehat H_S\) and \(H_U\) satisfy the same intertwining identity, so
\[
 (\1\otimes\langle e_V|)Y_t
 (\1\otimes|e_U\rangle)
 =\1_{\{U=V\}}\tau_t^{H_U}(A),
 \qquad U,V\subseteq S.
\]
For \(b\in\cM\) and \(V\subseteq\cM\), set
\[
 F_b:=\1\otimes\prod_{x\in Q_b}
       (|0\rangle\langle1|+|1\rangle\langle0|)_x,
 \qquad F_V:=\prod_{b\in V}F_b.
\]
The operators \(F_b\), \(b\in\cM\), are even, have norm one, are supported in \(Q_b\), and commute.
For \(U\subseteq S\) and \(V\subseteq\cM\),
\[
 F_V(\1\otimes|e_U\rangle)
 =\1\otimes|e_{U\mathbin\triangle V}\rangle.
\]
Define
\[
 \begin{split}
 \cD_{\cM}(Y)
 :=[F_{b_{N_{\rm sep}}},[\cdots,[F_{b_1},Y]\cdots]] =\sum_{V\subseteq\cM}(-1)^{N_{\rm sep}-|V|}
                       F_VYF_{\cM\setminus V},
 \qquad \cD_{\varnothing}(Y):=Y.
 \end{split}
\]
For \(S_0\subseteq U_0\subseteq S\setminus\cM\), the matrix entries of \(Y_t\) and the action of \(F_V\) give
\[
 \begin{split}
 \Delta_{U_0,\cM}(t)
 :=(\1\otimes\langle e_{U_0}|)
       \cD_{\cM}(Y_t)
       (\1\otimes|e_{U_0\cup\cM}\rangle) =\sum_{V\subseteq\cM}(-1)^{N_{\rm sep}-|V|}
                      \tau_t^{H_{U_0\cup V}}(A).
 \end{split}
\]
Define \(B_{\cM}:=\bigcup_{b\in\cM}\{x\in\Gamma\,|\,\dist(x,Q_b)\le\ell/3\}\).
The observable \(A\otimes\1\) is supported outside \(B_{\cM}\).
Set \(\Omega:=Q(S)\cup B_{\cM}\), and choose auxiliary spaces \(\C^2\) on \(Q(S)\) and \(\C\) on \(\Omega\setminus Q(S)\).
Applying Lemma~\ref{lem:card-separated} in this finite region to \(\widehat\Xi^S\) and the operators \(F_b\) yields
\[
 \|\Delta_{U_0,\cM}(t)\|
 \le\|\cD_{\cM}(Y_t)\|
 \le(2\varepsilon)^{N_{\rm sep}}\|A\|,\qquad |t|\le T.
\]
For \(N_{\rm sep}=0\), this inequality follows from \(\|Y_t\|=\|A\|\).
Regrouping \eqref{eq:card-mobius} as
\[
 A_X(S;t)=
 \sum_{S_0\subseteq U_0\subseteq S\setminus\cM}
       (-1)^{m-N_{\rm sep}-|U_0|}\Delta_{U_0,\cM}(t)
\]
gives \(\|A_X(S;t)\|\le2^{m-n_0}\varepsilon^{N_{\rm sep}}\|A\|\).
Since \(\varepsilon\le\E^{-\Kcell}<1\) and \(N_{\rm sep}\ge\max\{m/c_0-n_0,0\}\),
\begin{equation}
 \|A_X(S;t)\|
 \le 2^m\varepsilon^{\max\{m/c_0-n_0,0\}}\|A\|.
 \label{eq:card-mobius-bound}
\end{equation}
Moreover,
\[
 2^m\varepsilon^{\max\{m/c_0-n_0,0\}}
 \le\E^{\Kcell n_0-(\Kcell/c_0-\log2)m}
 \le\E^{\Kcell|X|-\mucell m}.
\]
We have proved
\begin{equation}
 \|A_X(S;t)\|
 \le\E^{\Kcell|X|-\mucell|S|}\|\Psi(X)\|.
 \label{eq:card-anchored-decay}
\end{equation}

It remains to count the cell sets in \(\scrC(S_0)\), which we do by encoding them as ordered trees.
Order \(S_0\) and fix neighbor lists \(n_b(1),\ldots,n_b(\deg b)\) in \(G_\ell\).
For \(S\in\scrC(S_0)\), let \(h_S(b):=\dist_{G_\ell[S]}(b,S_0)\), and define
\[
 \pi_S(b):=n_b\!\left(\min\{j\,|\,n_b(j)\in S,
                   \ h_S(n_b(j))=h_S(b)-1\}\right),
 \qquad b\in S\setminus S_0.
\]
The edges \(\pi_S(b)\to b\) form a forest with roots \(S_0\), since \(h_S\) decreases by one at each step towards a parent.
Order the children of each vertex by its neighbor list.
Let \(T_S\) be the ordered tree obtained by adjoining a vertex \(*\) whose ordered children are the ordered roots \(S_0\).
Encode the depth-first traversal of \(T_S\) by writing \(+1\) for each step away from \(*\) and \(-1\) for each return step.
This word in \(\{+1,-1\}^{2m}\) determines the tree shape, so there are at most \(4^m\) shapes.
For every edge \(a\to b\) not leaving \(*\), let \(j\) be the index with \(b=n_a(j)\).
The shape of \(T_S\), the ordered set \(S_0\), and the \(m-n_0\) indices determine all vertices of \(S\) recursively.
There are at most \(c_0\) choices per index, so
\[
 \bigl|\{S\in\scrC(S_0)\,|\,|S|=m\}\bigr|
 \le4^m c_0^{m-n_0}\le A_D^m,\qquad m\ge n_0.
\]
\smallskip
\noindent \emph{Diameter weights and absolute convergence.}
If \(S\in\scrC(S_0)\) and \(b\in S\), there is a path \(b=b_0,\ldots,b_k\) in \(G_\ell[S]\) with \(b_k\in S_0\) and \(k\le m-1\).
For \(z\in Q_b\) and \(x\in X\cap Q_{b_k}\), cell diameters and adjacency give
\[
 d(z,x)\le k\ell+(k+1)\Ccell\ell
          \le (\Ccell+1)\ell m.
\]
Thus
\[
 Q(S)\subseteq X^{((\Ccell+1)\ell m)},\qquad
 \diam Q(S)\le\diam X+2(\Ccell+1)\ell m,
\]
and, for each \(p\ge0\),
\[
 (1+\diam Q(S))^p
 \le C_p\bigl((1+\diam X)^p+\ell^p m^p\bigr).
\]
Fix \(b\in\cB_\ell\), choose \(y_b\in Q_b\), and set \(B_{b,m}:=\ball_{(\Ccell+1)\ell m}(y_b)\).
If \(b\in S\in\scrC(S_0(X))\), then \(X\cap B_{b,m}\ne\varnothing\).
The vanishing of terms outside \(\scrC(S_0(X))\), the count above, and \eqref{eq:card-anchored-decay} imply
\begin{align*}
 \sum_{S\ni b}\E^{\rho|S|}(1+\diam Q(S))^p&
       \sum_{\varnothing\ne X\subseteq Q(S)}\|A_X(S;t)\|\\*
 &\le C_p\sum_{m\ge1}\E^{-3m}
       \sum_{\substack{X\Subset\Gamma\\X\cap B_{b,m}\ne\varnothing}}
       \E^{\Kcell|X|}
       \bigl((1+\diam X)^p+\ell^p m^p\bigr)\|\Psi(X)\|\\
 &\le C_p\sum_{m\ge1}\E^{-3m}
       \sum_{y\in B_{b,m}}\sum_{X\ni y}
       \E^{\kappa_0|X|}
       \bigl((1+\diam X)^p+\ell^p m^p\bigr)\|\Psi(X)\|\\
 &\le C_p\sum_{m\ge1}\E^{-3m}|B_{b,m}|
       \left(\|\Psi\|^{\#}_{p;\kappa_0}
       +\ell^p m^p\|\Psi\|^{\#}_{0;\kappa_0}\right).
\end{align*}
The first inequality uses \(\E^{-\mucell m}\E^{\rho m}A_D^m=\E^{-3m}\).
The second inequality uses \(\1_{\{X\cap B_{b,m}\ne\varnothing\}} \le\sum_{y\in B_{b,m}}\1_{\{y\in X\}}\) and \(\Kcell<\kappa_0\).
Volume growth gives
\[
 |B_{b,m}|\le C_\Gamma(1+(\Ccell+1)\ell m)^{\dvol}
 \le C_\Gamma(\Ccell+2)^{\dvol}
                   \ell^{\dvol}m^{\dvol}.
\]
After this volume bound is substituted, the remaining exponential \(\E^{-3m}\) makes the polynomial factors \(m^{\dvol}\) and \(m^{\dvol+p}\) summable.
Taking the supremum over \(b\) proves \eqref{eq:card-short}.

For a fixed \(X\), the same count and diameter estimate give
\begin{align*}
 \sum_{S\supseteq S_0}\E^{\rho|S|}(1+\diam Q(S))^p\|A_X(S;t)\|
 &\le C_p\E^{\Kcell|X|}\|\Psi(X)\|
     \sum_{m\ge1}\E^{-3m}
       \bigl((1+\diam X)^p+\ell^p m^p\bigr).
\end{align*}
Summing over \(m\) and using \(\Kcell<\kappa_0\) proves \eqref{eq:finite-range-total-bound}.
Choose finite cell sets \(U_N\nearrow\cB_\ell\) with \(S_0\subseteq U_1\).
Fix \(q>\dvol\).
Finite range gives
\[
 \sup_{|u|\le T}\|\Xi_u\|^{\pol}_{q;0}\le(1+R)^qJ_{\rm fr}.
\]
Proposition~\ref{prop:polynomial-dynamics} therefore applies to the regions \(Q(U_N)\).
Using the absolute convergence above in \eqref{eq:card-telescope}, we obtain
\[
 \sum_{S\supseteq S_0}A_X(S;t)
 =\lim_{N\to\infty}\tau_t^{H_{U_N}}(A)
 =\tau_t^\Xi(\Psi(X)).
\]
\end{proof}

Using the coefficients \(A_X(S;t)\) constructed in Lemma~\ref{lem:card-short}, we define the linear map \(\cR_t^{\Xi,\ell}\) from interactions on \(\Gamma\) to cell interactions on \(\cB_\ell\) by
\[
 (\cR_t^{\Xi,\ell}\Psi)(S)
 :=\sum_{\varnothing\ne X\subseteq Q(S)}A_X(S;t).
\]
At \(t=0\), this map is the grouping of site interactions into cells defined in \eqref{eq:site-to-cell-grouping}: \(\cR_0^{\Xi,\ell}\Psi=\Psi^{(\ell)}\).

\section{Long-range propagation}
\label{sec:propagation}

We now use the finite-range estimates of Section~\ref{sec:finite-range} to treat Hamiltonians with polynomial spatial decay.
For a prescribed time horizon \(T\ge1\), split the Hamiltonian interaction at a range \(R\ge1\) as
\[
 \Phi_u=\Phi_u^{\le R}+\Phi_u^{>R},
 \qquad
 \Phi_u^{\le R}(Z):=\1_{\{\diam Z\le R\}}\Phi_u(Z),
 \qquad
 \Phi_u^{>R}:=\Phi_u-\Phi_u^{\le R}.
\]
We regard \(\Phi^{\le R}\) as the free interaction and include \(\Phi^{>R}\) through the interaction picture, following the range-decomposition strategy of \cite{FossFeigEtAl15,MKN17,EMNY20}.

We first establish the interaction-picture factorization in a finite region \(\Omega\), where the Hamiltonians are bounded operators.
The estimates below are independent of \(\Omega\) and permit passage to infinite volume.
For this calculation, take initial time zero.
Write \(H_{>R}^{\Omega}(u):=\sum_{\varnothing\ne Z\subseteq\Omega}\Phi_u^{>R}(Z)\).
Let \(\tau_{0,t}^{\mathrm I,\Omega}\) denote the dynamics generated by the time-dependent Hamiltonian
\[
 u\mapsto
 \tau_{0,u}^{\Phi^{\le R},\Omega}\bigl(H_{>R}^{\Omega}(u)\bigr).
\]
For \(A\in\A_\Omega\), the full evolution factors as
\[
 \tau_{0,t}^{\Phi,\Omega}(A)
 =\tau_{0,t}^{\mathrm I,\Omega}
       \bigl(\tau_{0,t}^{\Phi^{\le R},\Omega}(A)\bigr).
\]

We apply Lemma~\ref{lem:card-short} with \(\rho=4\) to the dynamics generated by \(\Phi^{\le R}\), first with input interaction \(\Phi_u^{>R}\) and then with input interaction \(\Psi\), using the same cell partition.
The short-time estimate below decreases this rate to \(3\), and the parameter-derivative estimates in Theorem~\ref{thm:smooth-hamiltonian-dependence} decrease it further to \(2\).
Write \(\cR_u:=\cR_u^{\Phi^{\le R},\ell}\) for the map defined after Lemma~\ref{lem:card-short}.
For a finite cell set \(\cS\) and \(\Omega:=Q(\cS)\), the two decompositions satisfy
\[
 \begin{aligned}
 \sum_{\varnothing\ne S\subseteq\cS}\bigl(\cR_u(\Phi_u^{>R})\bigr)(S)
 &=\tau_{0,u}^{\Phi^{\le R},\Omega}
      \bigl(H_{>R}^{\Omega}(u)\bigr),\\
 \sum_{\varnothing\ne S\subseteq\cS}(\cR_t\Psi)(S)
 &=\tau_{0,t}^{\Phi^{\le R},\Omega}
      \left(\sum_{\varnothing\ne X\subseteq\Omega}\Psi(X)\right).
 \end{aligned}
\]
To complete the factorization at the level of interactions, we construct a map \(\cC_t\) representing the dynamics generated by \(u\mapsto\cR_u(\Phi_u^{>R})\).
The composition
\[
 \Psi\mapsto\cC_t(\cR_t\Psi)
\]
then represents the full evolution.

In Subsection~\ref{subsec:card-conjugation}, we construct \(\cC_t\) and estimate its action in the weighted cell norms.
In Subsection~\ref{subsec:fixed-scale}, we choose the range cutoff and cell size so that these estimates and the finite-range estimates apply on the same time interval.
We then combine the estimates across time scales to prove Theorem~\ref{thm:card-propagation} in Subsection~\ref{subsec:propagation-proof}.
In Subsection~\ref{subsec:observable-extensions}, we extend the representation and its total estimates to arbitrary local observables and prove covariance under onsite automorphisms.

\subsection{Weighted bounds for short-time evolution on cells}
\label{subsec:card-conjugation}

We formulate the construction of \(\cC_t\) for a general self-adjoint cell interaction path \(\Xi\) on a fixed partition \((Q_b)_{b\in\cB_\ell}\), in the sense of Definition~\ref{def:cell-interactions}.
In the application above, we take \(\Xi_u=\cR_u(\Phi_u^{>R})\).

The estimate of Abanin--De Roeck--Ho--Huveneers \cite[Lemma~4.1]{ADHH17} controls short-time evolution by decreasing the rate \(\rho\) in the exponential weight \(\E^{\rho|S|}\) on the support size.
As observed by Yin--Lucas \cite[Lemma~28]{YinLucas25}, the estimate applies to arbitrary interaction supports, which need not be connected.
We use this principle for cell interactions and include a polynomial diameter weight.

\begin{figure}[htbp]
\centering
\begin{tikzpicture}[cell diagram]
 \begin{scope}[x=.68cm,y=.68cm]
  \node[fig title,anchor=south] at (3.3,3.25) {(a) Norms of the long-range part};
  \fill[figblue!20] (0,0) -- plot[domain=0:2.2,samples=30] (\x,{2.6/(1+\x)^1.5})
    -- (2.2,0) -- cycle;
  \fill[figred!22] (2.2,0) -- plot[domain=2.2:6.5,samples=40] (\x,{2.6/(1+\x)^1.5})
    -- (6.5,0) -- cycle;
  \draw[line width=.7pt] plot[domain=0:6.5,samples=60] (\x,{2.6/(1+\x)^1.5});
  \draw[->] (0,0) -- (6.9,0);
  \draw[->] (0,0) -- (0,3.0);
  \node[font=\footnotesize,anchor=north east] at (6.9,-.05) {$\diam Z$};
  \node[font=\footnotesize,rotate=90,anchor=south] at (-.05,1.5) {term norms};
  \draw[black!60,densely dashed] (2.2,0) -- (2.2,2.6);
  \node[font=\footnotesize,anchor=north] at (2.2,-.05) {$R$};
  \node[font=\footnotesize] at (.95,.5) {$\Phi^{\le R}$};
  \node[font=\footnotesize,anchor=south west] at (4.1,1.0) {$\Phi^{>R}$};
  \draw[->,black!60] (4.3,1.05) -- (3.7,.32);
  \node[fig note,align=left,text width=5.7cm,anchor=north west] at (-.3,-.75)
   {$\Phi^{\le R}$: range $\le R$, decomposed on cells.\\[1pt]
    $\|\Phi^{>R}_u\|^{\#}_{0;\kappa_0}\le R^{-a}L\to0$ as $R\to\infty$;\\
    $\|\Phi^{>R}_u\|^{\#}_{a;\kappa_0}\le L$ need not tend to zero.};
 \end{scope}
 \begin{scope}[shift={(6.6,.0)},x=.52cm,y=.52cm]
  \node[fig title,anchor=south] at (7,4.2) {(b) Overlapping cell supports};
  \fill[figblue!40] (1,1) rectangle (4,3);
  \fill[figred,opacity=.3] (3,2) rectangle (9,3);
  \fill[figred,opacity=.3] (8,0) rectangle (13,1);
  \fill[figred,opacity=.3] (8,1) rectangle (9,3);
  \fill[figblue!50!figred] (3,2) rectangle (4,3);
  \fill[figblue!50!figred] (8,2) rectangle (9,3);
  \foreach \u in {0,...,14} \draw[cell grid] (\u,-.0) -- (\u,4);
  \foreach \v in {0,...,4} \draw[cell grid] (0,\v) -- (14,\v);
  \draw[figblue!70!black,line width=.6pt] (1,1) rectangle (4,3);
  \draw[figred!80!black,line width=.6pt] (3,2) rectangle (9,3);
  \draw[figred!80!black,line width=.6pt] (8,0) -- (13,0) -- (13,1) -- (9,1) -- (9,3) -- (8,3) -- cycle;
  \node[font=\footnotesize] at (2,1.55) {$V$};
  \node[font=\footnotesize] at (6,2.5) {$U_1$};
  \node[font=\footnotesize] at (11,.5) {$U_2$};
  \draw[ext line] (1,1) -- (1,-.95);
  \draw[ext line] (13,0) -- (13,-.95);
  \draw[dim line] (1,-.7) -- (13,-.7)
    node[midway,fill=white,inner sep=1pt,font=\footnotesize] {$\asymp\diam Q(V\cup U_1\cup U_2)$};
  \node[fig note,align=left,text width=7.0cm,anchor=north west] at (0,-1.45)
   {$[\Xi_{u_2}(U_2),[\Xi_{u_1}(U_1),\Upsilon(V)]]$ is assigned to $V\cup U_1\cup U_2$.\\[2pt]
    $|V\cup U_1\cup U_2|\le|V|+|U_1|+|U_2|$.};
 \end{scope}
\end{tikzpicture}
\caption{Support size and spatial decay in the short-time estimate.
(a)~The schematic range split illustrates the bounds on the long-range part with and without the diameter weight, assuming \(\sup_u\|\Phi_u\|^{\#}_{a;\kappa_0}\le L\); see \eqref{eq:card-range-tail}.
(b)~Here \(\Xi\) is the generating cell interaction path and \(\Upsilon\) is the cell interaction being evolved.
The highlighted cells show \(V\cap U_1\) and \(U_1\cap U_2\).}
\label{fig:range-splitting}
\end{figure}
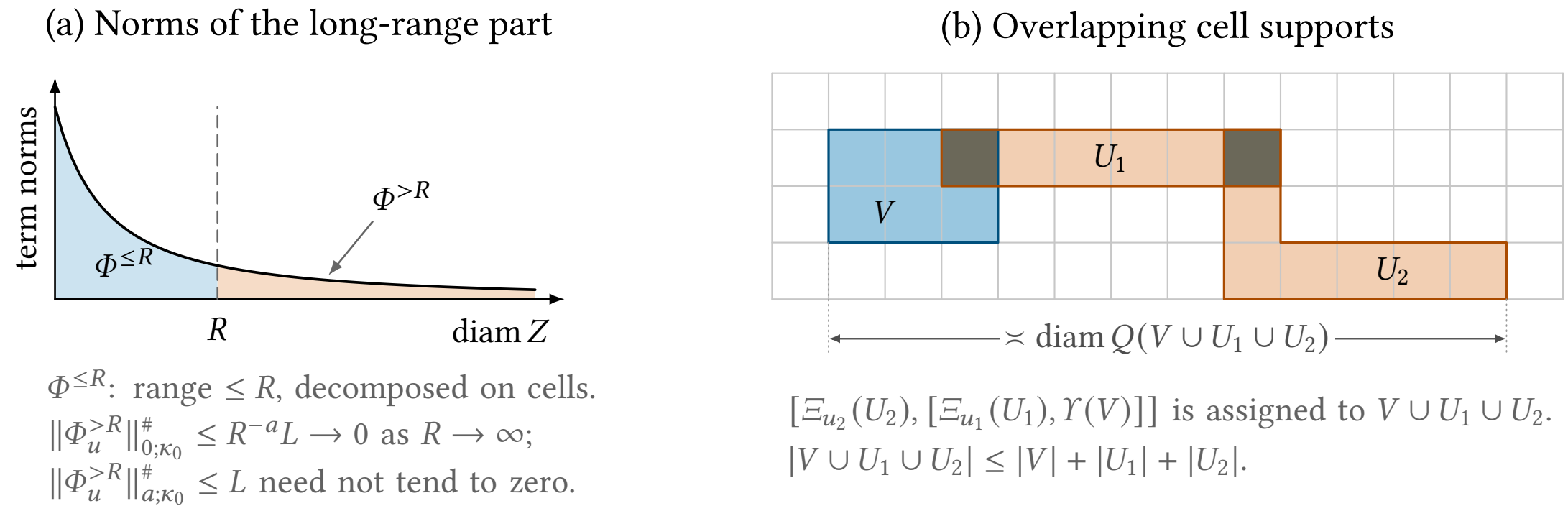

To construct \(\cC_t\), we use the ordinary Dyson expansion of the finite-volume dynamics generated by \(\Xi\) and assign each nested commutator to the union of its supports.
For an input cell interaction \(\Upsilon\), the integrands in this expansion are sums of terms of the form
\[
 \I^n\bigl[\Xi_{u_n}(U_n),[\cdots,[\Xi_{u_1}(U_1),\Upsilon(V)]\cdots]\bigr].
\]
For \(t\ge0\), the integration domain is \(0\le u_n\le\cdots\le u_1\le t\).
We number the time variables from the innermost commutator outwards, in reverse order to the labels in \(\Delta_n(T)\) of Lemma~\ref{lem:finite-volume-matrix-commutator}.
We assign each term to \(V\cup U_1\cup\cdots\cup U_n\).
Starting from \(\Upsilon(V)\), we add the supports \(U_1,\ldots,U_n\) in that order; each new support must meet the union already present for the nested commutator to be nonzero.
Figure~\ref{fig:range-splitting}(b) illustrates two such steps.
For fixed preceding supports, put \(S:=V\cup U_1\cup\cdots\cup U_{i-1}\) and sum over the supports \(U_i\) meeting \(S\).
We bound the sum of their weighted term norms by summing first over cells \(b\in S\) and then over supports \(U_i\ni b\).
The inner sum is at most \(\|\Xi_{u_i}\|^{\#,\ell}_{0;\rho}\) (see \eqref{eq:card-cell-norm}), uniformly in \(b\), which gives the factor \(|S|\).
Decreasing the exponential rate from \(\rho\) to \(\rho-\delta\) absorbs this factor through
\[
 |S|\E^{-\delta|S|}\le\frac1{\E\delta},\qquad \delta>0.
\]
The finite-range decomposition of Lemma~\ref{lem:card-short}, applied with \(\rho=4\), provides the exponential weight \(\E^{4|S|}\).
We estimate the evolved interaction with weight \(\E^{3|S|}\).
For a term with \(n\ge1\) commutators, we divide this decrease into \(n\) equal increments, taking \(\delta=1/n\) at each step.
Each commutator then contributes a factor proportional to \(n\).
The ordered time integral contributes \(|t|^n/n!\), and
\[
 \frac{n^n}{n!}\le\E^n.
\]
The resulting series converges when the time horizon multiplied by the generating interaction norm without a diameter weight is sufficiently small, with a condition independent of the volume.

It remains to estimate these nested commutators with the diameter weight of their output support \(V\cup U_1\cup\cdots\cup U_n\).
Since each \(U_i\) meets the union of the preceding supports, the diameter of \(Q(V\cup U_1\cup\cdots\cup U_n)\) is at most the sum of the diameters of \(Q(V),Q(U_1),\ldots,Q(U_n)\).
Repeated use of \((x+y)^p\le2^{\max\{p-1,0\}}(x^p+y^p)\), for \(x,y>0\), gives
\[
 (1+\diam Q(V\cup U_1\cup\cdots\cup U_n))^p\le 2^{n\max\{p-1,0\}}
 \left((1+\diam Q(V))^p
       +\sum_{i=1}^n(1+\diam Q(U_i))^p\right).
\]
We multiply this bound by \(\|\Upsilon(V)\|\prod_{i=1}^n\|\Xi_{u_i}(U_i)\|\) and sum over the supports as above.
The first term places the diameter weight on \(\Upsilon(V)\), so it is estimated using \(\|\Upsilon\|^{\#,\ell}_{p;4}\) and the generator norms \(\|\Xi_{u_i}\|^{\#,\ell}_{0;4}\).
In each of the other \(n\) terms, the weight falls on one \(\Xi_{u_i}(U_i)\), giving one factor \(\|\Xi_{u_i}\|^{\#,\ell}_{p;4}\); the input interaction and all other generating terms are measured without a diameter weight.
Thus the weighted generator norm occurs only once in each such term, regardless of the number of commutators.
The sum over the commutator orders converges under a smallness condition involving only the generator norm without a diameter weight.

For \(\Xi_u=\cR_u(\Phi_u^{>R})\), the resulting condition is
\[
 T\sup_{|u|\le T}\bigl\|\cR_u(\Phi_u^{>R})\bigr\|^{\#,\ell}_{0;4}
 \le\varepsilon_p,
\]
with \(\varepsilon_p\) as in Lemma~\ref{lem:card-conjugation}.
The norm \(\sup_{|u|\le T}\|\Xi_u\|^{\#,\ell}_{p;4}\) must be finite but need not be small.
This distinction permits \(p=a\): increasing the range cutoff makes \(\|\Phi_u^{>R}\|^{\#}_{0;\kappa_0}\) small, whereas \(\|\Phi_u^{>R}\|^{\#}_{a;\kappa_0}\) need not tend to zero, as illustrated in Figure~\ref{fig:range-splitting}(a).

To control the sum of local terms representing a single evolved observable, we use, in addition to the cell interaction norm \(\|\cdot\|^{\#,\ell}_{q;\rho}\), the weighted total sum
\[
 \Sigma_{q;\rho}^{(\ell)}(\Psi)
 :=\sum_{\varnothing\ne S\Subset\cB_\ell}
       \E^{\rho|S|}(1+\diam Q(S))^q\|\Psi(S)\|,
 \qquad \rho,q\ge0.
\]
For a nonnegative function \(f\) on the nonempty finite subsets of \(\cB_\ell\), \(\|f\|^{\#,\ell}_{q;\rho}\) and \(\Sigma_{q;\rho}^{(\ell)}(f)\) denote the same weighted sums with \(\|\Psi(S)\|\) replaced by \(f(S)\).

\begin{lemma}
\label{lem:card-conjugation}
There are constants \(\varepsilon_p,C_p>0\), depending only on \(p\ge0\), with \(\varepsilon_p\) nonincreasing in \(p\), such that the following holds.
Fix a cell partition, \(T>0\), and a componentwise norm-continuous path \((\Xi_u)_{|u|\le T}\) of self-adjoint cell interactions.
Assume that
\[
 L_j:=\sup_{|u|\le T}\|\Xi_u\|^{\#,\ell}_{j;4},
 \qquad j=0,p,
\]
are finite and \(TL_0\le\varepsilon_p\).
For each finite cell set \(\cS\Subset\cB_\ell\), define the restricted Hamiltonian by
\[
 H_\Xi^{\cS}(u):=\sum_{\varnothing\ne S\subseteq\cS}\Xi_u(S).
\]
Denote by \(\tau_{0,t}^{\Xi,\cS}\), \(|t|\le T\), the dynamics on \(\A_{Q(\cS)}\) generated by \(H_\Xi^{\cS}\).
There is a linear, adjoint-preserving interaction map \(\cC_t\), \(|t|\le T\), on the cell interactions of finite \(\|\cdot\|^{\#,\ell}_{p;4}\)-norm, with the following properties.
\begin{enumerate}[label=(\roman*),ref=(\roman*),leftmargin=*]
\item \emph{Representation and interaction estimates.}
\label{item:conjugation-ordinary}
For every interaction \(\Psi\) on the cells with \(\|\Psi\|^{\#,\ell}_{p;4}<\infty\), every finite cell set \(\cS\), and \(|t|\le T\),
\begin{equation}
 \sum_{\varnothing\ne S\subseteq\cS}(\cC_t\Psi)(S)
 =\tau_{0,t}^{\Xi,\cS}
       \left(\sum_{\varnothing\ne V\subseteq\cS}\Psi(V)\right).
 \label{eq:card-conjugation-finite}
\end{equation}
Moreover,
\begin{equation}
 \|\cC_t\Psi\|^{\#,\ell}_{p;3}
 \le C_p\bigl(\|\Psi\|^{\#,\ell}_{p;4}+TL_p\|\Psi\|^{\#,\ell}_{0;4}\bigr),
 \qquad |t|\le T.
 \label{eq:card-conjugation}
\end{equation}
If, in addition, \(\Sigma_{p;4}^{(\ell)}(\Psi)<\infty\), then the same map and constant satisfy
\begin{equation}
 \Sigma_{p;3}^{(\ell)}(\cC_t\Psi)
 \le C_p\left(
       \Sigma_{p;4}^{(\ell)}(\Psi)
       +TL_p\,\Sigma_{0;4}^{(\ell)}(\Psi)
       \right),
 \qquad |t|\le T.
 \label{eq:card-conjugation-total}
\end{equation}

\item \emph{Countable decompositions.}
\label{item:conjugation-decomposed}
Let \((\Psi_\alpha)_{\alpha\in\mathcal J}\) be a finite or countable family of interactions on the cells such that
\[
 g(S):=\sum_{\alpha\in\mathcal J}\|\Psi_\alpha(S)\|,
 \qquad \|g\|^{\#,\ell}_{p;4}<\infty,
\]
and define \(\Psi(S):=\sum_{\alpha\in\mathcal J}\Psi_\alpha(S)\).
For every nonempty finite cell set \(S\) and \(|t|\le T\),
\[
 (\cC_t\Psi)(S)
 =\sum_{\alpha\in\mathcal J}(\cC_t\Psi_\alpha)(S)
\]
with absolute convergence in norm, and
\begin{align}
 \sup_{b\in\cB_\ell}\sum_{S\ni b}
 \E^{3|S|}(1+\diam Q(S))^p
       \sum_{\alpha\in\mathcal J}\|(\cC_t\Psi_\alpha)(S)\|
 \le C_p\bigl(\|g\|^{\#,\ell}_{p;4}+TL_p\|g\|^{\#,\ell}_{0;4}\bigr).
 \label{eq:card-conjugation-decomposed}
\end{align}
If \(\Sigma_{p;4}^{(\ell)}(g)<\infty\), then also
\begin{align}
 \sum_S\E^{3|S|}(1+\diam Q(S))^p
       \sum_{\alpha\in\mathcal J}\|(\cC_t\Psi_\alpha)(S)\|
 \le C_p\bigl(\Sigma_{p;4}^{(\ell)}(g)
                   +TL_p\Sigma_{0;4}^{(\ell)}(g)\bigr).
 \label{eq:card-conjugation-decomposed-total}
\end{align}
\end{enumerate}
\end{lemma}

\begin{proof}
We expand the dynamics into a series of time integrals of nested commutators, assign each commutator to the union of its input supports, and control the series order by order through a recursion for the weighted cell norms.

\smallskip
\noindent \emph{Bounds for one commutator.}
For interactions \(\Xi,\Psi\), define
\begin{equation}
 [\Xi,\Psi](S):=
 \sum_{\substack{U\cup V=S\\U\cap V\ne\varnothing}}
 [\Xi(U),\Psi(V)].
 \label{eq:cell-interaction-bracket}
\end{equation}

The defining sum omits disjoint \(U,V\), because their interaction terms commute.
For every finite cell set \(\cS\), it follows that
\[
 \sum_{\varnothing\ne S\subseteq\cS}[\Xi,\Psi](S)
 =\left[
       \sum_{\varnothing\ne U\subseteq\cS}\Xi(U),
       \sum_{\varnothing\ne V\subseteq\cS}\Psi(V)
   \right].
\]
Put \(w_q(S):=(1+\diam Q(S))^q\).
The spatial weight of a union of overlapping supports is controlled by the weights of its parts: if \(U\cap V\ne\varnothing\), then
\[
 \diam Q(U\cup V)\le\diam Q(U)+\diam Q(V),
 \qquad
 w_p(U\cup V)\le 2^{\max\{p-1,0\}}\bigl(w_p(U)+w_p(V)\bigr).
\]
For nonnegative scalar families \(f,g\) on nonempty finite cell sets, every cell \(b\), and \(\delta>0\),
\begin{align*}
 \sum_{\substack{U\cap V\ne\varnothing\\b\in U\cup V}}
       \E^{-\delta(|U|+|V|)}f(U)g(V) &\le
 \sum_{U\ni b}\E^{-\delta|U|}f(U)
       \sum_{z\in U}\sum_{V\ni z}g(V)
 +\sum_{V\ni b}\E^{-\delta|V|}g(V)
       \sum_{z\in V}\sum_{U\ni z}f(U)\\
 & \le\frac{2}{\E\delta}
       \left(\sup_c\sum_{U\ni c}f(U)\right)
       \left(\sup_c\sum_{V\ni c}g(V)\right),
\end{align*}
using \(|U|\E^{-\delta|U|}\le1/(\E\delta)\).
Without the condition \(b\in U\cup V\), the same overlap sum satisfies
\[
 \sum_{U\cap V\ne\varnothing}
       \E^{-\delta(|U|+|V|)}f(U)g(V)
 \le\sum_V\E^{-\delta|V|}g(V)
       \sum_{z\in V}\sum_{U\ni z}f(U)
 \le\frac1{\E\delta}
       \left(\sup_c\sum_{U\ni c}f(U)\right)\sum_Vg(V).
\]
Apply these inequalities to each of the two terms in the spatial-weight bound, using
\[
 f(U)=\E^{\rho|U|}w_{q_1}(U)\|\Xi(U)\|,
 \qquad
 g(V)=\E^{\rho|V|}w_{q_2}(V)\|\Psi(V)\|,
 \qquad (q_1,q_2)=(p,0),(0,p).
\]
Using \(\|[\Xi(U),\Psi(V)]\|\le2\|\Xi(U)\|\|\Psi(V)\|\) and \(|U\cup V|\le|U|+|V|\), we obtain, with \(K_p:=8\cdot2^{\max\{p-1,0\}}\), for \(\delta>0\) and \(\rho\ge\delta\),
\begin{align}
 \|[\Xi,\Psi]\|^{\#,\ell}_{p;\rho-\delta}
 \le\frac{K_p}{\delta}\bigl(
 \|\Xi\|^{\#,\ell}_{p;\rho}\|\Psi\|^{\#,\ell}_{0;\rho}
 +\|\Xi\|^{\#,\ell}_{0;\rho}\|\Psi\|^{\#,\ell}_{p;\rho}\bigr),
 \label{eq:card-bracket}
\end{align}
and
\begin{equation}
\begin{split}
 \Sigma_{p;\rho-\delta}^{(\ell)}([\Xi,\Psi])
 \le\frac{K_p}{\delta}\bigl(
 \|\Xi\|^{\#,\ell}_{p;\rho}\Sigma_{0;\rho}^{(\ell)}(\Psi) +\|\Xi\|^{\#,\ell}_{0;\rho}\Sigma_{p;\rho}^{(\ell)}(\Psi)\bigr).
\end{split}
\label{eq:card-bracket-total}
\end{equation}
With \(w_0=1\), the overlap sums also give
\[
 \begin{aligned}
 \|[\Xi,\Psi]\|^{\#,\ell}_{0;\rho-\delta}
 &\le\frac{K_p}{\delta}
       \|\Xi\|^{\#,\ell}_{0;\rho}\|\Psi\|^{\#,\ell}_{0;\rho},\\
 \Sigma_{0;\rho-\delta}^{(\ell)}([\Xi,\Psi])
 &\le\frac{K_p}{\delta}
       \|\Xi\|^{\#,\ell}_{0;\rho}\Sigma_{0;\rho}^{(\ell)}(\Psi).
 \end{aligned}
\]
\smallskip
\noindent \emph{Bounds for the time-ordered commutator series.}
Define
\[
 (\scrL_u\Psi)(S):=\I[\Xi_u,\Psi](S),
 \qquad
 (\scrB_u f)(S):=
 2\sum_{\substack{U\cup V=S\\U\cap V\ne\varnothing}}
       \|\Xi_u(U)\|f(V).
\]
The operator \(\scrB_u\) acts on nonnegative families indexed by cell sets and assigns each pair \((U,V)\) to its union \(S\).
The one-commutator estimates apply to \(\scrB_u f\), since their proofs use the scalar bound \(2\|\Xi_u(U)\|f(V)\) for each pair.

Fix \(n\ge1\) and times \(v_1,\ldots,v_n\in[-T,T]\).
Set
\[
 \begin{aligned}
 \rho_j&:=4-\frac jn,\qquad 0\le j\le n,\\
 f_0(S)&:=\|\Psi(S)\|,\qquad
 f_j:=\scrB_{v_j}f_{j-1},\qquad 1\le j\le n.
 \end{aligned}
\]
For \(a_j:=\|f_j\|^{\#,\ell}_{0;\rho_j}\) and \(b_j:=\|f_j\|^{\#,\ell}_{p;\rho_j}\), the one-commutator bounds give
\[
 a_j\le K_p n L_0 a_{j-1},\qquad
 b_j\le K_p n\bigl(L_0b_{j-1}+L_pa_{j-1}\bigr).
\]
Starting from \(a_0=\|\Psi\|^{\#,\ell}_{0;4}\) and \(b_0=\|\Psi\|^{\#,\ell}_{p;4}\), induction yields, for \(1\le j\le n\),
\[
 \begin{aligned}
 a_j&\le(K_p n L_0)^j\|\Psi\|^{\#,\ell}_{0;4},\\
 b_j&\le(K_p n)^j
       \bigl(L_0^j\|\Psi\|^{\#,\ell}_{p;4}
             +jL_0^{j-1}L_p\|\Psi\|^{\#,\ell}_{0;4}\bigr).
 \end{aligned}
\]
By \eqref{eq:card-bracket-total}, the same recursion holds with \(a_j=\Sigma_{0;\rho_j}^{(\ell)}(f_j)\) and \(b_j=\Sigma_{p;\rho_j}^{(\ell)}(f_j)\).

For \(t=\sigma h\), \(0\le h\le T\), \(\sigma\in\{-1,1\}\), define the interaction terms and their scalar bounds by
\begin{equation}
 \begin{aligned}
 \cC_{n,t}\Psi
 &:=\sigma^n\int_{0\le u_n\le\cdots\le u_1\le h}
       \scrL_{\sigma u_n}\cdots\scrL_{\sigma u_1}\Psi\,
       \dd u_1\cdots\dd u_n,\\
 d_{n,t}
 &:=\int_{0\le u_n\le\cdots\le u_1\le h}
       \scrB_{\sigma u_n}\cdots\scrB_{\sigma u_1}f_0\,
       \dd u_1\cdots\dd u_n,
 \end{aligned}
 \label{eq:cell-conjugation-series}
\end{equation}
with \(\cC_{0,t}\Psi:=\Psi\) and \(d_{0,t}:=f_0\).
The integrals defining \(\cC_{n,t}\Psi\) are taken componentwise in the finite-dimensional algebras \(\A_{Q(S)}\); those defining \(d_{n,t}\) are scalar integrals for each \(S\).
Each component of these integrands is a finite sum of continuous functions, since every input support is contained in the output support.
The commutator norm bound gives
\[
 \|(\cC_{n,t}\Psi)(S)\|\le d_{n,t}(S).
\]
The integration domain in \eqref{eq:cell-conjugation-series} has volume \(h^n/n!\).
Using the recursion with \(v_j:=\sigma u_j\) and \(n^n/n!\le\E^n\), we obtain
\[
 \begin{aligned}
 \|d_{n,t}\|^{\#,\ell}_{p;3}
 &\le\frac{h^n(K_p n)^n}{n!}
       \bigl(L_0^n\|\Psi\|^{\#,\ell}_{p;4}
             +nL_0^{n-1}L_p\|\Psi\|^{\#,\ell}_{0;4}\bigr)\\
 &\le z^n\|\Psi\|^{\#,\ell}_{p;4}
       +\E K_pTL_p\|\Psi\|^{\#,\ell}_{0;4}\,n z^{n-1},
 \end{aligned}
\]
for \(z:=\E K_pTL_0\) and \(n\ge1\).
Choose
\[
 \varepsilon_p:=\frac1{2\E K_p},\qquad C_p:=2+4\E K_p.
\]
Then \(z\le1/2\), and
\[
 \sum_{n\ge0}\|d_{n,t}\|^{\#,\ell}_{p;3}
  \le\frac{\|\Psi\|^{\#,\ell}_{p;4}}{1-z}
       +\frac{\E K_pTL_p\|\Psi\|^{\#,\ell}_{0;4}}{(1-z)^2}\le C_p\bigl(\|\Psi\|^{\#,\ell}_{p;4}
                  +TL_p\|\Psi\|^{\#,\ell}_{0;4}\bigr).
\]
Thus the componentwise definition
\[
 (\cC_t\Psi)(S)
 :=\sum_{n\ge0}(\cC_{n,t}\Psi)(S)
\]
gives an absolutely convergent series in the \(\|\cdot\|^{\#,\ell}_{p;3}\)-norm and proves \eqref{eq:card-conjugation}.
Using the recursion for the weighted totals \(\Sigma_{q;\rho_j}^{(\ell)}(f_j)\), \(q=0,p\), and summing the resulting bounds proves \eqref{eq:card-conjugation-total}.
Each \(\scrL_u\) preserves adjoints, so \(\cC_t\) does as well.

\smallskip
\noindent \emph{Sums over a countable family of inputs.}
For the family in part~\ref{item:conjugation-decomposed}, each series \(\Psi(V)=\sum_\alpha\Psi_\alpha(V)\) converges absolutely, since its sum of norms is \(g(V)<\infty\).
Write \(d_{n,t}[g]\) for the scalar integral in \eqref{eq:cell-conjugation-series} with \(f_0\) replaced by \(g\), and set \(d_{0,t}[g]:=g\).
For \(n\ge1\) and a fixed output support \(S\), the integrand defining \((\cC_{n,t}\Psi)(S)\) contains only finitely many support sequences, all contained in \(S\).
For each sequence, the sum over \(\alpha\) of the norms of its nested commutators is bounded by the product of the Hamiltonian term norms, multiplied by \(2^n g(V)\) for its initial support \(V\).
Integrating these nonnegative bounds and using Tonelli's theorem gives
\[
 \sum_\alpha\int_{0\le u_n\le\cdots\le u_1\le |t|}
 \bigl\|(\scrL_{\sigma u_n}\cdots\scrL_{\sigma u_1}
                      \Psi_\alpha)(S)\bigr\|\,\dd u_1\cdots\dd u_n
 \le d_{n,t}[g](S)<\infty.
\]
Thus the sum over \(\alpha\) can be exchanged with the time integral, and
\[
 (\cC_{n,t}\Psi)(S)=\sum_\alpha(\cC_{n,t}\Psi_\alpha)(S),
 \qquad
 \sum_\alpha\|(\cC_{n,t}\Psi_\alpha)(S)\|\le d_{n,t}[g](S).
\]
The coefficient identity and bound also hold for \(n=0\) by definition.
The scalar recursion above, started at \(f_0=g\), gives
\[
 \sum_{n\ge0}\|d_{n,t}[g]\|^{\#,\ell}_{p;3}
 \le C_p\bigl(\|g\|^{\#,\ell}_{p;4}+TL_p\|g\|^{\#,\ell}_{0;4}\bigr).
\]
In particular, \(\sum_{n,\alpha}\|(\cC_{n,t}\Psi_\alpha)(S)\|<\infty\) for every nonempty \(S\), which permits exchanging the sums over \(n\) and \(\alpha\).
Their sum equals \((\cC_t\Psi)(S)\), and the last bound proves \eqref{eq:card-conjugation-decomposed}.
If \(\Sigma_{p;4}^{(\ell)}(g)<\infty\), the recursion for the weighted totals gives
\[
 \sum_{n\ge0}\Sigma_{p;3}^{(\ell)}(d_{n,t}[g])
 \le C_p\bigl(\Sigma_{p;4}^{(\ell)}(g)+TL_p\Sigma_{0;4}^{(\ell)}(g)\bigr).
\]
Summing the nonnegative bounds over \(S\) proves \eqref{eq:card-conjugation-decomposed-total}.

\smallskip
\noindent \emph{Finite-volume representation identity.}
Fix a finite cell set \(\cS\), and define \(\Xi_u^{\cS}(S):=\1_{\{S\subseteq\cS\}}\Xi_u(S)\) and \(\Psi^{\cS}(S):=\1_{\{S\subseteq\cS\}}\Psi(S)\).
Let \(\cC_t^{\cS}\) be the interaction series with generator \(\Xi^{\cS}\).
Every input support contributing to the coefficient at \(S\) is contained in \(S\), so
\[
 (\cC_t^{\cS}\Psi^{\cS})(S)
 =\1_{\{S\subseteq\cS\}}(\cC_t\Psi)(S).
\]
For \(A\in\A_{Q(\cS)}\), the dynamics satisfy
\[
 \partial_t\tau_{0,t}^{\Xi,\cS}(A)
 =\I\tau_{0,t}^{\Xi,\cS}\bigl([H_\Xi^{\cS}(t),A]\bigr),
 \qquad \tau_{0,0}^{\Xi,\cS}=\id.
\]
The Dyson series of \(\tau_{0,t}^{\Xi,\cS}\) has the ordering used in \(\cC_{n,t}\).
Summing the absolutely convergent interaction series over \(S\subseteq\cS\) therefore proves \eqref{eq:card-conjugation-finite}.
\end{proof}

\subsection{Propagation on a fixed cell partition}
\label{subsec:fixed-scale}

For a prescribed time horizon \(T\ge1\), we now estimate \(\cC_t(\cR_t\Psi)\) by applying Lemma~\ref{lem:card-conjugation} to the generator \(u\mapsto\cR_u(\Phi_u^{>R})\) and the input interaction \(\cR_t\Psi\).
The range cutoff \(R\) and cell size \(\ell\) must satisfy two requirements.
The cells must be large enough for the finite-range estimates of Lemma~\ref{lem:card-short} to hold throughout \(|t|\le T\).
At the same time, the norm of the interaction-picture generator without a diameter weight must satisfy the smallness condition of Lemma~\ref{lem:card-conjugation}.
The bound on this norm contains the cell-volume factor \(\ell^{\dvol}\), which competes with the decay factor \(R^{-a}\) of the long-range part.
We therefore choose \(R\) and \(\ell\) together.
The assumption \(a>\dvol\) permits choices that satisfy both requirements and grow at most polynomially in \(T\).

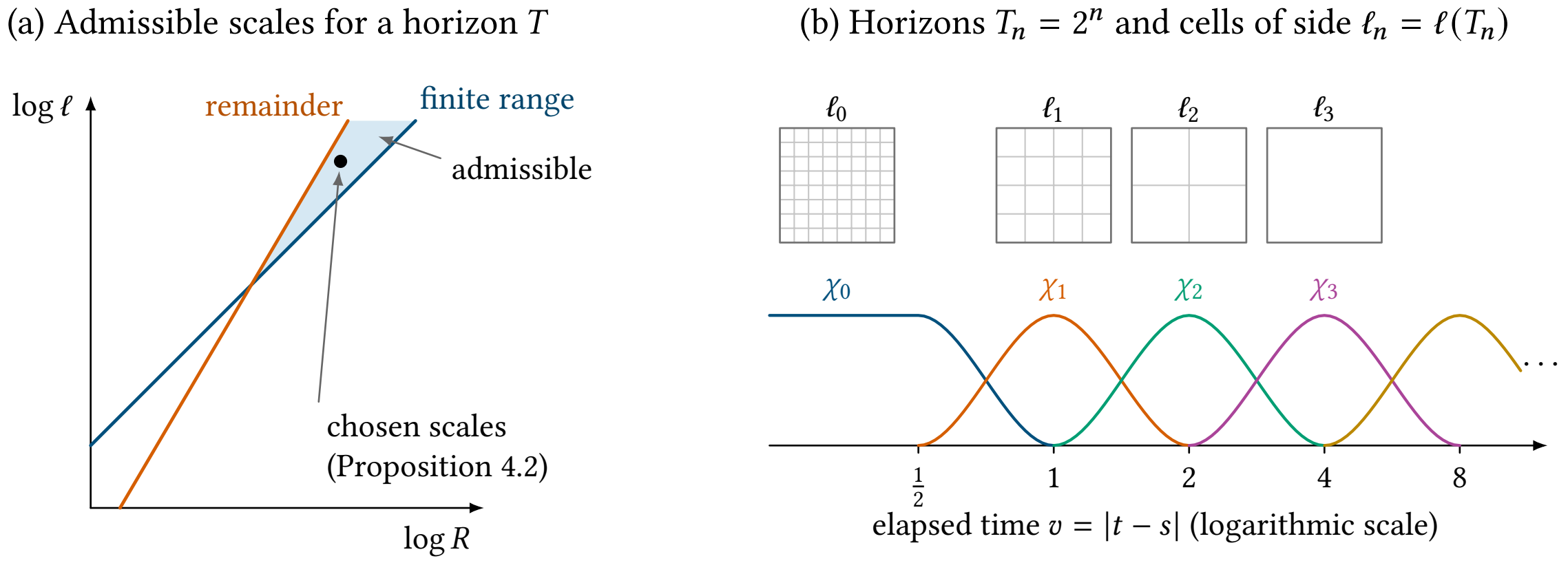
\begin{figure}[htbp]
\centering
\begin{tikzpicture}[cell diagram]
 \begin{scope}[x=.6cm,y=.6cm]
  \node[fig title,anchor=south] at (3,7.25) {(a) Admissible scales for a horizon $T$};
  \fill[figblue!16] (2.571,3.571) -- (4.118,6.2) -- (5.2,6.2) -- cycle;
  \draw[->] (0,0) -- (6.3,0);
  \draw[->] (0,0) -- (0,6.6);
  \node[font=\footnotesize,anchor=north east] at (6.3,-.05) {$\log R$};
  \node[font=\footnotesize,anchor=east] at (-.05,6.4) {$\log\ell$};
  \draw[figblue!70!black,line width=.9pt] (0,1) -- (5.2,6.2);
  \draw[figred,line width=.9pt] (.47,0) -- (4.118,6.2);
  \node[font=\footnotesize,text=figblue!60!black,anchor=south west,inner sep=1pt] at (5.2,6.2) {finite range};
  \node[font=\footnotesize,text=figred!85!black,anchor=south east,inner sep=1pt] at (4.118,6.2) {remainder};
  \fill (4.0,5.55) circle[radius=1.8pt];
  \node[font=\footnotesize,anchor=north west,align=left] at (3.55,1.75)
    {chosen scales\\ (Proposition~\ref{prop:fixed-scale})};
  \draw[->,black!60] (3.65,1.7) -- (3.98,5.4);
  \node[font=\footnotesize,anchor=west] at (5.55,5.45) {admissible};
  \draw[->,black!60] (5.6,5.6) -- (4.6,5.95);
 \end{scope}
 \begin{scope}[shift={(9.25,.6)},x=1.3cm,y=1cm]
  \definecolor{chigreen}{HTML}{009E73}
  \definecolor{chipurple}{HTML}{AA4499}
  \definecolor{chigold}{HTML}{BB8800}
  \node[fig title,anchor=south] at (.75,3.75)
    {(b) Horizons $T_n=2^n$ and cells of side $\ell_n=\ell(T_n)$};
  \def\h{1.25}
  \draw[->] (-2.1,0) -- (3.65,0);
  \foreach \u/\lab in {-1/{\tfrac12},0/1,1/2,2/4,3/8}{
   \draw (\u,0) -- (\u,-.07);
   \node[font=\footnotesize,anchor=north] at (\u,-.07) {$\lab$};}
  \node[font=\footnotesize,anchor=north] at (.75,-.5)
    {elapsed time $v=|t-s|$ (logarithmic scale)};
  \draw[figblue!70!black,line width=.8pt] (-2.1,\h) -- (-1,\h)
    -- plot[domain=-1:0,samples=25]
       (\x,{\h*(1-(3*(\x+1)*(\x+1)-2*(\x+1)*(\x+1)*(\x+1)))});
  \node[font=\footnotesize,text=figblue!70!black,anchor=south] at (-1.6,\h) {$\chi_0$};
  \foreach \n/\col in {1/figred,2/chigreen,3/chipurple}{
   \draw[\col,line width=.8pt]
     plot[domain=\n-2:\n-1,samples=25]
       (\x,{\h*(3*(\x-\n+2)*(\x-\n+2)-2*(\x-\n+2)*(\x-\n+2)*(\x-\n+2))})
     -- plot[domain=\n-1:\n,samples=25]
       (\x,{\h*(1-(3*(\x-\n+1)*(\x-\n+1)-2*(\x-\n+1)*(\x-\n+1)*(\x-\n+1)))});
   \node[font=\footnotesize,text=\col,anchor=south] at (\n-1,\h) {$\chi_{\n}$};}
  \draw[chigold,line width=.8pt]
     plot[domain=2:3,samples=25]
       (\x,{\h*(3*(\x-2)*(\x-2)-2*(\x-2)*(\x-2)*(\x-2))})
     -- plot[domain=3:3.45,samples=12]
       (\x,{\h*(1-(3*(\x-3)*(\x-3)-2*(\x-3)*(\x-3)*(\x-3)))});
  \node[font=\footnotesize] at (3.62,.75) {$\cdots$};
  \foreach \c/\N/\n in {-1.6/8/0,0/4/1,1/2/2,2/1/3}{
   \begin{scope}[shift={(\c-.4231,1.95)},x=1cm,y=1cm]
    \pgfmathsetmacro\st{1.1/\N}
    \draw[cell grid] (0,0) grid[step=\st] (1.1,1.1);
    \draw[black!55] (0,0) rectangle (1.1,1.1);
    \node[font=\footnotesize,anchor=south,inner sep=1.5pt] at (.55,1.1) {$\ell_{\n}$};
   \end{scope}}
 \end{scope}
\end{tikzpicture}
\caption{Choice of scales.
(a)~Schematic log--log plot for fixed \(T\) and \(L>0\).
The finite-range estimate of Lemma~\ref{lem:card-short} requires \(\ell\gtrsim R(T+\log(2+R))\); the blue line suppresses the logarithmic correction.
The remainder must satisfy \(CLT\ell^{\dvol}R^{-a}\le\varepsilon\), giving the red upper boundary.
For \(a>\dvol\), the choices \eqref{eq:fixed-scales} satisfy both constraints after the constants are fixed; the marked point illustrates these sufficient choices, not a sharp threshold.
For \(a\le\dvol\), these constraints cannot be satisfied uniformly over arbitrarily large horizons at fixed positive \(L\); the independent obstruction to extending the propagation theorem is proved in Proposition~\ref{prop:decay-optimality}.
(b)~The cutoffs \(\chi_n\) in \eqref{eq:time-cutoffs} combine coefficients for horizons \(T_n=2^n\).
At each elapsed time, at most two consecutive cutoffs are nonzero, and \(1+T_n\le4(1+|t-s|)\) on their supports.
The grids illustrate increasing cell scales only; the partitions need not be nested.
The bound \eqref{eq:scale-growth} converts powers of \(\ell_n\) into polynomial time growth.}
\label{fig:scales}
\end{figure}

In Proposition~\ref{prop:fixed-scale}, we bound the fixed-scale coefficients of Definition~\ref{def:interaction-representatives}.
For each even local observable \(A\), we also prove that these coefficients sum absolutely to \(\tau^\Phi_{s,t}(A)\).
The bounds on a fixed cell partition will be used to estimate parameter derivatives in Theorem~\ref{thm:smooth-hamiltonian-dependence}.
Figure~\ref{fig:scales}(a) illustrates the two scale requirements.

\FloatBarrier
\begin{proposition}
\label{prop:fixed-scale}
Assume \eqref{eq:volume-growth} and Hypothesis~\ref{hyp:cell-partitions}, and choose \(\kappa_0\) as in Lemma~\ref{lem:card-short} with \(\rho=4\).
For every \(a>\dvol\) and \(L\ge0\), there are constants \(c>4\), \(R_*\ge2\), \(C_{R_*}>0\), and \(C^{\rm fs}_p>0\), \(0\le p\le a\), depending only on \(a,L,\dvol,C_\Gamma,\Ccell,\Dcell\) and their indicated indices, with the following property.
With \(\gamma\) as in \eqref{eq:time-exponent}, set, for \(T\ge1\),
\begin{equation}
 R(T):=\left\lceil R_*(1+T)^\gamma\right\rceil,
 \qquad
 \ell(T):=\left\lceil cR(T)\bigl(T+\log(2+R(T))\bigr)\right\rceil.
 \label{eq:fixed-scales}
\end{equation}
These scales satisfy
\begin{equation}
 \ell(T)\le C_{R_*}(1+T)^{1+\gamma}.
 \label{eq:scale-growth}
\end{equation}
Choose arbitrary local multiplicities, an interval \(I\subseteq\R\), and a componentwise norm-continuous path \((\Phi_u)_{u\in I}\) of self-adjoint interactions such that
\[
 \sup_{u\in I}\|\Phi_u\|^{\#}_{a;\kappa_0}\le L.
\]
For every \(T\ge1\), choose a partition \((Q_b)_{b\in\cB_\ell}\) from Hypothesis~\ref{hyp:cell-partitions} at \(\ell=\ell(T)\), and use the fixed-scale coefficients of Definition~\ref{def:interaction-representatives}.
For every \(s,t\in I\) with \(|t-s|\le T\), we have:
\begin{enumerate}[label=(\roman*),ref=(\roman*),leftmargin=*]
\item \emph{Sum through a cell.}
\label{item:fixed-scale-cell-bound}
For every \(0\le p\le a\) and interaction \(\Psi\) of finite \(\|\cdot\|^{\#}_{p;\kappa_0}\)-norm,
\begin{align}
 \sup_{b\in\cB_\ell}\sum_{S\ni b}\E^{3|S|}(1+\diam Q(S))^p
 \sum_{\varnothing\ne X\subseteq Q(S)}
       \bigl\|(\cT^{\Phi,T}_{s,t;X}\Psi(X))(Q(S))\bigr\|
 \le C^{\rm fs}_p\ell^{\dvol+p}\|\Psi\|^{\#}_{p;\kappa_0}.
 \label{eq:fixed-scale-cell-bound}
\end{align}
\item \emph{Total sum for one input.}
\label{item:fixed-scale-total-bound}
For every nonempty \(X\Subset\Gamma\), \(A\in\A_X^+\), and \(0\le p\le a\),
\begin{align}
 \sum_{S\supseteq S_0(X)}\E^{3|S|}(1+\diam Q(S))^p
       \bigl\|(\cT^{\Phi,T}_{s,t;X}A)(Q(S))\bigr\|
 \le C^{\rm fs}_p\E^{\kappa_0|X|}\|A\|
       \bigl((1+\diam X)^p+\ell^p\bigr).
 \label{eq:fixed-scale-total-bound}
\end{align}
\item \emph{Representation.}
\label{item:fixed-scale-representation}
For every nonempty \(X\Subset\Gamma\) and \(A\in\A_X^+\),
\begin{equation}
 \tau^\Phi_{s,t}(A)=\sum_{Y\supseteq X}(\cT^{\Phi,T}_{s,t;X}A)(Y)
 \label{eq:fixed-scale-representation}
\end{equation}
with absolute convergence in norm, where \(\tau^\Phi\) is the dynamics obtained from Proposition~\ref{prop:polynomial-dynamics} on compact subintervals of \(I\).
\end{enumerate}
\end{proposition}

\begin{proof}
\smallskip
\noindent \emph{Decomposition under finite-range evolution.}
Choose \(\kappa_0\) from Lemma~\ref{lem:card-short} with \(\rho=4\), and fix \(a>\dvol\) and \(L\ge0\).
Put
\[
 J_{\rm fr}:=L\sup_{m\ge1}m\E^{-\kappa_0m},
\]
and choose the constant \(c>4\) of Lemma~\ref{lem:card-short} for this bound.
For each \(q\ge0\), choose \(C_q^{\rm fr}\) large enough to bound both constants in \eqref{eq:card-short} and \eqref{eq:finite-range-total-bound} at spatial exponent \(q\) and cell rate \(4\).
Fix these constants and set \(C_*:=2C_0^{\rm fr}\).
By Lemma~\ref{lem:card-conjugation}, \(\varepsilon_p\ge\varepsilon_a\) for \(0\le p\le a\).

Fix \(s,t\in I\) with \(|t-s|\le T\).
Restrict \(\Phi\) to \([\min(s,t),\max(s,t)]\) and extend it constantly beyond the two endpoints.
The extension preserves componentwise norm-continuity and the Hamiltonian norm bound.
Formula \eqref{eq:full-dynamics-mobius} is unchanged at \((s,t)\), and finite-volume convergence in Proposition~\ref{prop:polynomial-dynamics} shows that the infinite-volume evolution between \(s\) and \(t\) is unchanged as well.
For the extended path, put \(\Phi^{[s]}_u:=\Phi_{s+u}\).
The finite-volume dynamics in \eqref{eq:full-dynamics-mobius} give
\[
 \cT^{\Phi,T}_{s,t;X}
 =\cT^{\Phi^{[s]},T}_{0,t-s;X}.
\]
The translated path has the same norm bound, so it suffices to work at initial time zero with a path defined on \(\R\) and \(|t|\le T\).
For \(R\ge1\), define
\[
 \Phi_u^{\le R}(X):=\1_{\{\diam X\le R\}}\Phi_u(X),
 \qquad
 \Phi_u^{>R}(X):=\1_{\{\diam X>R\}}\Phi_u(X).
\]
For \(\diam X>R\), \((1+\diam X)^p\le R^{p-a}(1+\diam X)^a\) for \(0\le p\le a\).
Hence
\begin{equation}
 \sup_u\|\Phi_u^{>R}\|^{\#}_{0;\kappa_0}\le R^{-a}L,
 \qquad
 \sup_u\|\Phi_u^{>R}\|^{\#}_{p;\kappa_0}\le R^{p-a}L.
 \label{eq:card-range-tail}
\end{equation}
Also \(\sup_u\|\Phi_u^{\le R}\|^{\pol}_{0;1}\le J_{\rm fr}\).
Set
\[
 \ell:=\left\lceil cR\bigl(T+\log(2+R)\bigr)\right\rceil,
\]
choose a partition \((Q_b)_{b\in\cB_\ell}\) from Hypothesis~\ref{hyp:cell-partitions}, and write \(\cR_t:=\cR_t^{\Phi^{\le R},\ell}\) for \(|t|\le T\).
Use the single-support interactions \(\iota_X A\) defined in \eqref{eq:single-support-interaction}.

Fix \(0\le p\le a\) and an interaction \(\Psi\) with \(\|\Psi\|^{\#}_{p;\kappa_0}<\infty\).
Put
\[
 \widetilde\Phi_u:=\cR_u(\Phi_u^{>R}),
 \qquad
 g_t(S):=\sum_{\varnothing\ne X\subseteq Q(S)}
       \| (\cR_t(\iota_X\Psi(X)))(S)\|.
\]
Each \(\widetilde\Phi_u(S)\) is even and self-adjoint.
The finite sums in \eqref{eq:card-mobius} show that \(u\mapsto\widetilde\Phi_u(S)\) is norm-continuous.
By \eqref{eq:card-short}, \eqref{eq:card-range-tail}, and \(R\le\ell\),
\begin{equation}
\begin{aligned}
 L_0&:=\sup_{|u|\le T}\|\widetilde\Phi_u\|^{\#,\ell}_{0;4}
       \le C_*L\ell^{\dvol}R^{-a},\\
 L_p&:=\sup_{|u|\le T}\|\widetilde\Phi_u\|^{\#,\ell}_{p;4}
       \le C_p^{\rm fr}L\ell^{\dvol}
                     (R^{p-a}+\ell^pR^{-a})
       \le2C_p^{\rm fr}L\ell^{\dvol+p}R^{-a},\\
 \|g_t\|^{\#,\ell}_{0;4}&\le C_*\ell^{\dvol}\|\Psi\|^{\#}_{0;\kappa_0},\\
 \|g_t\|^{\#,\ell}_{p;4}&\le C_p^{\rm fr}\ell^{\dvol}
                      \bigl(\|\Psi\|^{\#}_{p;\kappa_0}
                            +\ell^p\|\Psi\|^{\#}_{0;\kappa_0}\bigr)\le2C_p^{\rm fr}\ell^{\dvol+p}\|\Psi\|^{\#}_{p;\kappa_0}.
\end{aligned}
 \label{eq:card-picture-bounds}
\end{equation}

\smallskip
\noindent \emph{The interaction picture and the choice of scales.}
For a finite site set \(\Omega\), define
\[
 H^\Omega(u):=\sum_{\varnothing\ne Z\subseteq\Omega}\Phi_u(Z),
 \qquad
 H_{\le R}^\Omega(u):=
       \sum_{\varnothing\ne Z\subseteq\Omega}\Phi_u^{\le R}(Z),
 \qquad H_{>R}^\Omega:=H^\Omega-H_{\le R}^\Omega.
\]
Let \(U^\Omega(0)=U_{\le R}^\Omega(0)=\1\) and
\[
 \I\partial_uU^\Omega(u)=H^\Omega(u)U^\Omega(u),
 \qquad
 \I\partial_uU_{\le R}^\Omega(u)
       =H_{\le R}^\Omega(u)U_{\le R}^\Omega(u).
\]
For \(C_u^\Omega:=U^\Omega(u)^*U_{\le R}^\Omega(u)\), differentiation gives
\begin{equation}
\begin{aligned}
 (C_u^\Omega)'
 &=\I U^\Omega(u)^*H_{>R}^\Omega(u)U_{\le R}^\Omega(u)
 =\I C_u^\Omega\,
       \tau_{0,u}^{\Phi^{\le R},\Omega}(H_{>R}^\Omega(u)),\\
 \tau_{0,t}^{\Phi,\Omega}(A)
 &=C_t^\Omega\,
       \tau_{0,t}^{\Phi^{\le R},\Omega}(A)(C_t^\Omega)^*.
\end{aligned}
 \label{eq:card-picture}
\end{equation}
The first identity and \(C_0^\Omega=\1\) show that \(C_t^\Omega B(C_t^\Omega)^*=\tau_{0,t}^{\mathrm I,\Omega}(B)\) for \(B\in\A_\Omega\).

Now take \(R=R(T)\) as in \eqref{eq:fixed-scales}, with \(R_*\ge2\) to be chosen.
Then
\[
 R_*(1+T)^\gamma\le R\le2R_*(1+T)^\gamma,
 \qquad
 \log(2+R)\le\log(4R_*)+\gamma\log(1+T).
\]
Since \(\log(1+T)\le T\), the definition of \(\ell\) gives \(\ell\le C(1+\log R_*)R(1+T)\), with \(C\) depending only on \(c,\gamma\).
Consequently, for a constant \(C_{\rm sc}=C_{\rm sc}(c,\gamma,\dvol)\),
\[
 \begin{aligned}
 T\ell^{\dvol}R^{-a}
 &\le C_{\rm sc}R_*^{\dvol-a}(1+\log R_*)^{\dvol}
       (1+T)^{\dvol+1-\gamma(a-\dvol)}\\
 &=C_{\rm sc}R_*^{\dvol-a}(1+\log R_*)^{\dvol}.
 \end{aligned}
\]
Choose \(R_*\ge2\) so that
\[
 C_*LC_{\rm sc}R_*^{\dvol-a}(1+\log R_*)^{\dvol}
 \le\varepsilon_a.
\]
This is where the assumption \(a>\dvol\) enters: the factor \(R_*^{\dvol-a}(1+\log R_*)^{\dvol}\) tends to zero as \(R_*\to\infty\), so this inequality can be satisfied.
The resulting scales obey
\begin{equation}
 TL_0\le C_*LT\ell^{\dvol}R^{-a}
 \le\varepsilon_a\le\varepsilon_p,
 \qquad 0\le p\le a,
 \label{eq:card-smallness}
\end{equation}
and the preceding bound on \(\ell\) proves \eqref{eq:scale-growth} with a constant \(C_{R_*}\) depending only on the stated data.

\smallskip
\noindent \emph{Absolute bounds for the interaction representatives.}
Let \(\cC_t\) be the interaction map of Lemma~\ref{lem:card-conjugation}, with generator \(\widetilde\Phi\), and define
\[
 \cW_t^{\Phi,T}\Psi:=\cC_t(\cR_t\Psi),
 \qquad |t|\le T.
\]
For the finite-volume identity, fix a finite cell set \(\cS\supseteq S_0(X)\) and put \(\Omega:=Q(\cS)\).
The telescoping identity \eqref{eq:card-telescope} gives
\[
 \begin{aligned}
 \sum_{S\subseteq\cS}\widetilde\Phi_u(S)
 &=\tau_{0,u}^{\Phi^{\le R},\Omega}(H_{>R}^{\Omega}(u)),\\
 \sum_{S\subseteq\cS}(\cR_t(\iota_XA))(S)
 &=\tau_{0,t}^{\Phi^{\le R},\Omega}(A).
 \end{aligned}
\]
The first identity identifies the cell dynamics \(\tau_{0,t}^{\widetilde\Phi,\cS}\) with \(\tau_{0,t}^{\mathrm I,\Omega}\).
Each commutator in \(\cC_t\) is assigned to the union of its input supports.
Its coefficients on subsets of \(\cS\) are therefore unchanged when the generator \(\widetilde\Phi\) and the interaction \(\cR_t(\iota_XA)\) are restricted to \(\cS\).
Equations \eqref{eq:card-conjugation-finite} and \eqref{eq:card-picture} yield
\begin{equation}
 \sum_{S\subseteq\cS}
       (\cW_t^{\Phi,T}(\iota_XA))(S)
 =\tau_{0,t}^{\mathrm I,\Omega}
       \bigl(\tau_{0,t}^{\Phi^{\le R},\Omega}(A)\bigr)
 =\tau_{0,t}^{\Phi,\Omega}(A).
 \label{eq:full-dynamics-finite-sum}
\end{equation}
The coefficients vanish unless \(S\supseteq S_0(X)\).
Since \eqref{eq:full-dynamics-finite-sum} holds for every finite \(\cS\supseteq S_0(X)\), M\"obius inversion gives
\begin{equation}
 (\cW_t^{\Phi,T}(\iota_XA))(S)
 =(\cT^{\Phi,T}_{0,t;X}A)(Q(S)).
 \label{eq:fixed-scale-identification}
\end{equation}

For each fixed \(t\), apply Lemma~\ref{lem:card-conjugation}\ref{item:conjugation-decomposed} to the countable decomposition
\[
 \cR_t\Psi=\sum_{\varnothing\ne X\Subset\Gamma}
                  \cR_t(\iota_X\Psi(X)).
\]
Its scalar majorant is \(g_t\), whose cell norms are bounded in \eqref{eq:card-picture-bounds}.
With \(C_p\) denoting the constant of Lemma~\ref{lem:card-conjugation}, \eqref{eq:card-conjugation-decomposed} gives
\[
 \begin{aligned}
 \sup_b\sum_{S\ni b}\E^{3|S|}(&1+\diam Q(S))^p
       \sum_{\varnothing\ne X\subseteq Q(S)}
       \|(\cW_t^{\Phi,T}(\iota_X\Psi(X)))(S)\|\\
 &\le C_p\bigl(\|g_t\|^{\#,\ell}_{p;4}
                   +TL_p\|g_t\|^{\#,\ell}_{0;4}\bigr)\\
 &\le 2C_pC_p^{\rm fr}\ell^{\dvol+p}
            \bigl(1+C_*LT\ell^{\dvol}R^{-a}\bigr)\|\Psi\|^{\#}_{p;\kappa_0}\\
 &\le C_p^{\rm fs}\ell^{\dvol+p}\|\Psi\|^{\#}_{p;\kappa_0},
 \qquad C_p^{\rm fs}:=2C_pC_p^{\rm fr}(1+\varepsilon_a).
 \end{aligned}
\]
The last inequality uses \eqref{eq:card-smallness}.
Together with \eqref{eq:fixed-scale-identification}, this proves \eqref{eq:fixed-scale-cell-bound}.

For a fixed input \(A\in\A_X^+\), the constants \(C_j^{\rm fr}\) chosen at the start of the proof give
\[
 \Sigma_{j;4}^{(\ell)}(\cR_t(\iota_XA))
 \le C_j^{\rm fr}\E^{\kappa_0|X|}\|A\|
                 \bigl((1+\diam X)^j+\ell^j\bigr),
 \qquad j=0,p.
\]
The total estimate \eqref{eq:card-conjugation-total} therefore yields
\[
 \begin{aligned}
 \Sigma_{p;3}^{(\ell)}(\cW_t^{\Phi,T}(\iota_XA))
 &\le C_p\E^{\kappa_0|X|}\|A\|
       \bigl(C_p^{\rm fr}((1+\diam X)^p+\ell^p)
                         +2C_0^{\rm fr}TL_p\bigr)\\
 &\le C_p^{\rm fs}\E^{\kappa_0|X|}\|A\|
                         \bigl((1+\diam X)^p+\ell^p\bigr).
 \end{aligned}
\]
Indeed, \eqref{eq:card-picture-bounds}, \(C_*=2C_0^{\rm fr}\), and \eqref{eq:card-smallness} give
\[
 2C_0^{\rm fr}TL_p
 \le 2C_p^{\rm fr}\ell^p
                  \bigl(C_*LT\ell^{\dvol}R^{-a}\bigr)
 \le 2C_p^{\rm fr}\varepsilon_a\ell^p.
\]
The definition of \(C_p^{\rm fs}\) thus proves \eqref{eq:fixed-scale-total-bound} with the same constant as in \eqref{eq:fixed-scale-cell-bound}.

\smallskip
\noindent \emph{Representation and continuity on a fixed time interval.}
Choose finite cell sets \(\cS_j\uparrow\cB_\ell\) containing \(S_0(X)\).
Equation~\eqref{eq:full-dynamics-finite-sum} gives
\[
 \sum_{Y\subseteq Q(\cS_j)}(\cT^{\Phi,T}_{0,t;X}A)(Y)
 =\tau_{0,t}^{\Phi,Q(\cS_j)}(A).
\]
By \eqref{eq:fixed-scale-total-bound} with \(p=0\), the series on the left converges absolutely as \(j\to\infty\).
The right side converges to \(\tau_{0,t}^\Phi(A)\) by Proposition~\ref{prop:polynomial-dynamics}.
This proves \eqref{eq:fixed-scale-representation} at initial time zero; the extension and time translation at the start of the proof give the identity for the original \(s,t\in I\).
In each finite region \(Q(U)\), the Hamiltonian is a finite sum of continuous matrix-valued functions.
Its propagator is continuous in both time endpoints.
The finite sum in \eqref{eq:full-dynamics-mobius} therefore shows that the coefficients are continuous in \((s,t)\).
That formula also shows that the maps are complex-linear and preserve adjoints.
\end{proof}

\subsection{Proof of the interaction propagation theorem}
\label{subsec:propagation-proof}

We now use the bounds on a fixed cell partition from Proposition~\ref{prop:fixed-scale} to prove Theorem~\ref{thm:card-propagation}.
The estimates in Proposition~\ref{prop:fixed-scale} weight a term supported on \(Q(S)\) by \(\E^{3|S|}\), where \(|S|\) counts cells.
To obtain a weight in the number of sites, we use the volume bound \(|Q(S)|\le C\ell^{\dvol}|S|\), with a geometric constant \(C\).
Consequently,
\[
 \E^{3|Q(S)|/(C\ell^{\dvol})}\le\E^{3|S|}.
\]
The same estimates therefore control an exponential weight in the number of sites, with a rate that decreases as the cell size grows.

We then combine the fixed-scale coefficients for the horizons \(T_n=2^n\), using the cutoffs of Definition~\ref{def:interaction-representatives}; see Figure~\ref{fig:scales}(b).
At each pair of times \(s,t\), these cutoffs select only horizons for which \(1+T_n\) is comparable to \(1+|t-s|\).
The growth bound \eqref{eq:scale-growth} therefore turns the powers of the cell size into polynomial time factors, while the diameter weight remains unchanged.

\begin{proof}[Proof of Theorem~\ref{thm:card-propagation}]
\phantomsection\label{proof:card-propagation}
\smallskip
\noindent\phantomsection\label{par:common-propagation-choices}\emph{Choice of cells and time cutoffs.}
Choose \(\kappa_0\) from Lemma~\ref{lem:card-short} with \(\rho=4\), fix \(a>\dvol\) and \(L\ge0\), and use the scales and constants of Proposition~\ref{prop:fixed-scale}.
For every \(T\ge1\), fix a cell partition at scale \(\ell(T)\).
For the horizons \(T_n=2^n\), use these partitions in Definition~\ref{def:interaction-representatives}, with one fixed function \(\chi\) satisfying its conditions.
These choices are independent of \(\Phi,\Psi,p,k\) and the local multiplicities.

Let \(\Phi\) satisfy the hypotheses of the theorem, fix \(0\le p\le a\), and let \(\Psi\) be an interaction with \(\|\Psi\|^{\#}_{p;\kappa_0}<\infty\).
For \(T\ge1\) and \(s,t\in I\) with \(|t-s|\le T\), write \(\ell=\ell(T)\).
For the unique cell \(b(x)\) containing \(x\), the sets \(Q(S)\ni x\) correspond to \(S\ni b(x)\).
Volume growth gives
\[
 |Q(S)|\le \Cvol\ell^{\dvol}|S|,
 \qquad \Cvol:=C_\Gamma(1+\Ccell)^{\dvol}.
\]
Equation~\eqref{eq:fixed-scale-cell-bound} therefore gives
\[
 \begin{aligned}
 \sup_x\sum_{Y\ni x}
       \E^{3|Y|/(\Cvol\ell^{\dvol})}(&1+\diam Y)^p
       \sum_{\varnothing\ne X\subseteq Y}
       \|(\cT^{\Phi,T}_{s,t;X}\Psi(X))(Y)\|\\
 &\le C_p^{\rm fs}\ell^{\dvol+p}\|\Psi\|^{\#}_{p;\kappa_0}\\
 &\le C_p^{\rm fs}C_{R_*}^{\dvol+p}(1+T)^{m_{p,0}}
                        \|\Psi\|^{\#}_{p;\kappa_0},
 \qquad |t-s|\le T.
 \end{aligned}
\]

For \(A\in\A_X^+\), \eqref{eq:fixed-scale-total-bound} and the same volume bound give
\[
 \begin{aligned}
 \sum_{Y\supseteq X}
       \E^{3|Y|/(\Cvol\ell^{\dvol})}(&1+\diam Y)^p
                 \|(\cT^{\Phi,T}_{s,t;X}A)(Y)\|\\
 &\le C_p^{\rm fs}\E^{\kappa_0|X|}\|A\|
                         \bigl((1+\diam X)^p+\ell^p\bigr)\\
 &\le C_p^{\rm fs}(1+C_{R_*}^{p})
                 \E^{\kappa_0|X|}(1+\diam X)^p\|A\|
                 (1+T)^{\widetilde m_{p,0}},
 \qquad |t-s|\le T,
 \end{aligned}
\]
where \(\widetilde m_{p,0}\) is defined in \eqref{eq:source-time-exponent}.

\smallskip
\noindent \emph{Interaction representatives for all times.}
We now estimate the sum over time scales in \eqref{eq:source-map-construction}.
The cutoffs \eqref{eq:time-cutoffs} satisfy \(1+T_n\le4(1+|v|)\) on \(\supp\chi_n\), including \(n=0\).
Only scales with \(|t-s|\le T_n\) contribute, so \eqref{eq:fixed-scale-representation} and \eqref{eq:fixed-scale-total-bound} apply to every summand.
Since \(\sum_n\chi_n=1\), they give \eqref{eq:source-representation} with absolute convergence.
The coefficientwise continuity, complex-linearity, preservation of adjoints, and support property pass from the fixed-scale maps to the locally finite sum.

On every scale contributing to \eqref{eq:source-map-construction},
\[
 \ell(T_n)^{\dvol}
  \le C_{R_*}^{\dvol}(1+T_n)^{(1+\gamma)\dvol} \le C_{R_*}^{\dvol}4^{(1+\gamma)\dvol}
                         (1+|t-s|)^{(1+\gamma)\dvol}.
\]
Choose
\[
 c_*:=\frac{3}{\Cvol C_{R_*}^{\dvol}4^{(1+\gamma)\dvol}}.
\]
Then \(\rho(s,t)\le3/(\Cvol\ell(T_n)^{\dvol})\) on each contributing scale.
Apply the triangle inequality to \eqref{eq:source-map-construction}, multiply by the output weights, and sum over \(X\) and \(Y\).
The fixed-scale estimate above gives
\[
 \begin{aligned}
 \sup_x\sum_{Y\ni x}\E^{\rho(s,t)|Y|}(1+\diam Y)^p
       \sum_{\varnothing\ne X\subseteq Y}
       \|(\cT^\Phi_{s,t;X}\Psi(X))(Y)\|  &\le\sum_{n\ge0}\chi_n(t-s)
       C_p^{\rm fs}\ell(T_n)^{\dvol+p}
                          \|\Psi\|^{\#}_{p;\kappa_0}\\
 & \le C_p^{\rm fs}C_{R_*}^{\dvol+p}4^{m_{p,0}}
       (1+|t-s|)^{m_{p,0}}\|\Psi\|^{\#}_{p;\kappa_0}.
 \end{aligned}
\]
The last inequality uses \(1+T_n\le4(1+|t-s|)\) on every contributing scale and \(\sum_n\chi_n(t-s)=1\).
For a fixed input \(A\in\A_X^+\), apply the same triangle inequality to the total sum over \(Y\supseteq X\), using the fixed-scale total estimate above in place of the sum through a site.
This gives \eqref{eq:exponential-source-total} with constant \(C_p^{\rm fs}(1+C_{R_*}^p)4^{\widetilde m_{p,0}}\).
Choose \(C_p\) to dominate this constant and \(C_p^{\rm fs}C_{R_*}^{\dvol+p}4^{m_{p,0}}\); then both \eqref{eq:exponential-support-propagation} and \eqref{eq:exponential-source-total} hold with \(C_p\).

For \(k>0\), the elementary inequality
\[
 h^k\le\left(\frac{k}{\E\rho}\right)^k\E^{\rho h},
 \qquad h\ge0,\quad \rho>0,
\]
applied with \(h=|Y|\) and \(\rho=\rho(s,t)\), proves \eqref{eq:card-propagation} and \eqref{eq:card-propagation-total}, since
\[
 m_{p,k}=m_{p,0}+k(1+\gamma)\dvol,\qquad
 \widetilde m_{p,k}=\widetilde m_{p,0}+k(1+\gamma)\dvol.
\]
For \(k=0\), discard the exponential weight.
We may therefore take \(C_{p,0}=C_p\) and \(C_{p,k}=C_p\max\{1,(k/(\E c_*))^k\}\) for \(k>0\).
The interaction-norm bounds follow by the triangle inequality.

If \(\sum_X\E^{\kappa_0|X|}\|\Psi(X)\|<\infty\), then \eqref{eq:card-propagation-total} with \(p=k=0\) gives
\[
 \sum_{X,Y}\| (\cT^\Phi_{s,t;X}\Psi(X))(Y)\|
 \le C_{0,0}\sum_X\E^{\kappa_0|X|}\|\Psi(X)\|<\infty.
\]
Interchanging the sums and using \eqref{eq:source-representation} proves \eqref{eq:card-propagation-finite}.
For a general interaction \(\Psi\) with finite \(\|\Psi\|^{\#}_{0;\kappa_0}\) and a local observable \(B\in\A_F\), \(F\Subset\Gamma\), evenness and \eqref{eq:card-propagation} give
\[
 \begin{aligned}
 \sum_{X,Y}\bigl\|[(\cT^\Phi_{s,t;X}\Psi(X))(Y),B]\bigr\|
 &\le2\|B\|\sum_{x\in F}\sum_{Y\ni x}
       \sum_{X\subseteq Y}\| (\cT^\Phi_{s,t;X}\Psi(X))(Y)\|\\
 &\le2|F|\|B\|C_{0,0}(1+|t-s|)^{m_{0,0}}
       \|\Psi\|^{\#}_{0;\kappa_0}.
 \end{aligned}
\]
Absolute convergence now gives
\[
 \sum_Y[(\cT_{s,t}^\Phi\Psi)(Y),B]
 =\sum_X\left[\sum_Y(\cT^\Phi_{s,t;X}\Psi(X))(Y),B\right]
 =\sum_X[\tau_{s,t}^\Phi(\Psi(X)),B],
\]
which proves \eqref{eq:card-propagation-derivation}.
For fixed \(\eta_X,\zeta_X\in\{0,1\}\), the interactions \(\Phi'_u(X):=\eta_X\Phi_u(X)\) and \(\Psi'(X):=\zeta_X\Psi(X)\) satisfy
\[
 \sup_u\|\Phi'_u\|^{\#}_{a;\kappa_0}\le L,
 \qquad
 \|\Psi'\|^{\#}_{p;\kappa_0}
       \le\|\Psi\|^{\#}_{p;\kappa_0}.
\]
The construction therefore gives the same constants after deletion.
Taking \(\eta_X=\zeta_X=\1_{\{X\subseteq\Lambda\}}\) gives the same estimates, with unchanged constants, for \(\Phi^\Lambda,\Psi^\Lambda\) viewed as interactions on \(\Gamma\).
If the underlying space is instead replaced by a subset \(\Gamma'\subseteq\Gamma\), the volume and cell bounds hold with the same constants after intersecting the cells with \(\Gamma'\) and discarding empty cells.
Applying the construction on \(\Gamma'\) therefore gives the same estimates, with representatives defined using these intersected cells.
\end{proof}

\subsection{Arbitrary local observables and onsite automorphisms}
\label{subsec:observable-extensions}

We now extend the local decomposition and its bounds to observables of arbitrary parity.
We then describe how the representing interactions transform under onsite automorphisms.

Bourne--Ogata \cite[Appendix~B]{BourneOgata21} obtain bounds for odd observables by multiplying them by a distant odd self-adjoint unitary.
Wilming--Werner \cite[Appendices~D.2--D.3]{WilmingWerner22} use stationary auxiliary fermions to obtain locality estimates by averaging over even unitaries.
We use a stationary auxiliary fermion to construct an explicit isometric embedding of the local algebra into the even part of an enlarged CAR algebra.
The embedding intertwines the dynamics and the tracial conditional expectations.

\begin{lemma}
\label{lem:auxiliary-even-embedding}
Let \(\Omega\Subset\Gamma\), and for \(Y\subseteq\Omega\) put
\[
 \widehat\A_Y:=\CAR\left(
       \bigoplus_{y\in Y}(\C^{r_y}\oplus\C)\right),
\]
with \(\A_Y\) identified with its canonical image.
Denote the even part of \(\widehat\A_Y\) under its natural parity automorphism by \(\widehat\A_Y^+\).
For \(x\in\Omega\), let \(b_x\) be the added annihilation operator and put \(v_x:=b_x+b_x^*\).
For \(B\in\A_\Omega\), put \(B_\pm:=\tfrac12(B\pm\Theta(B))\), where \(\Theta\) is the physical parity automorphism.
The map
\begin{equation}
 \jmath_x:\A_\Omega\rightarrow\widehat\A_\Omega^+,
 \qquad \jmath_x(B):=B_++\I B_-v_x,
 \label{eq:auxiliary-even-embedding}
\end{equation}
is an isometric unital \(*\)-homomorphism, and \(\jmath_x(\A_X)\subseteq\widehat\A_X^+\) whenever \(x\in X\subseteq\Omega\).

Let \(I\subseteq\R\) be an interval and \((\Xi_u)_{u\in I}\) a componentwise norm-continuous self-adjoint interaction path.
Regard each \(\Xi_u(Z)\) as an element of \(\widehat\A_Z\) through the canonical inclusion \(\A_Z\subseteq\widehat\A_Z\), and set \(\widehat\Xi_u(Z):=\Xi_u(Z)\).
Write \(\widehat\tau:=\tau^{\widehat\Xi,\Omega}\) and let \(\widehat{\F}^{\rm tr}_D\) be the tracial expectation onto \(\widehat\A_D\).
Then, for \(s,t\in I\) and \(x\in D\subseteq\Omega\),
\[
 \widehat\tau_{s,t}\circ\jmath_x
   =\jmath_x\circ\tau^{\Xi,\Omega}_{s,t},
 \qquad
 \widehat{\F}^{\rm tr}_D\circ\jmath_x
   =\jmath_x\circ\F_D^{\rm tr}.
\]
\end{lemma}

\begin{proof}
The Majorana \(v_x\) satisfies \(v_x^*=v_x\), \(v_x^2=\1\), and \(v_xB_\pm=\pm B_\pm v_x\) for \(B\in\A_\Omega\).
The unitary \(V_x:=\E^{-\I\pi v_x/4}=(\1-\I v_x)/\sqrt2\) satisfies
\[
 V_xBV_x^*=B_++\I B_-v_x=\jmath_x(B).
\]
Thus \(\jmath_x\) is the restriction of an inner automorphism of \(\widehat\A_\Omega\), hence an isometric unital \(*\)-homomorphism.
Both summands in \eqref{eq:auxiliary-even-embedding} are even and supported in \(X\) when \(B\in\A_X\) and \(x\in X\).
Every embedded Hamiltonian term commutes with \(V_x\), so \(\widehat\tau_{s,t}(V_x)=V_x\).
Since \(\widehat\tau\) agrees with \(\tau^{\Xi,\Omega}\) on \(\A_\Omega\), this proves the dynamics identity.
On a product of distinct physical Majoranas, both \(\widehat{\F}^{\rm tr}_D\) and \(\F_D^{\rm tr}\) act as the identity if its support is contained in \(D\), and give zero otherwise.
These monomials span \(\A_\Omega\), so \(\widehat{\F}^{\rm tr}_D|_{\A_\Omega} =\F_D^{\rm tr}\).
For \(x\in D\), we have \(V_x\in\widehat\A_D\), and bimodularity gives
\[
 \widehat{\F}^{\rm tr}_D(\jmath_x(B))
 =V_x\widehat{\F}^{\rm tr}_D(B)V_x^*
 =V_x\F_D^{\rm tr}(B)V_x^*
 =\jmath_x(\F_D^{\rm tr}(B)).\qedhere
\]
\end{proof}

The embedding also transfers the commutator criterion for local approximation to odd observables.
For \(x\in D\subseteq\Omega\) and \(C\in\A_\Omega\), isometry and compatibility with the conditional expectation give
\[
 \|C-\F_D^{\rm tr}(C)\|
 =\|\jmath_x(C)-\widehat{\F}_D^{\rm tr}(\jmath_x(C))\|\le\sup_{\substack{U\in\widehat\A_{\Omega\setminus D}\\
                    U^*U=UU^*=\1}}
                \|[\jmath_x(C),U]\|.
\]
The inequality follows by averaging over exterior unitaries, since \(\jmath_x(C)\) is even \cite[Lemma~4.3]{NSY18}.
For \(C=\tau_{s,t}^{\Xi,\Omega}(A)\) with \(A\in\A_X\) and \(x\in X\subseteq D\), Lemma~\ref{lem:auxiliary-even-embedding} expresses \(\jmath_x(C)\) as the evolution of the even observable \(\jmath_x(A)\).
We can therefore apply commutator bounds in the enlarged system and obtain local approximations of \(C\) with the same constants, provided those bounds are uniform under adjoining stationary fermionic modes.

\begin{corollary}[Representatives for arbitrary local observables]
\label{rem:odd-representatives}
Under the hypotheses of Theorem~\ref{thm:card-propagation}, fix the partitions, cutoffs, and constants provided there.
For every \(s,t\in I\), nonempty \(X\Subset\Gamma\), and \(A\in\A_X\), the coefficients of Definition~\ref{def:interaction-representatives} satisfy
\[
 \tau^\Phi_{s,t}(A)
 =\sum_{Y\supseteq X}(\cT^\Phi_{s,t;X}A)(Y),
\]
with absolute convergence in norm.
The polynomially and exponentially weighted estimates \eqref{eq:card-propagation-total} and \eqref{eq:exponential-source-total} hold for these coefficients, with the same constants, for every \(0\le p\le a\) and \(k\ge0\).
\end{corollary}

\begin{proof}
Adjoin one stationary auxiliary fermion at each site, as in Lemma~\ref{lem:auxiliary-even-embedding}, and denote the resulting quasi-local algebra by \(\widehat\A\).
Fix \(x\in X\).
The embeddings \(\jmath_x\) agree on local algebras as \(\Omega\) increases through finite regions containing \(x\), and extend by isometry to \(\jmath_x:\A\to\widehat\A^+\).
Let \(\widehat\Phi_u(Z):=\Phi_u(Z)\), and use the same cells and scales for \(\Phi\) and \(\widehat\Phi\).
In each finite region \(Q(U)\) appearing in \eqref{eq:full-dynamics-mobius}, Lemma~\ref{lem:auxiliary-even-embedding} gives \(\tau^{\widehat\Phi,Q(U)}_{s,t}\circ\jmath_x =\jmath_x\circ\tau^{\Phi,Q(U)}_{s,t}\), so that
\[
 (\cT^{\widehat\Phi}_{s,t;X}\jmath_x(A))(Y)
 =\jmath_x\bigl((\cT^\Phi_{s,t;X}A)(Y)\bigr),
 \qquad
 \|(\cT^\Phi_{s,t;X}A)(Y)\|
 =\|(\cT^{\widehat\Phi}_{s,t;X}\jmath_x(A))(Y)\|,
\]
by isometry of \(\jmath_x\).
The path \(\widehat\Phi\) satisfies the same Hamiltonian bound as \(\Phi\), with local multiplicities \(r_y+1\), and the constants in Theorem~\ref{thm:card-propagation} do not depend on the multiplicities.
Applying \eqref{eq:card-propagation-total} and \eqref{eq:exponential-source-total} to the even observable \(\jmath_x(A)\in\widehat\A_X^+\), with \(\|\jmath_x(A)\|=\|A\|\), gives both total estimates for \(A\).
The total bound at \(p=k=0\) permits exchanging the absolutely convergent sum with \(\jmath_x\).
Finite-volume convergence extends the dynamics identity in Lemma~\ref{lem:auxiliary-even-embedding} to the infinite-volume dynamics.
The coefficient identity above and the intertwining of the dynamics give
\[
 \jmath_x\!\left(\sum_Y
       (\cT^\Phi_{s,t;X}A)(Y)\right)
 =\tau^{\widehat\Phi}_{s,t}(\jmath_x(A))
 =\jmath_x(\tau^\Phi_{s,t}(A)).
\]
Injectivity of \(\jmath_x\) proves \eqref{eq:source-representation} for arbitrary \(A\).
\end{proof}

\begin{remark}[Covariance under onsite automorphisms]
\label{rem:onsite-covariance}
Let \(\Phi,\Psi\) satisfy the hypotheses of Theorem~\ref{thm:card-propagation}, and let \(\zeta\) be a parity-preserving \(*\)-automorphism of \(\A\) with \(\zeta(\A_X)=\A_X\) for every finite \(X\).
Write \((\zeta\Phi)_u(Z):=\zeta(\Phi_u(Z))\) and \((\zeta\Psi)(Z):=\zeta(\Psi(Z))\).
Use Definition~\ref{def:interaction-representatives} with the same partitions and cutoffs for \(\Phi\) and \(\zeta\Phi\).
For every \(s,t\in I\), finite nonempty \(X,Y\), and \(A\in\A_X\),
\[
 (\cT^{\zeta\Phi}_{s,t;X}\zeta(A))(Y)
 =\zeta\bigl((\cT^\Phi_{s,t;X}A)(Y)\bigr).
\]
Indeed, on each finite region \(\Omega\), \(\tau_{s,t}^{\zeta\Phi,\Omega}\circ\zeta =\zeta\circ\tau_{s,t}^{\Phi,\Omega}\).
Substitution in \eqref{eq:full-dynamics-mobius} and \eqref{eq:source-map-construction} proves the identity.
Summing over the finitely many input supports \(X\subseteq Y\) for each output support \(Y\) gives
\[
 \cT_{s,t}^{\zeta\Phi}(\zeta\Psi)
 =\zeta(\cT_{s,t}^\Phi\Psi).
\]
Since \(\zeta\) is isometric and preserves supports, the same norm bounds apply.
Boundedness of \(\zeta\) permits passing it through the filter integrals in Theorem~\ref{thm:card-limits}, so the filtered interactions obey the same covariance identity.
\end{remark}

\section{Parameter dependence, filtering, and observable locality}
\label{sec:filters}

This section proves Theorem~\ref{thm:card-limits}, the regularity results of Subsection~\ref{subsec:differentiability}, and the observable bounds of Corollary~\ref{cor:observable-LR} and Theorem~\ref{thm:polynomial-local-topology}.
Subsection~\ref{subsec:iterated-filtering} treats repeated filtering and commutators.
The continuity and regularity results use the explicit formulas \eqref{eq:full-dynamics-mobius} and \eqref{eq:source-map-construction}, the fixed-scale bounds of Proposition~\ref{prop:fixed-scale}, and the propagation bounds of Theorem~\ref{thm:card-propagation}.
We prove Theorem~\ref{thm:polynomial-local-topology} by applying Lemma~\ref{lem:finite-volume-matrix-commutator} and comparing dynamics in different volumes.
This proof is independent of Theorem~\ref{thm:card-propagation}.
Subsection~\ref{subsec:polynomial-time-locality} combines the same commutator estimate with a range-splitting iteration to prove Proposition~\ref{prop:polynomial-time-locality}, and then decomposes evolved interaction terms into dyadic shells in Corollary~\ref{cor:polynomial-interaction-shells}.

\subsection{Continuity, convergence, and filtering}
\label{subsec:continuity-filtering}

\begin{proof}[Proof of Theorem~\ref{thm:card-limits}]
\phantomsection\label{proof:card-limits}
\smallskip
\noindent \emph{\ref{item:card-limits-continuity} Parameter continuity and local convergence.}
Use Definition~\ref{def:interaction-representatives} with the partitions and cutoffs chosen in Theorem~\ref{thm:card-propagation} for the common values \(a,L\), and put \(\Phi^{(\infty)}:=\Phi\).
The finite-volume dynamics on \(I\) define the coefficients for every Hamiltonian family by \eqref{eq:full-dynamics-mobius} and \eqref{eq:source-map-construction}.

Fix a compact interval \(J\subseteq I\) and a nonempty finite output support \(Y\).
Put
\[
 \Delta_{n,J}(Y):=
 \sum_{\varnothing\ne Z\subseteq Y}
       \sup_{\xi\in\frakP}\int_J
       \|\Phi^{(n)}_{\xi,u}(Z)-\Phi_{\xi,u}(Z)\|\,\dd u.
\]
This finite sum tends to zero as \(n\to\infty\).
For every \(\Omega\subseteq Y\), Duhamel's formula and unitarity give
\[
 \|\tau^{\Phi_\xi^{(n)},\Omega}_{s,t}(A)
       -\tau^{\Phi_\xi,\Omega}_{s,t}(A)\|
 \le 2\Delta_{n,J}(Y)\|A\|,
 \qquad \xi\in\frakP,\quad s,t\in J,\quad A\in\A_\Omega.
\]
At a fixed scale, the coefficient at \(Y\) vanishes unless \(Y=Q(S)\), in which case the sum in \eqref{eq:full-dynamics-mobius} has at most \(2^{|S|}\le2^{|Y|}\) terms.
Since the cutoffs in \eqref{eq:source-map-construction} are nonnegative and sum to one,
\[
 \sup_{\xi\in\frakP}\sup_{s,t\in J}
 \sup_{A\in\A_X^+,\,\|A\|\le1}
 \| (\cT^{\Phi_\xi^{(n)}}_{s,t;X}A)(Y)
       -(\cT^{\Phi_\xi}_{s,t;X}A)(Y)\|
 \le 2^{|Y|+1}\Delta_{n,J}(Y)\xrightarrow{n\to\infty}0.
\]
Joint continuity and compactness of \(J\) make \(\xi\mapsto\Phi^{(n)}_{\xi,\cdot}(Z)\) continuous in \(C(J,\A_Z)\).
The same Duhamel estimate for two parameter values, together with continuity in the time endpoints, proves joint continuity of the finite-volume dynamics and of the fixed-scale coefficient maps.
Only finitely many cutoffs are nonzero on \(J^2\), so the full coefficient maps are continuous on \(\frakP\times J^2\).

Put \(\Psi^{(\infty)}:=\Psi\) and
\[
 \Upsilon^{(n)}_{\xi,s,t}:=
       \cT_{s,t}^{\Phi_\xi^{(n)}}\Psi_\xi^{(n)},
 \qquad n\in\N\cup\{\infty\}.
\]
For every finite nonempty \(Y\),
\[
 \Upsilon^{(n)}_{\xi,s,t}(Y)
 =\sum_{\varnothing\ne X\subseteq Y}
       (\cT^{\Phi_\xi^{(n)}}_{s,t;X}\Psi_\xi^{(n)}(X))(Y).
\]
The finite sum, the continuity of the coefficient maps, and the assumed continuity of \(\Psi_\xi^{(n)}(X)\) prove componentwise norm-continuity of \(\Upsilon^{(n)}\).
The operator-norm convergence of those maps and uniform convergence of \(\Psi_\xi^{(n)}(X)\) give
\[
 e_{n,J}(Y):=
 \sup_{\xi\in\frakP}\sup_{s,t\in J}
       \|\Upsilon^{(n)}_{\xi,s,t}(Y)
                 -\Upsilon^{(\infty)}_{\xi,s,t}(Y)\|
 \xrightarrow{n\to\infty}0.
\]
To pass from coefficient convergence to convergence in local interaction seminorms, we need a uniform bound on the sum over large output supports.
For each \(k\ge0\), Theorem~\ref{thm:card-propagation} gives
\[
 \sup_{n\in\N\cup\{\infty\}}\sup_{\xi\in\frakP}\sup_{s,t\in J}
       \|\Upsilon^{(n)}_{\xi,s,t}\|^{\pol}_{a;k}
 \le C_{a,k}M_\Psi(1+|J|)^{m_{a,k}}.
\]
Apply Lemma~\ref{lem:componentwise-local-continuity} to the families \(\Upsilon^{(n)}\) on the parameter space \(\frakP\times J^2\), with spatial exponent \(a\), support moment \(k\), and this uniform bound.
Their uniform componentwise convergence proves \eqref{eq:card-local-topology} for every nonempty finite \(K\) and every \(0\le q<a\).

\smallskip
\noindent \emph{\ref{item:card-limits-coefficients} Filtered coefficients.}
Assume \(I=\R\) and \(W\in L^1(\R)\).
For each nonempty finite \(Y\), the uniform total bound in \eqref{eq:card-propagation-total} gives
\[
 \begin{aligned}
 \|\Upsilon^{(n)}_{\xi,s,s+v}(Y)\|
 \le C_{0,0}\sum_{\varnothing\ne X\subseteq Y}
                    \E^{\kappa_0|X|}\|\Psi_\xi^{(n)}(X)\| \le C_{0,0}|Y|M_\Psi,
 \qquad n\in\N\cup\{\infty\}.
 \end{aligned}
\]
Thus each coefficient integral exists in norm.
Put \(\cK^{(\infty)}:=\cK\), and for \(H>0\) define
\[
 \cK^{(n),[H]}_{\xi,s}(Y)
 :=\int_{-H}^H W(v)\Upsilon^{(n)}_{\xi,s,s+v}(Y)\,\dd v.
\]
For a compact interval \(J\), set \(J_H:=[\min J-H,\max J+H]\).
Joint continuity on \(\frakP\times J_H^2\) makes
\[
 (\xi,s)\mapsto
       [v\mapsto\Upsilon^{(n)}_{\xi,s,s+v}(Y)]
\]
continuous into \(C([-H,H],\A_Y)\).
Integration against \(W\) is bounded on this space, so the truncated coefficients are continuous.
Their tails satisfy
\[
 \sup_{n\in\N\cup\{\infty\}}\sup_{\xi\in\frakP}\sup_{s\in J}
 \|\cK^{(n)}_{\xi,s}(Y)-\cK^{(n),[H]}_{\xi,s}(Y)\|
 \le C_{0,0}|Y|M_\Psi\int_{|v|>H}|W(v)|\,\dd v\xrightarrow{H\to\infty}0.
\]
Thus \(\cK^{(n)}_{\xi,s}(Y)\) is a uniform limit of continuous truncated integrals and is continuous.
To compare the approximating and limiting coefficients, we use
\[
  \sup_{\xi\in\frakP}\sup_{s\in J}
       \|\cK^{(n)}_{\xi,s}(Y)-\cK_{\xi,s}(Y)\| \le e_{n,J_H}(Y)\|W\|_{L^1}
       +2C_{0,0}|Y|M_\Psi\int_{|v|>H}|W(v)|\,\dd v\,.
\]
Letting \(n\to\infty\) and then \(H\to\infty\) proves componentwise convergence.
Parity preservation makes the filtered coefficients an interaction, and adjoint preservation gives self-adjointness when the input interaction is self-adjoint and the filter is real.

\smallskip
\noindent \emph{\ref{item:card-limits-estimates} Weighted estimates.}
If \(\mu_{p,k}(W)<\infty\), componentwise integration, Tonelli's theorem, and \eqref{eq:card-propagation} give
\[
 \|\cK_{\xi,s}\|^{\pol}_{p;k}
 \le\int_\R|W(v)|\,
       \|\Upsilon^{(\infty)}_{\xi,s,s+v}\|^{\pol}_{p;k}\,\dd v\ \le C_{p,k}\|\Psi_\xi\|^{\#}_{p;\kappa_0}\mu_{p,k}(W)\,.
\]
This proves \eqref{eq:filtered-norm}.
For a fixed term \(\Psi_\xi(X)\), the sharper total estimate \eqref{eq:card-propagation-total} gives
\[
 \begin{aligned}
 \int_\R|W(v)|\sum_{Y\supseteq X}|Y|^k(&1+\diam Y)^p
       \|(\cT^{\Phi_\xi}_{s,s+v;X}\Psi_\xi(X))(Y)\|\,\dd v\\
 & \le C_{p,k}\E^{\kappa_0|X|}(1+\diam X)^p
                     \|\Psi_\xi(X)\|\widetilde\mu_{p,k}(W).
 \end{aligned}
\]
Whenever \(\widetilde\mu_{p,k}(W)<\infty\), Tonelli's theorem proves \eqref{eq:filtered-source-total}.

\smallskip
\noindent \emph{\ref{item:card-limits-identities} Derivation and total-operator identities.}
Since \(\widetilde\mu_{0,0}(W)=\|W\|_{L^1}\), the representation \eqref{eq:source-representation} and the finite sum over input supports at each \(Y\) give
\[
 \sum_{Y\supseteq X}\cK_{\xi,s;X}(Y)
       =\cI_{\xi,s}(\Psi_\xi(X)),
 \qquad
 \cK_{\xi,s}(Y)
       =\sum_{\varnothing\ne X\subseteq Y}\cK_{\xi,s;X}(Y).
\]
The first sum is absolutely convergent, and the integral bound justifies interchanging it with the filter integral.

For the derivation identity, suppose \(\mu_{0,0}(W)<\infty\), and let \(B\in\A_F\) for a nonempty finite \(F\).
Define \(B_{X,Y}(v):=(\cT^{\Phi_\xi}_{s,s+v;X}\Psi_\xi(X))(Y)\).
This coefficient is even and supported in \(Y\), so its commutator with \(B\) vanishes when \(Y\cap F=\varnothing\).
The bound \eqref{eq:card-propagation}, with the sum of norms over \(X\), gives
\[
 \begin{aligned}
 \int_\R|W(v)|\sum_X\sum_{Y\supseteq X}
                         \|[B_{X,Y}(v),B]\|\,\dd v &\le 2\|B\|\sum_{x\in F}\int_\R|W(v)|
       \sum_{Y\ni x}\sum_{\varnothing\ne X\subseteq Y}
                            \|B_{X,Y}(v)\|\,\dd v\\
 & \le 2|F|\|B\|C_{0,0}
                 \|\Psi_\xi\|^{\#}_{0;\kappa_0}\mu_{0,0}(W).
 \end{aligned}
\]
Absolute convergence allows the two sums and the integral to be interchanged.
The preceding identity for each fixed \(X\) then proves \eqref{eq:filtered-derivation}.

If \(\sum_X\E^{\kappa_0|X|}\|\Psi_\xi(X)\|<\infty\), the integral bound for each term \(\Psi_\xi(X)\) instead gives
\[
 \sum_X\sum_{Y\supseteq X}\|\cK_{\xi,s;X}(Y)\|
 \le C_{0,0}\|W\|_{L^1}
       \sum_X\E^{\kappa_0|X|}\|\Psi_\xi(X)\|<\infty.
\]
Reordering this series and using \(\|\cI_{\xi,s}\|\le\|W\|_{L^1}\) proves \eqref{eq:filtered-total-operator}.

\smallskip
\noindent \emph{\ref{item:card-limits-filtered-convergence} Local convergence of filtered interactions.}
We use part~\ref{item:card-limits-continuity} on bounded time intervals and bound the remaining filter tails.
Fix \(0\le q<a\) and \(k\ge0\) with \(\mu_{q,k}(W)<\infty\), a nonempty \(K\Subset\Gamma\), and a compact interval \(J\subseteq\R\).
For \(H>0\), set
\[
 E_n(H):=
 \sup_{\xi\in\frakP}\sup_{s,t\in J_H}
 \|\Upsilon^{(n)}_{\xi,s,t}-\Upsilon^{(\infty)}_{\xi,s,t}
       \|^{\pol,\loc}_{q;k,K}.
\]
Equation~\eqref{eq:card-local-topology} gives \(E_n(H)\to0\) as \(n\to\infty\) for every fixed \(H\).
Since the local seminorm is bounded by the corresponding global norm, the propagation estimate and the common bound \(M_\Psi\) give
\[
 \sup_{\xi\in\frakP}\sup_{s\in J}
 \|\cK^{(n)}_{\xi,s}-\cK_{\xi,s}\|^{\pol,\loc}_{q;k,K}
 \le E_n(H)\int_{-H}^H|W(v)|\,\dd v+2C_{q,k}M_\Psi
       \int_{|v|>H}|W(v)|(1+|v|)^{m_{q,k}}\,\dd v.
\]
First let \(n\to\infty\) at fixed \(H\), and then \(H\to\infty\).
This proves \eqref{eq:filtered-local-topology}.

\smallskip
\noindent \emph{\ref{item:card-limits-stretched} Stretched-exponential support weights.}
Fix \(b>0\) and \(0<\alpha\le1\), and choose \(\theta,b'\) as in the theorem.
For \(h\ge0\) and \(r\ge1\), the weighted arithmetic--geometric mean inequality gives
\[
 b'h^\theta
 =(c_*hr^{-(1+\gamma)\dvol})^\theta
       \bigl((b/2)r^\alpha\bigr)^{1-\theta}
 \le c_*hr^{-(1+\gamma)\dvol}+(b/2)r^\alpha.
\]
For \(r=1+|v|\), apply this inequality with \(h=|Y|\).
The triangle inequality for the coefficient integrals, Tonelli's theorem, and \eqref{eq:exponential-support-propagation} give
\[
 \begin{aligned}
 \sup_x\sum_{Y\ni x}\E^{b'|Y|^\theta}(1+\diam Y)^p
       \sum_{\varnothing\ne X\subseteq Y}\|\cK_{\xi,s;X}(Y)\|
 &\le C_p\|\Psi_\xi\|^{\#}_{p;\kappa_0}
       \int_\R|W(v)|(1+|v|)^{m_{p,0}}
                     \E^{(b/2)(1+|v|)^\alpha}\,\dd v\\
 &\le C_{p,b,\alpha}F_{b,\alpha}(W)
                         \|\Psi_\xi\|^{\#}_{p;\kappa_0},
 \end{aligned}
\]
where \(C_{p,b,\alpha}:=C_p\sup_{r\ge1}r^{m_{p,0}}\E^{-(b/2)r^\alpha} <\infty\).
This proves \eqref{eq:filtered-stretched-support}.
For \(k>0\) and \(h\ge0\),
\[
 h^k\E^{b'h^\theta/2}
 \le\left(\frac{2k}{\E b'\theta}\right)^{k/\theta}
                  \E^{b'h^\theta}.
\]
Applying this inequality with \(h=|Y|\) proves \eqref{eq:filtered-stretched-moments} with \(C_{p,b,\alpha,k}=C_{p,b,\alpha} \max\{1,(2k/(\E b'\theta))^{k/\theta}\}\).
For \(k=0\), take \(C_{p,b,\alpha,0}=C_{p,b,\alpha}\).
\end{proof}

\subsection{Parameter derivatives of interactions}
\label{subsec:differentiability}

In perturbation theory one also needs derivatives of evolved and filtered interactions with respect to parameters.
The following theorem bounds these derivatives, starting from the explicit formula \eqref{eq:full-dynamics-mobius}, and proves their convergence under approximations.
It is used to prove the smooth parameter dependence of flows in Corollary~\ref{cor:filtered-flow}.

\begin{theorem}[Analytic propagation and smooth filtering]
\label{thm:smooth-hamiltonian-dependence}
Assume \eqref{eq:volume-growth} and Hypothesis~\ref{hyp:cell-partitions}, and choose \(\kappa_0\) as in Theorem~\ref{thm:card-propagation}.
Fix \(a>\dvol\) and \(L>0\), and use the propagation exponents for these data.
Set
\begin{equation}
 m^{(n)}_{p,k}:=m_{p,k}
       +n\bigl[1+(1+\gamma)\dvol\bigr],
 \qquad n\in\N_0,\quad 0\le p\le a,\quad k\ge0.
 \label{eq:derivative-time-exponent}
\end{equation}
There are constants \(C_{n,p,k},A_{p,k},B_p>0\), depending only on their indices, \(a,L\), and the geometric constants, with the following properties.
For every choice of local multiplicities, consider the real Banach space of componentwise norm-continuous paths \(\Phi=(\Phi_u)_{u\in\R}\) of self-adjoint interactions with finite norm
\[
 \|\Phi\|^{\#}_{\infty,a;\kappa_0}
 :=\sup_{u\in\R}\|\Phi_u\|^{\#}_{a;\kappa_0},
 \qquad
 B_L:=\{\Phi\,|\,\|\Phi\|^{\#}_{\infty,a;\kappa_0}<L\}.
\]
Choose the partitions and cutoffs in Theorem~\ref{thm:card-propagation} once for \(a,L\), independently of \(\Phi\), and use the resulting maps of Definition~\ref{def:interaction-representatives} for every \(\Phi\in B_L\).

\begin{enumerate}[label=(\roman*),ref=(\roman*),leftmargin=*]
\item \emph{Derivatives of propagated interactions.}
\label{item:smooth-propagation}
For every \(s,t\in\R\), \(0\le p\le a\), and \(k\ge0\), the map \(\Phi\mapsto\cT^\Phi_{s,t}\) is real analytic on \(B_L\) in the operator norm for bounded linear maps from interactions with norm \(\|\cdot\|^{\#}_{p;\kappa_0}\) to interactions with norm \(\|\cdot\|^{\pol}_{p;k}\).
The evaluation \((\Phi,\Psi)\mapsto\cT^\Phi_{s,t}\Psi\) is jointly real analytic in these norms.
For every interaction \(\Psi\) of finite \(\|\cdot\|^{\#}_{p;\kappa_0}\)-norm, \(n\ge0\), \(\Phi\in B_L\), and direction paths \(V_1,\ldots,V_n\) in the Hamiltonian path space, the Hamiltonian derivatives, taken with \(\Psi\) fixed, satisfy
\begin{align}
 \sup_{x\in\Gamma}\sum_{Y\ni x}|Y|^k(1+\diam Y)^p &
       \sum_{\varnothing\ne X\subseteq Y}
 \left\|D_\Phi^n\bigl[(\cT^\Phi_{s,t;X}\Psi(X))(Y)\bigr]
                  [V_1,\ldots,V_n]\right\|
 \nonumber\\*
 &\le C_{n,p,k}(1+|t-s|)^{m^{(n)}_{p,k}}
       \|\Psi\|^{\#}_{p;\kappa_0}
       \prod_{j=1}^n\|V_j\|^{\#}_{\infty,a;\kappa_0}.
 \label{eq:hamiltonian-derivative-bound}
\end{align}
For \(n=0\), the derivative denotes the coefficient itself and the empty product equals one.
The triangle inequality gives the same bound in \(\|\cdot\|^{\pol}_{p;k}\) for \(D_\Phi^n(\cT^\Phi_{s,t}\Psi)[V_1,\ldots,V_n]\).
For \(n\ge1\), the left-hand side of \eqref{eq:hamiltonian-derivative-bound} is also bounded by
\[
A_{p,k}(1+|t-s|)^{m_{p,k}}n!
       \bigl[B_p |t-s|(1+|t-s|)^{(1+\gamma)\dvol}\bigr]^n \|\Psi\|^{\#}_{p;\kappa_0}
       \prod_{j=1}^n\|V_j\|^{\#}_{\infty,a;\kappa_0}.
\]

\item \emph{Filtered derivatives.}
\label{item:smooth-filtering}
Fix \(s\in\R\) and \(0\le p\le a\).
For \(W\in L^1(\R;\C)\), use the map \(\cK_{W,s}^\Phi\) defined by \eqref{eq:card-filter-interaction} on interactions \(\Psi\) of finite \(\|\cdot\|^{\#}_{p;\kappa_0}\)-norm, and set
\[
 \mu^{(n)}_{p,k}(W)
 :=\int_\R|W(v)|(1+|v|)^{m^{(n)}_{p,k}}\,\dd v.
\]
Thus \(\mu^{(0)}_{p,k}(W)=\mu_{p,k}(W)\).
For \(n\ge0\) and \(k\ge0\) with \(\mu^{(n)}_{p,k}(W)<\infty\), \(\Phi\mapsto\cK_{W,s}^\Phi\) is real Fr\'echet \(C^n\) on \(B_L\) in the operator norm of part~\ref{item:smooth-propagation}.
Consequently, \((\Phi,\Psi)\mapsto\cK_{W,s}^\Phi[\Psi]\) is jointly \(C^n\) in the norms of part~\ref{item:smooth-propagation}.
For \(0\le j\le n\), the Hamiltonian derivatives satisfy
\begin{align}
  \bigl(D_\Phi^j\cK_{W,s}^\Phi[\Psi]
                       [V_1,\ldots,V_j]\bigr)(Y)
  =\int_\R W(v)
       \bigl(D_\Phi^j(\cT^\Phi_{s,s+v}\Psi)
                       [V_1,\ldots,V_j]\bigr)(Y)\,\dd v.
 \label{eq:filtered-hamiltonian-derivatives}
\end{align}
Formula~\eqref{eq:filtered-hamiltonian-derivatives} also holds for each input interaction \(\iota_X\Psi(X)\) (see~\eqref{eq:single-support-interaction}).
For every \(\Phi\in B_L\) and direction paths \(V_1,\ldots,V_j\), the individual input contributions satisfy
\begin{align}
 \sup_{x\in\Gamma}\sum_{Y\ni x}|Y|^k(1+\diam Y)^p&
       \sum_{\varnothing\ne X\subseteq Y}
 \left\|\int_\R W(v)
       D_\Phi^j\bigl[(\cT^\Phi_{s,s+v;X}\Psi(X))(Y)\bigr]
       [V_1,\ldots,V_j]\,\dd v\right\|
 \nonumber\\*
 &\le C_{j,p,k}\mu^{(j)}_{p,k}(W)
       \|\Psi\|^{\#}_{p;\kappa_0}
       \prod_{i=1}^j\|V_i\|^{\#}_{\infty,a;\kappa_0}.
 \label{eq:filtered-hamiltonian-derivative-bound}
\end{align}
By the triangle inequality, the same right-hand side bounds \(\|D_\Phi^j\cK_{W,s}^\Phi[\Psi][V_1,\ldots,V_j]\|^{\pol}_{p;k}\).

If \(W\) has finite absolute moments of every order, the joint map is \(C^\infty\) in \(\|\cdot\|^{\pol}_{p;k}\) for every \(k\ge0\).
If \(W\) has compact support, both \(\Phi\mapsto\cK_{W,s}^\Phi\) and \((\Phi,\Psi)\mapsto\cK_{W,s}^\Phi[\Psi]\) are real analytic in these norms.

\item \emph{Convergence of parameter derivatives.}
\label{item:smooth-approximation}
Fix \(r\in\N_0\) and a compact parameter interval \(J\).
Let \(\lambda\mapsto\Phi_\lambda^{(N)}\) and \(\lambda\mapsto\Psi_\lambda^{(N)}\) be \(C^r\) maps from a neighborhood of \(J\) into, respectively, the Hamiltonian path space and the interactions with norm \(\|\cdot\|^{\#}_{a;\kappa_0}\).
Include the limiting families \(\Phi_\lambda,\Psi_\lambda\) as the case \(N=\infty\).
Assume for all \(N\in\N\cup\{\infty\}\) and \(\lambda\in J\) that \( \Phi_\lambda^{(N)}\in B_L\), and
\[
 \begin{gathered}
 \sup_{N\in\N\cup\{\infty\}}\sup_{\lambda\in J}
       \|\partial_\lambda^j\Psi_\lambda^{(N)}
                         \|^{\#}_{a;\kappa_0}<\infty,
 \qquad 0\le j\le r,\\
 \sup_{N\in\N\cup\{\infty\}}\sup_{\lambda\in J}
       \|\partial_\lambda^j\Phi_\lambda^{(N)}
                   \|^{\#}_{\infty,a;\kappa_0}<\infty,
 \qquad 1\le j\le r.
 \end{gathered}
\]
Suppose that, for every \(0\le j\le r\), nonempty finite \(Z\), and compact physical-time interval \(I\subseteq\R\),
\[
 \begin{aligned}
 \sup_{\lambda\in J}\int_I
   \|\partial_\lambda^j\Phi_{\lambda,u}^{(N)}(Z)
                   -\partial_\lambda^j\Phi_{\lambda,u}(Z)\|\,\dd u
 &\xrightarrow{N\to\infty}0,\\
 \sup_{\lambda\in J}
   \|\partial_\lambda^j\Psi_\lambda^{(N)}(Z)
                    -\partial_\lambda^j\Psi_\lambda(Z)\|
 &\xrightarrow{N\to\infty}0.
 \end{aligned}
\]
For every \(0\le n\le r\), \(0\le q<a\), \(k\ge0\), nonempty finite \(K\), and compact physical-time interval \(I\subseteq\R\),
\begin{equation}
 \sup_{\lambda\in J}\sup_{s,t\in I}
 \left\|\partial_\lambda^n
   \bigl(\cT^{\Phi_\lambda^{(N)}}_{s,t}\Psi_\lambda^{(N)}
              -\cT^{\Phi_\lambda}_{s,t}\Psi_\lambda\bigr)
       \right\|^{\pol,\loc}_{q;k,K}\xrightarrow{N\to\infty}0.
 \label{eq:differentiated-local-convergence}
\end{equation}
For every \(W\in L^1(\R;\C)\) with \(\mu^{(n)}_{q,k}(W)<\infty\), one also has
\begin{equation}
 \sup_{\lambda\in J}\sup_{s\in I}
 \left\|\partial_\lambda^n
   \bigl(\cK_{W,s}^{\Phi_\lambda^{(N)}}[\Psi_\lambda^{(N)}]
              -\cK_{W,s}^{\Phi_\lambda}[\Psi_\lambda]\bigr)
       \right\|^{\pol,\loc}_{q;k,K}\xrightarrow{N\to\infty}0.
 \label{eq:differentiated-filtered-convergence}
\end{equation}
\end{enumerate}
\end{theorem}

\begin{proof}
We differentiate the finite-volume evolutions in \eqref{eq:full-dynamics-mobius} on a fixed cell partition and express the derivatives as time integrals of nested interaction brackets.
The estimates of Proposition~\ref{prop:fixed-scale} bound the evolved inputs to these brackets, and the one-commutator estimates \eqref{eq:card-bracket}--\eqref{eq:card-bracket-total} control each additional bracket.
Factorial bounds then give analyticity, and the filtered statements follow by differentiating under the filter integral.

\smallskip
\noindent \emph{Derivatives on a fixed cell partition.}
The translated path \(\Phi^{[s]}_u:=\Phi_{s+u}\) has the same path norm as \(\Phi\), and \eqref{eq:full-dynamics-mobius} and \eqref{eq:source-map-construction} give \(\cT^\Phi_{s,t}=\cT^{\Phi^{[s]}}_{0,t-s}\).

We prove parts~\ref{item:smooth-propagation} and~\ref{item:smooth-filtering} for \(s=0\).
Fix \(\Phi\in B_L\), \(0\le p\le a\), \(k\ge0\), and \(n\ge1\).
Use the partitions and cutoffs fixed for \(a,L\) in the statement, also for nearby Hamiltonian paths in \(B_L\).
At a fixed time scale \(T\ge1\), let \(\ell=\ell(T)\) be the scale of Proposition~\ref{prop:fixed-scale}.
For a fixed interaction \(\Psi\), every nonempty finite input support \(X\), and directions \(V_1,\ldots,V_n\), define the cell interactions
\[
 \begin{aligned}
 \Upsilon_{t;X}(S)&:=
       (\cT^{\Phi,T}_{0,t;X}\Psi(X))(Q(S)),\\
 \Xi_{j,u}(S)&:=
   \sum_{\varnothing\ne Z\subseteq Q(S)}
       (\cT^{\Phi,T}_{0,u;Z}V_{j,u}(Z))(Q(S)).
 \end{aligned}
\]
Put
\[
 M_{V_j}:=\|V_j\|^{\#}_{\infty,a;\kappa_0},
 \qquad
 g_0(S):=\sum_{\varnothing\ne X\subseteq Q(S)}
                       \|\Upsilon_{t;X}(S)\|.
\]
For nonnegative scalar families such as \(g_0\), use the weighted sums defined in Subsection~\ref{subsec:card-conjugation}.
The estimate \eqref{eq:fixed-scale-cell-bound} of Proposition~\ref{prop:fixed-scale} gives, for \(q=0,p\) and \(|t|,|u|\le T\),
\begin{equation}
 \begin{aligned}
 \|g_0\|^{\#,\ell}_{q;3}\le C_p\ell^{\dvol+q}\|\Psi\|^{\#}_{p;\kappa_0},\qquad
 \|\Xi_{j,u}\|^{\#,\ell}_{q;3}\le C_p\ell^{\dvol+q}M_{V_j}.
 \end{aligned}
 \label{eq:derivative-fixed-cell-inputs}
\end{equation}
For the finite-volume calculation, fix \(X\), choose a finite cell set \(\cS\supseteq S_0(X)\), and put
\[
 \Omega:=Q(\cS),\qquad
 H_{V_{j,u}}^\Omega:=\sum_{Z\subseteq\Omega}V_{j,u}(Z).
\]
We use the finite-dimensional form of the parameter-dependent Duhamel expansion; see Nachtergaele--Sims--Young \cite[Section~2.3.2, Proposition~2.6]{NSY19}.
For \(\varepsilon=(\varepsilon_1,\ldots,\varepsilon_n)\in\R^n\), let \(U_\varepsilon^\Omega(u,0)\) be the propagator generated by \(\Phi^\varepsilon_u:=\Phi_u+\sum_j\varepsilon_jV_{j,u}\), and put
\[
 C_\varepsilon^\Omega(u):=
          U_\varepsilon^\Omega(u,0)^*U_0^\Omega(u,0).
\]
Differentiation gives
\[
 (C_\varepsilon^\Omega)'(u)
 =\I C_\varepsilon^\Omega(u)
       \sum_j\varepsilon_j\tau_{0,u}^{\Phi,\Omega}(H_{V_{j,u}}^\Omega),
 \qquad C_\varepsilon^\Omega(0)=\1.
\]
Moreover,
\[
 \tau_{0,t}^{\Phi^\varepsilon,\Omega}(\Psi(X))
 =C_\varepsilon^\Omega(t)\,
    \tau_{0,t}^{\Phi,\Omega}(\Psi(X))\,
       C_\varepsilon^\Omega(t)^*.
\]
Summing \eqref{eq:card-telescope} over the direction supports \(Z\subseteq\Omega\) gives
\[
 \tau_{0,u}^{\Phi,\Omega}(H_{V_{j,u}}^\Omega)
       =\sum_{S\subseteq\cS}\Xi_{j,u}(S).
\]
For the fixed input support \(X\), the same identity gives
\[
 \tau_{0,t}^{\Phi,\Omega}(\Psi(X))
       =\sum_{S\subseteq\cS}\Upsilon_{t;X}(S).
\]
Thus the series \eqref{eq:cell-conjugation-series} gives the mixed derivative with the following time ordering.
Write \(t=\sigma h\), with \(h=|t|\) and \(\sigma\in\{-1,1\}\), and let \(\mathfrak S_n\) be the permutations of \(\{1,\ldots,n\}\).
With \(\ad_\Xi\Upsilon:=[\Xi,\Upsilon]\), define
\begin{equation}
 \cP_{n,t;X}^{\Phi,T}
 :=(\I\sigma)^n\sum_{\pi\in\mathfrak S_n}
   \int_{0\le u_n\le\cdots\le u_1\le h}
   \ad_{\Xi_{\pi(n),\sigma u_n}}\cdots
   \ad_{\Xi_{\pi(1),\sigma u_1}}\Upsilon_{t;X}
      \,\dd u_1\cdots\dd u_n.
 \label{eq:mixed-derivative-cell-formula}
\end{equation}
The bracket \([\Xi,\Upsilon]\) is defined in \eqref{eq:cell-interaction-bracket}.
For \(n\ge1\), set \(\cP_{n,0;X}^{\Phi,T}=0\).
Every coefficient of the integrand is a finite sum, and restriction to subsets of \(\cS\) commutes with the bracket.
Consequently,
\[
 \sum_{S\subseteq\cS}\cP_{n,t;X}^{\Phi,T}(S)
 =\left.\partial_{\varepsilon_1}\cdots\partial_{\varepsilon_n}
   \tau_{0,t}^{\Phi+\sum_j\varepsilon_jV_j,\Omega}(\Psi(X))
   \right|_{\varepsilon=0}.
\]
Applying M\"obius inversion on finite cell sets containing \(S_0(X)\) gives
\[
 \cP_{n,t;X}^{\Phi,T}(S)
 =D_\Phi^n\bigl[(\cT^{\Phi,T}_{0,t;X}\Psi(X))(Q(S))\bigr]
                  [V_1,\ldots,V_n].
\]

\smallskip
\noindent \emph{Derivative bounds and analyticity.}
Fix a permutation \(\pi\in\mathfrak S_n\) and times \(0\le u_n\le\cdots\le u_1\le h\).
For each input support \(X\), the nested commutator in \eqref{eq:mixed-derivative-cell-formula} is \(\Upsilon_X^{[n]}\) in the recursion
\[
 \begin{gathered}
 \rho_j:=3-j/n\quad(0\le j\le n),\qquad
 \Upsilon_X^{[0]}:=\Upsilon_{t;X}, \qquad
 \Upsilon_X^{[j]}:=[\Xi_{\pi(j),\sigma u_j},\Upsilon_X^{[j-1]}]
       \quad(1\le j\le n).
 \end{gathered}
\]
The bracket assigns terms to unions of their supports, so \(\Upsilon_X^{[j]}(S)=0\) unless \(X\subseteq Q(S)\).
Define, for \(1\le j\le n\),
\[
 g_j(S):=\sum_{\varnothing\ne X\subseteq Q(S)}
                          \|\Upsilon_X^{[j]}(S)\|,
\]
and, for \(0\le j\le n\), put
\[
 a_j:=\|g_j\|^{\#,\ell}_{0;\rho_j},\quad
 b_j:=\|g_j\|^{\#,\ell}_{p;\rho_j}.
\]
Bounding each commutator by twice the product of norms and then summing over \(X\) gives the pointwise inequality
\[
 g_j(S)\le
 2\sum_{\substack{U\cup V=S\\U\cap V\ne\varnothing}}
       \|\Xi_{\pi(j),\sigma u_j}(U)\|g_{j-1}(V).
\]
Apply the positive overlap estimates used in \eqref{eq:card-bracket} with \(\rho=\rho_{j-1}\) and \(\delta=1/n\).
Together with \eqref{eq:derivative-fixed-cell-inputs}, they give, after increasing \(C_p\),
\[
 a_j\le C_p n\ell^{\dvol}M_{V_{\pi(j)}}a_{j-1},\qquad
 b_j\le C_p nM_{V_{\pi(j)}}
      \bigl(\ell^{\dvol}b_{j-1}
             +\ell^{\dvol+p}a_{j-1}\bigr).
\]
Starting from \eqref{eq:derivative-fixed-cell-inputs}, induction yields
\[
 \begin{aligned}
 a_j&\le C_p(C_pn)^j\ell^{(j+1)\dvol}
       \|\Psi\|^{\#}_{p;\kappa_0}\prod_{r=1}^jM_{V_{\pi(r)}},\\
 b_j&\le(j+1)C_p(C_pn)^j
       \ell^{(j+1)\dvol+p}\|\Psi\|^{\#}_{p;\kappa_0}\prod_{r=1}^jM_{V_{\pi(r)}}.
 \end{aligned}
\]
There are \(n!\) permutations and the simplex has volume \(h^n/n!\).
The triangle inequality for each coefficient integral and Tonelli's theorem therefore give
\begin{equation}
  \sup_{b\in\cB_\ell}\sum_{S\ni b}\E^{2|S|}(1+\diam Q(S))^p
       \sum_{\varnothing\ne X\subseteq Q(S)}
                         \|\cP_{n,t;X}^{\Phi,T}(S)\|
 \le C_{n,p}|t|^n\ell^{(n+1)\dvol+p}
                         \|\Psi\|^{\#}_{p;\kappa_0}\prod_{j=1}^nM_{V_j}.
 \label{eq:derivative-fixed-cell-bound}
\end{equation}

For \(Y=Q(S)\), use \(|Y|^k\le C_k\ell^{k\dvol}|S|^k\) and \(\sup_{m\ge1}m^k\E^{-2m}<\infty\), as in \eqref{eq:cell-to-site-moments}.
This multiplies the right-hand side of \eqref{eq:derivative-fixed-cell-bound} by \(C_k\ell^{k\dvol}\) and keeps the sum of norms over \(X\).
The cutoffs \(\chi_\nu\) are independent of \(\Phi\), so coefficient differentiation commutes with the locally finite sum \eqref{eq:source-map-construction}.
For every input support \(X\) and output support \(Y\), the triangle inequality gives
\[
 \left\|D_\Phi^n\bigl[(\cT^\Phi_{0,t;X}\Psi(X))(Y)\bigr]
                  [V_1,\ldots,V_n]\right\| \le\sum_{\nu\ge0}\chi_\nu(t)
 \left\|D_\Phi^n\bigl[(\cT^{\Phi,T_\nu}_{0,t;X}\Psi(X))(Y)\bigr]
                  [V_1,\ldots,V_n]\right\|.
\]
Multiply this inequality by \(|Y|^k(1+\diam Y)^p\), sum over \(\varnothing\ne X\subseteq Y\) and \(Y\ni x\), and take the supremum over \(x\).
Using the bound on \(b_n\) in the derivation of \eqref{eq:derivative-fixed-cell-bound} and the conversion from cells to sites above, we bound the left-hand side of \eqref{eq:hamiltonian-derivative-bound} by
\[
 C_kC_p(n+1)(C_pn)^n|t|^n
 \sum_{\nu\ge0}\chi_\nu(t)\ell(T_\nu)^{(n+k+1)\dvol+p}
 \|\Psi\|^{\#}_{p;\kappa_0}\prod_{j=1}^n M_{V_j}.
\]
Here \(C_k\) includes the factor \(\sup_{m\ge1}m^k\E^{-2m}\) and is independent of \(n\).
For every index with \(\chi_\nu(t)>0\), \eqref{eq:scale-growth} and the support condition on \(\chi_\nu\) give
\[
 \ell(T_\nu)\le C_{R_*}(1+T_\nu)^{1+\gamma}
 \le 4^{1+\gamma}C_{R_*}(1+|t|)^{1+\gamma}.
\]
Since \(\sum_\nu\chi_\nu(t)=1\), substitution in the preceding bound and \(|t|^n\le(1+|t|)^n\) prove \eqref{eq:hamiltonian-derivative-bound} for the coefficientwise derivatives, with time exponent
\[
 n+(1+\gamma)((n+k+1)\dvol+p)=m^{(n)}_{p,k}.
\]
For \(n=0\), use \eqref{eq:card-propagation}.
For the sharper bound in part~\ref{item:smooth-propagation}, keep \(|t|^n\) and use
\[
 (n+1)(C_pn)^n\le(2\E C_p)^n n!,\qquad n\ge1.
\]
The power of \(1+|t|\) contributed by the cell scale is \(m_{p,k}+n(1+\gamma)\dvol\).
Thus one may choose, independently of \(n\),
\[
 A_{p,k}:=C_kC_p
       \bigl(4^{1+\gamma}C_{R_*}\bigr)^{(k+1)\dvol+p},\qquad
 B_p :=2\E C_p\bigl(4^{1+\gamma}C_{R_*}\bigr)^{\dvol}.
\]
These choices give the factorial bound for the same sum of norms over \(X\) and \(Y\) as in \eqref{eq:hamiltonian-derivative-bound}.

To prove analyticity in operator norm, we show that the Taylor series formed from the coefficientwise derivatives converges in that norm to the propagation map.
For \(n\ge1\), define the linear maps on interactions by
\[
 \bigl(Q_{n,t}^\Phi[V_1,\ldots,V_n]\Psi\bigr)(Y)
 :=\left.\partial_{\varepsilon_1}\cdots\partial_{\varepsilon_n}
       \bigl(\cT^{\Phi+\sum_j\varepsilon_jV_j}_{0,t}\Psi\bigr)(Y)
   \right|_{\varepsilon=0},
\]
and set \(Q_{0,t}^\Phi:=\cT^\Phi_{0,t}\).
The preceding estimates make each \(Q_{n,t}^\Phi\) a bounded symmetric \(n\)-linear map on the Hamiltonian path space, with values in the operator space of part~\ref{item:smooth-propagation}.
Write \(\|\cdot\|_{\rm op}\) for its operator norm and the induced multilinear norms.
The factorial bound proves absolute convergence of
\[
 \sum_{n=0}^\infty\frac1{n!}Q_{n,t}^\Phi[H,\ldots,H]
\]
in operator norm, locally uniformly for \(H\) satisfying
\[
 \Phi+H\in B_L,\qquad
 B_p|t|(1+|t|)^{(1+\gamma)\dvol}
                         \|H\|^{\#}_{\infty,a;\kappa_0}<1.
\]
For each interaction \(\Psi\) and output support \(Y\), formula \eqref{eq:full-dynamics-mobius}, summed over \(X\subseteq Y\) and the contributing scales, involves finitely many finite-volume evolutions.
Their parameter-dependent Dyson expansions give
\[
 \sum_{n=0}^\infty\frac1{n!}
       \bigl(Q_{n,t}^\Phi[H,\ldots,H]\Psi\bigr)(Y)
 =(\cT^{\Phi+H}_{0,t}\Psi)(Y).
\]
Equality of every output coefficient identifies the operator-norm sum with \(\cT^{\Phi+H}_{0,t}\).
This proves real analyticity in operator norm and identifies its Fr\'echet derivatives with \(Q_{n,t}^\Phi\).
Applying this conclusion to \(\iota_X\Psi(X)\) (see~\eqref{eq:single-support-interaction}) and evaluating each output coefficient shows that the derivatives in \eqref{eq:hamiltonian-derivative-bound} are real Fr\'echet derivatives.

Since evaluation of a bounded linear map at its argument is continuous and bilinear, \((\Phi,\Psi)\mapsto\cT^\Phi_{0,t}\Psi\) is jointly real analytic.

\smallskip
\noindent \emph{Differentiation under the filter integral.}
Let \(W\in L^1(\R;\C)\).
The uniform total bound \eqref{eq:card-propagation-total} at \(p=k=0\) gives
\[
 \sum_{\varnothing\ne X\subseteq Y}
       \|(\cT^\Phi_{0,t;X}\Psi(X))(Y)\|
 \le C_{0,0}\sum_{\varnothing\ne X\subseteq Y}
              \E^{\kappa_0|X|}\|\Psi(X)\|.
\]
Thus the coefficient integrals exist for individual input supports and for their finite sum at each \(Y\).

Fix \(n\in\N_0\) with \(\mu^{(n)}_{p,k}(W)<\infty\).
For \(0\le j\le n\), define the multilinear maps \(\overline Q_j^\Phi\) by
\[
 \bigl(\overline Q_j^\Phi[V_1,\ldots,V_j]\Psi\bigr)(Y)
 :=\int_\R W(t)
       \bigl(Q_{j,t}^\Phi[V_1,\ldots,V_j]\Psi\bigr)(Y)\,\dd t.
\]
For each fixed \(x\), bound the norm of every integral in \eqref{eq:filtered-hamiltonian-derivative-bound} by its integral of norms, and move the nonnegative sums over \(Y\ni x\) and \(X\subseteq Y\) inside the integral by Tonelli's theorem.
The bound \eqref{eq:hamiltonian-derivative-bound} at derivative order \(j\) is uniform in \(x\), so integration against \(|W(t)|\) gives \eqref{eq:filtered-hamiltonian-derivative-bound}.
The triangle inequality after summing the input contributions then gives \(\|\overline Q_j^\Phi\|_{\rm op}\le C_{j,p,k}\mu^{(j)}_{p,k}(W)\).

For continuity, the estimate of order \(j\) dominates the integrand, while the estimate of order \(j+1\) controls the difference at each fixed time.
For \(\Phi,\Phi+H\in B_L\), convexity of \(B_L\) gives \(\Phi+rH\in B_L\) for \(0\le r\le1\).
For \(0\le j\le n\), the first-order integral formula along this segment and the bounds for \(Q_{j,t}\) and \(Q_{j+1,t}\) give
\[
 \begin{aligned}
 \|\overline Q_j^{\Phi+H}-\overline Q_j^\Phi\|_{\rm op} \le \int_\R |W(t)|\min\!\left\{
       2C_{j,p,k}(1+|t|)^{m^{(j)}_{p,k}},
       C_{j+1,p,k}(1+|t|)^{m^{(j+1)}_{p,k}}
                          \|H\|^{\#}_{\infty,a;\kappa_0}\right\}\,\dd t.
 \end{aligned}
\]
Dominated convergence, using only \(\mu^{(j)}_{p,k}(W)<\infty\), proves continuity of \(\overline Q_j\) in multilinear operator norm.
For \(1\le j\le n\), the coefficientwise identity
\[
 (Q_{j-1,t}^{\Phi+H}-Q_{j-1,t}^\Phi)[V_1,\ldots,V_{j-1}]
 =\int_0^1Q_{j,t}^{\Phi+rH}[V_1,\ldots,V_{j-1},H]\,\dd r
\]
can be integrated against \(W\), giving
\[
 \|\overline Q_{j-1}^{\Phi+H}-\overline Q_{j-1}^\Phi
                     -\overline Q_j^\Phi[\,\cdot\,,H]\|_{\rm op} \le \|H\|^{\#}_{\infty,a;\kappa_0}
       \sup_{0\le r\le1}
         \|\overline Q_j^{\Phi+rH}-\overline Q_j^\Phi\|_{\rm op}
       =o(\|H\|^{\#}_{\infty,a;\kappa_0}).
\]
Thus \(D_\Phi\overline Q_{j-1}=\overline Q_j\), proving \eqref{eq:filtered-hamiltonian-derivatives} using only \(\mu^{(j)}_{p,k}(W)\).
Applying this argument to \(\iota_X\Psi(X)\) proves the coefficient derivative identity for each input support separately.

Since evaluation is continuous and bilinear, \((\Phi,\Psi)\mapsto\cK^\Phi_{W,0}[\Psi]\) is jointly \(C^n\).
If \(W\) is supported in \([-R,R]\), the maps \(\overline Q_j^\Phi\) are defined for every \(j\ge0\).
The factorial bound permits termwise integration of the propagation series whenever \(\Phi+H\in B_L\) and \(B_pR(1+R)^{(1+\gamma)\dvol} \|H\|^{\#}_{\infty,a;\kappa_0}<1\).
Termwise integration gives
\[
 \cK_{W,0}^{\Phi+H}
   =\sum_{j=0}^\infty\frac1{j!}\overline Q_j^\Phi[H,\ldots,H]
\]
with convergence in operator norm.
This proves analyticity and, by evaluation at \(\Psi\), joint analyticity.

\smallskip
\noindent \emph{Convergence of parameter derivatives.}
We first obtain a uniform bound at spatial exponent \(a\).
We then prove convergence on each finite region and use the bound at exponent \(a\) to estimate the contributions from supports of large diameter at exponents \(q<a\).
Choose \(M_{\rm der}\ge1\) bounding the derivatives of \(\Psi_\lambda^{(N)}\) of orders \(0,\ldots,r\) and of \(\Phi_\lambda^{(N)}\) of orders \(1,\ldots,r\) in the norms of part~\ref{item:smooth-approximation}.
For \(0\le n\le r\), repeated product and chain rules express \(\partial_\lambda^n (\cT^{\Phi_\lambda^{(N)}}_{s,t}\Psi_\lambda^{(N)})\) as a finite linear combination, with coefficients depending only on \(n\), of terms
\[
 \left.
 D_\Phi^\nu
   (\cT^\Phi_{s,t}\partial_\lambda^j\Psi_\lambda^{(N)})
   [\partial_\lambda^{r_1}\Phi_\lambda^{(N)},\ldots,
    \partial_\lambda^{r_\nu}\Phi_\lambda^{(N)}]
 \right|_{\Phi=\Phi_\lambda^{(N)}},
 \quad
 j,\nu\in\N_0,\quad j+\sum_{i=1}^\nu r_i=n,
 \quad r_i\in\N.
\]
For \(\nu=0\), this term is \(\cT^{\Phi_\lambda^{(N)}}_{s,t} \partial_\lambda^n\Psi_\lambda^{(N)}\).
For \(n<r\), differentiating any such term increases \(j\), increases one \(r_i\), or introduces a new direction \(\partial_\lambda\Phi_\lambda^{(N)}\), which proves the displayed expansion by induction on \(n\).
For \(0\le p\le a\) and \(k\ge0\), \eqref{eq:hamiltonian-derivative-bound} bounds each such term by
\[
 C_{\nu,p,k}(1+|t-s|)^{m^{(\nu)}_{p,k}}
 \|\partial_\lambda^j\Psi_\lambda^{(N)}\|^{\#}_{p;\kappa_0}
 \prod_{i=1}^\nu
 \|\partial_\lambda^{r_i}\Phi_\lambda^{(N)}\|^{\#}_{\infty,a;\kappa_0}
 \le C_{\nu,p,k}(1+|t-s|)^{m^{(n)}_{p,k}}M_{\rm der}^{n+1},
\]
using \(\|\cdot\|^{\#}_{p;\kappa_0} \le\|\cdot\|^{\#}_{a;\kappa_0}\), \(\nu\le n\), \(m^{(\nu)}_{p,k}\le m^{(n)}_{p,k}\), and \(M_{\rm der}\ge1\).
Since the number of terms and their coefficients depend only on \(n\), there are constants \(C'_{n,p,k}\), depending only on \(n,p,k,a,L\), and the geometric constants, such that, for \(0\le n\le r\), \(0\le p\le a\), and \(k\ge0\),
\[
 \sup_{N\in\N\cup\{\infty\}}\sup_{\lambda\in J}
 \|\partial_\lambda^n
       (\cT^{\Phi_\lambda^{(N)}}_{s,t}\Psi_\lambda^{(N)})
       \|^{\pol}_{p;k}
 \le C'_{n,p,k}M_{\rm der}^{n+1}(1+|t-s|)^{m^{(n)}_{p,k}}.
\]
Fix a finite region \(\Omega\) and a compact physical-time interval \(I\).
Let \(U_\lambda^{(N),\Omega}(t,s)\) be the finite-volume propagator of \(\Phi_\lambda^{(N)}\).
For \(0\le j\le r\), define
\[
 H_{\lambda,j}^{(N)}(u)
 :=\sum_{Z\subseteq\Omega}
                \partial_\lambda^j\Phi_{\lambda,u}^{(N)}(Z),
 \qquad
 U_{\lambda,j}^{(N)}(t,s)
 :=\partial_\lambda^jU_\lambda^{(N),\Omega}(t,s).
\]
Here \(N=\infty\) denotes the limiting Hamiltonian.
Differentiating the finite-dimensional evolution equation and using variation of constants gives, for \(1\le j\le r\),
\[
 U_{\lambda,j}^{(N)}(t,s)
 =-\I\sum_{i=1}^j\binom ji\int_s^t
 U_{\lambda,0}^{(N)}(t,u)H_{\lambda,i}^{(N)}(u)
 U_{\lambda,j-i}^{(N)}(u,s)\,\dd u.
\]
The initial values are \(U_{\lambda,0}^{(N)}(s,s)=\1\) and \(U_{\lambda,j}^{(N)}(s,s)=0\) for \(j\ge1\); the oriented integral also applies when \(t<s\).
Put
\[
 \begin{aligned}
 b_j&:=\sup_{N\in\N\cup\{\infty\}}\sup_{\lambda\in J}\int_I
                         \|H_{\lambda,j}^{(N)}(u)\|\,\dd u<\infty,\\
 e_{N,j}&:=\sup_{\lambda\in J}\int_I
       \|H_{\lambda,j}^{(N)}(u)-H_{\lambda,j}^{(\infty)}(u)\|\,\dd u
                    \xrightarrow{N\to\infty}0,\\
 D_{N,j}&:=\sup_{\lambda\in J}\sup_{s,t\in I}
       \|U_{\lambda,j}^{(N)}(t,s)-U_{\lambda,j}^{(\infty)}(t,s)\|.
 \end{aligned}
\]
Finiteness of \(b_j\) and convergence of \(e_{N,j}\) follow from the uniform norm bounds and the finite sum over \(Z\subseteq\Omega\).
Unitarity and the recursion bound \(\|U_{\lambda,j}^{(N)}(t,s)\|\) by \(c_j\), where
\[
 c_0:=1,\qquad c_j:=\sum_{i=1}^j\binom ji b_i c_{j-i}.
\]
Duhamel's formula gives \(D_{N,0}\le e_{N,0}\).
Subtracting the two recursions and expanding each product difference gives
\[
 D_{N,j}\le\sum_{i=1}^j\binom ji
 \bigl(b_i c_{j-i}D_{N,0}
       +c_{j-i}e_{N,i}+b_i D_{N,j-i}\bigr).
\]
Induction proves \(D_{N,j}\to0\) as \(N\to\infty\) for \(0\le j\le r\).
The finite sums \(\partial_\lambda^jH_{\Psi_\lambda^{(N)}}^\Omega =\sum_{Z\subseteq\Omega}\partial_\lambda^j\Psi_\lambda^{(N)}(Z)\) converge in norm to \(\partial_\lambda^jH_{\Psi_\lambda}^\Omega\), uniformly on \(J\), for \(0\le j\le r\).
Leibniz' rule gives
\[
 \partial_\lambda^n
   \tau_{s,t}^{\Phi_\lambda^{(N)},\Omega}(H_{\Psi_\lambda^{(N)}}^\Omega)
 =\sum_{i_1+i_2+i_3=n}\frac{n!}{i_1!i_2!i_3!}
       U_{\lambda,i_1}^{(N)}(t,s)^*
       \bigl(\partial_\lambda^{i_2}H_{\Psi_\lambda^{(N)}}^\Omega\bigr)
                                      U_{\lambda,i_3}^{(N)}(t,s).
\]
Every factor is uniformly bounded and converges uniformly, so the sum converges uniformly on \(J\times I^2\) for \(0\le n\le r\).
For a fixed output support \(Y\), formula \eqref{eq:full-dynamics-mobius} uses only input supports \(X\subseteq Y\) and finite regions \(Q(U)\subseteq Y\).
Apply the preceding argument to the interaction families \(\iota_X\Psi_\lambda^{(N)}(X)\) (see~\eqref{eq:single-support-interaction}) in each of these regions.
There are finitely many such terms and, on \(I^2\), finitely many contributing scales in \eqref{eq:source-map-construction}.
Summing their differentiated coefficients proves uniform convergence for the output support \(Y\).
The bound obtained from the chain rule with \(p=a\) gives a common \(\|\cdot\|^{\pol}_{a;k}\)-bound on \(J\times I^2\).
Lemma~\ref{lem:componentwise-local-continuity} then proves \eqref{eq:differentiated-local-convergence}.
For filtered derivatives, split the integral at \(|v|=T\).
Apply \eqref{eq:differentiated-local-convergence} on \(I+[-T,T]\) to control the part with \(|v|\le T\), uniformly in \(\lambda\in J\) and \(s\in I\).
The difference of the tails is bounded, uniformly in \(\lambda,s,N\), by
\[
 2C'_{n,q,k}M_{\rm der}^{n+1}
 \int_{|v|>T}|W(v)|(1+|v|)^{m^{(n)}_{q,k}}\,\dd v.
\]
The tail bound tends to zero as \(T\to\infty\) under the assumption \(\mu^{(n)}_{q,k}(W)<\infty\), proving \eqref{eq:differentiated-filtered-convergence}.
\end{proof}

\subsection{Iterated filtering and commutators}
\label{subsec:iterated-filtering}

Higher-order adiabatic constructions involve repeated applications of quasi-adiabatic filters.
Teufel--Wessel \cite[Section~4, after Theorem~9]{TeufelWessel25} show that their estimate requires one additional polynomial support-size moment at each filtering step, improving earlier bounds.
Repeated application nevertheless requires stronger spatial decay of the Hamiltonian as the number of steps increases.
Under the exponential support-size assumptions of Theorem~\ref{thm:card-propagation}, the next proposition constructs interactions representing any finite expression formed by time filtering and commutators, without strengthening the Hamiltonian decay assumption or losing the input spatial exponent.
For example, consider the formula
\[
 E(A_1,A_2,A_3)
 :=\cI_f\!\left([A_1,\cI_g([A_2,A_3])]\right),
\]
where \(\cI_f\) denotes time filtering with kernel \(f\).

\begin{proposition}
\label{prop:iterated-filters}
Assume \eqref{eq:volume-growth} and Hypothesis~\ref{hyp:cell-partitions}, and choose \(\kappa_0\) as in Theorem~\ref{thm:card-propagation}.
Fix \(a>\dvol\), \(L\ge0\), and \(0\le p\le a\).
Let \(m\ge1\), and let \(E=E(A_1,\ldots,A_m)\) be a finite formula built from the symbols \(A_1,\ldots,A_m\) by the operations \([E_1,E_2]\) and \(\cI_f(E_1)\), with each symbol \(A_j\) appearing exactly once.
Assume that every filter kernel occurring in \(E\) belongs to \(L^1(\R;\C)\).
For every \(k\ge0\) such that
\begin{equation}
 \int_\R|f(t)|(1+|t|)^{\,m\,m_{p,k+m-1}}\,\dd t<\infty
 \quad\text{for every filter in }E,
 \label{eq:iterated-filter-moment}
\end{equation}
there is a constant \(C_{E,p,k}>0\) for the estimate in part~\ref{item:iterated-estimates}, depending only on the arrangement of filters and commutators in \(E\), the displayed absolute moments, and \(a,L,p,k,\dvol,C_\Gamma,\Ccell,\Dcell\).
Expressions without filters are included, with the moment conditions understood as vacuous.
Fix the partitions and cutoffs of Definition~\ref{def:interaction-representatives} as in Theorem~\ref{thm:card-propagation} for \(a,L\).

Choose local multiplicities, an autonomous Hamiltonian interaction \(\Phi\) with \(\|\Phi\|^{\#}_{a;\kappa_0}\le L\), and interactions \(\Psi_1,\ldots,\Psi_m\), not necessarily distinct, with finite \(\|\cdot\|^{\#}_{p;\kappa_0}\)-norms.
Interpret the filters on operators by
\[
 \cI_f(B):=\cI^\Phi_{f,0}(B)=\int_\R f(t)\tau_t^\Phi(B)\,\dd t,
 \qquad B\in\A.
\]
For \(\Lambda\Subset\Gamma\), define
\[
 \Psi_j^\Lambda(X):=\1_{\{X\subseteq\Lambda\}}\Psi_j(X),
 \qquad
 A_j^\Lambda:=\sum_{\varnothing\ne X\subseteq\Lambda}\Psi_j(X),
 \qquad 1\le j\le m,
\]
and set \(E_\Lambda:=E(A_1^\Lambda,\ldots,A_m^\Lambda)\), using the full dynamics \(\tau^\Phi\) in every filter.
\begin{enumerate}[label=(\roman*),ref=(\roman*),leftmargin=*]
\item \emph{Coefficient construction and total-operator identities.}
\label{item:iterated-coefficients}
Formula \eqref{eq:iterated-filter-construction} below defines an interaction \(\Xi_E\), depending complex-multilinearly on \((\Psi_1,\ldots,\Psi_m)\).
Each coefficient integral exists in norm under the \(L^1\) assumptions alone, including for unrestricted extensive inputs.
The same construction on the restricted interactions \(\Psi_j^\Lambda\) gives interactions \(\Xi_E^\Lambda\) satisfying
\[
 \sum_Y\Xi_E^\Lambda(Y)=E_\Lambda
\]
with absolute convergence in norm and the sum over all output supports.
If
\[
 S_j:=\sum_X\E^{\kappa_0|X|}\|\Psi_j(X)\|<\infty,
 \qquad 1\le j\le m,
\]
then \(\sum_Y\Xi_E(Y)\) also converges absolutely and equals \(E\) evaluated at \(A_j=\sum_X\Psi_j(X)\).

\item \emph{Weighted interaction estimates.}
\label{item:iterated-estimates}
For every \(k\ge0\) satisfying \eqref{eq:iterated-filter-moment},
\begin{equation}
 \|\Xi_E\|^{\pol}_{p;k}
 \le C_{E,p,k}\prod_{j=1}^m
                    \|\Psi_j\|^{\#}_{p;\kappa_0}.
 \label{eq:iterated-filter-bound}
\end{equation}
The construction of \(\Xi_E\) is the same for every admissible \(k\).
If every filter has finite absolute moments of every order, this estimate holds for every \(k\ge0\).

\item \emph{Derivation identity for extensive inputs.}
\label{item:iterated-derivation}
Assume that
\[
 \int_\R|f(t)|(1+|t|)^{\,m\,m_{0,m-1}}\,\dd t<\infty
 \quad\text{for every filter in }E.
\]
Then, for every local observable \(A\),
\begin{equation}
 \lim_{\Lambda\nearrow\Gamma}[E_\Lambda,A]
       =\sum_Y[\Xi_E(Y),A]
 \quad\text{in norm},
 \label{eq:iterated-filter-derivation}
\end{equation}
where the series is absolutely convergent.
\end{enumerate}
\end{proposition}

\begin{proof}
We move the time evolutions onto the original inputs, represent these evolved inputs by Theorem~\ref{thm:card-propagation}, and assign each commutator to the union of its supports.

For an interaction \(\Xi\) and a family of local operators \(\Upsilon(V)\in\A_V\), use the bracket \eqref{eq:cell-interaction-bracket} on site supports:
\begin{equation}
 [\Xi,\Upsilon](Z)
 :=\sum_{\substack{U\cup V=Z\\U\cap V\ne\varnothing}}
                         [\Xi(U),\Upsilon(V)].
 \label{eq:site-interaction-bracket}
\end{equation}

For a family \(\Upsilon(Z)\in\A_Z\) and \(p,k\ge0\), also put
\begin{equation}
 \frakt_{p;k}(\Upsilon)
 :=\sum_{\varnothing\ne Z\Subset\Gamma}
                  |Z|^k(1+\diam Z)^p\|\Upsilon(Z)\|.
 \label{eq:total-support-norm}
\end{equation}

Since \(\Xi\) is even, \([\Xi(U),\Upsilon(V)]=0\) for disjoint \(U,V\) and arbitrary \(\Upsilon(V)\in\A_V\).
Thus the bracket includes all nonzero commutators.

\smallskip
\noindent \emph{\ref{item:iterated-coefficients} Coefficient construction and total-operator identities.}
Number the filter applications in \(E\) by \(1,\ldots,r\), with kernels \(f_1,\ldots,f_r\), assigning a separate index to each application even when kernels coincide.
Let \(P_j\subseteq\{1,\ldots,r\}\) consist of the indices of the filters whose arguments contain the symbol \(A_j\), and set
\[
 v_j(\mathbf t):=\sum_{b\in P_j}t_b,\qquad
 F(\mathbf t):=\prod_{b=1}^rf_b(t_b),
 \qquad \mathbf t\in\R^r.
\]
For an \(m\)-tuple \(\mathbf B=(B_1,\ldots,B_m)\) of operators or interactions, define \(\Br_E(\mathbf B)\) recursively by
\[
 \begin{aligned}
 \Br_{A_j}(\mathbf B)&:=B_j,\\
 \Br_{[E_1,E_2]}(\mathbf B)
   &:=[\Br_{E_1}(\mathbf B),\Br_{E_2}(\mathbf B)],\\
 \Br_{\cI_f(E_1)}(\mathbf B)&:=\Br_{E_1}(\mathbf B),
 \end{aligned}
\]
using the operator or interaction bracket, respectively.
The expression \(\Br_E\) contains \(m-1\) commutators.
The automorphism identity \(\tau_t^\Phi([B,C])=[\tau_t^\Phi(B),\tau_t^\Phi(C)]\) and the group law give
\[
 E_\Lambda
 =\int_{\R^r}F(\mathbf t)\,
   \Br_E\bigl(\tau_{v_1(\mathbf t)}^\Phi(A_1^\Lambda),
                 \ldots,\tau_{v_m(\mathbf t)}^\Phi(A_m^\Lambda)\bigr)
       \,\dd\mathbf t.
\]
The integrand norm is bounded by
\[
 2^{m-1}|F(\mathbf t)|\prod_{j=1}^m\|A_j^\Lambda\|,
\]
which is integrable.
When \(r=0\), omit the integral, set every \(v_j=0\), and interpret products over filter indices as \(1\).

Use one family of maps \(\cT^\Phi_{s,t}\) from Theorem~\ref{thm:card-propagation} for all inputs \(\Psi_j\), and define componentwise
\begin{equation}
 \Xi_E(Y):=\int_{\R^r}F(\mathbf t)\,
 \Bigl(\Br_E\bigl(
       \cT^\Phi_{0,v_1(\mathbf t)}\Psi_1,\ldots,
       \cT^\Phi_{0,v_m(\mathbf t)}\Psi_m\bigr)\Bigr)(Y)
       \,\dd\mathbf t.
 \label{eq:iterated-filter-construction}
\end{equation}
For an \(m\)-tuple \(\mathbf X=(X_1,\ldots,X_m)\) of nonempty finite subsets of \(\Gamma\), let \(\Xi_{E;\mathbf X}\) be given by the same formula with \(\iota_{X_j}\Psi_j(X_j)\) (see~\eqref{eq:single-support-interaction}) assigned to label \(j\).
For the example preceding the proposition, \eqref{eq:iterated-filter-construction} reads
\[
 \Xi_E(Y)=\int_{\R^2}f(u)g(v)
 \Bigl([\cT^\Phi_{0,u}\Psi_1,
       [\cT^\Phi_{0,u+v}\Psi_2,\cT^\Phi_{0,u+v}\Psi_3]]\Bigr)(Y)
       \,\dd u\,\dd v.
\]
The brackets in this formula are interaction brackets.

Suppose first that \(S_j<\infty\) for every \(j\).
The total estimate \eqref{eq:card-propagation-total} at \(p=k=0\), summed over the terms \(\Psi_j(X)\), gives
\[
 \sum_X\frakt_{0;0}(\cT^\Phi_{0,t;X}\Psi_j(X))
 \le C_{0,0}S_j,\qquad t\in\R.
\]
For absolutely summable families, the bracket definition gives
\[
 \frakt_{0;0}([\Xi,\Upsilon])
 \le2\sum_{U\cap V\ne\varnothing}\|\Xi(U)\|\|\Upsilon(V)\|
 \le2\frakt_{0;0}(\Xi)\frakt_{0;0}(\Upsilon).
\]
Applying this inequality to the \(m-1\) commutators in \eqref{eq:iterated-filter-construction} and using Tonelli's theorem yields
\[
 \frakt_{0;0}(\Xi_E)
 \le\sum_{\mathbf X}\frakt_{0;0}(\Xi_{E;\mathbf X})
 \le2^{m-1}C_{0,0}^{\,m}
       \prod_{b=1}^r\|f_b\|_{L^1}
       \prod_{j=1}^m S_j.
\]
These bounds prove existence of the coefficient integrals and absolute total summability for \(L^1\) filters.
The identity \eqref{eq:source-representation} and absolute convergence permit rearranging the sums and integrals.
This gives \(\sum_Y\Xi_E(Y)=E(A_1,\ldots,A_m)\).
Applying the same argument to the restricted interactions \(\Psi_j^\Lambda\) gives \(\sum_Y\Xi_E^\Lambda(Y)=E_\Lambda\).

For arbitrary input interactions, fix a nonempty finite output support \(Y\).
Every propagated term has an output support containing its input support, and each interaction bracket assigns terms to the union of their supports.
Consequently, the coefficient at \(Y\) in \eqref{eq:iterated-filter-construction} is unchanged when every \(\Psi_j\) is replaced by \(\Psi_j^Y\).
The preceding total estimates applied to these restricted interactions give
\[
 \begin{aligned}
 \int_{\R^r}|F(\mathbf t)|&
 \left\|
 \Bigl(\Br_E\bigl(
 \cT^\Phi_{0,v_1(\mathbf t)}\Psi_1,\ldots,
 \cT^\Phi_{0,v_m(\mathbf t)}\Psi_m
 \bigr)\Bigr)(Y)
 \right\|\,\dd\mathbf t\\
 & \le
 2^{m-1}C_{0,0}^{\,m}
 \prod_{j=1}^m\left(
       \sum_{\varnothing\ne X\subseteq Y}
              \E^{\kappa_0|X|}\|\Psi_j(X)\|\right)
 \prod_{b=1}^r\|f_b\|_{L^1}<\infty.
 \end{aligned}
\]

Thus the coefficient integrals exist for arbitrary input interactions under \(L^1\) integrability alone.
Complex multilinearity and evenness follow from \eqref{eq:iterated-filter-construction}.
At each output support \(Y\), the sum over input labels is finite and gives
\[
 \Xi_E(Y)=\sum_{\varnothing\ne X_1,\ldots,X_m\subseteq Y}
                         \Xi_{E;\mathbf X}(Y).
\]

\smallskip
\noindent \emph{\ref{item:iterated-estimates} Weighted interaction estimates.}
Write \(w_p(U)=(1+\diam U)^p\).
For overlapping nonempty \(U,V\),
\[
 |U\cup V|^k\le2^k|U|^k|V|^k,\qquad
 w_p(U\cup V)\le2^{\max\{p-1,0\}}\bigl(w_p(U)+w_p(V)\bigr).
\]
For example, the part of the overlap sum with \(x\in U\) and the diameter weight on \(U\) satisfies
\[
 \sum_{\substack{U\cap V\ne\varnothing\\x\in U}}
 |U|^k|V|^kw_p(U)\|\Xi(U)\|\|\Upsilon(V)\|\le
 \sum_{U\ni x}|U|^{k+1}w_p(U)\|\Xi(U)\|
                      \|\Upsilon\|^{\pol}_{0;k}.
\]
The other choices of the support containing \(x\) and the support carrying \(w_p\), together with the commutator norm bound, give
\begin{equation}
 \|[\Xi,\Upsilon]\|^{\pol}_{p;k}
 \le C^{\rm br}_{p,k}\bigl(
 \|\Xi\|^{\pol}_{p;k+1}\|\Upsilon\|^{\pol}_{0;k+1}\ +\|\Xi\|^{\pol}_{0;k+1}\|\Upsilon\|^{\pol}_{p;k+1}\bigr)\,.
 \label{eq:polynomial-interaction-bracket}
\end{equation}
Summing over all overlapping pairs \(U,V\), without fixing a site, gives
\[
 \sum_{U\cap V\ne\varnothing}
 |U|^k|V|^kw_p(U)\|\Xi(U)\|\|\Upsilon(V)\|
 \le\|\Xi\|^{\pol}_{p;k}\frakt_{0;k+1}(\Upsilon).
\]
Putting the diameter weight on \(V\) instead proves
\begin{equation}
 \frakt_{p;k}([\Xi,\Upsilon])
 \le C^{\rm br}_{p,k}\bigl(
 \|\Xi\|^{\pol}_{p;k+1}\frakt_{0;k+1}(\Upsilon) +\|\Xi\|^{\pol}_{0;k+1}\frakt_{p;k+1}(\Upsilon)\bigr).
 \label{eq:polynomial-total-bracket}
\end{equation}
The overlap estimates also bound the sum of the norms of the individual commutators assigned to each union \(U\cup V\).

Fix \(k\ge0\) satisfying \eqref{eq:iterated-filter-moment}.
We first track the support moments required by the \(m-1\) brackets in \(\Br_E\).
We prove, for arbitrary interactions \(\Upsilon_1,\ldots,\Upsilon_m\), that
\[
 \|\Br_E(\Upsilon_1,\ldots,\Upsilon_m)\|^{\pol}_{p;k}
 \le C\prod_{j=1}^m\|\Upsilon_j\|^{\pol}_{p;k+m-1}.
\]
For \(m=1\), the expression is its sole input, and support moment \(k\) suffices.
For \(m\ge2\), the outermost bracket divides the inputs into two groups of sizes \(m_1,m_2\ge1\), with \(m_1+m_2=m\).
Apply \eqref{eq:polynomial-interaction-bracket} at moment \(k\), and apply the inductive estimates to its two factors at moment \(k+1\).
The inputs in group \(i\) then require moment
\[
 k+1+(m_i-1)=k+m_i\le k+m-1.
\]
Since the norms increase with the support moment and \(\|\cdot\|^{\pol}_{0;k'}\le\|\cdot\|^{\pol}_{p;k'}\), this proves the displayed bound by induction on \(m\).

To keep the original input labels separate, write the coefficient integrand for \(\mathbf X=(X_1,\ldots,X_m)\) as
\[
 B_{E;\mathbf X}(\mathbf t;Y):=
 \Bigl(\Br_E\bigl(
       \cT^\Phi_{0,v_1(\mathbf t);X_1}\Psi_1(X_1),\ldots,
       \cT^\Phi_{0,v_m(\mathbf t);X_m}\Psi_m(X_m)
                         \bigr)\Bigr)(Y).
\]
Thus \(\Xi_{E;\mathbf X}(Y)=\int_{\R^r}F(\mathbf t)B_{E;\mathbf X}(\mathbf t;Y)\,\dd\mathbf t\).
For two finite or countable labelled families of interactions \((\Xi_\alpha)_\alpha\) and \((\Upsilon_\beta)_\beta\), bounding each commutator separately gives
\[
 \sum_{\alpha,\beta}\|[\Xi_\alpha,\Upsilon_\beta](Y)\|
 \le2\sum_{\substack{U\cup V=Y\\U\cap V\ne\varnothing}}
       \left(\sum_\alpha\|\Xi_\alpha(U)\|\right)
       \left(\sum_\beta\|\Upsilon_\beta(V)\|\right).
\]
Apply the positive overlap estimates proving \eqref{eq:polynomial-interaction-bracket} to these scalar sums at each bracket.
The preceding induction on \(m\), together with the input-separated propagation estimate \eqref{eq:card-propagation}, gives
\[
\sup_x\sum_{Y\ni x}|Y|^k(1+\diam Y)^p
       \sum_{\mathbf X}\|B_{E;\mathbf X}(\mathbf t;Y)\| \le C\prod_{j=1}^m
       (1+|v_j(\mathbf t)|)^{m_{p,k+m-1}}
                         \|\Psi_j\|^{\#}_{p;\kappa_0}.
\]
Here \(C\) depends only on the arrangement of the commutators in \(E\), the propagation data, and \(p,k\).

For each input \(j\), \(v_j(\mathbf t)=\sum_{b\in P_j}t_b\), so \(1+|v_j(\mathbf t)|\le\prod_{b\in P_j}(1+|t_b|)\).
Consequently,
\[
 \prod_{j=1}^m(1+|v_j(\mathbf t)|)^{m_{p,k+m-1}}
 \le\prod_{b=1}^r
       (1+|t_b|)^{m_{p,k+m-1}|\{j\in\{1,\ldots,m\}\mid b\in P_j\}|}\le\prod_{b=1}^r(1+|t_b|)^{m\,m_{p,k+m-1}}.
\]
The second inequality uses the fact that each filter acts on at most \(m\) inputs.
Multiplying the bound for \(B_{E;\mathbf X}\) by \(|F(\mathbf t)|\), integrating, and applying Tonelli's theorem therefore give
\[
 \begin{aligned}
 \sup_x\sum_{Y\ni x}|Y|^k(1+\diam Y)^p &
       \sum_{\mathbf X}\int_{\R^r}|F(\mathbf t)|
                    \|B_{E;\mathbf X}(\mathbf t;Y)\|\,\dd\mathbf t\\
 & \le C
       \prod_{b=1}^r\left(
       \int_\R|f_b(t)|(1+|t|)^{m\,m_{p,k+m-1}}\,\dd t\right)
       \prod_{j=1}^m\|\Psi_j\|^{\#}_{p;\kappa_0}.
 \end{aligned}
\]
Choose \(C_{E,p,k}>0\) no smaller than \(C\) times the product of filter moments on the right.
The triangle inequality for the coefficient integrals and the sums over \(\mathbf X\) proves \eqref{eq:iterated-filter-bound} with this constant.

\smallskip
\noindent \emph{\ref{item:iterated-derivation} Derivation identity for extensive inputs.}
Assume the filter moment of part~\ref{item:iterated-derivation}.
We prove convergence of the commutators with local observables by summing the contributions indexed by the original input supports.
Formula~\eqref{eq:iterated-filter-construction} does not depend on the spatial exponent used to estimate it, and \(\|\Psi_j\|^{\#}_{0;\kappa_0}\le\|\Psi_j\|^{\#}_{p;\kappa_0}<\infty\).
We may therefore apply part~\ref{item:iterated-estimates} at \(p=k=0\) to the same coefficients.
Keeping the sum of norms over input labels in those estimates gives
\[
 \sup_x\sum_{Y\ni x}\sum_{\mathbf X}
          \|\Xi_{E;\mathbf X}(Y)\|
 \le C_{E,0,0}\prod_{j=1}^m\|\Psi_j\|^{\#}_{0;\kappa_0}.
\]
The same bound holds with the absolute time integrals in place of the coefficient norms.
For \(A\in\A_K\), evenness makes the commutator vanish unless the output support meets \(K\).
Summing first over \(x\in K\) bounds the absolute sum of all commutator contributions by
\[
 2|K|\|A\|C_{E,0,0}
                  \prod_{j=1}^m\|\Psi_j\|^{\#}_{0;\kappa_0}.
\]
Restricting each interaction \(\Psi_j\) to \(\Lambda\) keeps exactly the tuples with \(X_j\subseteq\Lambda\) for every \(j\).
These tuples exhaust the countable sum as \(\Lambda\nearrow\Gamma\).
Dominated convergence therefore proves \eqref{eq:iterated-filter-derivation}.
\end{proof}

\subsection{Commutator sums and observable propagation bounds}
\label{subsec:observable-locality}

We now derive local-approximation bounds for evolved observables and estimate how their evolution changes when Hamiltonian terms are removed or added.
We begin with the commutator sums and comparison estimates of Proposition~\ref{prop:commutator-comparison}.
We then prove the observable bounds of Corollary~\ref{cor:observable-LR} and Theorem~\ref{thm:polynomial-local-topology}.

\begin{proposition}
\label{prop:commutator-comparison}
Assume \eqref{eq:volume-growth} and Hypothesis~\ref{hyp:cell-partitions}, and choose \(\kappa_0\) as in Theorem~\ref{thm:card-propagation}.
Fix \(a>\dvol\) and \(0\le L<\infty\), and use the constants \(C_{p,0}\) and exponents \(m_{p,0}\), \(0\le p\le a\), from that theorem.
For every choice of local multiplicities \((r_x)_{x\in\Gamma}\) and every componentwise norm-continuous path \((\Phi_u)_{u\in\R}\) of self-adjoint interactions with
\[
 \sup_{u\in\R}\|\Phi_u\|^{\#}_{a;\kappa_0}\le L,
\]
the following conclusions hold for every nonempty \(X\Subset\Gamma\) and every \(A\in\A_X\).
\begin{enumerate}[label=(\roman*),leftmargin=*]
\item \emph{Sums of commutators.}
\label{item:commutator-sum}
For every interaction \(\Psi\) with \(\|\Psi\|^{\#}_{0;\kappa_0}<\infty\) and every \(s,t\in\R\),
\begin{equation}
 \sum_Z\|[\Psi(Z),\tau_{s,t}^\Phi(A)]\|
 \le 2C_{0,0}|X|\|A\|(1+|t-s|)^{m_{0,0}}
             \|\Psi\|^{\#}_{0;\kappa_0}.
 \label{eq:evolved-commutator-sum}
\end{equation}

\item \emph{Boundary sums.}
\label{item:commutator-boundary}
For a finite \(\Lambda\) with \(X\subseteq\Lambda\subsetneq\Gamma\), put
\[
 D:=\dist(X,\Lambda^c),\qquad
 \scrE_\partial(\Lambda):=\{Z\Subset\Gamma\,|\,
       Z\cap\Lambda\ne\varnothing,\ Z\cap\Lambda^c\ne\varnothing\}.
\]
For every \(0\le p\le a\), every interaction \(\Psi\) with \(\|\Psi\|^{\#}_{p;\kappa_0}<\infty\), and every \(s,t\in\R\),
\begin{equation}
 \sum_{Z\in\scrE_\partial(\Lambda)}
       \|[\Psi(Z),\tau_{s,t}^{\Phi,\Lambda}(A)]\|
 \le 2C_{p,0}|X|\|A\|
       \frac{(1+|t-s|)^{m_{p,0}}}{(1+D)^p}
       \|\Psi\|^{\#}_{p;\kappa_0}.
 \label{eq:boundary-crossing-sum}
\end{equation}
For a measurable filter \(f:\R\to\C\), define
\[
 \frakm_{p}(f):=\int_\R |f(t)|\,|t|(1+|t|)^{m_{p,0}}\,\dd t.
\]
If \(\Phi\) is autonomous and \(\frakm_{p}(f)<\infty\), then
\begin{align}
 \int_\R |f(t)|
     \int_{\min(0,t)}^{\max(0,t)}
       \sum_{Z\in\scrE_\partial(\Lambda)}
          \|[\Psi(Z),\tau_u^{\Phi,\Lambda}(A)]\|\,\dd u\,\dd t \le 2C_{p,0}\frakm_{p}(f)|X|\|A\|(1+D)^{-p}
                   \|\Psi\|^{\#}_{p;\kappa_0}.
 \label{eq:filtered-boundary-sum}
\end{align}
For interactions \(\Psi_1,\ldots,\Psi_N\) with \(\|\Psi_j\|^{\#}_{p;\kappa_0}<\infty\), both estimates hold with the commutator sum taken also over \(j=1,\ldots,N\) and \(\|\Psi\|^{\#}_{p;\kappa_0}\) replaced by \(\sum_{j=1}^N\|\Psi_j\|^{\#}_{p;\kappa_0}\).

\item \emph{Perturbative comparison.}
\label{item:commutator-perturbation}
Let \(\Phi\) be autonomous, let \(\cV\) be a self-adjoint interaction with \(\|\cV\|^{\#}_{a;\kappa_0}<\infty\), and set \(\Phi_\lambda:=\Phi+\lambda \cV\) for \(\lambda\in\R\).
For every \(\lambda,t\in\R\),
\begin{equation}
 \|\tau_t^{\Phi_\lambda}(A)-\tau_t^\Phi(A)\|
 \le 2C_{0,0}|\lambda|\,|X|\|A\|
       \|\cV\|^{\#}_{0;\kappa_0}
       \int_0^{|t|}(1+u)^{m_{0,0}}\,\dd u.
 \label{eq:perturbative-dynamics-comparison}
\end{equation}
For every measurable \(f:\R\to\C\) with \(\frakm_{0}(f)<\infty\), the following integral converges absolutely in norm and satisfies
\begin{align}
 \left\|\int_\R f(t)
      \bigl(\tau_t^{\Phi_\lambda}(A)-\tau_t^\Phi(A)\bigr)\,\dd t\right\|
  \le 2C_{0,0}\frakm_{0}(f)|\lambda|\,|X|\|A\|
                       \|\cV\|^{\#}_{0;\kappa_0}.
 \label{eq:perturbative-filter-comparison}
\end{align}
If also \(f\in L^1(\R)\), this integral is the difference of the two separately defined filtered observables.
\end{enumerate}
All bounds hold uniformly under finite restrictions and deletion of terms of \(\Phi\), \(\Psi\), or \(\cV\), whenever the resulting Hamiltonian paths remain componentwise norm-continuous.
Parts~\ref{item:commutator-sum} and~\ref{item:commutator-perturbation} also apply to expansions \(A=\sum_{j=1}^\infty A_j\), with \(A_j\in\A_{X_j}\), nonempty finite \(X_j\), and \(\sum_{j=1}^\infty|X_j|\|A_j\|<\infty\), replacing \(|X|\|A\|\) by this sum.
\end{proposition}

\begin{proof}
\smallskip
\noindent \emph{Sums of commutators and boundary sums.}
Fix \(\Psi,s,t,X,A\) as in part~\ref{item:commutator-sum}, and set
\[
 B_Z(Y):=(\cT^\Phi_{t,s;Z}\Psi(Z))(Y).
\]
By Theorem~\ref{thm:card-propagation}, the operators \(B_Z(Y)\in\A_Y\) are even, vanish unless \(Z\subseteq Y\), and satisfy \(\tau_{t,s}^\Phi(\Psi(Z))=\sum_{Y\supseteq Z}B_Z(Y)\) absolutely in norm.
Isometry and the inverse cocycle relation give
\[
 \sum_Z\|[\Psi(Z),\tau_{s,t}^\Phi(A)]\|
 =\sum_Z\left\|\sum_{Y\supseteq Z}[B_Z(Y),A]\right\| \le2\|A\|\sum_{x\in X}\sum_{Y\ni x}
                \sum_{\varnothing\ne Z\subseteq Y}\|B_Z(Y)\|.
\]
The inequality uses \([B_Z(Y),A]=0\) when \(Y\cap X=\varnothing\), which holds for arbitrary \(A\) because \(B_Z(Y)\) is even.
Equation~\eqref{eq:card-propagation} with \(p=k=0\) bounds the last expression by the right-hand side of \eqref{eq:evolved-commutator-sum}.

For part~\ref{item:commutator-boundary}, set
\[
 \Phi_u^\Lambda(Z):=\1_{\{Z\subseteq\Lambda\}}\Phi_u(Z),
 \qquad
 \Psi^\partial(Z):=\1_{\{Z\in\scrE_\partial(\Lambda)\}}\Psi(Z),\qquad
 B_Z^\partial(Y):=
       (\cT^{\Phi^\Lambda}_{t,s;Z}\Psi^\partial(Z))(Y).
\]
If \([B_Z^\partial(Y),A]\ne0\), choose \(x\in X\cap Y\) and \(z\in Z\cap\Lambda^c\).
Since \(Z\subseteq Y\), \(\diam Y\ge d(x,z)\ge D\).
The commutator expansion therefore gives
\[
\sum_{Z\in\scrE_\partial(\Lambda)}
       \|[\Psi(Z),\tau_{s,t}^{\Phi,\Lambda}(A)]\| \le 2\|A\|(1+D)^{-p}
   \sum_{x\in X}\sum_{Y\ni x}(1+\diam Y)^p
          \sum_{\varnothing\ne Z\subseteq Y}\|B_Z^\partial(Y)\|\,.
\]
The norms of \(\Phi_u^\Lambda\) and \(\Psi^\partial\) are bounded by those of \(\Phi_u\) and \(\Psi\), respectively.
Applying \eqref{eq:card-propagation} to \(\Phi^\Lambda,\Psi^\partial\) with \(k=0\) proves \eqref{eq:boundary-crossing-sum}.
For an autonomous \(\Phi\), use this estimate with \(s=0\) and \(t=u\), and integrate over \(u\in[\min(0,t),\max(0,t)]\).
Since
\[
 \int_{\min(0,t)}^{\max(0,t)}(1+|u|)^{m_{p,0}}\,\dd u
 \le |t|(1+|t|)^{m_{p,0}},
\]
integration against \(|f(t)|\) and Tonelli's theorem give \eqref{eq:filtered-boundary-sum}.

\smallskip
\noindent \emph{Perturbative comparison.}
Proposition~\ref{prop:polynomial-dynamics} constructs the dynamics of \(\Phi_\lambda\), since
\[
 \|\Phi_\lambda\|^{\pol}_{a;0}
 \le L+|\lambda|\|\cV\|^{\#}_{a;\kappa_0}<\infty.
\]
For a finite \(\Omega\supseteq X\), define
\[
 F^\Omega(u):=\tau_{t-u}^{\Phi_\lambda,\Omega}
                    (\tau_u^{\Phi,\Omega}(A)).
\]
Differentiation gives
\[
 (F^\Omega)'(u)=-\I\lambda\tau_{t-u}^{\Phi_\lambda,\Omega}\!\left(
         \sum_{Z\subseteq\Omega}
                [\cV(Z),\tau_u^{\Phi,\Omega}(A)]\right).
\]
Since \(F^\Omega(0)=\tau_t^{\Phi_\lambda,\Omega}(A)\) and \(F^\Omega(t)=\tau_t^{\Phi,\Omega}(A)\), integration and isometry give
\[
\|\tau_t^{\Phi_\lambda,\Omega}(A)-\tau_t^{\Phi,\Omega}(A)\|\le |\lambda|\int_{\min(0,t)}^{\max(0,t)}
         \sum_{Z\subseteq\Omega}
                \|[\cV(Z),\tau_u^{\Phi,\Omega}(A)]\|\,\dd u\,.
\]
Applying \eqref{eq:evolved-commutator-sum} to the restrictions of \(\Phi\) and \(\cV\) to \(\Omega\) gives \eqref{eq:perturbative-dynamics-comparison} in finite volume, for both signs of \(t\) and \(\lambda\).
By Proposition~\ref{prop:polynomial-dynamics}, both finite-volume evolutions of \(A\) converge in norm as \(\Omega\nearrow\Gamma\), so \eqref{eq:perturbative-dynamics-comparison} holds in infinite volume.
Using \(\int_0^{|t|}(1+u)^{m_{0,0}}\,\dd u \le |t|(1+|t|)^{m_{0,0}}\), we obtain
\[
 \int_\R |f(t)|\,
       \|\tau_t^{\Phi_\lambda}(A)-\tau_t^\Phi(A)\|\,\dd t
 \le 2C_{0,0}\frakm_{0}(f)|\lambda|\,|X|\|A\|
                \|\cV\|^{\#}_{0;\kappa_0}.
\]
This proves absolute convergence and \eqref{eq:perturbative-filter-comparison}.
If \(f\in L^1(\R)\), isometry also gives absolute convergence of \(\int_\R f(t)\tau_t^{\Phi_\lambda}(A)\,\dd t\) and \(\int_\R f(t)\tau_t^\Phi(A)\,\dd t\).

Restrictions and deletions do not increase the interaction norms used above, so the same constants apply to the restricted interactions and to interactions obtained by deleting terms.

\smallskip
\noindent \emph{Absolutely summable expansions.}
For \(A=\sum_{j=1}^\infty A_j\) as in the statement and \(N\ge0\), put
\[
 A^{(N)}:=\sum_{j=1}^N A_j,\qquad
 R_N:=\sum_{j>N}|X_j|\|A_j\|.
\]
Then \(A^{(N)}\to A\) in norm and \(R_N\to0\) as \(N\to\infty\).
Apply \eqref{eq:evolved-commutator-sum} to each \(A_j\), sum over \(N<j\le M\), and let \(M\to\infty\).
Norm convergence for each \(Z\) and Fatou's lemma give
\[
 \sum_Z\|[\Psi(Z),\tau_{s,t}^\Phi(A-A^{(N)})]\|
 \le 2C_{0,0}(1+|t-s|)^{m_{0,0}}
                 \|\Psi\|^{\#}_{0;\kappa_0}R_N.
\]
Summing \eqref{eq:perturbative-dynamics-comparison} over \(N<j\le M\) and letting \(M\to\infty\) gives that estimate for \(A-A^{(N)}\), with \(|X|\|A\|\) replaced by \(R_N\).
Integrating the bound for \(A^{(M)}-A^{(N)}\) against \(|f(t)|\) and using Fatou's lemma yields
\[
 \int_\R |f(t)|\,
   \|\tau_t^{\Phi_\lambda}(A-A^{(N)})
                      -\tau_t^\Phi(A-A^{(N)})\|\,\dd t\ \le 2C_{0,0}\frakm_{0}(f)|\lambda|
                         \|\cV\|^{\#}_{0;\kappa_0}R_N\,.
\]
Taking \(N=0\) proves \eqref{eq:evolved-commutator-sum}, \eqref{eq:perturbative-dynamics-comparison}, and \eqref{eq:perturbative-filter-comparison} for \(A\), with \(|X|\|A\|\) replaced by \(\sum_j|X_j|\|A_j\|\).
Letting \(N\to\infty\) proves convergence of the commutator series and of the filtered differences of the partial sums.
\end{proof}

The proof of the conditional-expectation bound in Corollary~\ref{cor:observable-LR} is organized as follows.
We estimate the approximation error by a sum of commutators with single-site test operators outside the approximation region, evolve these test operators backwards, and apply Theorem~\ref{thm:card-propagation}.
Lemma~\ref{lem:auxiliary-even-embedding} then reduces observables of arbitrary parity to even ones.

\smallskip
\noindent \emph{Single-site averaging.}
For this averaging argument, fix a finite region \(\Omega\) and a subset \(D\subseteq\Omega\).
Use the auxiliary algebras, Majoranas \(v_x\), and tracial expectations from Lemma~\ref{lem:auxiliary-even-embedding}.
For \(y\in\Omega\), let \(\varsigma_{y,1},\ldots,\varsigma_{y,2r_y}\) be the physical Majorana fields at \(y\), normalized by \(\varsigma_{y,j}^*=\varsigma_{y,j}\) and \(\{\varsigma_{y,j},\varsigma_{y,k}\}=2\delta_{jk}\1\).
Define
\[
 \scrV_y:=
 \left\{\varsigma_{y,1}^{\varepsilon_1}\cdots
        \varsigma_{y,2r_y}^{\varepsilon_{2r_y}}\,|\,
                      \varepsilon_j\in\{0,1\}\right\},
 \qquad
 \cP_y(C):=\frac1{|\scrV_y|}
                  \sum_{V\in\scrV_y}V^*CV.
\]
Each \(V\in\scrV_y\) is a parity-homogeneous unitary.
Majorana monomials at distinct sites commute or anticommute, and the sign disappears under conjugation.
Hence the conjugation averages \(\cP_y\) commute; each is contractive as an average of isometries.
For an even Majorana monomial \(M\), averaging at \(y\) fixes \(M\) if no Majorana at \(y\) occurs and sends \(M\) to zero otherwise.
The product over \(y\in\Omega\setminus D\) fixes precisely the even Majorana monomials supported in \(D\) and sends all others to zero.
Since this is also the action of \(\F_D^{\rm tr}\) on a basis of \(\A_\Omega^+\), we obtain
\[
 \prod_{y\in\Omega\setminus D}\cP_y(C)=\F_D^{\rm tr}(C),
 \qquad C\in\A_\Omega^+.
\]
This is the mode-by-mode averaging construction of Nachtergaele--Sims--Young \cite[Eq.~(4.6) and Lemmas~4.1--4.2]{NSY18}, with the \(r_y\) modes at each site grouped together.

Enumerate \(\Omega\setminus D=\{y_1,\ldots,y_m\}\).
The identity
\[
 C-\F_D^{\rm tr}(C)
 =\sum_{j=1}^m\left(\prod_{i=1}^{j-1}\cP_{y_i}\right)
                 (C-\cP_{y_j}(C)),
\]
with the empty product interpreted as the identity, therefore gives
\begin{equation}
 \|C-\F_D^{\rm tr}(C)\|
 \le\sum_{y\in\Omega\setminus D}\|C-\cP_y(C)\|
 \le\sum_{y\in\Omega\setminus D}
              \max_{V\in\scrV_y}\|[C,V]\|,
 \qquad C\in\A_\Omega^+.
 \label{eq:single-site-averaging}
\end{equation}

\begin{proof}[Proof of Corollary~\ref{cor:observable-LR}]
\phantomsection\label{proof:observable-LR}
Fix a finite region \(\Omega\supseteq X\), \(s,t\in\R\), and \(R\ge0\).
First take \(A\in\A_X^+\), and put \(\Omega_R:=\Omega\cap X^{(R)}\).
For each \(y\in\Omega\setminus \Omega_R\), choose
\[
 V_y\in\scrV_y,\qquad
 \|[\tau_{s,t}^{\Phi,\Omega}(A),V_y]\|
 =\max_{V\in\scrV_y}\|[\tau_{s,t}^{\Phi,\Omega}(A),V]\|.
\]
Use the enlarged algebra and the auxiliary Majoranas \(v_y\) defined above, and define an even interaction by
\[
 \Psi(\{y\}):=
 \begin{cases}
 V_y,&V_y\text{ even},\\
 V_yv_y,&V_y\text{ odd},
 \end{cases}
 \quad y\in\Omega\setminus \Omega_R,\qquad
 \Psi(Z):=0\quad\text{for all other }Z.
\]
It satisfies \(\|\Psi\|^{\#}_{a;\kappa_0}\le\E^{\kappa_0}\).
Let \(\widehat\Phi\) be the embedded Hamiltonian path and \(\widehat\tau:=\tau^{\widehat\Phi,\Omega}\).
The inverse relation for the finite-volume dynamics and the stationarity of \(v_y\) give
\[
 \|[\tau_{s,t}^{\Phi,\Omega}(A),V_y]\|
 =\|[A,\tau_{t,s}^{\Phi,\Omega}(V_y)]\|
 =\|[A,\widehat\tau_{t,s}(\Psi(\{y\}))]\|.
\]
For odd \(V_y\), the last equality uses \([A,\tau_{t,s}^{\Phi,\Omega}(V_y)v_y] =[A,\tau_{t,s}^{\Phi,\Omega}(V_y)]v_y\) and the unitarity of \(v_y\).

Apply Theorem~\ref{thm:card-propagation} with underlying metric space \(\Omega\), equipped with the induced metric and cell partitions, and with the enlarged CAR algebra over \(\Omega\).
Define
\[
 B_{y,Z}:=
       (\cT^{\widehat\Phi}_{t,s;\{y\}}\Psi(\{y\}))(Z),
 \qquad
 \widehat\tau_{t,s}(\Psi(\{y\}))=\sum_{Z\ni y}B_{y,Z}.
\]
The operators \(B_{y,Z}\) are even and supported in \(Z\).
If \(Z\cap X=\varnothing\), then \([A,B_{y,Z}]=0\).
If \(Z\cap X\ne\varnothing\) and \(y\in Z\setminus \Omega_R\), then \(\diam Z>R\).
Consequently,
\begin{align}
 \sum_{y\in\Omega\setminus \Omega_R}
       \|[\tau_{s,t}^{\Phi,\Omega}(A),V_y]\|
 &\le 2\|A\|\sum_{x\in X}
       \sum_{\substack{Z\subseteq\Omega\,|\,x\in Z\\\diam Z>R}}
       \sum_{y\in Z\setminus \Omega_R}\|B_{y,Z}\|
 \nonumber\\
 & \le 2\E^{\kappa_0}C_{a,0}|X|\|A\|
       (1+|t-s|)^{m_{a,0}}(1+R)^{-a}.
 \label{eq:backward-single-site-sum}
\end{align}
The last step uses \eqref{eq:card-propagation} with \(p=a\), \(k=0\), and the norm bound for \(\Psi\).
Combining \eqref{eq:single-site-averaging} and \eqref{eq:backward-single-site-sum} proves \eqref{eq:observable-LR} for even \(A\), uniformly in \(\Omega\).
For arbitrary \(A\in\A_X\), choose \(x\in X\) and apply Lemma~\ref{lem:auxiliary-even-embedding}.
Its isometry and intertwining identities give
\[
 \bigl\|\tau_{s,t}^{\Phi,\Omega}(A)
       -\F_{\Omega_R}^{\rm tr}(\tau_{s,t}^{\Phi,\Omega}(A))\bigr\|
 =\bigl\|\widehat\tau_{s,t}(\jmath_x(A))
       -\widehat{\F}^{\rm tr}_{\Omega_R}
          (\widehat\tau_{s,t}(\jmath_x(A)))\bigr\|.
\]
The even-observable estimate is uniform in the local multiplicities, so it applies to \(\jmath_x(A)\) in the enlarged CAR algebra, whose multiplicities are \(\widehat r_y=r_y+1\).
In this application, the single-site averages use all Majoranas of the enlarged site algebra, and the proof adjoins fresh auxiliary modes for its odd test operators.
The embedded Hamiltonian has the same interaction norm.
Thus the bound for arbitrary \(A\) holds with the same constant \(C\ge2\E^{\kappa_0}C_{a,0}\).

For \(B\in\A_\Omega\), consistency gives \(\F_{\Omega_R}^{\rm tr}(B)=\F_{X^{(R)}}^{\rm tr}(B)\).
Proposition~\ref{prop:polynomial-dynamics} and contractivity of \(\F_{X^{(R)}}^{\rm tr}\) therefore pass the bound to the limit \(\Omega\nearrow\Gamma\).
Deleting Hamiltonian terms does not increase the interaction norm, so the same argument gives uniformity under deletion.

For homogeneous \(A\), put \(A_R:=\F_{X^{(R)}}^{\rm tr}(\tau_{s,t}^\Phi(A))\).
Parity preservation gives \(|A_R|=|A|\) whenever \(A_R\ne0\).
For homogeneous \(B\) supported outside \(X^{(R)}\), \(A_RB=(-1)^{|A||B|}BA_R\), and hence
\[
 \|\tau_{s,t}^\Phi(A)B-(-1)^{|A||B|}B\tau_{s,t}^\Phi(A)\|
 \le 2\|B\|\,\|\tau_{s,t}^\Phi(A)-A_R\|.
\]
Equation~\eqref{eq:observable-LR} gives the bound with prefactor \(2C|X|\).
If \(Y\) is finite and nonempty, isometry and the inverse cocycle relation give
\[
\|\tau_{s,t}^\Phi(A)B-(-1)^{|A||B|}B\tau_{s,t}^\Phi(A)\| =
 \|\tau_{t,s}^\Phi(B)A-(-1)^{|A||B|}A\tau_{t,s}^\Phi(B)\|\,.
\]
Since \(X\subseteq\Gamma\setminus Y^{(R)}\), the same argument with \((A,X,s,t)\) and \((B,Y,t,s)\) exchanged gives the prefactor \(2C|Y|\).
Taking the smaller bound gives the prefactor \(2C\min\{|X|,|Y|\}\) in the graded commutator estimate.
If \(Y\) is empty, \(B\) is scalar and the graded commutator vanishes.

\smallskip
\noindent \emph{Comparison with restricted dynamics.}
Fix a finite \(\Lambda\) with \(X\subseteq\Lambda\subsetneq\Gamma\), and put \(D:=\dist(X,\Lambda^c)\).
For each \(u\) between \(s\) and \(t\), apply Proposition~\ref{prop:commutator-comparison}\ref{item:commutator-boundary} with \(\Psi=\Phi_u\), exponent \(p=a\), and evolution \(\tau_{u,t}^{\Phi,\Lambda}\).
Here \(\|\Phi_u\|^{\#}_{a;\kappa_0}\le L\).
For every finite \(\Omega\supseteq\Lambda\), differentiating \(\tau_{s,u}^{\Phi,\Omega}(\tau_{u,t}^{\Phi,\Lambda}(A))\) and integrating in \(u\) gives
\[
\tau_{s,t}^{\Phi,\Omega}(A)-\tau_{s,t}^{\Phi,\Lambda}(A) =\I\int_s^t\tau_{s,u}^{\Phi,\Omega}\!\Big(
       \sum_{\substack{Z\in\scrE_\partial(\Lambda)\\Z\subseteq\Omega}}
       [\Phi_u(Z),\tau_{u,t}^{\Phi,\Lambda}(A)]\Big)\dd u\,.
\]
Terms contained in \(\Omega\setminus\Lambda\) do not contribute, by evenness.
Taking norms and applying \eqref{eq:boundary-crossing-sum} gives \eqref{eq:polynomial-time-finite-approximation} with \(\tau_{s,t}^{\Phi,\Omega}\) in place of \(\tau_{s,t}^\Phi\), for both signs of \(t-s\).
By Proposition~\ref{prop:polynomial-dynamics}, \(\tau_{s,t}^{\Phi,\Omega}(A)\to\tau_{s,t}^\Phi(A)\) in norm, so the bound holds in infinite volume.
All conclusions therefore hold with \(C=2C_{a,0}\max\{\E^{\kappa_0},L\}\).
\end{proof}

\begin{proof}[Proof of Theorem~\ref{thm:polynomial-local-topology}]
\phantomsection\label{proof:polynomial-local-topology}
We apply Lemma~\ref{lem:finite-volume-matrix-commutator} to the finite-volume dynamics and estimate its matrix series with the weight \((1+d(x,y))^p\); see also Nachtergaele--Sims--Young \cite[Theorem~3.1]{NSY18} for fermionic commutator estimates.
We then compare finite-volume dynamics and pass to infinite volume.

\smallskip
\noindent \emph{Finite-volume commutator estimate.}
Fix a finite region \(\Omega\), and write \(\tau^\Omega:=\tau^{\Xi,\Omega}\).
The function \(u\mapsto\|\Xi_u^\Omega\|^{\pol}_{p;1}\) is continuous and bounded on \(J\).
Let \(K_u^\Omega\) and \(\mathsf U_{s,t}^\Omega\) be the matrices of Lemma~\ref{lem:finite-volume-matrix-commutator} for \(\Xi^\Omega\).
For every \(x\in\Omega\), the definition of \(K_u^\Omega\) gives
\begin{equation}
 \begin{aligned}
 \sum_{y\in\Omega}(1+d(x,y))^p K_u^\Omega(x,y)
 &=2\sum_{\substack{Z\subseteq\Omega\\Z\ni x}}
       \left(1+\sum_{y\in Z\setminus\{x\}}(1+d(x,y))^p\right)
          \|\Xi_u(Z)\|\\
 &\le2\|\Xi_u^\Omega\|^{\pol}_{p;1}
          \le2\|\Xi_u\|^{\pol}_{p;1}.
 \end{aligned}
 \label{eq:polynomial-matrix-row}
\end{equation}
We can therefore apply \eqref{eq:finite-volume-weighted-matrix-bound} with \(w(r)=(1+r)^p\) to obtain
\[
 \sup_{x\in\Omega}\sum_{y\in\Omega}(1+d(x,y))^p
       \mathsf U_{s,t}^\Omega(x,y)
 \le\E^{2M(s,t)}.
\]
Together with \eqref{eq:polynomial-matrix-commutator}, this bounds the commutators needed to compare two finite volumes.

\smallskip
\noindent \emph{Comparison of finite-volume dynamics.}
Let \(X^{(R)}\subseteq\Lambda\subseteq\Omega\) be finite regions.
For \(A\in\A_X\) and \(s\le t\), differentiating \(u\mapsto\tau_{s,u}^\Omega(\tau_{u,t}^\Lambda(A))\) gives
\[
 \tau_{s,t}^\Omega(A)-\tau_{s,t}^\Lambda(A)
 =\I\int_s^t\tau_{s,u}^\Omega
       \Big(\sum_{\substack{Z\subseteq\Omega\\Z\not\subseteq\Lambda}}
                 [\Xi_u(Z),\tau_{u,t}^\Lambda(A)]\Big)\dd u.
\]
Terms with \(Z\cap\Lambda=\varnothing\) vanish because \(\Xi_u(Z)\) is even.
Apply the commutator estimate to the interaction \(\Xi^\Lambda\), regarded as an interaction on \(\Omega\), and the even test operator \(\Xi_u(Z)\).
The matrix series for this interaction agrees with \(\mathsf U_{u,t}^\Lambda(x,y)\) for \(x,y\in\Lambda\) and vanishes for \(x\in\Lambda\), \(y\notin\Lambda\): the matrices in the series are block diagonal, with restriction \(K_v^\Lambda\) to \(\Lambda\times\Lambda\).
The norm of the integrand is consequently at most
\[
 2\|A\|\sum_{x\in X}\sum_{y\in\Lambda}
       \mathsf U_{u,t}^\Lambda(x,y)
       \sum_{\substack{Z\subseteq\Omega\,|\,y\in Z\\
                       Z\not\subseteq\Lambda}}\|\Xi_u(Z)\|.
\]
Every \(Z\) in the inner sum contains a site in \(\Omega\setminus\Lambda\), so the inner sum is at most \(\frac12\sum_{z\in\Omega\setminus\Lambda}K_u^\Omega(y,z)\).
For \(x\in X\), \(y\in\Lambda\), and \(z\notin\Lambda\),
\[
 (1+R)^p\le(1+d(x,y))^p(1+d(y,z))^p.
\]
For each \(x\in X\), insert the distance weights and sum first over \(z\):
\[
 \begin{aligned}
 \sum_{y\in\Lambda}\sum_{z\in\Omega\setminus\Lambda}
       \mathsf U_{u,t}^\Lambda(x,y)K_u^\Omega(y,z) &\le(1+R)^{-p}
       \sum_{y\in\Lambda}(1+d(x,y))^p\mathsf U_{u,t}^\Lambda(x,y)
       \sum_{z\in\Omega}(1+d(y,z))^pK_u^\Omega(y,z)\\
 & \le2(1+R)^{-p}\|\Xi_u\|^{\pol}_{p;1}
       \sum_{y\in\Lambda}(1+d(x,y))^p\mathsf U_{u,t}^\Lambda(x,y)\\
 & \le2(1+R)^{-p}\E^{2M(u,t)}\|\Xi_u\|^{\pol}_{p;1}.
 \end{aligned}
\]
The last two inequalities use \eqref{eq:polynomial-matrix-row} and the weighted row bound for \(\mathsf U^\Lambda\), respectively.
Multiplying by \(\|A\|\) and summing over \(x\in X\) bounds the integrand by \(2|X|\|A\|(1+R)^{-p}\E^{2M(u,t)}\|\Xi_u\|^{\pol}_{p;1}\).

Integrating gives
\begin{equation}
 \begin{split}
 \|\tau_{s,t}^\Omega(A)-\tau_{s,t}^\Lambda(A)\|
 &\le2|X|\|A\|(1+R)^{-p}
       \int_s^t\E^{2M(u,t)}\|\Xi_u\|^{\pol}_{p;1}\,\dd u\\
 &=|X|\|A\|\bigl(\E^{2M(s,t)}-1\bigr)(1+R)^{-p}.
 \end{split}
 \label{eq:polynomial-matrix-cutoff}
\end{equation}
The substitution \(\check\Xi_v=-\Xi_{-v}\) proves the last bound for \(t<s\) as well.

\smallskip
\noindent \emph{Infinite-volume dynamics and its differential equation.}
For finite \(\Omega_1,\Omega_2\supseteq X^{(R)}\), compare both dynamics with the restriction to \(X^{(R)}\).
Equation \eqref{eq:polynomial-matrix-cutoff} gives
\[
 \sup_{s,t\in J}
 \|\tau_{s,t}^{\Omega_1}(A)-\tau_{s,t}^{\Omega_2}(A)\|
 \le2|X|\|A\|\bigl(\E^{2M(\min J,\max J)}-1\bigr)(1+R)^{-p}.
\]
Since \(X^{(R)}\) is finite and \(p>0\), these estimates make the finite-volume dynamics a Cauchy net on each local observable, uniformly on \(J^2\).
Isometry and density extend the limit to every observable in \(\A\).
The standard limit argument in \cite[proof of Theorem~3.5]{NSY19} then gives a strongly continuous automorphism cocycle: multiplication, adjoints, and the cocycle law pass to the limit, and \(\tau_{t,s}\) is the inverse of \(\tau_{s,t}\).
Parity preservation follows because every finite-volume dynamics commutes with \(\Theta\).

Put \(B_0:=\sup_{u\in J}\|\Xi_u\|^{\pol}_{p;0}\).
For local \(A\in\A_X\), the sum \(\delta_{\Xi_u}(A):=\sum_Z[\Xi_u(Z),A]\) is absolutely convergent.
If \(\Omega\supseteq X^{(R)}\), every omitted term that meets \(X\) has diameter greater than \(R\), and therefore
\[
 \sup_{u\in J}
 \|\delta_{\Xi_u}(A)-\delta_{\Xi_u^\Omega}(A)\|
 \le2|X|\|A\|B_0(1+R)^{-p}.
\]
The finite sums \(u\mapsto\delta_{\Xi_u^\Omega}(A)\) are continuous, so their uniform limit \(u\mapsto\delta_{\Xi_u}(A)\) is continuous.
Its image is compact, so isometry extends the uniform convergence of the finite-volume dynamics from fixed observables to this image.

We may consequently pass to the limit in the finite-volume integral equation and obtain
\[
 \tau_{s,t}(A)=A+\I\int_s^t
                       \tau_{s,u}(\delta_{\Xi_u}(A))\,\dd u.
\]
The integrand is norm-continuous, so differentiation in \(t\) proves \eqref{eq:dynamics}.

\smallskip
\noindent \emph{Uniqueness and local approximation.}
Let \(\alpha\) be another strongly continuous automorphism cocycle satisfying \eqref{eq:dynamics}, and fix \(\Lambda=X^{(R)}\).
For \(s\le t\), put \(B(u):=\tau_{u,t}^{\Lambda}(A)\).
This path lies in the fixed finite-dimensional algebra \(\A_\Lambda\) and satisfies \(B'(u)=-\I\delta_{\Xi_u^\Lambda}(B(u))\).
The product rule, applied in a basis of \(\A_\Lambda\), gives
\[
 \alpha_{s,t}(A)-\tau_{s,t}^\Lambda(A)
 =\I\int_s^t\alpha_{s,u}
       \left(\sum_{Z\not\subseteq\Lambda}[\Xi_u(Z),B(u)]\right)
       \dd u.
\]
The sum converges absolutely and uniformly in \(u\), since
\[
 \sup_{u\in J}\sum_{Z\not\subseteq\Lambda^{(H)}}
       \|[\Xi_u(Z),B(u)]\|
 \le2|\Lambda|B_0\|A\|(1+H)^{-p},\qquad H\ge0.
\]
The weighted matrix estimate used in \eqref{eq:polynomial-matrix-cutoff} bounds every finite partial sum of the integrand by
\(2|X|\|A\|(1+R)^{-p}\E^{2M(u,t)}\|\Xi_u\|^{\pol}_{p;1}\).
Taking the norm limit and integrating gives \eqref{eq:polynomial-matrix-cutoff} with \(\alpha_{s,t}\) in place of \(\tau_{s,t}^\Omega\).
Time reflection gives the bound for \(t<s\).
Letting \(R\to\infty\) proves \(\alpha=\tau\) on local observables and hence on \(\A\).

Letting \(\Omega\nearrow\Gamma\) in \eqref{eq:polynomial-matrix-cutoff} proves \eqref{eq:polynomial-finite-approximation}.
For \(\Lambda=X^{(R)}\), the finite-volume observable \(\tau_{s,t}^{\Xi,\Lambda}(A)\) belongs to \(\A_\Lambda\).
The conditional expectation fixes this algebra and is contractive, so
\[
 \|\tau_{s,t}^\Xi(A)
       -\F_\Lambda^{\rm tr}(\tau_{s,t}^\Xi(A))\|=\bigl\|(\id-\F_\Lambda^{\rm tr})
       (\tau_{s,t}^\Xi(A)-\tau_{s,t}^{\Xi,\Lambda}(A))\bigr\| \le2\|\tau_{s,t}^\Xi(A)-\tau_{s,t}^{\Xi,\Lambda}(A)\|.
\]
Equation~\eqref{eq:polynomial-finite-approximation} gives \eqref{eq:polynomial-cocycle-locality}.

For parity-homogeneous \(A\), put \(A_R:=\tau_{s,t}^{\Xi,X^{(R)}}(A)\).
The restricted dynamics preserve parity, so \(A_RB=(-1)^{|A||B|}BA_R\) for parity-homogeneous \(B\in\A_Y\) with \(Y\subseteq\Gamma\setminus X^{(R)}\).
Consequently,
\[
 \bigl\|\tau_{s,t}^\Xi(A)B-(-1)^{|A||B|}B\tau_{s,t}^\Xi(A)\bigr\|
 \le2\|B\|\,\|\tau_{s,t}^\Xi(A)-A_R\|.
\]
Equation~\eqref{eq:polynomial-finite-approximation} gives \eqref{eq:integrated-graded-commutator} with \(|X|\) in place of the minimum.
If \(Y\) is finite and nonempty, isometry and the inverse cocycle relation exchange \((A,X,s,t)\) and \((B,Y,t,s)\), as in the proof of Corollary~\ref{cor:observable-LR}.
Since \(M(s,t)=M(t,s)\), the same bound holds with \(|Y|\).
For empty \(Y\), the graded commutator vanishes because \(B\) is scalar.
\end{proof}

\begin{proof}[Proof of Remark~\ref{rem:general-spatial-weights}]
\phantomsection\label{proof:general-spatial-weights}
For a finite region \(\Omega\), the calculation in \eqref{eq:polynomial-matrix-row}, with \(w(d(x,y))\) in place of \((1+d(x,y))^p\), gives
\[
 \sup_{x\in\Omega}\sum_{y\in\Omega}w(d(x,y))K_u^\Omega(x,y)
 \le2\|\Xi_u^\Omega\|_{w;1}.
\]
Applying \eqref{eq:finite-volume-weighted-matrix-bound} yields
\[
 \sup_{x\in\Omega}\sum_{y\in\Omega}w(d(x,y))\mathsf U_{s,t}^\Omega(x,y)
 \le\E^{2M_w(s,t)}.
\]

If \(X^{(R)}\subseteq\Lambda\), \(x\in X\), \(y\in\Lambda\), and \(z\notin\Lambda\), then
\[
 w(R)\le w(d(x,y))w(d(y,z)).
\]
The comparison of finite-volume dynamics leading to \eqref{eq:polynomial-matrix-cutoff} therefore gives that estimate with \(M_w(s,t)\) in place of \(M(s,t)\) and \(w(R)^{-1}\) in place of \((1+R)^{-p}\).
For \(A\in\A_X\) and \(\Omega\supseteq X^{(R)}\), every interaction term omitted by restriction to \(\Omega\) that meets \(X\) has diameter greater than \(R\).
Since \(w\) is nondecreasing, the derivation tails are bounded by \(2|X|\|A\|\sup_{u\in J}\|\Xi_u\|_{w;0}/w(R)\), and the corresponding tail bound in the uniqueness argument is \(2|\Lambda|\|A\|\sup_{u\in J}\|\Xi_u\|_{w;0}/w(H)\).
Since \(w(R)\to\infty\), the convergence, differentiability, and uniqueness arguments in the preceding proof apply with these replacements.
The conditional-expectation and graded-commutator bounds follow from the finite-volume approximation as in that proof, since these steps do not involve the weight.
\end{proof}

\subsection{Polynomial bounds in time without exponential support-size weights}
\label{subsec:polynomial-time-locality}

Under a uniform bound on \(\|\Xi_u\|^{\pol}_{a;0}\), the next proposition bounds the local-approximation error for evolved observables by a polynomial in time.
It requires no cell partitions, but gives spatial decay at every exponent \(q<a\), rather than at \(a\).

\begin{proposition}
\label{prop:polynomial-time-locality}
Assume \eqref{eq:volume-growth}, and fix \(a>\dvol\) and \(0<q<a\).
There are \(C_q>0\) and an integer \(m_q\ge1\), depending only on \(a,q,\dvol,C_\Gamma\), with the following property.
Let \((r_x)_{x\in\Gamma}\) be arbitrary local multiplicities, let \(I\subseteq\R\) be an interval, let \(0\le L<\infty\), and let \((\Xi_u)_{u\in I}\) be a componentwise norm-continuous path of self-adjoint interactions satisfying
\[
 \sup_{u\in I}\|\Xi_u\|^{\pol}_{a;0}\le L.
\]
The cocycle \(\tau^\Xi\) of Proposition~\ref{prop:polynomial-dynamics} satisfies
\begin{equation}
 \|\tau^\Xi_{s,t}(A)
       -\F^{\rm tr}_{X^{(R)}}(\tau^\Xi_{s,t}(A))\|
 \le C_q|X|\|A\|(1+L|t-s|)^{m_q}(1+R)^{-q}
 \label{eq:polynomial-time-locality}
\end{equation}
for every nonempty \(X\Subset\Gamma\), \(A\in\A_X\), \(s,t\in I\), and \(R\ge0\).
The same constants apply after finite restrictions or deletion of interaction terms, whenever the resulting paths remain componentwise norm-continuous.
\end{proposition}

\begin{proof}
If \(L=0\), every interaction term vanishes and the cocycle is the identity, so the left-hand side of \eqref{eq:polynomial-time-locality} is zero.
Assume \(L>0\), work first in a finite region \(\Omega\), and write \(\tau^\Omega:=\tau^{\Xi,\Omega}\).
Fix \(s,t\in I\) and put \(T:=|t-s|\).
We prove the intermediate bounds on every subinterval between \(s\) and \(t\), using the common time horizon \(T\).
Constants denoted by \(C\) depend only on \(a,q,\dvol,C_\Gamma\); indexed constants may also depend on their indicated indices.
We adapt the range-splitting iteration of Else--Machado--Nayak--Yao \cite{EMNY20}, in the time-dependent form developed by Teufel--Wessel \cite[Proposition~1, Lemmas~2--3, and Appendix~C]{TeufelWessel25}, to one-site observables and polynomial volume growth.

\smallskip
\noindent \emph{Finite-range commutators.}
For a nonnegative nonincreasing function \(f:[0,\infty)\to[0,\infty)\), define
\[
 \mathfrak r(f):=\sup_{x\in\Gamma}\sum_{z\in\Gamma}f(d(x,z)).
\]
Splitting the distances below one and into the annuli \([2^j,2^{j+1})\), \(j\ge0\), gives
\begin{equation}
 \mathfrak r(f)
 \le C\left(f(0)+\sum_{j\ge0}2^{j\dvol}f(2^j)\right).
 \label{eq:polynomial-time-radial-sum}
\end{equation}
Indeed, the first region lies in the unit ball, and the \(j\)-th annulus contains at most \(C_\Gamma3^{\dvol}2^{j\dvol}\) sites.
In particular, \(F_a(r):=(1+r)^{-a}\) satisfies \(\mathfrak r(F_a)<\infty\).

For \(h\ge1\), define \(\Xi_u^{<h}(Z):=\1_{\{\diam Z<h\}}\Xi_u(Z)\), and let \(\tau^{<h}\) be its dynamics in \(\Omega\).
Use the matrices \(K_u^{<h}\) of Lemma~\ref{lem:finite-volume-matrix-commutator}, including their diagonal entries.
The interaction bound gives \(K_u^{<h}(x,y)\le2LF_a(d(x,y))\), and these entries vanish for \(d(x,y)\ge h\).
Thus
\[
 \sup_{x\in\Omega}\sum_{y\in\Omega}
       \E^{d(x,y)/h}K_u^{<h}(x,y)
 \le2\E L\,\mathfrak r(F_a).
\]
Choose \(c\ge\max\{1,2\E\mathfrak r(F_a)\}\), depending only on \(a,\dvol,C_\Gamma\), and put \(v:=cL\).
Applying \eqref{eq:finite-volume-weighted-matrix-bound} with \(w(r)=\E^{r/h}\), then \eqref{eq:polynomial-matrix-commutator} and the trivial commutator bound, gives
\[
 \|[\tau^{<h}_{s,t}(A_x),B]\|
 \le\|A_x\|\|B\|\lambda_h^{(0)}(\dist(x,Y)),
 \qquad
 \lambda_h^{(0)}(r):=\min\{2,2\E^{vT-r/h}\},
\]
for \(x\in\Omega\), \(A_x\in\A_{\{x\}}\), and \(B\in\A_Y^+\) with nonempty \(Y\subseteq\Omega\).

\smallskip
\noindent \emph{Range splitting and iteration.}
For \(1\le h'<h\), the Duhamel comparison of the two range cuts gives
\[
 \|[\tau^{<h}_{s,t}(A_x),B]\|
 \le\|[\tau^{<h'}_{s,t}(A_x),B]\| +2\|B\|\int_{\min(s,t)}^{\max(s,t)}
       \sum_{\substack{Z\subseteq\Omega\\h'\le\diam Z<h}}
       \|[\tau^{<h'}_{u,t}(A_x),\Xi_u(Z)]\|\,\dd u,
\]
as in Teufel--Wessel \cite[Lemma~2]{TeufelWessel25}.
This comparison holds for either ordering of \(s,t\).

Suppose that, for every \(h\ge1\), a nonnegative nonincreasing function \(\lambda_h\) satisfies
\[
 \|[\tau^{<h}_{u,w}(A_x),B]\|
 \le\|A_x\|\|B\|\lambda_h(\dist(x,Y))
\]
for all \(u,w\) between \(s\) and \(t\), all \(x\in\Omega\), \(A_x\in\A_{\{x\}}\), and \(B\in\A_Y^+\).
Since \(\Xi_u(Z)\) is even, this bound applies to every term in the Duhamel sum.
Choosing a site of each \(Z\) nearest to \(x\) and enlarging to a sum over all sites gives
\begin{align*}
 \sum_{\substack{Z\subseteq\Omega\\h'\le\diam Z<h}}
       \lambda_{h'}(\dist(x,Z))\|\Xi_u(Z)\| \le\sum_{z\in\Omega}\lambda_{h'}(d(x,z))
       \sum_{\substack{Z\ni z\\\diam Z\ge h'}}\|\Xi_u(Z)\|
 \le L(1+h')^{-a}\mathfrak r(\lambda_{h'}).
\end{align*}
Bound the first commutator in the Duhamel estimate by \(\lambda_{h'}^{(0)}\).
For any choice \(h'=h'(r)\ge1\), the resulting bound is
\begin{equation}
 \widetilde\lambda_h(r):=
 \min\left\{2,\,
       2\E^{vT-r/h'}+
       2TL\1_{\{h>h'\}}(1+h')^{-a}
                                      \mathfrak r(\lambda_{h'})\right\}.
 \label{eq:polynomial-time-range-recurrence}
\end{equation}
When \(h'\ge h\), the finite-range estimate alone gives this bound, since \(\E^{-r/h}\le\E^{-r/h'}\).

Choose
\[
 \alpha:=\frac12\left(1+
       \max\left\{\frac{\dvol}{a},\frac qa\right\}\right),
 \qquad \upsilon:=1-\alpha,\qquad
 \varepsilon:=\alpha a-\dvol>0,\qquad
 \widehat T:=vT.
\]
Then \(\alpha<1\) and \(\alpha a>\max\{\dvol,q\}\).
For \(r\ge1\), take \(h'=r^\alpha\) in \eqref{eq:polynomial-time-range-recurrence}.
We will construct successive bounds of the form
\begin{equation}
 \lambda_h^{(n)}(r):=
 \min\left\{2,\,
  2\E^{\widehat T-r^\upsilon}
  +C_n\1_{\{h>r^\alpha\}}
       (1+\widehat T)^{\dvol/\upsilon+n}(1+r)^{\mu_n}\right\},
 \qquad n\ge1,
 \label{eq:polynomial-time-iteration}
\end{equation}
and set \(\lambda_h^{(n)}(r):=2\) for \(0\le r<1\).
The exponents \(\mu_n\) below are negative; hence these functions are nonincreasing in \(r\) and nondecreasing in \(\widehat T\).
The latter property makes each bound valid on all the subintervals used in the recurrence.

Equation~\eqref{eq:polynomial-time-radial-sum} gives
\[
 \begin{aligned}
 \mathfrak r\bigl(r\mapsto\min\{2,2\E^{\widehat T-r/h}\}\bigr)
 \le Ch^{\dvol}(1+\widehat T)^{\dvol}, \quad
 \mathfrak r\bigl(r\mapsto\min\{2,2\E^{\widehat T-r^\upsilon}\}\bigr)
 \le C(1+\widehat T)^{\dvol/\upsilon}.
 \end{aligned}
\]
For the first inequality, split the radial sum at \(2h(1+\widehat T)\); outside this radius, \(\widehat T-r/h\le-r/(2h)\).
For the second, split at \([2(1+\widehat T)]^{1/\upsilon}\); outside this radius, \(\widehat T-r^\upsilon\le-r^\upsilon/2\).
Volume growth bounds the sums inside these radii by their \(\dvol\)-th powers.
On successive dyadic annuli outside them, the volume factors grow as \(2^{j\dvol}\), whereas the exponential factors decay at least as \(\E^{-2^j}\), respectively \(\E^{-2^{j\upsilon}}\); this proves both bounds.
Substituting the first radial bound into \eqref{eq:polynomial-time-range-recurrence}, with \(TL\le\widehat T\), proves \eqref{eq:polynomial-time-iteration} for
\[
 \mu_1:=\alpha(\dvol-a)<0.
\]
Here the time power \(\dvol+1\) is bounded by \(\dvol/\upsilon+1\).

For every \(\mu<0\), \(\varepsilon'>0\), and \(U\ge1\), dyadic summation also gives
\begin{equation}
 \sup_{x\in\Gamma}\sum_{\substack{z\in\Gamma\\d(x,z)<U}}
                       (1+d(x,z))^\mu
 \le C_{\mu,\varepsilon'}
       (1+U)^{\max\{\dvol+\mu,0\}+\varepsilon'}.
 \label{eq:polynomial-time-truncated-sum}
\end{equation}
The annular contributions are bounded by \(C2^{j(\dvol+\mu)}\).
Their sum is bounded independently of \(U\) when \(\mu<-\dvol\), grows at most logarithmically when \(\mu=-\dvol\), and is bounded by \(C(1+U)^{\dvol+\mu}\) when \(\mu>-\dvol\).
These three bounds imply \eqref{eq:polynomial-time-truncated-sum}.
For \(\mu<-\dvol\), we will use the uniform bound instead.

In \(\mathfrak r(\lambda_{h'}^{(n)})\), the indicator in \eqref{eq:polynomial-time-iteration} restricts the algebraic term to distances less than \((h')^{1/\alpha}\).
For \(h'=r^\alpha\), this radius is \(r\).
The distances below one contribute a uniform constant.
Using \(\min\{2,A+B\}\le\min\{2,A\}+B\) for \(A,B\ge0\) and \eqref{eq:polynomial-time-truncated-sum} with \(\varepsilon'=\varepsilon/2\), we obtain
\[
 \mathfrak r(\lambda_{h'}^{(n)})
 \le C(1+\widehat T)^{\dvol/\upsilon} +C_n'(1+\widehat T)^{\dvol/\upsilon+n}
       (1+r)^{\max\{\dvol+\mu_n,0\}+\varepsilon/2}\,.
\]
As long as \(\mu_n\ge-\dvol\), substitution in \eqref{eq:polynomial-time-range-recurrence} gives
\[
 \widetilde\lambda_h(r)\le
 \min\left\{2,\,
 2\E^{\widehat T-r^\upsilon}
 +C_{n+1}\1_{\{h>r^\alpha\}}
 (1+\widehat T)^{\dvol/\upsilon+n+1}
 (1+r)^{\mu_n-\varepsilon/2}\right\}.
\]
Indeed, \(-\alpha a+\dvol+\mu_n+\varepsilon/2=\mu_n-\varepsilon/2\), and the first term in the radial bound contributes the faster spatial decay \((1+r)^{-\alpha a}\).
The factor \(TL\le1+\widehat T\) increases the time power by at most one.
Thus \eqref{eq:polynomial-time-iteration} holds at the next step with \(\mu_{n+1}:=\mu_n-\varepsilon/2\).
For \(r\ge1\), replacing \(r\) by \(1+r\) in these power bounds changes only constants.

Set
\[
 n_*:=\max\left\{1,\,
       2+\left\lfloor\frac{2(\mu_1+\dvol)}{\varepsilon}
                                                \right\rfloor\right\}.
\]
Then \(\mu_{n_*}=\mu_1-(n_*-1)\varepsilon/2<-\dvol\).
The uniform bound for the truncated power sum now gives
\[
 \mathfrak r(\lambda_{h'}^{(n_*)})
 \le C(1+\widehat T)^{\dvol/\upsilon+n_*}.
\]
A final application of \eqref{eq:polynomial-time-range-recurrence}, with \(h'=r^\alpha\) and the remaining indicator bounded by one, therefore yields
\[
 \widetilde\lambda_h(r)\le
 \min\left\{2,\,
 2\E^{\widehat T-r^\upsilon}
 +C(1+\widehat T)^{\dvol/\upsilon+n_*+1}
          (1+r)^{-\alpha a}\right\}.
\]
For \(r\le[2(1+\widehat T)]^{1/\upsilon}\), the trivial bound is at most \(C(1+\widehat T)^{q/\upsilon}(1+r)^{-q}\).
For larger \(r\), use \(\widehat T-r^\upsilon\le-r^\upsilon/2\) and the boundedness of \((1+r)^q\E^{-r^\upsilon/2}\).
Since \(\alpha a>q\), we may choose
\[
 m_q:=\left\lceil\max\left\{\frac q\upsilon,\,
                   \frac{\dvol}\upsilon+n_*+1\right\}\right\rceil.
\]
Taking \(h>\diam\Omega\) gives the full finite-volume dynamics and proves
\begin{equation}
 \|[\tau^\Omega_{s,t}(A_x),B]\|
 \le C\|A_x\|\|B\|(1+LT)^{m_q}
                         (1+\dist(x,Y))^{-q},
 \quad A_x\in\A_{\{x\}},\quad B\in\A_Y^+.
 \label{eq:polynomial-time-one-site}
\end{equation}
We used \(1+\widehat T\le c(1+LT)\) and increased \(C\) to cover radii below one.

\smallskip
\noindent \emph{General supports and arbitrary parity.}
For nonempty \(X,Y\subseteq\Omega\), let \(A\in\A_X^+\), let \(B\in\A_Y\), and put \(C_A:=\tau^\Omega_{s,t}(A)\).
The observable \(\F^{\rm tr}_{\Omega\setminus Y}(C_A)\) is even and commutes with \(B\).
The single-site averaging bound \eqref{eq:single-site-averaging} therefore gives
\[
 \|[C_A,B]\|
  \le2\|B\|\,\|C_A-\F^{\rm tr}_{\Omega\setminus Y}(C_A)\| \le2\|B\|\sum_{y\in Y}
       \sup_{D_y\in\A_{\{y\}},\,\|D_y\|\le1}
                       \|[C_A,D_y]\|\,.
\]
Isometry and the inverse cocycle relation imply
\[
 \|[C_A,D_y]\|
 =\|[A,\tau^\Omega_{t,s}(D_y)]\|.
\]
Apply \eqref{eq:polynomial-time-one-site} with the one-site observable \(D_y\), the even test observable \(A\), and the reversed parameters.
Since \(\dist(y,X)\ge\dist(X,Y)\), we obtain
\[
 \|[\tau^\Omega_{s,t}(A),B]\|
 \le C|Y|\|A\|\|B\|(1+LT)^{m_q}
                         (1+\dist(X,Y))^{-q}.
\]
When \(B\) is also even, reversing the dynamics and interchanging \(A,B\) gives the same bound with \(|X|\) in place of \(|Y|\).
For arbitrary \(B\), choose \(y\in Y\) and apply Lemma~\ref{lem:auxiliary-even-embedding}.
Denote by \(\widehat\tau^\Omega\) the dynamics of the same interaction terms on the enlarged algebra.
The embedding \(\jmath_y\) fixes \(A\), maps \(B\) into the even algebra on \(Y\), and intertwines the dynamics.
It is isometric, so
\[
 \|[\tau^\Omega_{s,t}(A),B]\|
 =\|[\widehat\tau^\Omega_{s,t}(A),\jmath_y(B)]\|
 \le C\min\{|X|,|Y|\}\|A\|\|B\|
       (1+LT)^{m_q}(1+\dist(X,Y))^{-q}.
\]
The enlarged interaction has the same norm bound, and the preceding argument is independent of the local multiplicities.

Put \(\Omega_R:=\Omega\cap X^{(R)}\) and \(Y:=\Omega\setminus \Omega_R\).
For even \(C_A\), the product \(\prod_{y\in Y}\cP_y(C_A)=\F_{\Omega_R}^{\rm tr}(C_A)\) from the single-site averaging argument is an average of conjugations by products of the unitaries \(V_y\in\scrV_y\).
Every such product is a unitary in \(\A_Y\).
If \(Y\ne\varnothing\), the last commutator bound therefore gives
\[
 \|C_A-\F_{\Omega_R}^{\rm tr}(C_A)\|
 \le\sup_{U\in\A_Y,\,U^*U=UU^*=\1}\|[C_A,U]\|
 \le C|X|\|A\|(1+LT)^{m_q}(1+R)^{-q}.
\]
If \(Y=\varnothing\), the approximation error is zero.
For arbitrary \(A\in\A_X\), choose \(x\in X\) and apply Lemma~\ref{lem:auxiliary-even-embedding} once more.
Its intertwining identities and isometry give
\[
 \|\tau^\Omega_{s,t}(A)-\F_{\Omega_R}^{\rm tr}(\tau^\Omega_{s,t}(A))\|
 =
 \|\widehat\tau^\Omega_{s,t}(\jmath_x(A))
    -\widehat\F_{\Omega_R}^{\rm tr}(\widehat\tau^\Omega_{s,t}(\jmath_x(A)))\|,
\]
so the estimate for even observables proves the same bound for \(A\).

\smallskip
\noindent \emph{Infinite volume and uniformity.}
Proposition~\ref{prop:polynomial-dynamics}, with \(p=a>\dvol\), constructs the dynamics and gives norm convergence of finite-volume evolutions on every compact subinterval of \(I\).
The conditional expectations are contractive and agree on local algebras, so the finite-volume approximation errors converge to the left side of \eqref{eq:polynomial-time-locality}.
This proves that estimate for all \(s,t\in I\).
Every bound above depends only on \(a,q,\dvol,C_\Gamma\) and the common interaction bound \(L\), which is unchanged or decreased by finite restrictions and deletion of terms.
This proves the uniformity under finite restrictions and deletion of terms.
\end{proof}

Proposition~\ref{prop:polynomial-time-locality} also controls time integrals of individual observables.
Assume \(I=\R\), and let \(f:\R\to\C\) be measurable with
\[
 \int_\R |f(v)|(1+L|v|)^{m_q}\,\dd v<\infty.
\]
Isometry and \(f\in L^1(\R)\) give absolute convergence in norm of \(\int_\R f(v)\tau^\Xi_{s,s+v}(A)\,\dd v\).
Applying the bounded map \(\id-\F^{\rm tr}_{X^{(R)}}\) under the integral and using \eqref{eq:polynomial-time-locality} gives
\begin{align}
 \left\|(\id-\F^{\rm tr}_{X^{(R)}})
       \int_\R f(v)\tau^\Xi_{s,s+v}(A)\,\dd v\right\| \le C_q|X|\|A\|(1+R)^{-q}
       \int_\R |f(v)|(1+L|v|)^{m_q}\,\dd v.
 \label{eq:polynomial-time-filtered-observable}
\end{align}
The estimate is uniform in \(s\in\R\) and also holds for time-dependent \(\Xi\).

The local-approximation estimate of Proposition~\ref{prop:polynomial-time-locality} also gives a decomposition of each evolved interaction term.
The next corollary controls the weighted sum of the norms of all terms whose assigned support contains a fixed site.
Keeping the original support \(Z\) as a label allows us to sum over selected inputs, such as interaction terms crossing a boundary.

\begin{corollary}
\label{cor:polynomial-interaction-shells}
Assume \eqref{eq:volume-growth}, and fix \(a>\dvol\) and \(0\le p<a-\dvol\).
There are \(C>0\) and an integer \(m\ge1\), depending only on \(a,p,\dvol,C_\Gamma\), with the following property.
Let \((r_x)_{x\in\Gamma}\) be arbitrary local multiplicities, let \(I\subseteq\R\) be an interval, and let \(0\le L<\infty\).
Let \((\Xi_u)_{u\in I}\) be a componentwise norm-continuous path of self-adjoint interactions, and let \(\Upsilon\) be an interaction, with
\[
 \sup_{u\in I}\|\Xi_u\|^{\pol}_{a;0}\le L,
 \qquad \|\Upsilon\|^{\pol}_{p;1}<\infty.
\]
Define the dyadic radii by \(R_0:=0\) and \(R_n:=2^{n-1}\) for \(n\ge1\).
For every nonempty finite \(Z\), define the shell maps by
\[
 \begin{aligned}
 \Delta_{Z,0}^{\rm tr}&:=\F_Z^{\rm tr},\\
 \Delta_{Z,n}^{\rm tr}
 &:=\F_{Z^{(R_n)}}^{\rm tr}
       -\F_{Z^{(R_{n-1})}}^{\rm tr}\quad(n\ge1),
 \end{aligned}
\]
and, for the cocycle \(\tau^\Xi\) of Proposition~\ref{prop:polynomial-dynamics}, set
\[
 \Upsilon_{s,t;Z,n}:=\Delta_{Z,n}^{\rm tr}
                       \tau_{s,t}^\Xi(\Upsilon(Z)),
 \qquad s,t\in I,\quad n\ge0.
\]
The shell terms satisfy
\begin{equation}
 \sup_{y\in\Gamma}
 \sum_{\substack{\varnothing\ne Z\Subset\Gamma,\,n\ge0\\
                  y\in Z^{(R_n)}}}
 (1+\diam Z^{(R_n)})^p\|\Upsilon_{s,t;Z,n}\|
 \le C(1+L|t-s|)^m\|\Upsilon\|^{\pol}_{p;1},
 \qquad s,t\in I.
 \label{eq:polynomial-shell-rooted-bound}
\end{equation}
Each shell term is even and supported in \(Z^{(R_n)}\), and
\[
 \tau_{s,t}^\Xi(\Upsilon(Z))=\sum_{n\ge0}\Upsilon_{s,t;Z,n}
 \quad\text{absolutely in norm}.
\]
If \(\Upsilon\) is self-adjoint, every shell term is self-adjoint.
The same constants apply after finite restrictions or deletion of terms of either interaction, provided the Hamiltonian path remains componentwise norm-continuous.
\end{corollary}

\begin{proof}
Fix \(s,t\in I\).
Set \(q:=(a+p+\dvol)/2\), so that \(p+\dvol<q<a\), and take \(m=m_q\) from Proposition~\ref{prop:polynomial-time-locality}.
For \(n\ge1\), the conditional-expectation identities give
\[
 \Delta_{Z,n}^{\rm tr}
 =\F_{Z^{(R_n)}}^{\rm tr}
       \bigl(\id-\F_{Z^{(R_{n-1})}}^{\rm tr}\bigr).
\]
Contractivity, Proposition~\ref{prop:polynomial-time-locality}, and \(1+R_n\le2(1+R_{n-1})\) therefore give
\[
 \|\Upsilon_{s,t;Z,n}\|
 \le C|Z|\|\Upsilon(Z)\|
        (1+L|t-s|)^m(1+R_n)^{-q}.
\]
For \(n=0\), the same bound follows from contractivity of \(\F_Z^{\rm tr}\), after increasing \(C\) if necessary.
For every \(R\ge0\),
\[
 1+\diam Z^{(R)}
 \le 2(1+\diam Z)(1+R).
\]
For every \(y\in\Gamma\), the volume bound and the definition of \(\|\Upsilon\|^{\pol}_{p;1}\) give
\[
 \begin{aligned}
 \sum_{\dist(y,Z)\le R_n}
       |Z|(1+\diam Z)^p\|\Upsilon(Z)\|&\le\sum_{x\in\ball_{R_n}(y)}\sum_{Z\ni x}
       |Z|(1+\diam Z)^p\|\Upsilon(Z)\|\\
 &\le C_\Gamma(1+R_n)^{\dvol}
       \|\Upsilon\|^{\pol}_{p;1}.
 \end{aligned}
\]
Combining these estimates yields
\[
 \sum_{\substack{\varnothing\ne Z\Subset\Gamma,\,n\ge0\\
                   y\in Z^{(R_n)}}}
 (1+\diam Z^{(R_n)})^p\|\Upsilon_{s,t;Z,n}\|\le C(1+L|t-s|)^m\|\Upsilon\|^{\pol}_{p;1}
       \sum_{n\ge0}(1+R_n)^{p+\dvol-q}.
\]
The final dyadic series converges because \(q>p+\dvol\); this proves \eqref{eq:polynomial-shell-rooted-bound}.
The dynamics and the shell maps preserve parity and commute with taking adjoints.
The shell estimate gives \(\sum_{n\ge0}\|\Upsilon_{s,t;Z,n}\|<\infty\) for each \(Z\).
The identity
\[
 \sum_{n=0}^N \Upsilon_{s,t;Z,n}
 =\F_{Z^{(R_N)}}^{\rm tr}\bigl(\tau_{s,t}^\Xi(\Upsilon(Z))\bigr)
\]
and norm convergence of the local conditional expectations prove the expansion.
Finite restrictions and deletion of terms do not increase the interaction norms and preserve the local-approximation estimate of Proposition~\ref{prop:polynomial-time-locality}, so the same constants apply.
\end{proof}

\section{Applications and extensions of the propagation bounds}
\label{sec:applications}

In this section, time-filtered interactions serve as generators of flows in a real parameter.
We establish the locality, convergence, and smoothness properties of these flows needed for gap stability \cite{Stability} and LPPL \cite{LPPLBdG}.
With quasi-adiabatic filters, the flows transport isolated spectral sectors and their infinite-volume limit states.
We show that the filtered generators and their flows converge when finite-region cutoffs on an extensive perturbation are removed, as needed, for example, in the gap-stability proof \cite{Stability}.
The final subsection treats the extensions to unbounded onsite terms and spin systems.

\subsection{Flows generated by filtered interactions}
\label{subsec:coupling-flows}

The construction starts from two paths of self-adjoint interactions, \((\Phi_\lambda)_{\lambda\in J}\) and \((\Psi_\lambda)_{\lambda\in J}\), indexed by a real parameter \(\lambda\) in an interval \(J\).
For fixed \(\lambda\), we evolve \(\Psi_\lambda\) in physical time \(t\) with the autonomous dynamics \(\tau_t^{\Phi_\lambda}\) generated by \(\Phi_\lambda\), and integrate against a real filter \(W\).
The resulting interaction \(\eta_\lambda\) generates a flow \(\beta_{\lambda,\mu}\) in the parameter \(\lambda\).
The filter assumptions below give polynomial support-size bounds on \(\eta_\lambda\), but do not in general give the exponential support-size bound required by Theorem~\ref{thm:card-propagation}.

\begin{corollary}
\label{cor:filtered-flow}
Assume \eqref{eq:volume-growth} and Hypothesis~\ref{hyp:cell-partitions}, and choose \(\kappa_0\) as in Theorem~\ref{thm:card-propagation}.
Fix \(a>\dvol\), \(0\le L<\infty\), and a measurable real filter \(W\) satisfying
\[
 \int_\R |W(t)|(1+|t|)^{m_{a,1}}\,\dd t<\infty.
\]
Use the constants \(C_{a,k}\) from Theorem~\ref{thm:card-propagation} for the Hamiltonian bound \(L\), and set
\[
 C_{W,k}:=C_{a,k}\mu_{a,k}(W),
 \qquad k\ge0.
\]
If \(W\) has finite absolute moments of every order, there are also constants \(C_n^{\rm sm}>0\), \(n\ge0\), depending only on \(n,a,L\), finitely many absolute moments of \(W\), and the geometric constants, for the smoothness bound \eqref{eq:smooth-coupling-bound} below.
Let \(J\subseteq\R\) be a compact interval and \((\Phi_\lambda)_{\lambda\in J}\) a componentwise norm-continuous path of self-adjoint interactions satisfying
\[
 \sup_{\lambda\in J}\|\Phi_\lambda\|^{\#}_{a;\kappa_0}\le L.
\]
Regard each \(\Phi_\lambda\) as a time-independent Hamiltonian, and use Definition~\ref{def:interaction-representatives} with the partitions and cutoffs chosen in Theorem~\ref{thm:card-propagation} for \(a,L\).
Let \((\Psi_\lambda)_{\lambda\in J}\) be any componentwise norm-continuous path of self-adjoint interactions with
\[
 \sup_{\lambda\in J}
       \|\Psi_\lambda\|^{\#}_{a;\kappa_0}<\infty.
\]
Using the componentwise definition \eqref{eq:card-filter-interaction} of \(\cK\), set
\[
 \eta_\lambda(Y):=\cK^{\Phi_\lambda}_{W,0}[\Psi_\lambda](Y)
 =\int_\R W(t)(\cT^{\Phi_\lambda}_{0,t}\Psi_\lambda)(Y)\,\dd t
\]
for every nonempty \(Y\Subset\Gamma\).
\begin{enumerate}[label=(\roman*),ref=(\roman*),leftmargin=*]
\item \emph{Filtered generators and locality.}
\label{item:filtered-flow-continuity}
For every \(k\ge0\) with \(C_{W,k}<\infty\),
\begin{equation}
 \sup_{\lambda\in J}\|\eta_\lambda\|^{\pol}_{a;k}
 \le C_{W,k}\sup_{\lambda\in J}
                  \|\Psi_\lambda\|^{\#}_{a;\kappa_0}.
 \label{eq:application-generator}
\end{equation}
The interaction path \(\eta\) is componentwise norm-continuous and self-adjoint.
It generates a unique strongly continuous parity-preserving automorphism cocycle \(\beta_{\lambda,\mu}\), with
\begin{equation}
 \sup_{\lambda,\mu\in J}
 \|\beta_{\lambda,\mu}(A)-
       \F_{X^{(R)}}^{\rm tr}(\beta_{\lambda,\mu}(A))\|
 \le 2\left[\exp\!\left(2C_{W,1}
       \int_J\|\Psi_u\|^{\#}_{a;\kappa_0}\,\dd u\right)-1\right]
       |X|\|A\|(1+R)^{-a}
 \label{eq:application-flow}
\end{equation}
for every nonempty finite \(X\), \(A\in\A_X\), and \(R\ge0\).
The graded commutator bound \eqref{eq:integrated-graded-commutator} also holds for \(\beta=\tau^\eta\), with \(p=a\); the integrated norm of \(\eta\) satisfies
\[
 \int_{\min(\lambda,\mu)}^{\max(\lambda,\mu)}
       \|\eta_u\|^{\pol}_{a;1}\,\dd u
 \le C_{W,1}\int_{\min(\lambda,\mu)}^{\max(\lambda,\mu)}
          \|\Psi_u\|^{\#}_{a;\kappa_0}\,\dd u.
\]
The representation identities \eqref{eq:filtered-derivation}--\eqref{eq:filtered-total-operator} apply with \(\xi=\lambda\) and \(s=0\), under the summability assumptions stated there.

\item \emph{Convergence of filtered interactions and their flows.}
\label{item:filtered-flow-approximation}
Suppose that \((\Phi^{(N)}_\lambda)_{\lambda\in J}\), \(N\in\N\), are componentwise norm-continuous paths of self-adjoint interactions, with
\[
 \sup_{N\in\N}\sup_{\lambda\in J}
       \|\Phi^{(N)}_\lambda\|^{\#}_{a;\kappa_0}\le L,
 \qquad
 \sup_{\lambda\in J}
       \|\Phi^{(N)}_\lambda(Y)-\Phi_\lambda(Y)\|\xrightarrow{N\to\infty}0
\]
for every nonempty finite \(Y\).
Use the same partitions and cutoffs for \(\Phi\) and every \(\Phi^{(N)}\).
Then the following holds for every sequence of componentwise norm-continuous paths of self-adjoint interactions \((\Psi^{(N)}_\lambda)_{\lambda\in J}\) satisfying
\[
 \sup_{N\in\N}\sup_{\lambda\in J}
       \|\Psi^{(N)}_\lambda\|^{\#}_{a;\kappa_0}<\infty,
 \qquad
 \sup_{\lambda\in J}
       \|\Psi^{(N)}_\lambda(Y)-\Psi_\lambda(Y)\|\xrightarrow{N\to\infty}0
\]
for every nonempty finite \(Y\).
The filtered interactions
\[
 \eta^{(N)}_\lambda:=\cK^{\Phi^{(N)}_\lambda}_{W,0}[\Psi^{(N)}_\lambda]
\]
satisfy, for every nonempty finite \(K\), \(0\le q<a\), and \(k\ge0\) with \(\mu_{q,k}(W)<\infty\),
\[
 \sup_{\lambda\in J}
 \|\eta_\lambda^{(N)}-\eta_\lambda\|^{\pol,\loc}_{q;k,K}
 \xrightarrow{N\to\infty}0.
\]
Writing \(\beta^{(N)}:=\tau^{\eta^{(N)}}\), one also has
\[
 \sup_{\lambda,\mu\in J}
 \|\beta^{(N)}_{\lambda,\mu}(A)-\beta_{\lambda,\mu}(A)\|
 \xrightarrow{N\to\infty}0,\qquad A\in\A.
\]

\item \emph{Smooth dependence on the perturbation parameter.}
\label{item:filtered-flow-smoothness}
Suppose additionally that \(L>0\), \(\lambda\mapsto\Phi_\lambda\) and \(\lambda\mapsto\Psi_\lambda\) are \(C^\infty\) in \(\|\cdot\|^{\#}_{a;\kappa_0}\) on a neighborhood of \(J\), \(\|\Phi_\lambda\|^{\#}_{a;\kappa_0}<L\) for every \(\lambda\in J\), and \(W\) has finite absolute moments of every order.
Then \(\lambda\mapsto\eta_\lambda\) is \(C^\infty\) in every \(\|\cdot\|^{\pol}_{a;k}\)-norm.
For every nonempty finite \(X\), \(A\in\A_X\), and fixed \(\lambda_0\in J\), \(\lambda\mapsto\beta_{\lambda_0,\lambda}(A)\) is norm-\(C^\infty\).
Both maps are smooth on the interior of \(J\), and all their derivatives have continuous one-sided extensions to the endpoints.
For affine paths \(\Phi_\lambda=\Phi_0+\lambda \cV\) with \(\Psi_\lambda=\cV\), the following bound holds for every \(n\ge0\):
\begin{equation}
 \sup_{\lambda_0,\lambda\in J}
 \bigl\|\partial_\lambda^n\beta_{\lambda_0,\lambda}(A)\bigr\|
 \le C_n^{\rm sm}|X|^n\|A\|
             \bigl(\|\cV\|^{\#}_{a;\kappa_0}\bigr)^n.
 \label{eq:smooth-coupling-bound}
\end{equation}
For \(n=0\), take \(C_0^{\rm sm}=1\) and interpret the zeroth powers as one.
If the self-adjoint approximating families \(\Phi_\lambda^{(N)},\Psi_\lambda^{(N)}\) of part~\ref{item:filtered-flow-approximation} satisfy the approximation hypotheses of Theorem~\ref{thm:smooth-hamiltonian-dependence}\ref{item:smooth-approximation} for every \(r\), every derivative of \(\eta^{(N)}\) converges in the local seminorms stated there, and
\begin{equation}
 \sup_{\lambda_0,\lambda\in J}
 \left\|\partial_\lambda^n
   \bigl(\beta^{(N)}_{\lambda_0,\lambda}(A)
                -\beta_{\lambda_0,\lambda}(A)\bigr)\right\|
 \xrightarrow{N\to\infty}0,\qquad n\ge0.
 \label{eq:smooth-coupling-convergence}
\end{equation}
For every \(n\ge0\) and \(k\ge0\), the norms \(\|\partial_\lambda^n\eta_\lambda^{(N)}\|^{\pol}_{a;k}\) and \(\|\partial_\lambda^n\beta^{(N)}_{\lambda_0,\lambda}(A)\|\) are bounded uniformly in \(N\) and \(\lambda_0,\lambda\in J\).
For affine families \(\Phi_\lambda^{(N)}=\Phi_0^{(N)}+\lambda\cV^{(N)}\) with \(\Psi_\lambda^{(N)}=\cV^{(N)}\), \eqref{eq:smooth-coupling-bound} holds uniformly under the common bounds on \(\|\cV^{(N)}\|^{\#}_{a;\kappa_0}\).
\end{enumerate}
\end{corollary}

\begin{proof}
\smallskip
\noindent \emph{\ref{item:filtered-flow-continuity} Filtered generators and locality.}
Apply Theorem~\ref{thm:card-limits}\ref{item:card-limits-coefficients}--\ref{item:card-limits-filtered-convergence} with \(\frakP=J\), \(I=\R\), \(\Phi_{\lambda,u}:=\Phi_\lambda\), and \(s=0\).
Then \(\cK_{\lambda,0}=\eta_\lambda\), so parts \ref{item:card-limits-coefficients}--\ref{item:card-limits-identities} give componentwise norm-continuity, parity, self-adjointness, \eqref{eq:application-generator}, and the representation identities under their summability assumptions.
Since \(m_{a,0}\le m_{a,1}\), one has \(C_{W,0}<\infty\), and part~\ref{item:card-limits-estimates} gives
\[
 \sup_{\lambda\in J}\|\eta_\lambda\|^{\pol}_{a;0}<\infty,
 \qquad
 \int_J\|\eta_u\|^{\pol}_{a;1}\,\dd u
 \le C_{W,1}\int_J
            \|\Psi_u\|^{\#}_{a;\kappa_0}\,\dd u<\infty.
\]
Theorem~\ref{thm:polynomial-local-topology}, with \(p=a\), therefore gives the unique flow \(\beta:=\tau^\eta\) generated by \(\eta\).
Substituting the integral bound into \eqref{eq:polynomial-cocycle-locality} proves \eqref{eq:application-flow}.
Applying the pointwise generator bound on the interval between \(\lambda\) and \(\mu\) also gives the graded commutator estimate of part~\ref{item:filtered-flow-continuity}.

\smallskip
\noindent \emph{\ref{item:filtered-flow-approximation} Convergence of filtered interactions and their flows.}
For the approximating Hamiltonians, put \(\Phi^{(N)}_{\lambda,u}:=\Phi^{(N)}_\lambda\).
Apply Theorem~\ref{thm:card-limits} to the approximating and limiting families with the common partitions and cutoffs.
Part~\ref{item:card-limits-filtered-convergence} gives local interaction convergence whenever \(\mu_{q,k}(W)<\infty\).
Part~\ref{item:card-limits-coefficients} gives componentwise norm-continuity and uniform componentwise convergence of \(\eta^{(N)}\), and part~\ref{item:card-limits-estimates} gives a common \(\|\cdot\|^{\pol}_{a;0}\)-bound.
Uniform componentwise convergence on the compact interval \(J\) implies componentwise \(L^1(J)\)-convergence.
Applying Proposition~\ref{prop:polynomial-dynamics} with \(p=a>\dvol\) proves convergence of \(\beta^{(N)}_{\lambda,\mu}(A)\) to \(\beta_{\lambda,\mu}(A)\), uniformly for \(\lambda,\mu\in J\), for every \(A\in\A\).

\smallskip
\noindent \emph{\ref{item:filtered-flow-smoothness} Smooth dependence on the perturbation parameter.}
Restrict Theorem~\ref{thm:smooth-hamiltonian-dependence}\ref{item:smooth-filtering} to autonomous Hamiltonians and apply the chain rule to \((\Phi_\lambda,\Psi_\lambda)\).
This gives \(C^\infty\) dependence of \(\eta_\lambda\).
Compactness of \(J\) then gives
\[
 \sup_{\lambda\in J}
 \|\partial_\lambda^j\eta_\lambda\|^{\pol}_{p;k}
 <\infty,
 \qquad j\ge0,\quad 0\le p\le a,\quad k\ge0.
\]
To differentiate \(\beta_{\lambda_0,\lambda}(A)\) repeatedly, we construct the observables on which each successive generator acts.
For \(A\in\A_X\), define families of local terms recursively by
\[
 \frakb_{\lambda,0}(Y):=\1_{\{Y=X\}}A,
 \qquad
 \frakb_{\lambda,r+1}
 :=\partial_\lambda\frakb_{\lambda,r}
                 +\I[\eta_\lambda,\frakb_{\lambda,r}].
\]
Here the bracket is defined in \eqref{eq:site-interaction-bracket}; by evenness of \(\eta_\lambda\), it accounts for all nonzero commutators, whatever the parity of \(A\).
Use the total norms \(\frakt_{p;k}\) defined in \eqref{eq:total-support-norm}.

Induction on \(r\) gives the explicit form
\begin{equation}
 \frakb_{\lambda,r}
 =\sum_{m=0}^r\ \sum_{\substack{j\in\N_0^m\\ m+|j|=r}}
   c_{j}\,\I^m
   [\partial_\lambda^{j_1}\eta_\lambda,
    [\partial_\lambda^{j_2}\eta_\lambda,\ldots,
     [\partial_\lambda^{j_m}\eta_\lambda,\frakb_{\lambda,0}]\cdots]],
 \qquad |j|:=j_1+\cdots+j_m,
 \label{eq:coupling-derivative-structure}
\end{equation}
where the empty nested commutator is \(\frakb_{\lambda,0}\) and \(c_\varnothing=1\).
The nonnegative integer coefficients \(c_j\) are determined by the recursion: differentiating a term raises one \(j_i\) by one, and the commutator with \(\eta_\lambda\) prepends a factor with \(j_1=0\).
For interactions \(\Xi_1,\ldots,\Xi_m\), iterating \eqref{eq:polynomial-total-bracket} \(m\) times, with \(\|\cdot\|^{\pol}_{0;k'}\le\|\cdot\|^{\pol}_{p;k'}\), \(\frakt_{0;k'}\le\frakt_{p;k'}\), and \(\frakt_{p;k'}(\frakb_{\lambda,0}) =|X|^{k'}(1+\diam X)^p\|A\|\), gives
\begin{equation}
 \frakt_{p;k}\bigl([\Xi_1,[\Xi_2,\ldots,
                       [\Xi_m,\frakb_{\lambda,0}]\cdots]]\bigr)
 \le\Bigl(\prod_{i=1}^m2C^{\rm br}_{p,k+i-1}\Bigr)
   |X|^{k+m}(1+\diam X)^p\|A\|
   \prod_{i=1}^m\|\Xi_i\|^{\pol}_{p;k+m}.
 \label{eq:nested-bracket-total}
\end{equation}
Starting from the constant family \(\frakb_{\lambda,0}\), the bounded bilinear map in \eqref{eq:polynomial-total-bracket} proves by induction that each \(\frakb_{\lambda,r}\) is \(C^\infty\) in \(\frakt_{p;k}\) for \(0\le p\le a\) and \(k\ge0\).
For \(m\ge1\) and \(q\in\N_0\), Leibniz' rule gives
\[
\partial_\lambda^q
   [\partial_\lambda^{j_1}\eta_\lambda,
    [\ldots,[\partial_\lambda^{j_m}\eta_\lambda,
                    \frakb_{\lambda,0}]\ldots]]
 =\hspace{-10pt}\sum_{\substack{h\in\N_0^m\\h_1+\cdots+h_m=q}}\hspace{-10pt}
   \frac{q!}{h_1!\cdots h_m!}
   [\partial_\lambda^{j_1+h_1}\eta_\lambda,
    [\ldots,[\partial_\lambda^{j_m+h_m}\eta_\lambda,
                    \frakb_{\lambda,0}]\ldots]].
\]
Apply this identity to \eqref{eq:coupling-derivative-structure}.
Since \(m\le r\) and \(|X|\ge1\), \eqref{eq:nested-bracket-total} gives
\[
 \frakt_{p;k}(\partial_\lambda^q\frakb_{\lambda,r})
 \le C_{r,q,p,k}|X|^{k+r}(1+\diam X)^p\|A\|,
\]
where the constant depends only on \(r,q,p,k\) and the uniform \(\|\cdot\|^{\pol}_{p;k+r}\)-bounds for the derivatives of \(\eta\) of orders at most \(r+q\).
Every nonzero term \(\partial_\lambda^q\frakb_{\lambda,r}(Y)\) has \(Y\supseteq X\).

For \(\Phi_\lambda=\Phi_0+\lambda\cV\) and \(\Psi_\lambda=\cV\), we obtain from Theorem~\ref{thm:smooth-hamiltonian-dependence}\ref{item:smooth-filtering} the more precise bound
\[
 \sup_{\lambda\in J}\|\partial_\lambda^j\eta_\lambda\|^{\pol}_{p;k}
 \le C_{j,p,k}\mu^{(j)}_{p,k}(W)
       \bigl(\|\cV\|^{\#}_{a;\kappa_0}\bigr)^{j+1}.
\]
In \eqref{eq:coupling-derivative-structure}, each term carries \(m\) factors \(\partial_\lambda^{j_i}\eta_\lambda\) with \(m+|j|=r\), so the exponents of \(\|\cV\|^{\#}_{a;\kappa_0}\) add to \(r\).
Equation~\eqref{eq:nested-bracket-total} therefore gives
\begin{equation}
 \sup_{\lambda\in J}\frakt_{p;k}(\frakb_{\lambda,r})
 \le C_{r,p,k,W}|X|^{k+r}(1+\diam X)^p\|A\|
       \bigl(\|\cV\|^{\#}_{a;\kappa_0}\bigr)^r.
 \label{eq:coupling-derivative-total}
\end{equation}

The sums \(B_r(\lambda):=\sum_Y\frakb_{\lambda,r}(Y)\) converge absolutely.
The commutator series needed for their evolution also converges absolutely, since
\[
 \sum_Z\sum_Y
      \|[\eta_\lambda(Z),\frakb_{\lambda,r}(Y)]\|
 \le2\|\eta_\lambda\|^{\pol}_{0;0}
                   \frakt_{0;1}(\frakb_{\lambda,r})<\infty.
\]
The norm derivatives and the absolutely convergent sums give
\[
 B_{r+1}(\lambda)=\partial_\lambda B_r(\lambda)
       +\I\sum_Y[\eta_\lambda(Y),B_r(\lambda)].
\]
For finite \(\Lambda\supseteq X\), put
\[
 B_r^\Lambda(\lambda):=\sum_{Y\subseteq\Lambda}
                                  \frakb_{\lambda,r}(Y).
\]
Apply the differential equation for \(\beta\) and the product rule to \(\beta_{\lambda_0,\lambda}(B_r^\Lambda(\lambda))\), and integrate in \(\lambda\).
The norm smoothness proved above implies that \(\lambda\mapsto\frakb_{\lambda,r}\) and \(\lambda\mapsto\partial_\lambda\frakb_{\lambda,r}\) are continuous in \(\frakt_{0;1}\).
Their images on the compact interval \(J\) are therefore compact in this norm.
Truncation to \(Y\subseteq\Lambda\) is contractive and converges to the identity on each family of finite \(\frakt_{0;1}\)-norm.
Covering each compact image by finitely many norm balls and choosing \(\Lambda\) large enough for their centers proves uniform convergence of the truncations on \(J\).
The absolute commutator estimate above therefore permits passage to the limit and gives
\[
 \beta_{\lambda_0,\lambda+h}(B_r(\lambda+h))
       -\beta_{\lambda_0,\lambda}(B_r(\lambda))
 =\int_\lambda^{\lambda+h}
       \beta_{\lambda_0,u}(B_{r+1}(u))\,\dd u
\]
whenever \(\lambda,\lambda+h\in J\).
The integrand is norm-continuous, so induction yields
\[
 \partial_\lambda^r\beta_{\lambda_0,\lambda}(A)
       =\beta_{\lambda_0,\lambda}(B_r(\lambda)).
\]
This proves norm-\(C^\infty\) dependence, including the endpoint extensions.
For \(\Phi_\lambda=\Phi_0+\lambda\cV\) and \(\Psi_\lambda=\cV\), isometry and \eqref{eq:coupling-derivative-total} with \(p=k=0\) prove \eqref{eq:smooth-coupling-bound}.

Under the approximation hypotheses, Theorem~\ref{thm:smooth-hamiltonian-dependence}\ref{item:smooth-approximation} gives uniform componentwise convergence of every derivative of \(\eta^{(N)}_\lambda\).
By \eqref{eq:coupling-derivative-structure}, each coefficient \(\frakb^{(N)}_{\lambda,r}(Y)\) is a finite sum of nested commutators of coefficients \(\partial_\lambda^{j_i}\eta^{(N)}_\lambda(Y_i)\), \(Y_i\subseteq Y\), with \(A\), so it converges uniformly on \(J\).
The common bounds on the derivatives of \(\Psi^{(N)}\) and \(\Phi^{(N)}\) give common bounds on the derivatives of \(\eta^{(N)}\), and hence on the total norms of \(\frakb^{(N)}_{\lambda,r}\) just estimated.
In particular,
\[
 \sum_{\diam Y>R}\|\frakb^{(N)}_{\lambda,r}(Y)\|
 \le(1+R)^{-a}\frakt_{a;0}
                         (\frakb^{(N)}_{\lambda,r}),
\]
uniformly in \(N\) and \(\lambda\).
Since \(Y\supseteq X\), the remaining supports lie in the finite set \(X^{(R)}\).
Thus \(B_r^{(N)}\to B_r\) in norm, uniformly on \(J\).
The previously proved convergence of \(\beta^{(N)}\), together with isometry, is uniform on the compact set \(\{B_r(\lambda)\,|\,\lambda\in J\}\).
Applying
\[
 \|\beta^{(N)}_{\lambda_0,\lambda}(B_r^{(N)}(\lambda))
                     -\beta_{\lambda_0,\lambda}(B_r(\lambda))\|
 \le \|B_r^{(N)}(\lambda)-B_r(\lambda)\|
      +\|(\beta^{(N)}_{\lambda_0,\lambda}
                 -\beta_{\lambda_0,\lambda})(B_r(\lambda))\|
\]
proves \eqref{eq:smooth-coupling-convergence}.
\end{proof}

For the transported states \(\omega_\lambda=\omega_0\circ\beta_{0,\lambda}\) in \cite[Theorem~3.3 and Section~9]{Stability}, Corollary~\ref{cor:filtered-flow}\ref{item:filtered-flow-smoothness} gives \(C^\infty\) dependence of \(\omega_\lambda(A)\) for each local observable \(A\).

Symmetries are inherited as well.
For the paths in Corollary~\ref{cor:filtered-flow}, suppose that a parity-preserving onsite automorphism \(\zeta\) fixes every term of \(\Phi_\lambda\) and \(\Psi_\lambda\), for all \(\lambda\in J\).
Remark~\ref{rem:onsite-covariance} gives \(\zeta(\eta_\lambda(Y))=\eta_\lambda(Y)\) for every \(Y\).
Uniqueness of the generated dynamics then gives \(\zeta\circ\beta_{\lambda,\mu} =\beta_{\lambda,\mu}\circ\zeta\).
In particular, gauge-invariant \(\Phi_\lambda\) and \(\Psi_\lambda\) give generator terms fixed by the gauge transformations, and the flow \(\beta\) commutes with those transformations; see also \cite[Remark~3.8]{Stability}.

\paragraph{Application to perturbation cutoffs.}
\label{par:perturbation-limits}
In gap-stability proofs, cutting off an extensive perturbation allows one to work with a bounded perturbing operator while keeping the background Hamiltonian in infinite volume.
The following application of Corollary~\ref{cor:filtered-flow} explains how the filtered generators and their flows converge when these cutoffs are removed.
Use the geometric hypotheses and \(a,\kappa_0\) of that corollary, and choose \(W\) by the quasi-adiabatic construction in Subsection~\ref{subsec:quasi-adiabatic-filters} from a real even Schwartz function \(w\), so that \(W\) has finite absolute moments of every order.
Fix a self-adjoint interaction \(\Phi_0\) and self-adjoint interactions \(\cV^{(n)}\) converging componentwise to \(\cV\), with
\[
 \|\Phi_0\|^{\#}_{a;\kappa_0}<\infty,
 \qquad
 \sup_n\|\cV^{(n)}\|^{\#}_{a;\kappa_0}<\infty,
 \qquad
 \sum_X\E^{\kappa_0|X|}\|\cV^{(n)}(X)\|<\infty
 \quad(n\in\N).
\]
For example, one may take \(\cV^{(n)}(X):=\1_{\{X\subseteq\Lambda_n\}}\cV(X)\) along an exhaustion \(\Lambda_n\nearrow\Gamma\), for any self-adjoint \(\cV\) of finite \(\|\cdot\|^{\#}_{a;\kappa_0}\)-norm.
For a compact parameter interval \(J\), set
\[
 \Phi_\lambda^{(n)}:=\Phi_0+\lambda\cV^{(n)},
 \qquad \Phi_\lambda:=\Phi_0+\lambda\cV.
\]
By Fatou's lemma, \(\|\cV\|^{\#}_{a;\kappa_0}\le\sup_n\|\cV^{(n)}\|^{\#}_{a;\kappa_0}\).
The Hamiltonian norms on \(J\) are therefore uniformly bounded by
\[
 \|\Phi_0\|^{\#}_{a;\kappa_0}
 +\left(\sup_n\|\cV^{(n)}\|^{\#}_{a;\kappa_0}\right)
       \sup_{\lambda\in J}|\lambda|.
\]
Apply Corollary~\ref{cor:filtered-flow} with \(\Psi_\lambda^{(n)}=\cV^{(n)}\) and \(\Psi_\lambda=\cV\), and denote the filtered interactions by \(\eta_\lambda^{(n)}\) and \(\eta_\lambda\).
The total-operator identity is
\[
 \sum_Y\eta_\lambda^{(n)}(Y)
 =\int_\R W(t)\tau_t^{\Phi_\lambda^{(n)}}
       \left(\sum_X\cV^{(n)}(X)\right)\,\dd t.
\]
The corollary also gives convergence of the filtered interactions in every local seminorm with spatial exponent \(q<a\) and fixed support moment \(k\ge0\), uniformly on \(J\), and convergence of the flows generated by \(\eta^{(n)}\) to the flow generated by \(\eta\) on every observable, uniformly on \(J^2\).
The limiting perturbation \(\cV\) need not have a norm-convergent total sum: componentwise convergence and the uniform interaction bounds suffice for convergence of these flows.
The physical perturbation cutoffs in the polynomial stability proof satisfy these hypotheses; see \cite[Section~9]{Stability}.

The locality constants can also be independent of the perturbation strength if the parameter interval shrinks accordingly.
To see this, fix \(c,C_\eta>0\).
For each \(b_{\rm pert}>0\), let \(I_{b_{\rm pert}}\) be a compact interval and \(\eta^{(b_{\rm pert})}\) a componentwise norm-continuous path of self-adjoint interactions on \(I_{b_{\rm pert}}\).
The bounds
\[
 \sup_{\lambda\in I_{b_{\rm pert}}}
       \|\eta^{(b_{\rm pert})}_\lambda\|^{\pol}_{a;1}
       \le C_\eta b_{\rm pert},
 \qquad |I_{b_{\rm pert}}|\le\frac c{b_{\rm pert}}
\]
give \(\int_{I_{b_{\rm pert}}} \|\eta^{(b_{\rm pert})}_\lambda\|^{\pol}_{a;1}\,\dd\lambda \le cC_\eta\).
Theorem~\ref{thm:polynomial-local-topology} therefore gives local-approximation bounds for the flows generated by \(\eta^{(b_{\rm pert})}\), with constants independent of \(b_{\rm pert}\).

\subsection{Quasi-adiabatic filters and spectral transport}
\label{subsec:quasi-adiabatic-filters}
\label{subsec:spectral-transport}

Quasi-adiabatic continuation was introduced by Hastings \cite{Hastings04} and developed further by Hastings--Wen \cite{HastingsWen05}.
We use the exact-filter construction of Bachmann--Michalakis--Nachtergaele--Sims \cite[Proposition~2.4 and Eq.~(2.12)]{BMNS12}; see also \cite{NSY18,NSY19}.
Our propagation bounds extend the associated spectral-flow construction to polynomially decaying fermionic interactions without loss of the spatial exponent.

Fix \(g>0\), use the Fourier convention \(\widehat f(\omega):=\int_\R\E^{\I t\omega}f(t)\,\dd t\), and choose a real even Schwartz function \(w\) with \(\widehat w\in C_c^\infty((-g,g))\) and \(\widehat w(0)=1\).

Define
\[
 W(t):=-\int_t^\infty w(v)\,\dd v\quad(t>0),
 \qquad W(-t):=-W(t),\qquad W(0):=0.
\]
The sign is opposite to that in \cite[Eq.~(2.12)]{BMNS12}, in accordance with our evolution convention \eqref{eq:dynamics}.
Since \(\int_\R w=1\), distributional differentiation gives \(W'=w-\delta_0\), and hence
\[
 \widehat W(\omega)=\frac{1-\widehat w(\omega)}{\I\omega},
 \qquad
 \widehat W(\omega)=-\frac{\I}{\omega}\quad(|\omega|\ge g).
\]
The quotient is set to zero at \(\omega=0\).
Moreover, \(W\) has finite absolute moments of every order: for every \(m\ge0\)
\[
 \int_\R |t|^m|W(t)|\,\dd t
 \le\frac1{m+1}\int_\R|v|^{m+1}|w(v)|\,\dd v<\infty.
\]

The relation \(W'=w-\delta_0\) yields an identity for the generator of a strongly continuous automorphism group.
We state this identity of Nachtergaele--Sims--Young \cite[Proposition~6.9]{NSY19} in our sign convention and prove the domain inclusion under integrability of the two filters.
In infinite volume, the dynamics may have no total Hamiltonian in \(\A\).
The identity still allows one to apply the closed generator to a filtered observable and compute the result, as needed in perturbative and adiabatic constructions.

\begin{lemma}
\label{lem:filter-generator}
Let \((\tau_t)_{t\in\R}\) be a strongly continuous group of \(*\)-automorphisms of \(\A\), and let \(G\) be its closed generator.
Let \(w,W:\R\to\R\) be integrable filters satisfying
\[
 W'=w-\delta_0
\]
in the sense of distributions.
Define
\[
 \cI(A):=\int_\R W(t)\tau_t(A)\,\dd t,
 \qquad
 \cD(A):=\int_\R w(t)\tau_t(A)\,\dd t,
 \qquad A\in\A.
\]
Then, for every \(A\in\A\),
\begin{equation}
 \cI(A)\in\operatorname{Dom}G,
 \qquad
 G\cI(A)=A-\cD(A).
 \label{eq:filter-generator-identity}
\end{equation}
\end{lemma}

\begin{proof}
Fix \(A\in\A\) and \(r>0\).
The relation \(W'=w-\delta_0\) gives, for almost every \(t\),
\[
 W(t-r)-W(t)
 =\1_{(0,r)}(t)-\int_0^r w(t-u)\,\dd u.
\]
Since
\[
 \int_0^r\int_\R |w(t-u)|\|\tau_t(A)\|\,\dd t\,\dd u
 =r\|w\|_{L^1}\|A\|,
\]
Fubini's theorem applies.
A change of variables and the group law give
\[
 \tau_r(\cI(A))-\cI(A)
 =\int_\R[W(t-r)-W(t)]\tau_t(A)\,\dd t\ =\int_0^r\tau_u
                  (A-\cD(A))\,\dd u.
\]
For \(r<0\), apply \(\tau_r\) to the identity with increment \(-r>0\) to obtain the same formula.
Dividing by \(r\) and using strong continuity as \(r\to0\) proves \eqref{eq:filter-generator-identity} by the definition of \(G\).
\end{proof}

This lemma applies to the autonomous dynamics of Propositions~\ref{prop:polynomial-dynamics} and~\ref{prop:onsite-terms}.
For \(\tau=\tau^\Phi\), its map \(\cI\) is \(\cI^\Phi_{W,0}\).

Under the hypotheses below, the Fourier condition on \(W\) gives a filtered generator whose flow transports the isolated spectral sector.
The next lemma recalls this fact in finite dimension, following Bachmann--Michalakis--Nachtergaele--Sims \cite[Proposition~2.4]{BMNS12} and Nachtergaele--Sims--Young \cite[Theorem~6.3]{NSY19}.

\begin{lemma}
\label{lem:finite-spectral-transport}
Fix \(g>0\) and a real filter \(W\in L^1(\R)\) such that \(\widehat W(\omega)=-\I/\omega\) for \(|\omega|\ge g\).
Let \(J\subseteq\R\) be an interval and \(J\ni \lambda\mapsto H_\lambda\) a \(C^1\) family of self-adjoint Hamiltonians on a finite-dimensional Hilbert space, and let \(P_\lambda\) be a continuously chosen spectral projection with
\[
 \dist\bigl(\spec(H_\lambda|_{\Ran P_\lambda}),
            \spec(H_\lambda|_{\Ran(\1-P_\lambda)})\bigr)
 \ge g,\qquad \lambda\in J.
\]
For each \(\lambda\), let \(\cI_\lambda\) denote time filtering with \(W\) for the autonomous dynamics generated by \(H_\lambda\).
Define
\[
 D_\lambda:=\cI_\lambda(\partial_\lambda H_\lambda)
 =\int_\R W(t)\E^{\I tH_\lambda}
          (\partial_\lambda H_\lambda)\E^{-\I tH_\lambda}\,\dd t.
\]
The path \(P_\lambda\) is \(C^1\) and satisfies
\begin{equation}
 \partial_\lambda P_\lambda=-\I[D_\lambda,P_\lambda].
 \label{eq:finite-spectral-transport}
\end{equation}
For the automorphism cocycle \(\beta\), defined by \(\partial_\lambda\beta_{\lambda_0,\lambda}(B) =\I\beta_{\lambda_0,\lambda}([D_\lambda,B])\) for every operator \(B\) and \(\beta_{\lambda_0,\lambda_0}=\id\), one has
\[
 \beta_{\lambda_0,\lambda}(P_\lambda)=P_{\lambda_0},
 \qquad \lambda_0,\lambda\in J.
\]
\end{lemma}

\begin{proof}
The local Riesz formula makes \(P_\lambda\) a \(C^1\) path.
We write the spectral-flow calculation of Bachmann--Michalakis--Nachtergaele--Sims \cite[Proposition~2.4]{BMNS12} and Nachtergaele--Sims--Young \cite[Section~6.4]{NSY19} in our sign convention \eqref{eq:dynamics}.
Fix \(\lambda\) and choose an eigenbasis of \(H_\lambda\) adapted to \(P_\lambda\), with eigenvalues \(E_j\) for \(H_\lambda\) and \(\pi_j\in\{0,1\}\) for \(P_\lambda\).
For indices \(j,k\) in opposite sectors, the Fourier condition and differentiation of \([H_\lambda,P_\lambda]=0\) give, respectively,
\[
 \begin{aligned}
 (D_\lambda)_{jk}
 &=-\frac{\I(\partial_\lambda H_\lambda)_{jk}}{E_j-E_k},\\
 (\partial_\lambda P_\lambda)_{jk}
 &=\frac{(\pi_j-\pi_k)(\partial_\lambda H_\lambda)_{jk}}
         {E_j-E_k}.
 \end{aligned}
\]
Differentiating \(P_\lambda^2=P_\lambda\) shows that
\[
 P_\lambda(\partial_\lambda P_\lambda)P_\lambda
 =(\1-P_\lambda)(\partial_\lambda P_\lambda)(\1-P_\lambda)=0.
\]
The commutator \([D_\lambda,P_\lambda]\) also has zero diagonal blocks relative to \(P_\lambda\).
The two matrix formulas therefore prove \eqref{eq:finite-spectral-transport} on every block.
Since \(W\) is real and integrable, \(D_\lambda\) is self-adjoint and continuous, and hence generates \(\beta\).
Finally,
\[
 \partial_\lambda\bigl(\beta_{\lambda_0,\lambda}(P_\lambda)\bigr)
 =\beta_{\lambda_0,\lambda}
       (\I[D_\lambda,P_\lambda]+\partial_\lambda P_\lambda)=0.
\]
The initial value gives the transport identity.
\end{proof}

We now apply this transport identity to isolated spectral sectors of finite-volume Hamiltonians.
We will prove that the infinite-volume flow transports the sets of weak* limits of states supported in these sectors.
To pass to infinite volume, we represent the finite-volume filtered generators by interactions using Theorem~\ref{thm:card-propagation}.
Let \(\Phi\) satisfy the Hamiltonian assumptions of Corollary~\ref{cor:filtered-flow}, and suppose that each component \(\lambda\mapsto\Phi_\lambda(X)\) is continuously differentiable and that
\[
 \Psi_\lambda(X):=\partial_\lambda\Phi_\lambda(X)
 \quad\mbox{satisfies}\quad
 \sup_{\lambda\in J}
       \|\Psi_\lambda\|^{\#}_{a;\kappa_0}<\infty.
\]
Let \(W\) be a real filter with finite absolute moment of order \(m_{a,1}\).
For a finite region \(\Lambda\), set \(H_\lambda^\Lambda:=\sum_{Z\subseteq\Lambda}\Phi_\lambda(Z)\).
Using common cells and scales for \(\Phi_\lambda\) and \(\Phi_\lambda^\Lambda\), define
\[
 \eta_\lambda:=\cK^{\Phi_\lambda}_{W,0}[\Psi_\lambda],
 \qquad
 \eta_\lambda^{[\Lambda]}
 :=\cK^{\Phi_\lambda^\Lambda}_{W,0}[\Psi_\lambda^\Lambda],
\]
where \(\cK\) is defined in \eqref{eq:card-filter-interaction}.
The interaction \(\Psi_\lambda^\Lambda\) satisfies the total summability assumption in Theorem~\ref{thm:card-limits}\ref{item:card-limits-identities}.
The total-operator identity \eqref{eq:filtered-total-operator} gives
\begin{equation}
 \sum_Y\eta_\lambda^{[\Lambda]}(Y)
 =\int_\R W(t)\tau_{0,t}^{\Phi_\lambda,\Lambda}
                  (\partial_\lambda H_\lambda^\Lambda)\,\dd t
 =:D_\lambda^\Lambda.
 \label{eq:finite-input-filter-identity}
\end{equation}
The sum converges absolutely in norm and runs over all supports \(Y\), including those extending beyond \(\Lambda\).
Each coefficient belongs to \(\A_\Lambda\), and \(\eta_\lambda^{[\Lambda]}(Y)=\eta_\lambda(Y)\) for \(Y\subseteq\Lambda\).

The following corollary extends the transport of weak* limits of finite-volume states in Bachmann--Michalakis--Nachtergaele--Sims \cite[Theorem~5.5]{BMNS12} to our setting, with a filtered generator bounded at the same spatial exponent as the Hamiltonian.
Spectral transport for polynomially decaying fermionic interactions, with a loss of spatial decay, was already obtained by Teufel--Wessel \cite[Section~4]{TeufelWessel25}.

\begin{corollary}
\label{cor:finite-sector-transport}
Assume \eqref{eq:volume-growth} and Hypothesis~\ref{hyp:cell-partitions}, and choose \(\kappa_0\) as in Theorem~\ref{thm:card-propagation}.
Fix \(a>\dvol\), \(0\le L<\infty\), \(g>0\), and a measurable filter \(W:\R\to\R\) satisfying
\[
 \int_\R |W(t)|(1+|t|)^{m_{a,1}}\,\dd t<\infty,
 \qquad
 \widehat W(\omega)=-\frac{\I}{\omega}\quad(|\omega|\ge g).
\]
Use \(C_{W,1}\) from Corollary~\ref{cor:filtered-flow} for these choices of \(a,L,W\).
Let \(J\subseteq\R\) be a compact interval and \((\Phi_\lambda)_{\lambda\in J}\) a componentwise \(C^1\) path of self-adjoint interactions satisfying
\[
 \sup_{\lambda\in J}\|\Phi_\lambda\|^{\#}_{a;\kappa_0}\le L,
 \qquad
 \sup_{\lambda\in J}\|\partial_\lambda\Phi_\lambda\|^{\#}_{a;\kappa_0}<\infty.
\]
Let \(\Lambda_n\nearrow\Gamma\).
Suppose \(P_\lambda^{(n)}\) are spectral projections of \(H_\lambda^{\Lambda_n}\), continuous in \(\lambda\), and satisfying
\[
 \dist\!\left(
       \spec(H_\lambda^{\Lambda_n}|_{\Ran P_\lambda^{(n)}}),
       \spec(H_\lambda^{\Lambda_n}|_{\Ran(\1-P_\lambda^{(n)})})
       \right)\ge g,
 \qquad n\in\N,\quad\lambda\in J.
\]
Extend finite-volume states to \(\A\) using the tracial conditional expectations, and define
\[
 \begin{aligned}
 \cS_\lambda^{(n)}
 &:=\{\nu\circ\F_{\Lambda_n}^{\rm tr}\,|\,
       \nu\text{ is a state on }\A_{\Lambda_n},\
       \nu(P_\lambda^{(n)})=1\},\\
\mathcal G_\lambda
&:=\left\{
\mathrm{w}^*\!-\!\lim_{j\to\infty}\phi_j
\;\middle|\;
n_j\nearrow\infty,\quad
\phi_j\in\mathcal S_\lambda^{(n_j)}
\right\}.
 \end{aligned}
\]
Apply Corollary~\ref{cor:filtered-flow} with \(\Psi_\lambda=\partial_\lambda\Phi_\lambda\), and denote the resulting flow by \(\beta\).
Then
\begin{equation}
 \cG_\mu
 =\{\phi\circ\beta_{\lambda,\mu}\,|\,\phi\in\cG_\lambda\},
 \qquad \lambda,\mu\in J,
 \label{eq:thermodynamic-sector-transport}
\end{equation}
and \(\beta\) satisfies the locality bound \eqref{eq:application-flow} at spatial exponent \(a\).
\end{corollary}

\begin{proof}
Lemma~\ref{lem:finite-spectral-transport} applies to each \(H_\lambda^{\Lambda_n}\) and \(P_\lambda^{(n)}\).
Define \(\eta^{[\Lambda_n]}\) by the construction above for \(\Lambda=\Lambda_n\) with this \(W\), and set \(\beta^{(n)}:=\tau^{\eta^{[\Lambda_n]}}\).
The absolute total convergence in \eqref{eq:finite-input-filter-identity} gives, for every \(B\in\A\),
\[
 \sum_Y[\eta_\lambda^{[\Lambda_n]}(Y),B]
 =\left[\sum_Y\eta_\lambda^{[\Lambda_n]}(Y),B\right]
 =[D_\lambda^{\Lambda_n},B],
\]
with absolute convergence of the commutator series.
The inner cocycle generated by \(D_\lambda^{\Lambda_n}\) therefore satisfies the same differential equation as \(\beta^{(n)}\), and uniqueness makes the two cocycles equal.
Its implementing unitaries belong to \(\A_{\Lambda_n}\), so bimodularity of \(\F_{\Lambda_n}^{\rm tr}\) makes this conditional expectation commute with \(\beta^{(n)}\).
Finite-volume spectral transport then gives

\[
 \cS_\mu^{(n)}
 =\{\phi\circ\beta_{\lambda,\mu}^{(n)}\,|\,
                            \phi\in\cS_\lambda^{(n)}\}.
\]
The restricted Hamiltonian and derivative interactions converge componentwise and satisfy the common norm bounds.
Corollary~\ref{cor:filtered-flow} therefore gives \(\beta_{\lambda,\mu}^{(n)}(A)\to\beta_{\lambda,\mu}(A)\) in norm for every \(A\in\A\), uniformly on \(J^2\).
Passage to the limit uses the argument of Bachmann--Michalakis--Nachtergaele--Sims \cite[Theorem~5.5 and Lemma~5.6]{BMNS12}.
For \(\phi_j\in\cS_\lambda^{(n_j)}\) converging weak* to \(\phi\), and every \(A\in\A\),
\[
|\phi_j(\beta_{\lambda,\mu}^{(n_j)}(A))
                         -\phi(\beta_{\lambda,\mu}(A))| \le\|\beta_{\lambda,\mu}^{(n_j)}(A)
                         -\beta_{\lambda,\mu}(A)\|
       +|(\phi_j-\phi)(\beta_{\lambda,\mu}(A))|
       \xrightarrow{j\to\infty}0.
\]
Thus \(\{\phi\circ\beta_{\lambda,\mu}\,|\,\phi\in\cG_\lambda\} \subseteq\cG_\mu\).
Interchanging \(\lambda\) and \(\mu\) and using the inverse cocycle gives the reverse inclusion.
\end{proof}

Corollary~\ref{cor:finite-sector-transport} applies to exact ground-state sectors and to isolated low-energy bands containing several eigenvalues.
For ground-state sectors, identifying the limit sets \(\cG_\lambda\) with all infinite-volume ground states requires additional assumptions.
For spectral transport under a gap in infinite-volume ground-state representations, see Moon--Ogata \cite[Assumption~1.2 and Theorem~1.3]{MoonOgata20} and Becker--Teufel--Wesle \cite[Definition~3.1 and Theorem~3.2]{BTW26}, whose hypotheses differ from the finite-volume separation used here.

The transport results above require only finitely many absolute filter moments.
We now choose filters that also give stretched-exponential decay in the sizes of the interaction supports.
The filter \(W\) cannot decay exponentially: otherwise \(\widehat W\) would extend analytically to a strip around the real axis and, by the identity theorem, coincide there with \(-\I/\omega\), which is singular at \(\omega=0\).
This is one reason why polynomial rather than exponential time growth matters in the applications.
For the stronger filter assumption in Theorem~\ref{thm:card-limits}\ref{item:card-limits-stretched}, choose \(w\) from Bachmann--Michalakis--Nachtergaele--Sims \cite[Lemmas~2.3 and~2.6]{BMNS12}, with Fourier support in \([-g/2,g/2]\).
Their estimates for \(w\) and its tail \(W\) give constants \(C,c>0\), depending on \(g\), such that
\[
 |w(t)|+|W(t)|
 \le C\exp\!\left(-\frac{c|t|}{\log^2(2+|t|)}\right),
 \qquad t\in\R.
\]
For every \(0<\alpha<1\), the ratio of \(|t|/\log^2(2+|t|)\) to \((1+|t|)^\alpha\) tends to infinity as \(|t|\to\infty\).
Consequently,
\[
 F_{b,\alpha}(W)
 =\int_\R |W(v)|\E^{b(1+|v|)^\alpha}\,\dd v<\infty,
 \qquad b>0,\quad 0<\alpha<1.
\]

For the Hamiltonian and derivative interactions \(\Phi_\lambda,\Psi_\lambda\) above, use this filter in \(\eta_\lambda=\cK^{\Phi_\lambda}_{W,0}[\Psi_\lambda]\).
With \(\theta=\alpha/(\alpha+(1+\gamma)\dvol)\) and \(b'=c_*^\theta(b/2)^{1-\theta}\), Theorem~\ref{thm:card-limits}\ref{item:card-limits-stretched} therefore gives, for \(0\le p\le a\),
\[
  \sup_{\lambda\in J}\sup_x
       \sum_{Y\ni x}\E^{b'|Y|^\theta}(1+\diam Y)^p
                         \|\eta_\lambda(Y)\| \le C_{p,b,\alpha}F_{b,\alpha}(W)
       \sup_{\lambda\in J}\|\Psi_\lambda\|^{\#}_{p;\kappa_0}.
\]
The polynomial stability proof \cite[Subsection~5.2 and Section~9]{Stability} uses \(\alpha=1/2\), for which \(\theta=1/(1+2(1+\gamma)\dvol)\).

\subsection{Unbounded onsite terms and spin systems}
\label{subsec:onsite-spin}

Onsite terms, such as external potentials, need not be uniformly bounded in space.
The following proposition covers, for example, onsite potentials \(\Phi_u(\{x\})=V_u(x)n_x\).
Here \(n_x:=\sum_{j=1}^{r_x}a_{x,j}^*a_{x,j}\) is the local number operator, with \(a_{x,j}\) the annihilation operators for an orthonormal basis of \(\C^{r_x}\), and each \(u\mapsto V_u(x)\) is real and continuous.

\begin{proposition}
\label{prop:onsite-terms}
For an interaction \(\Xi\), write \(\Xi^{\ge2}(Z):=\1_{\{|Z|\ge2\}}\Xi(Z)\).
In each part, the onsite terms of the Hamiltonian interaction must be even and self-adjoint and satisfy the componentwise continuity and, for approximation families, convergence assumptions of the cited results.
No uniform spatial bound on the onsite terms is required.
Under the hypotheses of each part, the full interaction generates a unique strongly continuous parity-preserving automorphism cocycle satisfying \eqref{eq:dynamics}, and its finite-volume approximations converge in norm on every observable, uniformly in compact intervals.

\begin{enumerate}[label=(\roman*),ref=(\roman*),leftmargin=*]
\item \emph{Exponential support-size bounds.}
\label{item:onsite-exponential}
The conclusions of Theorems~\ref{thm:card-propagation}--\ref{thm:card-limits} and Corollary~\ref{cor:observable-LR} hold when their Hamiltonian norm bounds are imposed only on \(\Phi^{\ge2}\).
The single-input estimates \eqref{eq:card-propagation-total} and \eqref{eq:exponential-source-total} also hold under these offsite assumptions.
The constants, cell partitions, and cutoffs are chosen as in those results, using only the offsite Hamiltonian norm bound.

\item \emph{Bounds from an integrated interaction norm.}
\label{item:onsite-integrated}
The conclusions of Theorem~\ref{thm:polynomial-local-topology} and Remark~\ref{rem:general-spatial-weights} hold when their Hamiltonian norm assumptions are imposed only on \(\Xi^{\ge2}\).
The integrated norms are also computed from \(\Xi^{\ge2}\); in particular, replace \(M(s,t)\) by
\[
 M_{\ge2}(s,t):=\int_{\min(s,t)}^{\max(s,t)}
                   \|\Xi_u^{\ge2}\|^{\pol}_{p;1}\,\dd u,
 \qquad s,t\in J.
\]

\item \emph{Polynomial time bounds.}
\label{item:onsite-polynomial-time}
The conclusions of Proposition~\ref{prop:polynomial-time-locality} and Corollary~\ref{cor:polynomial-interaction-shells} hold when their Hamiltonian norm bound is imposed only on \(\Xi^{\ge2}\).
In their estimates, \(L\) denotes this offsite bound, and the same constants apply.

\end{enumerate}
\end{proposition}

\begin{proof}
We remove the onsite terms by the interaction picture of Nachtergaele--Sims--Young \cite[Section~3.2]{NSY19}.
Use \(\Phi\) for the Hamiltonian path in all three parts, with time domain \(I\), taking \(I=J\) in part~\ref{item:onsite-integrated}.

\smallskip
\noindent \emph{Onsite dynamics and factorization.}
For each \(x\), let \(U_x^{\rm on}(t,s)\) solve
\[
 \partial_tU_x^{\rm on}(t,s)=-\I\Phi_t(\{x\})U_x^{\rm on}(t,s),
 \qquad U_x^{\rm on}(s,s)=\1,
\]
and, for finite \(X\), put
\[
 U_X^{\rm on}(t,s):=\prod_{x\in X}U_x^{\rm on}(t,s),
 \qquad
 \alpha_{s,t}(A):=U_X^{\rm on}(t,s)^*AU_X^{\rm on}(t,s),\quad A\in\A_X.
\]
The propagators are even, so the definitions agree when \(X\) is enlarged.
Continuity on each local algebra and isometry extend \(\alpha\) to a strongly continuous parity-preserving automorphism cocycle on \(\A\).

Fix \(r\in I\), and define
\[
 \widetilde\Phi_u(Z):=\alpha_{r,u}(\Phi_u^{\ge2}(Z)).
\]
The onsite maps preserve supports, adjoints, parity, and term norms.
In parts~\ref{item:onsite-exponential} and~\ref{item:onsite-polynomial-time}, apply Proposition~\ref{prop:polynomial-dynamics} to \(\widetilde\Phi\) on compact subintervals of \(I\).
In part~\ref{item:onsite-integrated}, apply Theorem~\ref{thm:polynomial-local-topology} or, for general spatial weights, Remark~\ref{rem:general-spatial-weights}.
Write \(\widetilde\tau:=\tau^{\widetilde\Phi}\) for the resulting dynamics.
Differentiating the finite-volume propagators gives
\begin{equation}
 \tau^{\Phi,\Omega}_{s,t}
 =\alpha_{s,r}\circ\tau^{\widetilde\Phi,\Omega}_{s,t}
                    \circ\alpha_{r,t}
 \quad\text{on }\A_\Omega.
 \label{eq:onsite-finite-factorization}
\end{equation}
Define the full dynamics by
\begin{equation}
 \tau_{s,t}:=\alpha_{s,r}\circ\widetilde\tau_{s,t}
                              \circ\alpha_{r,t}.
 \label{eq:onsite-factorization}
\end{equation}
The middle onsite maps cancel in compositions, so \(\tau\) is an automorphism cocycle.
For local \(A\), the set \(\{\alpha_{r,t}(A)\mid t\in J\}\) is compact in its local algebra for each compact \(J\subseteq I\).
Uniform convergence of the transformed finite-volume dynamics on this set proves convergence of \(\tau^{\Phi,\Omega}_{s,t}(A)\), uniformly on \(J^2\).
Density and isometry extend this convergence to all \(A\in\A\).

The onsite derivation of \(A\in\A_F\) is a finite sum, and the offsite sum converges absolutely, with
\[
 \|\delta_{\Phi_t^{\ge2}}(A)\|
 \le2|F|\|\Phi_t^{\ge2}\|^{\pol}_{0;0}\|A\|.
\]
The norm on the right is uniformly bounded on compact time intervals in each part.
The path \(t\mapsto\alpha_{r,t}(A)\) lies in \(\A_F\), and differentiation there gives
\[
 \partial_t\alpha_{r,t}(A)
 =\I\alpha_{r,t}\!\left(
       \sum_{x\in F}[\Phi_t(\{x\}),A]\right).
\]
The definition of \(\widetilde\Phi_t\) and absolute convergence of the offsite derivation also give
\[
 \delta_{\widetilde\Phi_t}(\alpha_{r,t}(A))
 =\alpha_{r,t}(\delta_{\Phi_t^{\ge2}}(A)).
\]
Apply the product rule to \eqref{eq:onsite-factorization}, using a finite basis of \(\A_F\) for the varying input \(\alpha_{r,t}(A)\).
The two preceding identities yield
\[
 \partial_t\tau_{s,t}(A)
  =\alpha_{s,r}\!\left(
     \widetilde\tau_{s,t}\bigl(
       \I\delta_{\widetilde\Phi_t}(\alpha_{r,t}(A))
           +\partial_t\alpha_{r,t}(A)\bigr)\right) =\I\tau_{s,t}(\delta_{\Phi_t}(A)),
\]
which proves \eqref{eq:dynamics} for \(\tau\).
For uniqueness, let \(\varphi\) be another strongly continuous automorphism cocycle satisfying \eqref{eq:dynamics}.
Define the transformed cocycle by
\[
 \widetilde\varphi_{s,t}
 :=\alpha_{r,s}\circ\varphi_{s,t}\circ\alpha_{t,r}.
\]
For \(A\in\A_F\), differentiation with respect to the initial time of \(\alpha\) gives
\[
 \partial_t\alpha_{t,r}(A)
 =-\I\sum_{x\in F}[\Phi_t(\{x\}),\alpha_{t,r}(A)].
\]
Since \(\alpha_{t,r}(A)\in\A_F\), apply the product rule to \(\varphi_{s,t}(\alpha_{t,r}(A))\) using a finite basis of \(\A_F\).
The onsite terms cancel explicitly:
\[
 \partial_t\widetilde\varphi_{s,t}(A)
 =\alpha_{r,s}\!\left(\varphi_{s,t}\bigl(
       \I\delta_{\Phi_t}(\alpha_{t,r}(A))
                    +\partial_t\alpha_{t,r}(A)\bigr)\right) =\I\alpha_{r,s}\!\left(\varphi_{s,t}
       (\delta_{\Phi_t^{\ge2}}(\alpha_{t,r}(A)))\right) =\I\widetilde\varphi_{s,t}(\delta_{\widetilde\Phi_t}(A)).
\]
Uniqueness for \(\widetilde\Phi\) gives \(\widetilde\varphi=\widetilde\tau\), hence \(\varphi=\tau\).

\smallskip
\noindent \emph{Transfer of the observable estimates.}
The onsite maps also commute with tracial conditional expectations:
\begin{equation}
 \F_D^{\rm tr}\circ\alpha_{s,r}
       =\alpha_{s,r}\circ\F_D^{\rm tr},
 \qquad D\subseteq\Gamma.
 \label{eq:onsite-expectation-commutation}
\end{equation}
Indeed, on each finite algebra \(\A_\Omega\), the onsite automorphism preserves the trace and \(\A_{D\cap\Omega}\).
Uniqueness of the tracial conditional expectation \cite[Theorem~4.7]{ArakiMoriya03} gives
\(\alpha_{r,s}\circ\F_{D\cap\Omega}^{\rm tr}\circ\alpha_{s,r}
=\F_{D\cap\Omega}^{\rm tr}\).
Consistency of the conditional expectations and density extend this identity to \(\A\).

For \(A\in\A_X\), put \(B:=\alpha_{r,t}(A)\).
Then \(\|B\|=\|A\|\), and, for \(X\subseteq\Lambda\Subset\Gamma\),
\[
 \begin{aligned}
 \|(\id-\F_D^{\rm tr})(\tau_{s,t}(A))\|
 &=\|(\id-\F_D^{\rm tr})(\widetilde\tau_{s,t}(B))\|,\\
 \|\tau_{s,t}(A)-\tau^{\Phi,\Lambda}_{s,t}(A)\|
 &=\|\widetilde\tau_{s,t}(B)
                   -\tau^{\widetilde\Phi,\Lambda}_{s,t}(B)\|.
 \end{aligned}
\]
These identities transfer the conditional-expectation and finite-volume approximation bounds.
For the graded commutator bounds in parts~\ref{item:onsite-exponential} and~\ref{item:onsite-integrated}, take parity-homogeneous \(A\in\A_X\) and \(B'\in\A_Y\) and apply the transformed bound with \(B=\alpha_{r,t}(A)\) and \(\alpha_{r,s}(B')\).
The observable \(\alpha_{r,s}(B')\) has the same support, parity, and norm as \(B'\), so \eqref{eq:onsite-factorization} gives the full graded commutator bound.

\smallskip
\noindent \emph{Interaction representatives and parameter dependence.}
For part~\ref{item:onsite-exponential}, use the same partitions and cutoffs for \(\Phi\) and \(\widetilde\Phi\), chosen for the common offsite Hamiltonian norm bound.
Substitution of \eqref{eq:onsite-finite-factorization} into \eqref{eq:full-dynamics-mobius} and \eqref{eq:source-map-construction} gives
\[
 (\cT^\Phi_{s,t;X}A)(Y)=
 \alpha_{s,r}\!\left(
     (\cT^{\widetilde\Phi}_{s,t;X}(\alpha_{r,t}(A)))(Y)
                    \right).
\]
The input interaction \((\alpha_{r,t}\Psi)(X):=\alpha_{r,t}(\Psi(X))\) has the same supports and term norms as \(\Psi\).
Applying Theorem~\ref{thm:card-propagation} and \eqref{eq:card-propagation-total}--\eqref{eq:exponential-source-total} to this input proves the propagation bounds.
The representation identities follow from \eqref{eq:onsite-factorization} and the absolutely convergent sums, since \(\alpha_{s,r}\) is a bounded automorphism.
Apply Corollary~\ref{cor:observable-LR} to \(\widetilde\tau\) and use the approximation identities and graded commutator bound above to obtain its conclusions for the full dynamics.

For the parameter families, put \(\Phi^{(\infty)}:=\Phi\) and
\[
 H_{\xi,u}^{(n),\Omega}
 :=\sum_{\varnothing\ne Z\subseteq\Omega}\Phi_{\xi,u}^{(n)}(Z),
 \qquad n\in\N\cup\{\infty\}.
\]
These finite-volume Hamiltonians are norm-continuous, and for every compact \(J\subseteq I\),
\[
 \sup_\xi\int_J
 \|H_{\xi,u}^{(n),\Omega}-H_{\xi,u}^{(\infty),\Omega}\|\,\dd u
 \le\sum_{\varnothing\ne Z\subseteq\Omega}
       \sup_\xi\int_J
       \|\Phi_{\xi,u}^{(n)}(Z)-\Phi_{\xi,u}(Z)\|\,\dd u
 \xrightarrow{n\to\infty}0.
\]
Duhamel's formula and \eqref{eq:full-dynamics-mobius} and \eqref{eq:source-map-construction} give coefficient continuity and convergence.
The propagation bounds already proved are uniform over these families.
Together with Lemma~\ref{lem:componentwise-local-continuity}, they give convergence in local interaction seminorms at every spatial exponent \(q<a\).
For \(I=\R\), the total bound at \(p=k=0\) gives the \(L^1\) majorant for filtered coefficients.
The proof of Theorem~\ref{thm:card-limits} therefore applies with these coefficient limits and absolute bounds, giving its filtering, representation, and local convergence conclusions under the respective moment assumptions.

\smallskip
\noindent \emph{Polynomial support-size assumptions.}
In part~\ref{item:onsite-integrated}, the transformed integrated norm equals \(M_{\ge2}(s,t)\).
The two approximation identities above and the transformation of the test observable give \eqref{eq:polynomial-finite-approximation}--\eqref{eq:integrated-graded-commutator} with this integrated norm.
The same reduction applies to Remark~\ref{rem:general-spatial-weights}, since the onsite maps preserve both norms of the offsite interaction appearing there.

In part~\ref{item:onsite-polynomial-time}, apply Proposition~\ref{prop:polynomial-time-locality} to \(\widetilde\Phi\).
The conditional-expectation identity transfers its bound to the full dynamics with \(L\) given by the offsite norm bound.
For the shell bounds, fix \(s,t\) and put
\[
 \widetilde\Upsilon_t(Z):=\alpha_{r,t}(\Upsilon(Z)),
 \qquad
 \|\widetilde\Upsilon_t\|^{\pol}_{p;1}
 =\|\Upsilon\|^{\pol}_{p;1}.
\]
Commutation with conditional expectations gives
\[
 \Delta_{Z,n}^{\rm tr}\tau^\Phi_{s,t}(\Upsilon(Z))
 =\alpha_{s,r}\!\left(
   \Delta_{Z,n}^{\rm tr}\tau^{\widetilde\Phi}_{s,t}
                      (\widetilde\Upsilon_t(Z))\right),
 \qquad n\ge0.
\]
Applying Corollary~\ref{cor:polynomial-interaction-shells} to \(\widetilde\Phi\) and \(\widetilde\Upsilon_t\) gives the shell expansion and weighted bounds with the same constants and support labels \(Z^{(R_n)}\).
The shell terms are even and are self-adjoint when \(\Upsilon\) is self-adjoint.
If \(L=0\), the dynamics preserve every \(\A_Z\), so all shells with \(n\ge1\) vanish.
Finite restrictions and deletion of terms preserve the offsite norm bounds, which proves the stated uniformity.
\end{proof}

\begin{remark}[Spin systems and additional site factors]
\label{rem:spin-systems}
The interaction and observable propagation results also hold for spin systems, with parity conditions omitted and with the same constants, independent of the local dimensions.
For nonzero finite-dimensional Hilbert spaces \(\cH_x\), define
\[
 \cB_X:=\cB\!\left(\bigotimes_{x\in X}\cH_x\right),
 \qquad
 \cB:=\overline{\bigcup_{X\Subset\Gamma}\cB_X}.
\]
The inclusions use \(B\mapsto B\otimes\1\), and the overline denotes completion in the norm induced by the local norms.
The dimensions \(\dim\cH_x\) need not be uniformly bounded.
For finite \(\Omega\Subset\Gamma\) and \(D\subseteq\Omega\), define the conditional expectation by
\[
 \EE_D^\Omega(B):=
       (\id_{\cB_D}\otimes
                    \operatorname{tr}_{\Omega\setminus D})(B),
 \qquad B\in\cB_\Omega,
\]
where \(\operatorname{tr}_{\Omega\setminus D}\) is the normalized trace and \(\cB_D\) is identified with its image in \(\cB_\Omega\).
For \(D\subseteq\Gamma\) and finite \(\Omega\subseteq\Omega'\), these maps satisfy
\[
 \left.\EE_{D\cap\Omega'}^{\Omega'}\right|_{\cB_\Omega}
 =\EE_{D\cap\Omega}^{\Omega}.
\]
Their common action on local observables therefore extends by norm continuity to a conditional expectation \(\EE_D:\cB\to\cB_D\), where \(\cB_D\) is the norm completion of the local algebras in \(D\).
The proofs of Theorems~\ref{thm:card-propagation} and~\ref{thm:card-limits} use only the following properties of the local algebras, and each of them holds for \(\cB_X\) and \(\EE_D^\Omega\):
\begin{enumerate}[label=(\alph*),leftmargin=*]
\item \(\cB_X\subseteq\cB_Y\) for \(X\subseteq Y\), and \([B,C]=0\) for \(B\in\cB_X\), \(C\in\cB_Y\) with \(X\cap Y=\varnothing\); this replaces the evenness of the fermionic terms and observables, and gives \(\|[B,C]\|\le2\|B\|\|C\|\) in all cases.
\item The finite-volume dynamics generated by a componentwise norm-continuous self-adjoint interaction with terms in \(\cB_Z\), \(Z\subseteq Y\), are isometric automorphisms of \(\cB_\Omega\) that preserve \(\cB_Y\).
For Hamiltonian paths supported in fixed disjoint regions, the Hamiltonians commute at all pairs of times, and the propagator of their sum is the product of their propagators.
These properties are used in the M\"obius formulas \eqref{eq:card-mobius} and \eqref{eq:full-dynamics-mobius}.
The interaction-picture factorization \eqref{eq:card-picture} follows from the finite-dimensional evolution equations and requires no commutation assumption between the finite-range and long-range Hamiltonians.
\item Adjoining the auxiliary factors \(\cB(\C^2)\) of Lemma~\ref{lem:card-short} replaces \(\cH_x\) by \(\cH_x\otimes\C^2\) and stays within the same class of algebras.
The commutator iteration in Lemma~\ref{lem:finite-volume-matrix-commutator} and the boundary argument in Lemma~\ref{lem:card-separated} use commutation on disjoint supports and isometry of the dynamics, so both proofs apply to these algebras without parity restrictions.

\item \(\EE_D^\Omega\) is contractive, restricts to the identity on \(\cB_D\), and is \(\cB_D\)-bimodular.
\end{enumerate}
The constants depend only on the geometric constants and the interaction norms, not on the dimensions \(\dim\cH_x\).

For the observable bounds, let \(\mu_x\) be normalized Haar measure on the unitary group \(\mathrm U(\cH_x)\), and put
\[
 \cP_x(B):=
       \int_{\mathrm U(\cH_x)}U^*BU\,\dd\mu_x(U).
\]
The maps \(\cP_x\) commute, are contractive, and satisfy \(\EE_D^\Omega=\prod_{x\in\Omega\setminus D}\cP_x\).
The telescoping argument for \eqref{eq:single-site-averaging} gives
\[
 \|B-\EE_D^\Omega(B)\|
 \le\sum_{x\in\Omega\setminus D}
       \sup_{U\in\mathrm U(\cH_x)}\|[B,U]\|,
 \qquad B\in\cB_\Omega.
\]
Apply this bound to the evolved observable and evolve each single-site unitary backwards, as in the proof of Corollary~\ref{cor:observable-LR}, to obtain the same local-approximation estimate for spin systems.
The matrix estimate of Lemma~\ref{lem:finite-volume-matrix-commutator} and the comparison of volumes in the proofs of Theorem~\ref{thm:polynomial-local-topology} and Remark~\ref{rem:general-spatial-weights} also apply to arbitrary spin observables.
Proposition~\ref{prop:polynomial-time-locality} holds for spin systems as well: its matrix estimate and range-splitting identity are unchanged, and the site averages above give its local-approximation bound.
Corollary~\ref{cor:polynomial-interaction-shells} follows with the same constants by using the contractive spin conditional expectations in its dyadic shells.
The finite-volume derivative formula \eqref{eq:mixed-derivative-cell-formula} and the interaction bracket bounds use only properties (a) and (b), so Theorem~\ref{thm:smooth-hamiltonian-dependence} and Proposition~\ref{prop:iterated-filters} also hold for spin systems, and Corollary~\ref{cor:filtered-flow}, including part~\ref{item:filtered-flow-smoothness}, follows as before.

The perturbative comparison in Proposition~\ref{prop:commutator-comparison} uses the same commutator bounds and Duhamel identities, which apply to spin interactions.
For Proposition~\ref{prop:onsite-terms}, the single-site unitary propagators preserve supports, adjoints, and term norms; they also preserve the normalized traces and hence commute with \(\EE_D\).
The onsite factorization and its convergence arguments therefore apply without a uniform bound on the onsite terms.
Finally, the proof of Corollary~\ref{cor:finite-sector-transport} uses finite-dimensional spectral identities and norm limits of the generated flows, so it also gives transport of limit states for spin systems.

One may also replace \(\A_X\) by \(\A_X\otimes\cB_X\), with parity \(\Theta\otimes\id\) and conditional expectations \(\F_D^{\rm tr}\otimes\EE_D\).
Even operators on disjoint supports still commute, and the tensor product of the two site averages is contractive.
The fermionic proofs therefore apply to even interactions on these enlarged algebras.
For the parity reductions in Corollaries~\ref{rem:odd-representatives} and~\ref{cor:observable-LR} and Proposition~\ref{prop:polynomial-time-locality}, use \(\jmath_x\otimes\id_{\cB_\Omega}\) in each finite region~\(\Omega\).
This embedding has the support, isometry, and intertwining properties of Lemma~\ref{lem:auxiliary-even-embedding}, so the same constants apply.
Taking \(\cH_x=\C^2\) gives the additional site factors used in Lemma~\ref{lem:card-short}.
\end{remark}
\appendix
\section{Convergence of interactions and infinite-volume dynamics}
\label{app:dynamics}

This appendix proves the criterion for convergence in local interaction seminorms used above (Lemma~\ref{lem:componentwise-local-continuity}) and deduces Proposition~\ref{prop:polynomial-dynamics} from Theorem~\ref{thm:polynomial-local-topology}.

\subsection{Local continuity and convergence of interactions}

The convergence criterion uses finiteness of the neighborhoods of finite sets: by \eqref{eq:volume-growth},
\[
 |K^{(R)}|\le\sum_{x\in K}|\ball_R(x)|
       \le C_\Gamma|K|(1+R)^{\dvol}<\infty,
 \qquad K\Subset\Gamma,\quad R\ge0.
\]

\begin{lemma}
\label{lem:componentwise-local-continuity}
Fix \(p>0\), \(k\ge0\), \(0\le B<\infty\), \(0\le q<p\), and a nonempty finite \(K\subseteq\Gamma\).
Let \(\frakP\) be a topological parameter space and \((\Psi_\xi)_{\xi\in\frakP}\) a family of interactions satisfying \(\sup_{\xi\in\frakP}\|\Psi_\xi\|^{\pol}_{p;k}\le B\).
If every map \(\xi\mapsto\Psi_\xi(X)\) is norm-continuous, then \(\xi\mapsto\Psi_\xi\) is continuous in \(\|\cdot\|^{\pol,\loc}_{q;k,K}\).

Let \((\Psi_\xi^{(n)})_{\xi\in\frakP}\), \(n\in\N\), be families satisfying \(\sup_{n,\xi}\|\Psi_\xi^{(n)}\|^{\pol}_{p;k}\le B\).
If
\[
 \sup_{\xi\in\frakP}\|\Psi_\xi^{(n)}(X)-\Psi_\xi(X)\|
       \xrightarrow{n\to\infty}0
 \quad\text{for every nonempty finite }X\subseteq\Gamma,
\]
then
\[
 \sup_{\xi\in\frakP}
       \|\Psi_\xi^{(n)}-\Psi_\xi\|^{\pol,\loc}_{q;k,K}
       \xrightarrow{n\to\infty}0.
\]
\end{lemma}

\begin{proof}
For \(R\ge0\), define the finite family
\[
 \scrF_{K,R}:=\{X\Subset\Gamma\,|\,
               X\cap K\ne\varnothing,\ \diam X\le R\}.
\]
Every \(X\in\scrF_{K,R}\) is contained in the finite set \(K^{(R)}\).
If \(\Psi\) and \(\widetilde\Psi\) have \(\|\cdot\|^{\pol}_{p;k}\)-norm at most \(B\), then
\[
 \|\Psi-\widetilde\Psi\|^{\pol,\loc}_{q;k,K}
 \le\sum_{X\in\scrF_{K,R}}|X|^k(1+\diam X)^q
                    \|\Psi(X)-\widetilde\Psi(X)\| +2B(1+R)^{q-p}.
\]
Indeed, in each site sum the terms with \(\diam X>R\) satisfy \((1+\diam X)^q\le(1+R)^{q-p}(1+\diam X)^p\).

For continuity at \(\xi_0\), take \(\Psi=\Psi_\xi\) and \(\widetilde\Psi=\Psi_{\xi_0}\); the finite sum tends to zero as \(\xi\to\xi_0\).
For the sequence \(\Psi^{(n)}_\xi\), componentwise uniform convergence makes the finite sum tend to zero uniformly in \(\xi\).
In both cases, choose \(R\) first to make the tail arbitrarily small, which is possible because \(q<p\), and then use convergence of the finite sum.
\end{proof}

The strict decrease of the spatial exponent in this convergence criterion is necessary in general, even under an exponential support-size bound.
For example, fix \(a>0\) and \(\kappa_0>0\) and, on \(\Gamma=\Z\) with its usual distance, choose \(A_n\in\A_{\{0,n\}}^+\) with \(\|A_n\|=1\), for \(n\ge1\).
Define \(\Psi^{(n)}(\{0,n\}):=(1+n)^{-a}A_n\) and set all other terms to zero.
Then \(\Psi^{(n)}\) converges componentwise to zero, whereas, for \(0\le q\le a\),
\[
 \|\Psi^{(n)}\|^{\#}_{a;\kappa_0}=\E^{2\kappa_0},
 \qquad
 \|\Psi^{(n)}\|^{\pol,\loc}_{q;0,\{0\}}=(1+n)^{q-a}.
\]
The local seminorm tends to zero as \(n\to\infty\) for \(q<a\) but equals one for \(q=a\).
Preservation of the spatial exponent in the propagation estimates therefore does not imply convergence at that exponent.

To apply this example to Theorem~\ref{thm:card-limits}, take \(a>1\) and the zero Hamiltonian interaction \(\Phi=0\).
At \(s=t\), Definition~\ref{def:interaction-representatives} assigns the only nonzero input term to the union of the scale-\(T_0\) cells meeting \(\{0,n\}\).
That union contains \(\{0,n\}\), so
\[
 \|\cT^0_{s,s}\Psi^{(n)}\|^{\pol,\loc}_{a;0,\{0\}}\ge1.
\]
For the filtered conclusion, choose \(W\in C_c^\infty(\R)\) with \(W\ge0\) and \(\int_\R W=1\).
Every nonzero output coefficient of \(\cK^0_{W,s}[\Psi^{(n)}]\) has a support containing \(\{0,n\}\), and the total-operator identity in Theorem~\ref{thm:card-limits}\ref{item:card-limits-identities} gives
\[
 \sum_Y\bigl(\cK^0_{W,s}[\Psi^{(n)}]\bigr)(Y)
 =(1+n)^{-a}A_n.
\]
The triangle inequality therefore yields
\[
 \|\cK^0_{W,s}[\Psi^{(n)}]\|^{\pol,\loc}_{a;0,\{0\}}
 \ge (1+n)^a
       \left\|\sum_Y\bigl(\cK^0_{W,s}[\Psi^{(n)}]\bigr)(Y)\right\|
 =1.
\]
Thus the restriction \(q<a\) is necessary in general for both the propagated and the filtered convergence statements.
This example uses a filter with nonzero integral and does not establish the same obstruction for the odd quasi-adiabatic filters used in Section~\ref{sec:applications}.

\subsection{Existence and convergence of dynamics}
\label{subsec:dynamics-proof}

\begin{proof}[Proof of Proposition~\ref{prop:polynomial-dynamics}]
\phantomsection\label{proof:polynomial-dynamics}
Put \(q:=p-\dvol>0\).
For every nonempty finite \(Z\) and \(z\in Z\), volume growth gives
\[
 |Z|\le|\ball_{\diam Z}(z)|
       \le C_\Gamma(1+\diam Z)^{\dvol}.
\]
Consequently, for every interaction \(\Xi\),
\[
 \|\Xi\|^{\pol}_{q;1}
       \le C_\Gamma\|\Xi\|^{\pol}_{p;0}.
\]
Theorem~\ref{thm:polynomial-local-topology}, applied with spatial exponent \(q\), therefore gives existence and uniqueness of the cocycle and uniform convergence of its finite-volume dynamics on \(J^2\).

To prove convergence of the dynamics generated by \(\Phi^{(n)}\) to those generated by \(\Phi\), put \(\Phi^{(\infty)}:=\Phi\) and choose \(L<\infty\) so that
\[
 \sup_{n\in\N\cup\{\infty\}}\sup_{u\in J}
       \|\Phi_u^{(n)}\|^{\pol}_{p;0}\le L.
\]
Fix a nonempty finite \(X\), \(A\in\A_X\), \(R\ge0\), and \(\Lambda:=X^{(R)}\).
The preceding norm comparison bounds the integrated \(\|\cdot\|^{\pol}_{q;1}\)-norm of every path by \(C_\Gamma L|J|\).
Compare both infinite-volume dynamics with their restrictions to \(\Lambda\).
Equation~\eqref{eq:polynomial-finite-approximation} bounds the two restriction errors, and the finite-volume Duhamel formula gives
\[
 \begin{aligned}
 \sup_{s,t\in J}
 \|\tau_{s,t}^{\Phi^{(n)}}(A)-\tau_{s,t}^{\Phi}(A)\|
 &\le2\bigl(\E^{2C_\Gamma L|J|}-1\bigr)
       |X|\|A\|(1+R)^{-q}\\
 &\quad+2\|A\|
       \sum_{\varnothing\ne Z\subseteq\Lambda}
          \int_J\|\Phi_u^{(n)}(Z)-\Phi_u(Z)\|\,\dd u.
 \end{aligned}
\]
For fixed \(R\), the finite sum tends to zero as \(n\to\infty\).
Then letting \(R\to\infty\) proves convergence for local \(A\).
For arbitrary \(B\in\A\), the estimate
\[
 \|(\tau_{s,t}^{\Phi^{(n)}}-\tau_{s,t}^\Phi)(B)\|
 \le2\|B-A\|+
       \|(\tau_{s,t}^{\Phi^{(n)}}-\tau_{s,t}^\Phi)(A)\|
\]
and density prove convergence uniformly on \(J^2\).
\end{proof}

\section{Optimality of the decay assumption}
\label{app:decay-optimality}

This appendix shows that the representation \eqref{eq:source-representation} and the polynomial bound \eqref{eq:card-propagation} of Theorem~\ref{thm:card-propagation} cannot both hold when \(0<a\le\dvol\) with constants uniform over Hamiltonian paths and finite regions under a common interaction-norm bound and fixed geometric data.
The obstruction already occurs for smooth paths of two-site interactions.
To show this, we use the rapid state-transfer protocol of Tran et al.\ \cite[Eq.~(3) and Section~III]{TranStateTransfer21} to obtain a large commutator between distant single-site observables, whereas a representation with this polynomial bound would force this commutator to tend to zero as their distance grows.

\begin{proposition}
\label{prop:decay-optimality}
Fix an integer \(n\ge1\) and set \(\Gamma:=\Z^n\), with the Euclidean distance and \(\dvol=n\).
Assume that each site carries two fermionic modes.
Let \(0<a\le n\) and \(\kappa_0>0\).
There are an unbounded set \(\cR\subseteq\N\) and, for every \(r\in\cR\), a smooth path \((\Phi^{(r)}_u)_{u\in\R}\) of self-adjoint interactions with only two-site terms, supported in a finite set \(\Lambda_r\subseteq\Z^n\), sites \(x_r,y_r\in\Lambda_r\) with \(|x_r-y_r|=r\), norm-one observables \(\widetilde A_r\in\A^+_{\{y_r\}}\) and \(\widetilde B_r\in\A^+_{\{x_r\}}\), and times \(t_r>0\) such that
\begin{equation}
 \sup_{r\in\cR,\,u\in\R}\|\Phi_u^{(r)}\|^{\#}_{a;\kappa_0}<\infty,
 \qquad
 t_r=o(r^\varepsilon)\ \text{for every }\varepsilon>0,
 \qquad
 \bigl\|[\tau_{0,t_r}^{\Phi^{(r)}}(\widetilde A_r),\widetilde B_r]\bigr\|=2.
 \label{eq:decay-optimality-transfer}
\end{equation}
Consequently, there are no constants \(m\ge0\), \(0<C<\infty\), and maps
\[
 \cT^{\Phi^{(r)}}_{0,t;\{y_r\}}:
 \A^+_{\{y_r\}}\rightarrow\{\text{interactions on }\Gamma\},
 \qquad r\in\cR,\quad t\in\R,
\]
such that, for every \(r\in\cR\), \(t\in\R\), and \(A\in\A^+_{\{y_r\}}\), one has \((\cT^{\Phi^{(r)}}_{0,t;\{y_r\}}A)(Y)=0\) whenever \(y_r\notin Y\), the decomposition \eqref{eq:source-representation} holds in norm, and
\begin{equation}
 \sup_{x\in\Gamma}\sum_{Y\ni x}(1+\diam Y)^a
 \bigl\|(\cT^{\Phi^{(r)}}_{0,t;\{y_r\}}A)(Y)\bigr\|
 \le C(1+|t|)^m\E^{\kappa_0}\|A\|.
 \label{eq:decay-optimality-hypothetical}
\end{equation}
In particular, Theorem~\ref{thm:card-propagation} does not extend to \(0<a\le\dvol\) with uniform polynomial time bounds.
\end{proposition}

\begin{proof}
\smallskip
\noindent \emph{The spin protocol.}
We use the state-transfer protocol of Tran et al.\ \cite[Section~III and Eq.~(3)]{TranStateTransfer21} at pointwise decay exponent \(2n\).
We first specify the boxes and the properties of that protocol needed below; the constants in these properties depend only on \(n\).
To choose boxes compatible with its recursive subdivision, set
\[
 \ell_0:=\left\lceil\E^{8/n}\right\rceil,
 \qquad
 m_j:=\left\lceil\exp\!\left(
        \frac{3}{2\sqrt n}\sqrt{\log\ell_j}\right)\right\rceil,
 \qquad
 \ell_{j+1}:=m_j\ell_j,
 \qquad j\ge0.
\]
Each box with \(\ell_{j+1}\) sites along each coordinate splits into \(m_j^n\) disjoint boxes with \(\ell_j\) sites along each coordinate, as required by the protocol.
Define
\[
 \cR:=\{\ell_j-1\,|\,j\ge0\},
 \qquad
 \Lambda_r:=\{0,\ldots,r\}^n,
 \qquad x_r:=0,
 \qquad y_r:=r e_1,
 \qquad r\in\cR,
\]
where \(e_1\) is the first coordinate vector.
Thus \(|x_r-y_r|=r\), \(\diam\Lambda_r=\sqrt n\,r\), and \(\cR\) is unbounded because \(m_j\ge2\).
For each \(r\in\cR\), the protocol in \cite[Section~III]{TranStateTransfer21} on \(\Lambda_r\) consists of a finite sequence of instantaneous onsite unitaries and constant Hamiltonian pulses with two-site terms satisfying \(\|H_{x,y}\|\le C(1+|x-y|)^{-2n}\), with \(C\) independent of \(r\).
Let \(T_r\) be the sum of the durations of these Hamiltonian pulses and \(U_r\) the unitary implemented by the full sequence.
The choice of \(m_j\) satisfies the subdivision condition preceding \cite[Eqs.~(19)--(20)]{TranStateTransfer21}.
The recursive time estimate in those equations gives constants \(C,c>0\), independent of \(r\), such that
\[
 T_r\le C\exp\!\bigl(c\sqrt{\log(2+r)}\bigr).
\]
The protocol permits arbitrary source and target sites in the box, and the unitary \(U_r\) transfers an arbitrary qubit state exactly from \(x_r\) to \(y_r\) \cite[Sections~II--III]{TranStateTransfer21}:
\begin{equation}
 U_r\bigl(|\psi\rangle_{x_r}\otimes|0\cdots0\rangle\bigr)
 =|\chi_r\rangle\otimes|\psi\rangle_{y_r}
 \qquad\text{for every }|\psi\rangle\in\C^2,
 \label{eq:decay-optimality-transfer-map}
\end{equation}
where \(|0\cdots0\rangle\) is the state of the sites other than \(x_r\), and the state \(|\chi_r\rangle\) of the sites other than \(y_r\) does not depend on \(|\psi\rangle\).

\smallskip
\noindent \emph{Commutator lower bound.}
Let \(A_r:=\sigma^z_{y_r}\), \(B_r:=\sigma^x_{x_r}\), with \(\sigma^z,\sigma^x\) the Pauli matrices, and let \(P_r\) be the projection onto the states in which every site except \(x_r\) is in \(|0\rangle\).
Applying \eqref{eq:decay-optimality-transfer-map} to \(|\psi\rangle\) and to \(\sigma^z|\psi\rangle\) gives \(A_rU_rP_r=U_r\sigma^z_{x_r}P_r\), hence \(U_r^*A_rU_rP_r=\sigma^z_{x_r}P_r\).
Since \(P_r\) acts as the identity on site \(x_r\), it commutes with \(B_r\) and with \(\sigma^z_{x_r}\), and therefore
\[
 \|[U_r^*A_rU_r,B_r]\|
 \ge\|P_r[U_r^*A_rU_r,B_r]P_r\|
 =\|[\sigma^z_{x_r},\sigma^x_{x_r}]P_r\|
 =\|2\I\sigma^y_{x_r}P_r\|=2.
\]

\smallskip
\noindent \emph{Removing the onsite unitaries.}
The identity \(\E^{-\I hH}G=G\E^{-\I hG^*HG}\) moves an onsite unitary \(G\) to the left of a pulse, and conjugation by \(G\) preserves the support and the norm of every two-site term.
Repeating this operation gives \(U_r=G_rV_r\), where \(G_r\) is a product of onsite unitaries and \(V_r\) is generated by two-site pulses alone.
Since \(U_r^*A_rU_r=V_r^*(G_r^*A_rG_r)V_r\), the observable \(G_r^*A_rG_r\in M_2(\C)_{y_r}\) has norm one and satisfies the same commutator lower bound with respect to \(V_r\).

\smallskip
\noindent \emph{Smoothing in time.}
Choose \(f\in C_c^\infty((0,2))\) with \(0\le f\le1\) and \(\int_0^2f(v)\,\dd v=1\).
Replace each constant pulse \(H\) of duration \(h>0\) by the generator \(f(u/h)H\), \(0\le u\le2h\), whose propagator at time \(2h\) is \(\E^{-\I hH}\).
The modified generators vanish near the endpoints of their time intervals, so their concatenation, extended by zero, defines a smooth path \(\Xi^{(r)}\) of two-site spin interactions without increasing any term norm.
Let \(W_r(t)\) be its propagator with \(W_r(0)=\1\).
Then \(W_r(2T_r)=V_r\).

\smallskip
\noindent \emph{Rescaling the interaction and time.}
Set
\[
 b_r:=\log(2+r),\qquad
 \widehat\Xi_u^{(r)}(Z):=b_r^{-1}\Xi_{u/b_r}^{(r)}(Z),
 \qquad t_r:=2b_rT_r.
\]
Differentiating \(\widehat W_r(u):=W_r(u/b_r)\) shows that it is the propagator of \(\widehat\Xi^{(r)}\), so \(\widehat W_r(t_r)=V_r\).
The rescaled path is smooth, its two-site terms satisfy
\[
 \|\widehat\Xi_u^{(r)}(\{x,y\})\|
 \le\frac{C}{b_r}(1+|x-y|)^{-2n},
\]
and the transfer time satisfies, for every \(\varepsilon>0\),
\[
 t_r\le 2C\log(2+r)\exp\!\bigl(c\sqrt{\log(2+r)}\bigr)
      =o(r^\varepsilon).
\]

\smallskip
\noindent \emph{Fermionic embedding.}
With two fermionic modes per site, \(\A^+_{\{x\}}\cong M_2(\C)\oplus M_2(\C)\), and
\[
 \jmath_x^{\mathrm{spin}}:M_2(\C)\rightarrow\A_{\{x\}}^+,
 \qquad \jmath_x^{\mathrm{spin}}(B):=B\oplus B,
\]
is a unital \(*\)-homomorphism.
Even operators on disjoint sites commute, so the maps \(\jmath_x^{\mathrm{spin}}\), \(x\in\Lambda\), define a unital \(*\)-homomorphism \(\jmath_\Lambda^{\mathrm{spin}}\) from the spin algebra of every finite \(\Lambda\subseteq\Z^n\) into \(\A^+_\Lambda\), compatible under enlargement of \(\Lambda\).
Each \(\jmath_\Lambda^{\mathrm{spin}}\) is injective, since the spin algebra of \(\Lambda\) is a full matrix algebra and \(\jmath_\Lambda^{\mathrm{spin}}\) is unital; hence it is isometric.
For every nonempty finite \(Z\subseteq\Gamma\), define
\[
 \Phi_u^{(r)}(Z):=
 \begin{cases}
 \jmath_Z^{\mathrm{spin}}(\widehat\Xi_u^{(r)}(Z)),&Z\subseteq\Lambda_r,\\
 0,&Z\not\subseteq\Lambda_r.
 \end{cases}
\]
Applying \(\jmath_{\Lambda_r}^{\mathrm{spin}}\) to the Schr\"odinger equation for \(\widehat W_r(t)\) shows that \(\jmath_{\Lambda_r}^{\mathrm{spin}}(\widehat W_r(t))\) is the propagator generated by \(\sum_{Z\subseteq\Lambda_r}\Phi_t^{(r)}(Z)\).
Thus, for every spin observable \(B\) on \(\Lambda_r\),
\[
 \tau^{\Phi^{(r)}}_{0,t}(\jmath_{\Lambda_r}^{\mathrm{spin}}(B))
 =\jmath_{\Lambda_r}^{\mathrm{spin}}(\widehat W_r(t)^*B\widehat W_r(t)).
\]
Put \(\widetilde A_r:=\jmath_{\Lambda_r}^{\mathrm{spin}}(G_r^*A_rG_r)\in\A^+_{\{y_r\}}\) and \(\widetilde B_r:=\jmath_{\Lambda_r}^{\mathrm{spin}}(B_r)\in\A^+_{\{x_r\}}\).
Both have norm one, and \(\widehat W_r(t_r)=V_r\) together with the commutator lower bound gives the last relation in \eqref{eq:decay-optimality-transfer}.
For \(0<a\le n\), the two-site decay bound and \(\diam\Lambda_r\le C_nr\) give
\[
 \sup_{u\in\R}\|\Phi_u^{(r)}\|^{\#}_{a;\kappa_0}
 \le\sup_{u\in\R}\|\Phi_u^{(r)}\|^{\#}_{n;\kappa_0} \le\frac{C\E^{2\kappa_0}}{b_r}
       \sum_{\substack{z\in\Z^n\\0<|z|\le C_nr}}
                    (1+|z|)^{-n}
 \le C'\E^{2\kappa_0},
\]
uniformly in \(r\).
Indeed, each annulus \(\{z\in\Z^n\,|\,2^j\le|z|<2^{j+1}\}\), \(j\ge0\), contains at most \(C_n'2^{jn}\) sites and contributes at most \(C_n'\) to the sum, while at most \(C_n''b_r\) such annuli meet \(\{z\,|\,0<|z|\le C_nr\}\).

\smallskip
\noindent \emph{The hypothetical bound.}
Suppose that maps \(\cT^{\Phi^{(r)}}_{0,t;\{y_r\}}\) with \eqref{eq:source-representation} and \eqref{eq:decay-optimality-hypothetical} existed, and apply them to \(A=\widetilde A_r\) and \(t=t_r\).
Equation~\eqref{eq:decay-optimality-hypothetical}, applied at \(x=y_r\), implies absolute norm convergence of the decomposition.
By the assumed locality and support properties, every nonzero interaction term \((\cT^{\Phi^{(r)}}_{0,t_r;\{y_r\}}\widetilde A_r)(Y)\) belongs to \(\A_Y^+\) with \(y_r\in Y\), and commutes with \(\widetilde B_r\) unless \(x_r\in Y\).
The remaining supports satisfy \(\diam Y\ge|x_r-y_r|=r\), so the site sum at \(x_r\) in \eqref{eq:decay-optimality-hypothetical} gives
\begin{align*}
 \bigl\|[\tau_{0,t_r}^{\Phi^{(r)}}(\widetilde A_r),\widetilde B_r]\bigr\|
 \le2\sum_{Y\ni x_r,y_r}
   \bigl\|(\cT^{\Phi^{(r)}}_{0,t_r;\{y_r\}}\widetilde A_r)(Y)\bigr\| \le\frac{2C\E^{\kappa_0}(1+t_r)^m}{(1+r)^a}
 \rightarrow0
 \qquad(r\to\infty,\ r\in\cR),
\end{align*}
since \(t_r=o(r^{\varepsilon})\) for every \(\varepsilon>0\).
This contradicts \eqref{eq:decay-optimality-transfer}.

The two-mode construction suffices because the constants in Theorem~\ref{thm:card-propagation} are uniform in the local multiplicities, so any extension with this uniformity must apply when \(r_x=2\) at every site.
As a consequence, Theorem~\ref{thm:card-propagation} cannot be extended to \(0<a\le\dvol\) with uniform polynomial time bounds, since such an extension would imply \eqref{eq:decay-optimality-hypothetical} by taking \(p=a\), \(k=0\), and an interaction whose only nonzero term is supported on \(\{y_r\}\).
\end{proof}

\section*{Acknowledgments}

The author profited from numerous discussions with Marius Wesle and Tom Wessel on long-range Lieb--Robinson and propagation bounds.
He also thanks Domenico Monaco and Gianluca Panati at Sapienza University of Rome and Massimo Moscolari at Politecnico di Milano for their hospitality during his two-month sabbatical visits to their respective institutions.
This work was supported by the Deutsche Forschungsgemeinschaft (DFG, German Research Foundation) -- Project numbers 470903074 and 465199066.

\section*{Declaration on the use of generative AI}

During the preparation of this manuscript, the author used ChatGPT Pro (GPT-5.6 Sol) to assist in exploring and critically examining mathematical arguments and literature.
The final version was checked for consistency and polished using GPT-6 Astra and Anthropic's Opus 5.5.
The author reviewed and revised all AI-assisted material.
The author takes full responsibility for the content of the manuscript.


\begin{thebibliography}{99}
\small
\setlength{\itemsep}{3pt plus 1pt minus 1pt}

\bibitem{ADHH17}
D.~Abanin, W.~De Roeck, W.~W.~Ho, and F.~Huveneers, \href{https://doi.org/10.1007/s00220-017-2930-x}{\emph{A rigorous theory of many-body prethermalization for periodically driven and closed quantum systems}}, Commun. Math. Phys. \textbf{354}, 809--827 (2017), \href{https://arxiv.org/abs/1509.05386}{arXiv:\allowbreak 1509.05386}.

\bibitem{ArakiMoriya03}
H.~Araki and H.~Moriya, \href{https://doi.org/10.1142/S0129055X03001606}{\emph{Equilibrium statistical mechanics of fermion lattice systems}}, Rev. Math. Phys. \textbf{15}, 93--198 (2003), \href{https://arxiv.org/abs/math-ph/0211016}{arXiv:\allowbreak math-ph/0211016}.

\bibitem{BMNS12}
S.~Bachmann, S.~Michalakis, B.~Nachtergaele, and R.~Sims, \href{https://doi.org/10.1007/s00220-011-1380-0}{\emph{Automorphic equivalence within gapped phases of quantum lattice systems}}, Commun. Math. Phys. \textbf{309}, 835--871 (2012), \href{https://arxiv.org/abs/1102.0842}{arXiv:\allowbreak 1102.0842}.

\bibitem{BeckerQBP26}
L.~Becker, \href{https://arxiv.org/abs/2609.31858}{\emph{Locality of KMS states for infinite fermionic lattice systems}}, \href{https://arxiv.org/abs/2609.31858}{arXiv:\allowbreak 2609.31858} (2026).

\bibitem{BTW26}
L.~Becker, S.~Teufel, and M.~Wesle, \href{https://doi.org/10.1016/j.jfa.2026.111421}{\emph{Automorphic equivalence within gapped phases of infinitely extended fermion systems}}, J. Funct. Anal. \textbf{290}, 111421 (2026), \href{https://arxiv.org/abs/2507.13321}{arXiv:\allowbreak 2507.13321}.

\bibitem{BTWAd25}
L.~Becker, S.~Teufel, and M.~Wesle, \href{https://arxiv.org/abs/2510.20914}{\emph{The generalized adiabatic theorem for extended lattice systems}}, \href{https://arxiv.org/abs/2510.20914}{arXiv:\allowbreak 2510.20914} (2025).

\bibitem{BourneOgata21}
C.~Bourne and Y.~Ogata, \href{https://doi.org/10.1017/fms.2021.19}{\emph{The classification of symmetry protected topological phases of one-dimensional fermion systems}}, Forum Math. Sigma \textbf{9}, e25 (2021), \href{https://arxiv.org/abs/2006.15232}{arXiv:\allowbreak 2006.15232}.

\bibitem{BR97}
O.~Bratteli and D.~W.~Robinson, \href{https://doi.org/10.1007/978-3-662-03444-6}{\emph{Operator Algebras and Quantum Statistical Mechanics 2: Equilibrium States, Models in Quantum Statistical Mechanics}}, 2nd ed., Springer, Berlin (1997).

\bibitem{BruPedra17}
J.-B.~Bru and W.~de~Siqueira Pedra, \href{https://doi.org/10.1007/978-3-319-45784-0}{\emph{Lieb--Robinson Bounds for Multi-Commutators and Applications to Response Theory}}, SpringerBriefs in Mathematical Physics, vol.~13, Springer (2017), \href{https://arxiv.org/abs/1702.08869}{arXiv:\allowbreak 1702.08869}.

\bibitem{CMTW25}
\'{A}.~Capel, M.~Moscolari, S.~Teufel, and T.~Wessel, \href{https://doi.org/10.1007/s00220-024-05198-x}{\emph{From decay of correlations to locality and stability of the Gibbs state}}, Commun. Math. Phys. \textbf{406}, 43 (2025), \href{https://arxiv.org/abs/2310.09182}{arXiv:\allowbreak 2310.09182}.

\bibitem{ChenLucas19}
C.-F.~Chen and A.~Lucas, \href{https://doi.org/10.1103/PhysRevLett.123.250605}{\emph{Finite speed of quantum scrambling with long range interactions}}, Phys. Rev. Lett. \textbf{123}, 250605 (2019), \href{https://arxiv.org/abs/1907.07637}{arXiv:\allowbreak 1907.07637}.

\bibitem{EMNY20}
D.~V.~Else, F.~Machado, C.~Nayak, and N.~Y.~Yao, \href{https://doi.org/10.1103/physreva.101.022333}{\emph{Improved Lieb--Robinson bound for many-body Hamiltonians with power-law interactions}}, Phys. Rev. A \textbf{101}, 022333 (2020), \href{https://arxiv.org/abs/1809.06369}{arXiv:\allowbreak 1809.06369}.

\bibitem{FossFeigEtAl15}
M.~Foss-Feig, Z.-X.~Gong, C.~W.~Clark, and A.~V.~Gorshkov, \href{https://doi.org/10.1103/PhysRevLett.114.157201}{\emph{Nearly linear light cones in long-range interacting quantum systems}}, Phys. Rev. Lett. \textbf{114}, 157201 (2015), \href{https://arxiv.org/abs/1410.3466}{arXiv:\allowbreak 1410.3466}.

\bibitem{Hastings04}
M.~B.~Hastings, \href{https://doi.org/10.1103/PhysRevB.69.104431}{\emph{Lieb--Schultz--Mattis in higher dimensions}}, Phys. Rev. B \textbf{69}, 104431 (2004), \href{https://arxiv.org/abs/cond-mat/0305505}{arXiv:\allowbreak cond-mat/0305505}.

\bibitem{Hastings07}
M.~B.~Hastings, \href{https://doi.org/10.1103/PhysRevB.76.201102}{\emph{Quantum belief propagation: An algorithm for thermal quantum systems}}, Phys. Rev. B \textbf{76}, 201102(R) (2007), \href{https://arxiv.org/abs/0706.4094}{arXiv:\allowbreak 0706.4094}.

\bibitem{HastingsKoma06}
M.~B.~Hastings and T.~Koma, \href{https://doi.org/10.1007/s00220-006-0030-4}{\emph{Spectral gap and exponential decay of correlations}}, Commun. Math. Phys. \textbf{265}, 781--804 (2006), \href{https://arxiv.org/abs/math-ph/0507008}{arXiv:\allowbreak math-ph/0507008}.

\bibitem{HastingsWen05}
M.~B.~Hastings and X.-G.~Wen, \href{https://doi.org/10.1103/physrevb.72.045141}{\emph{Quasiadiabatic continuation of quantum states: The stability of topological ground-state degeneracy and emergent gauge invariance}}, Phys. Rev. B \textbf{72}, 045141 (2005), \href{https://arxiv.org/abs/cond-mat/0503554}{arXiv:\allowbreak cond-mat/0503554}.

\bibitem{HernandezSantanaEtAl17}
S.~Hern\'andez-Santana, C.~Gogolin, J.~I.~Cirac, and A.~Ac\'in, \href{https://doi.org/10.1103/PhysRevLett.119.110601}{\emph{Correlation decay in fermionic lattice systems with power-law interactions at nonzero temperature}}, Phys. Rev. Lett. \textbf{119}, 110601 (2017), \href{https://arxiv.org/abs/1702.00371}{arXiv:\allowbreak 1702.00371}.

\bibitem{HytonenKairema12}
T.~Hyt\"onen and A.~Kairema, \href{https://doi.org/10.4064/cm126-1-1}{\emph{Systems of dyadic cubes in a doubling metric space}}, Colloq. Math. \textbf{126}, 1--33 (2012), \href{https://arxiv.org/abs/1012.1985}{arXiv:\allowbreak 1012.1985}.

\bibitem{KuwaharaSaito20}
T.~Kuwahara and K.~Saito, \href{https://doi.org/10.1103/PhysRevX.10.031010}{\emph{Strictly linear light cones in long-range interacting systems of arbitrary dimensions}}, Phys. Rev. X \textbf{10}, 031010 (2020), \href{https://arxiv.org/abs/1910.14477}{arXiv:\allowbreak 1910.14477}.

\bibitem{LemmWessel25}
M.~Lemm and T.~Wessel, \href{https://doi.org/10.1063/5.0252675}{\emph{Enhanced Lieb--Robinson bounds for commuting long-range interactions}}, J. Math. Phys. \textbf{66}, 091901 (2025), \href{https://arxiv.org/abs/2411.19241}{arXiv:\allowbreak 2411.19241}.

\bibitem{LR72}
E.~H.~Lieb and D.~W.~Robinson, \href{https://doi.org/10.1007/BF01645779}{\emph{The finite group velocity of quantum spin systems}}, Commun. Math. Phys. \textbf{28}, 251--257 (1972).

\bibitem{MKN17}
T.~Matsuta, T.~Koma, and S.~Nakamura, \href{https://doi.org/10.1007/s00023-016-0526-1}{\emph{Improving the Lieb--Robinson bound for long-range interactions}}, Ann. Henri Poincar\'e \textbf{18}, 519--528 (2017), \href{https://arxiv.org/abs/1604.05809}{arXiv:\allowbreak 1604.05809}.

\bibitem{McDonoughEtAl25}
B.~T.~McDonough, C.~Yin, A.~Lucas, and C.~Zhang, \href{https://doi.org/10.1103/n242-m1l2}{\emph{Lieb--Robinson bounds with exponential-in-volume tails}}, PRX Quantum \textbf{6}, 040322 (2025), \href{https://arxiv.org/abs/2502.02652}{arXiv:\allowbreak 2502.02652}.
Numbered results and equations are cited according to the published version.

\bibitem{MoonOgata20}
A.~Moon and Y.~Ogata, \href{https://doi.org/10.1016/j.jfa.2019.108422}{\emph{Automorphic equivalence within gapped phases in the bulk}}, J. Funct. Anal. \textbf{278}, 108422 (2020), \href{https://arxiv.org/abs/1906.05479}{arXiv:\allowbreak 1906.05479}.

\bibitem{NOS06}
B.~Nachtergaele, Y.~Ogata, and R.~Sims, \href{https://doi.org/10.1007/s10955-006-9143-6}{\emph{Propagation of correlations in quantum lattice systems}}, J. Stat. Phys. \textbf{124}, 1--13 (2006), \href{https://arxiv.org/abs/math-ph/0603064}{arXiv:\allowbreak math-ph/0603064}.

\bibitem{NSY18}
B.~Nachtergaele, R.~Sims, and A.~Young, \href{https://doi.org/10.1090/conm/717/14443}{\emph{Lieb--Robinson bounds, the spectral flow, and stability of the spectral gap for lattice fermion systems}}, in \emph{Mathematical Problems in Quantum Physics}, F.~Bonetto, D.~Borthwick, E.~Harrell, and M.~Loss (eds.), Contemp. Math. \textbf{717}, American Mathematical Society, 93--115 (2018), \href{https://arxiv.org/abs/1705.08553}{arXiv:\allowbreak 1705.08553}.

\bibitem{NSY19}
B.~Nachtergaele, R.~Sims, and A.~Young, \href{https://doi.org/10.1063/1.5095769}{\emph{Quasi-locality bounds for quantum lattice systems. I. Lieb--Robinson bounds, quasi-local maps, and spectral flow automorphisms}}, J. Math. Phys. \textbf{60}, 061101 (2019), \href{https://arxiv.org/abs/1810.02428}{arXiv:\allowbreak 1810.02428}.

\bibitem{RubilianiWesleWessel26}
C.~Rubiliani, M.~Wesle, and T.~Wessel, \href{https://arxiv.org/abs/2610.12731}{\emph{Lieb--Robinson bounds for irreversible dynamics via adaptive Dyson expansion}}, \href{https://arxiv.org/abs/2610.12731}{arXiv:\allowbreak 2610.12731} (2026).

\bibitem{Stability}
S.~Teufel, \href{https://arxiv.org/abs/2610.40299}{\emph{Bulk gap stability and spectral flow for fermions with weak long-range interactions}}, \href{https://arxiv.org/abs/2610.40299}{arXiv:\allowbreak 2610.40299} (2026).

\bibitem{LPPLBdG}
S.~Teufel, \href{https://arxiv.org/abs/2610.40337}{\emph{Local determination of low-energy states in gapped fermion systems with weak long-range interactions}}, \href{https://arxiv.org/abs/2610.40337}{arXiv:\allowbreak 2610.40337} (2026).

\bibitem{TeufelWessel25}
S.~Teufel and T.~Wessel, \href{https://doi.org/10.1007/s11040-026-09584-w}{\emph{Lieb--Robinson bounds, automorphic equivalence and LPPL for long-range interacting fermions}}, Math. Phys. Anal. Geom. \textbf{29}, 56 (2026), \href{https://arxiv.org/abs/2507.03319v2}{arXiv:\allowbreak 2507.03319}.

\bibitem{TranEtAl19}
M.~C.~Tran, A.~Y.~Guo, Y.~Su, J.~R.~Garrison, Z.~Eldredge, M.~Foss-Feig, A.~M.~Childs, and A.~V.~Gorshkov, \href{https://doi.org/10.1103/PhysRevX.9.031006}{\emph{Locality and digital quantum simulation of power-law interactions}}, Phys. Rev. X \textbf{9}, 031006 (2019), \href{https://arxiv.org/abs/1808.05225}{arXiv:\allowbreak 1808.05225}.

\bibitem{TranHierarchy20}
M.~C.~Tran, C.-F.~Chen, A.~Ehrenberg, A.~Y.~Guo, A.~Deshpande, Y.~Hong, Z.-X.~Gong, A.~V.~Gorshkov, and A.~Lucas, \href{https://doi.org/10.1103/PhysRevX.10.031009}{\emph{Hierarchy of linear light cones with long-range interactions}}, Phys. Rev. X \textbf{10}, 031009 (2020), \href{https://arxiv.org/abs/2001.11509}{arXiv:\allowbreak 2001.11509}.

\bibitem{TranEtAl21}
M.~C.~Tran, A.~Y.~Guo, C.~L.~Baldwin, A.~Ehrenberg, A.~V.~Gorshkov, and A.~Lucas, \href{https://doi.org/10.1103/physrevlett.127.160401}{\emph{Lieb--Robinson light cone for power-law interactions}}, Phys. Rev. Lett. \textbf{127}, 160401 (2021), \href{https://arxiv.org/abs/2103.15828}{arXiv:\allowbreak 2103.15828}.

\bibitem{TranStateTransfer21}
M.~C.~Tran, A.~Y.~Guo, A.~Deshpande, A.~Lucas, and A.~V.~Gorshkov, \href{https://doi.org/10.1103/PhysRevX.11.031016}{\emph{Optimal state transfer and entanglement generation in power-law interacting systems}}, Phys. Rev. X \textbf{11}, 031016 (2021), \href{https://arxiv.org/abs/2010.02930}{arXiv:\allowbreak 2010.02930}.

\bibitem{WilmingWerner22}
H.~Wilming and A.~H.~Werner, \href{https://doi.org/10.1103/PhysRevB.105.125101}{\emph{Lieb--Robinson bounds imply locality of interactions}}, Phys. Rev. B \textbf{105}, 125101 (2022), \href{https://arxiv.org/abs/2006.10062}{arXiv:\allowbreak 2006.10062}.

\bibitem{YinLucas25}
C.~Yin and A.~Lucas, \href{https://doi.org/10.1103/361k-nj4b}{\emph{Low-density parity-check codes as stable phases of quantum matter}}, PRX Quantum \textbf{6}, 030329 (2025), \href{https://arxiv.org/abs/2411.01002}{arXiv:\allowbreak 2411.01002}.

\end{thebibliography}
\end{document}